\documentclass[]{article}
\usepackage [english] {babel}
\usepackage{amsmath,amsfonts,amssymb,amsthm,epsfig,epstopdf,titling,array}
\usepackage [utf8]{inputenc}
\usepackage[T1]{fontenc}
\usepackage{csquotes}
\usepackage{textcomp}
\usepackage{geometry}
\usepackage{bm}
\usepackage[margin=10pt, font=scriptsize, labelfont=bf]{caption}
\usepackage{subcaption}
\usepackage{braket}
\usepackage{environ}
\usepackage{verbatim}
\usepackage{dsfont}
\usepackage{tikz}
\usetikzlibrary{shapes}
\usetikzlibrary{positioning}
\usepackage[colorinlistoftodos]{todonotes}
\usepackage{newunicodechar}
\usepackage{authblk}
\usepackage[colorlinks]{hyperref}
\usepackage[capitalize]{cleveref}

\usepackage[style=alphabetic,url=false,date=year,giveninits=true,isbn=false,sorting=nyt]{biblatex}
\AtEveryBibitem{\clearfield{issue} \clearfield{number}}
\DeclareFieldFormat
  [article,inbook,incollection,inproceedings,patent,thesis,unpublished]
  {titlecase:title}{\MakeSentenceCase*{#1}}

\newunicodechar{ȩ}{\c{e}}

\newcommand{\orcid}[1]{\href{https://orcid.org/#1}{\textcolor[HTML]{A6CE39}{\aiOrcid}}}

\usepackage[colorlinks]{hyperref}
\usepackage[capitalize]{cleveref}

\usetikzlibrary{trees}
\usetikzlibrary{calc}
\usetikzlibrary{decorations.pathmorphing}
\usetikzlibrary{decorations.markings}
\usetikzlibrary{shapes}
\usetikzlibrary{fit}

\makeatletter
\newsavebox{\measure@tikzpicture}
\NewEnviron{scaletikz}[1]{%
  \def\tikz@width{#1}%
  \begin{lrbox}{\measure@tikzpicture}%
  \BODY
  \end{lrbox}%
  \pgfmathparse{#1/\wd\measure@tikzpicture}%
  \BODY
}
\makeatother

\theoremstyle{plain}

\newtheorem{theorem}{Theorem}
\numberwithin{theorem}{section}
\newtheorem{lemma}[theorem]{Lemma}

\newtheorem{proposition}[theorem]{Proposition}

\newtheorem{remark}[theorem]{Remark}

\newtheorem{corollary}[theorem]{Corollary}

\numberwithin{equation}{section}
\numberwithin{equation}{subsection}
\renewcommand*{\theequation}{%
  \ifnum\value{subsection}=0 %
    \thesection
  \else
    \thesubsection
  \fi
  .\arabic{equation}%
}

\theoremstyle{definition}

\definecolor{darkGreen}{rgb}{0,0.5,0}
\hypersetup{
    colorlinks,
    citecolor=darkGreen,
    filecolor=red,
    linkcolor=blue}

\newcommand{\condset}[2]{\left\{ {#1} \middle| {#2} \right\}}

\DeclareMathOperator{\dist}{dist}
\DeclareMathOperator{\diam}{diam}

\newcommand{\ul}[1]{{\ensuremath{\underline{#1}}}}
\newcommand{\lis}[1]{{\ensuremath{\overline{#1}}}}

\newcommand{\dd}{\,\text{\rm d}}     
\newcommand{\hdd}{\mkern7mu\hat{\mkern-7mu\,\text{\rm d}}}

\newcommand{\cyl}{{\ensuremath{{\rm cyl}}}}
\newcommand{\scal}{{\ensuremath{{\rm scal}}}}
\newcommand{\per}{{\ensuremath{{\rm per}}}}
\newcommand{\free}{{\ensuremath{{\rm free}}}}

\newcommand{\black}{{\ensuremath{{\rm black}}}}

\newcommand{\Min}{{\ensuremath{{\rm min}}}}

\newcommand{\B}{\ensuremath{\textup{B}} }
\newcommand{\D}{\ensuremath{\textup{D}} }
\newcommand{\E}{\ensuremath{\textup{E}} }

\newcommand{\M}{\ensuremath{\textup{M}} }
\renewcommand{\P}{\ensuremath{\textup{P}} }
\newcommand{\R}{\ensuremath{\textup{R}} }

\newcommand{\qf}{\ensuremath{\textup{qf}}}
\newcommand{\ct}{\ensuremath{\textup{ct}}}

\newcommand{\f}{\ensuremath{\textup{f}}}
\newcommand{\p}{\ensuremath{\textup{p}}}

\newcommand{\cA}{\ensuremath{\mathcal{A}}}
\newcommand{\cB}{\ensuremath{\mathcal{B}}}

\newcommand{\cD}{\ensuremath{\mathcal{D}}}
\newcommand{\cE}{\ensuremath{\mathcal{E}}}
\newcommand{\cF}{\ensuremath{\mathcal{F}}}

\newcommand{\cH}{\ensuremath{\mathcal{H}}}
\newcommand{\cL}{\ensuremath{\mathcal{L}}}
\newcommand{\cO}{\ensuremath{\mathcal{O}}}
\newcommand{\cP}{\ensuremath{\mathcal{P}}}
\newcommand{\cQ}{\ensuremath{\mathcal{Q}}}
\newcommand{\cR}{\ensuremath{\mathcal{R}}}
\newcommand{\cS}{\ensuremath{\mathcal{S}}}
\newcommand{\cT}{\ensuremath{\mathcal{T}}}
\newcommand{\cV}{\ensuremath{\mathcal{V}}}

\newcommand{\cX}{\ensuremath{\mathcal{X}}}
\newcommand{\cW}{\ensuremath{\mathcal{W}}}
\newcommand{\cM}{\ensuremath{\mathcal{M}}}

\newcommand{\tcL}{\ensuremath{\widetilde \cL}}
\newcommand{\tcR}{\ensuremath{\widetilde \cR}}

\newcommand{\bbR}{\ensuremath{\mathbb{R}}}

\newcommand{\bZ}{\ensuremath{\mathbb{Z}}}

\newcommand{\fB}{\ensuremath{\mathfrak{B}}}
\newcommand{\fG}{\ensuremath{\mathfrak{G}}}

\newcommand{\fT}{\ensuremath{\mathfrak{T}}}

\newcommand{\fg}{\ensuremath{\mathfrak{g}}}

\newcommand{\bs}[1]{{\ensuremath{\boldsymbol{#1}}}}

\newcommand{\piecewise}[1]{\left\{ \begin{array}{*2{>{\displaystyle}l}}  #1 \end{array} \right. }

\newcommand{\floor}[1]{\left\lfloor {#1} \right\rfloor}

\newcommand{\successor}{\ensuremath{\rhd}}

\newcommand{\allct}{\ul{{\upsilon}}}

\DeclareMathOperator{\INT}{INT}

\newcommand{\media}[1]{ { \left\langle #1 \right\rangle}}

\tikzset{vertex/.style={circle,fill=black,inner sep=2pt},
ctVertex/.style={diamond,fill=black,inner sep=2pt},
bigvertex/.style={circle,fill=black,inner sep=4pt},
E/.append style={fill=white,draw},
probeEP/.style={circle,fill=black,draw,inner sep=2pt,
  prefix after command= {\pgfextra{\tikzset{every pin/.style = {pin edge={decorate,decoration={snake,amplitude=2pt,segment length =4pt}}}}}}
},
bareProbeEP/.style={rectangle,fill=black,draw,inner sep=3pt,
  prefix after command= {\pgfextra{\tikzset{every pin/.style = {pin edge={decorate,decoration={snake,amplitude=2pt,segment length =4pt}}}}}}
},
nuEP/.style={circle,fill=white,draw, inner sep=2pt},
linelabel/.style={sloped,above,very near start, inner sep=1pt,execute at begin node=$\scriptstyle,execute at end node=$},
baseline=(current  bounding  box.center),doubled/.style={double distance= 1pt,line width=1.5pt}
}
\pgfdeclarelayer{background}
\pgfsetlayers{background,main}

\newcommand{\tikzvertex}[1]{\tikz[baseline=default]{ \node [#1] {};}}

\begin{document}
\title{Energy correlations of non-integrable Ising models: \\ The scaling limit in the cylinder}
\author[1]{Giovanni Antinucci}
\author[2,3]{Alessandro Giuliani}
\author[2,*]{Rafael L. Greenblatt}
\affil[1]{\small{Universit\'e de Gen\`eve, Section de math\'ematiques, 2-4 rue du Li\`evre, 1211 Gen\`eve 4, Switzerland}}
\affil[2]{\small{Universit\`a degli Studi Roma Tre, Dipartimento di Matematica e Fisica, L.go S. L. Murialdo 1, 00146 Roma, Italy}}
\affil[3]{\small{Centro Linceo Interdisciplinare {\it Beniamino Segre}, Accademia Nazionale dei Lincei, Palazzo Corsini, Via della Lungara 10,
00165 Roma, Italy.}}
\affil[*]{\small{Present affiliation: 
		Universit\`a degli Studi di Roma ``Tor Vergata'', Dipartimento di Matematica, via della Ricerca Scientifica 1, 00133 Roma, Italy
}}
\date{\today}

\maketitle

\begin{abstract} 
	We consider a class of non-integrable $2D$ Ising models whose Hamiltonian, in addition to the standard nearest neighbor 
	couplings, includes additional weak multi-spin interactions which are even under spin flip. 
We study the model in cylindrical domains of arbitrary aspect ratio 
and compute the multipoint energy correlations at the critical temperature via a multiscale expansion, uniformly convergent in the domain size and in the lattice spacing. 
We prove that, in the scaling limit, the multipoint energy correlations converge to the same limiting correlations as those of the {\it nearest neighbor} Ising model in a finite cylinder 
with renormalized horizontal and vertical couplings,
up to an overall multiplicative constant independent of the shape of the domain. The proof is based on a representation of 
the generating function of correlations in terms of a non-Gaussian Grassmann integral, and a constructive Renormalization Group (RG) analysis thereof. 

A key technical novelty compared with previous works is a systematic analysis of the effect of the boundary corrections to the RG flow, in particular a proof that the scaling 
dimension of boundary operators is better by one dimension than their bulk counterparts.  In addition, a cancellation mechanism based on an approximate image rule for the fermionic Green's function is of crucial importance for controlling the flow of the (superficially) marginal boundary terms under RG iterations. 
\end{abstract}

\tableofcontents

\section{Introduction}\label{sec:intro}

The Ising model may very well be the most studied model in statistical mechanics; partly because it is the simplest finite-dimensional model with bounded local interactions 
which can be shown to exhibit a phase transition, but probably more so because the two-dimensional (2D) Ising model with nearest-neighbor interactions on a locally 
planar graph is exactly solvable in a very strong sense.
Exact solutions and related methods have been used over decades to generate a remarkably rich picture of the model, from the algebraic transfer matrix approach 
used by Lars Onsager to calculate the free energy and spontaneous magnetization on the square lattice with periodic boundary conditions \cite{Onsager} through the 
mapping into free Fermions and Grassmann integrals \cite{SML, Hurst1966, KC, S80a} which give rise to many relationships among the correlation functions, to the 
set of quadratic difference equations used to determine the lattice correlation functions at any separation and any temperature \cite{MWP}, 
just to mention a few. Among the byproducts of these methods are the identification of the critical point of the model, of the corresponding set of critical 
exponents and the computation of the scaling limit of the critical correlations \cite{WMTB}. More recently, using the methods of discrete 
holomorphicity introduced in \cite{Ken01}, and later developed in \cite{SMI, LSW, Smi10}, 
the set of correlation functions in the scaling limit has been proved to be conformally covariant 
\cite{CS, duminil2012conformal, HS13, CHI}, 
in agreement with the predictions based on Wilsonian Renormalization Group (RG) \cite{WILi, WILii} and on Conformal Field Theory (CFT) \cite{BPZ}.

The predominant part of the mathematical results on the behavior of the 2D Ising model, in particular nearly all the results on 
the theory at the critical point (including the scaling limit of the critical correlations) are based on the exact solvability of the model with nearest-neighbor interactions; 
and, even more specifically, on the fact that the exact solution exhibits a simple determinant structure \cite{KW}. This is quite unsatisfactory: the predictions on the structure and properties of 
the scaling limit of the critical theory, based on RG and CFT, are expected to be robust under a large class of perturbations of the microscopic Hamiltonian, 
said to be {\it irrelevant} in the RG jargon, which break the exact solvability of the model. 
Furthermore, the robustness of the limit under such class of perturbations is the very content of the {\it universality} 
hypothesis, which is one of the cornerstones of modern statistical mechanics and is a key hypothesis that one would like to test. Therefore, 
it would be highly desirable to prove stability of the scaling limit of the critical theory under a large class of perturbations 
of the microscopic Hamiltonian. 

Motivated by this, we introduce the following deformation of the standard nearest neighbor Ising model: 
\begin{equation}
\label{eq:HM}
H_{\Lambda}(\sigma)=-\sum_{l=1}^2J_l\sum_{z\in \Lambda}\sigma_{z}\sigma_{z+\hat{e}_l} -\lambda \sum_{X\subset \Lambda} V(X) \sigma_X,
\end{equation}
where: $\Lambda$ is a finite portion of $\mathbb Z^2$, with appropriate boundary conditions, to be specified below; 
$J_1,J_2$ are two positive constants, representing the couplings in the horizontal and vertical directions, and $\hat e_1,\hat e_2$ are the unit vectors in the 
two coordinate directions; the local spins $\sigma_x$ take values in $\{+1,-1\}$, and $\sigma_X:=\prod_{z\in X}\sigma_z$; $V(X)$ is a finite range, translationally 
invariant, even interaction; finally, $\lambda$ is a parameter controlling the strength of the interaction, which can be 
of either sign and, for most of the discussion below, the reader can think of as being small, compared to $J_1,J_2$, but independent of the system size. In the following, we shall refer to 
model \eqref{eq:HM} with $\lambda\neq0$ as to the `interacting' model, in contrast with the standard nearest-neighbor model, which we will refer to as the `non-interacting', one of several terminological conventions motivated by analogy with quantum field theory (an anology motivated, in particular, by the Grassmann representation of the model, discussed in 
\cref{sec:gen} below). 

Even though, physically, the presence of the additional interaction is not expected to change the macroscopic behavior of the system, 
particularly at or close to the critical point, rigorous results in this sense are rare. Among the very few available results, let us mention the recent proof 
of Pfaffian behavior of the boundary spin correlations \cite{ADTW} for general (non-planar) pair interactions, based on the use of random currents and on a 
generalization of Russo-Seymour-Welsh theory \cite{Russo,SW78}; remarkably, this result does not require a smallness assumption on $\lambda$, 
even though it requires the additional interaction beyond the nearest neighbor one to be a pair interaction of ferromagnetic type. 
The second result that we mention, and the only one where the scaling limit of the critical correlation functions of \eqref{eq:HM} has been fully computed, is 
\cite{GGM}, where the infinite plane multipoint energy correlations have been considered, and their scaling limit proved to coincide with that of the nearest-neighbor 
model, for $\lambda$ small enough\footnote{This result was stated for pair interactions (of either sign) only, but the generalization to other finite-range interactions is straightforward.}, up to a finite multiplicative renormalization of the energy observable. The proof of \cite{GGM} is based on constructive RG 
methods, and follows an earlier proposal \cite{PS, S00}. One of the limitations of the result in \cite{GGM} is that it only concerns infinite plane 
observables\footnote{More precisely, in \cite{GGM} the authors considered model \eqref{eq:HM} on a discrete torus $\Lambda$ of side $L$ and performed 
the following special scaling limit: first $L\to\infty$ for $\beta\neq\beta_c(\lambda)$, with $\beta_c(\lambda)$ the critical inverse temperature of the 
interacting model;  then $\beta\to \beta_c(\lambda)$ and, simultaneously, after appropriate rescaling of the energy observable, lattice mesh to zero.}: the method of 
proof is based on translational invariance and is not able to accomodate the presence of a boundary. In particular, the result in \cite{GGM} is not 
strong enough to allow one to check conformal covariance of the scaling limit under geometric deformations of the domain, in the spirit of 
\cite{Smi10, CS, duminil2012conformal, HS13, chelkak2014convergence, CHI}.
In this paper we make a first step towards the longer term goal of understanding scaling limits of critical, non-integrable, statistical mechanics models in domains 
of arbitrary shape and their conformal covariance \cite{G21}: we consider the multipoint energy correlations of \eqref{eq:HM} in cylindrical domains of arbitrary aspect ratio and prove that the scaling limit coincides 
with the one of the non-interacting model, up to a finite multiplicative renormalization of the energy observable, independent of the shape of the domain. 
The major technical novelty required for proving this result is the control of the boundary effects under iterations of the RG map; the new technical 
tools introduced for this purpose may have an impact in several related problems, as discussed in more details below. 

\paragraph{Main results} We consider the model \eqref{eq:HM} in cylindrical geometry with free boundary conditions; that is, we let $\Lambda=\mathbb Z_L \times \left(\mathbb Z\cap [1, M]\right)$, 
where\footnote{We use the convention that $\mathbb N$ is the set of positive integers, and $\mathbb N_0$ the set of non-negative integers.} $L\in 2\mathbb N$,  $M\in \mathbb N$, and $\mathbb Z_L:=\mathbb Z/L\mathbb Z$ is the set of integers modulo $L$ (in the following, we shall identify the elements of $\mathbb Z_L$ with the integers $1,\ldots, L$); moreover, if $z=((z)_1,(z)_2)$\footnote{In the following we shall denote the components of $z\in \Lambda$ by $(z)_1, (z)_2$, rather than by $z_1,z_2$, 
in order to avoid confusion with the first two elements of an $n$-tuple $\bs z\in\Lambda^n$, for which we will use the notation $\bs z=(z_1,\ldots, z_n)$.}
is on the upper boundary of $\Lambda$, that is, if $z=((z)_1,M)$ for some $(z)_1\in\{1,\ldots,L\}$, 
then we interpret $\sigma_{z+\hat e_2}$ as being equal to zero; if $z=(L,(z)_2)$ for some $(z)_2\in\{1,\ldots,M\}$, then we interpret $\sigma_{z+\hat e_1}$ as 
being equal to $\sigma_{(1,(z)_2)}$.

The set $\Lambda$ can be naturally thought of as the vertex set of the graph $G_\Lambda=(\Lambda,\mathfrak B_\Lambda)$, whose edge set is the set 
$\mathfrak B_\Lambda$ of nearest neighbor edges (or `bonds') of $\Lambda$, including, of course, those connecting a point $z=(L,(z)_2)$ with $z+\hat e_1\equiv (1,(z)_2)$. 
The edges in $\mathfrak B_\Lambda$ are in one-to-one correspondence with their midpoints; therefore, in the following, we shall identify 
the elements of $\fB_\Lambda$ with the midpoints of the nearest-neighbor edges of $\Lambda$ and, with some 
abuse of notation, we shall indiscriminately refer to $x\in\fB_\Lambda$ as an edge, or as the midpoint of an edge. For $x\in \fB_\Lambda$ and $\sigma\in \Omega_\Lambda:=\{\pm1\}^{\Lambda}$, we let 
$\epsilon_x= \epsilon_x(\sigma):=\sigma_z\sigma_{z'}$, with $z,z'$ the two endpoints of the edge $x$. 
We are interested in computing the multipoint energy correlations $\media{\epsilon_{x_1}\cdots\epsilon_{x_n}}_{\beta,\Lambda}$, where
$\media{\cdot}_{\beta,\Lambda}$ is the average with respect to the Gibbs measure associated with $H_\Lambda$ at inverse temperature $\beta$; that is, 
given an observable $F:\Omega_\Lambda\to\mathbb R$, 
\begin{equation} 
\media{F}_{\beta,\Lambda}:= \frac{\sum_{\sigma\in\Omega_{\Lambda}}e^{-\beta H_{\Lambda}(\sigma)}F(\sigma)}{\sum_{\sigma\in\Omega_{\Lambda}} e^{-\beta H_{\Lambda}(\sigma)}}.\end{equation}
Our results are more straightforward to derive and state in terms of the {\it truncated} correlations $\media{\epsilon_{x_1};\cdots;\epsilon_{x_n}}_{\beta,\Lambda}$, or cumulants, 
defined, for any $n>1$, as
\begin{equation} \media{\epsilon_{x_1};\cdots;\epsilon_{x_n}}_{\beta,\Lambda}:=\frac{\partial^n}{\partial A_1 \cdots\partial A_n} \log\media{e^{A_1\epsilon_{x_1}+\cdots
A_n\epsilon_{x_n}}}_{\beta,\Lambda}\Big|_{A_1=\cdots=A_n=0};\end{equation}
the ordinary correlation functions can easily be reconstructed from them.
More precisely, we are interested in these truncated correlations at the critical temperature, in the limit $L,M\to\infty$ with fixed aspect ratio $L/M$. 

We fix once and for all an interaction $V$ with the properties spelled out after \eqref{eq:HM}. 
We also assume that $J_1/J_2$ and $L/M$ belong to a compact 
$K\subset (0,+\infty)$. We let $t_l:=t_l(\beta):=\tanh \beta J_l$, with $l=1,2$, and recall that in the non-interacting case, $\lambda=0$, 
the critical temperature $\beta_c(J_1,J_2)$ is the unique solution of $t_2(\beta)=(1-t_1(\beta))/(1+t_1(\beta))$. Note that 
there exists a suitable compact $K'\subset (0,1)$ such that whenever $J_1/J_2 \in K$ and $\beta \in [\tfrac12 \beta_c(J_1,J_2),2 \beta_c(J_1,J_2)]$, 
then $t_1,t_2 \in K'$. From now on, we will think of $K,K'$ as being fixed once and for all. In order to emphasize the dependence of the Gibbs measure upon $\lambda, t_1, t_2$, we add labels 
to the Gibbs measure, and denote
\begin{equation}\label{media}\media{\cdot}_{\beta,\Lambda}\equiv \media{\cdot}_{\lambda,t_1,t_2;\Lambda}.\end{equation}
We are now ready to state our main result. 
\begin{theorem}
\label{prop:main} Fix $V$ as discussed above. Fix $J_1,J_2$ so that $J_1/J_2$ belongs to the compact $K$ introduced above.  
There exists $\lambda_0>0$ and analytic functions
 $\beta_c(\lambda)$, $t_1^*(\lambda)$, $Z_1(\lambda)$, $Z_2(\lambda)$, defined for $|\lambda| \le \lambda_0$, 
such that, for any finite cylinder $\Lambda$ with $L/M\in K$ and any 
$m$-tuple $\bs x=(x_1,\ldots x_m)$ of distinct elements of $\fB_\Lambda$, with $m_1$ horizontal elements, $m_2$ vertical elements, and $m=m_1+m_2\ge 2$, 
	\begin{equation}
		\media{\epsilon_{x_1}; \dots ; \epsilon_{x_m}}_{\lambda,t_1(\lambda),t_2(\lambda);\Lambda}
		=
		\big(Z_1(\lambda)\big)^{m_1}\big(Z_2(\lambda)\big)^{m_2}
		\media{\epsilon_{x_1}; \dots ; \epsilon_{x_m}}_{0,t^*_1(\lambda),t^*_2(\lambda);\Lambda}
		+ R_\Lambda(\bs x), 
		\label{eq:corr_main_statement}
	\end{equation}
where $t_1(\lambda):=\tanh(\beta_c(\lambda)J_1)$, $t_2(\lambda):=\tanh(\beta_c(\lambda)J_2)$ and $t_2^*(\lambda):=(1-t_1^*(\lambda))/(1+t_1^*(\lambda))$. 
Moreover, denoting by $\delta(\bs x)$ the tree distance of $\bs x$, i.e., the cardinality of the smallest connected subset of $\fB_\Lambda$ containing the elements of 
$\bs x$, by $d=d(\bs x)=\min_{1\le i<j\le m}\delta(x_i,x_j)$, and by $d_\partial=d_\partial(\bs x)=\min_{i=1}^m\delta_\E(x_i)$, with $\delta_\E(x)$ the $L^1$ distance between the midpoint of $x$ and the boundary of $\Lambda$, for all $\theta\in(0,1)$ and $\varepsilon\in(0,1/2)$ and a
suitable $C_{\theta,\varepsilon}>0$, the remainder $R_\Lambda$ can be
bounded as
\begin{equation} |R_\Lambda(\bs x)|\le  C_{\theta,\varepsilon}^m  |\lambda| m!
\frac1{d^{m}}\left(\frac{d}{\delta(\bs x)}\right)^{2-2\varepsilon}(\min\{d,d_\partial\})^{-\theta}\;.\label{10b}\end{equation}
\end{theorem}

In this theorem, $\beta_c(\lambda)$ has the interpretation of an interacting inverse critical temperature, and $\media{\cdot}_{0,t_1^*(\lambda),t_2^*(\lambda);\Lambda}$ 
plays the role of the reference non-interacting critical Gibbs measure. In fact, \cref{eq:corr_main_statement} tells us that, 
for the purpose of computing the multipoint energy correlations, we can use this non-interacting measure instead of the interacting one, 
up to the finite multiplicative renormalization constants $Z_1(\lambda), Z_2(\lambda)$ and the remainder term $R_\Lambda$. 
The non-interacting correlation function $\media{\epsilon_{x_1}; \dots ; \epsilon_{x_m}}_{0,t^*_1(\lambda),t^*_2(\lambda);\Lambda}$ in the right side of \eqref{eq:corr_main_statement}
is explicitly known from Onsager's solution of the nearest neighbor Ising model. In particular, the \textit{non-truncated} energy correlation 
$\media{(\epsilon_{x_1}-\media{\epsilon_{x_1}}_{0,t^*_1(\lambda),t^*_2(\lambda);\Lambda})\cdots
(\epsilon_{x_m}-\media{\epsilon_{x_m}}_{0,t^*_1(\lambda),t^*_2(\lambda);\Lambda})}_{0,t^*_1(\lambda),t^*_2(\lambda);\Lambda}$
is the Pfaffian of a suitable $2m\times 2m$ anti-symmetric matrix. In the limit that $L$, $M$, $d$, $d_\partial$ all become large compared to the lattice spacing, with e.g. their ratios all kept bounded, 
this Pfaffian goes to zero like $d^{-m}$, which indicates that the remainder term $R_\Lambda$ is subdominant with 
respect to $\media{\epsilon_{x_1}; \dots ; \epsilon_{x_m}}_{0,t^*_1(\lambda),t^*_2(\lambda);\Lambda}$ at large distances. Note that, since we consider the 
{\it truncated} energy correlations, the remainder also goes to zero as $\delta(\bs x)$ becomes large at $d$ fixed, as expected. Notably, the 
analyticity radius $\lambda_0$ and the functions $\beta_c(\lambda)$, $t_1^*(\lambda)$, $Z_1(\lambda)$, $Z_2(\lambda)$ are all independent of $L,M$, as long as 
$L/M\in K$. That is, the result is uniform in $L,M\to \infty$, as long as the aspect ratio is bounded from above and below.

\medskip

As a corollary of \cref{prop:main} (thanks in particular to the uniformity in $L,M$), we can compute the scaling limit of the energy correlations as follows.
Fix two positive constants $\ell_1,\ell_2$ with $\ell_1/\ell_2\in K$, and let $L=2\lfloor a^{-1}\ell_1/2\rfloor$, $M=\lfloor a^{-1}\ell_2\rfloor$ for $a>0$ sufficiently small. 
Let $\Lambda^a:=a\Lambda$ and let $\fB_{\Lambda^a}$ be the corresponding set of nearest neighbor bonds. As in the case $a=1$, for $x\in \fB_{\Lambda^a}$,
we let $\epsilon_x$ be the product of the spins at the vertices of the edge $x$. We 
denote by $\Lambda_{\ell_1,\ell_2}:=(\mathbb R/\ell_1\mathbb R)\times[0,\ell_2]$ the continuous cylinder 
which $\Lambda^a$ reduces to in the limit $a\to 0$. Moreover, for any $z$ in the interior of $\Lambda_{\ell_1,\ell_2}$, we define the rescaled energy observable as follows: 
\begin{equation} \varepsilon_l^a(z):= a^{-1}\big(\epsilon_{x(z,l)}-e^a_{z,l}(\lambda)\big),\label{def_varepsilon}\end{equation}
where $l\in\{1,2\}$, $x(z,l)$ is the edge of $\fB_{\Lambda^a}$ with vertices $a\lfloor z/a\rfloor$ and $a\lfloor z/a\rfloor+a\hat e_l$, and 
\begin{equation} e^a_{z,l}(\lambda):=
\media{\epsilon_{x(z,l)}}_{\lambda,t_1(\lambda),t_2(\lambda);\Lambda^a},\end{equation}
where $\media{\cdot}_{\lambda,t_1(\lambda),t_2(\lambda);\Lambda^a}$ 
denotes the interacting critical Gibbs measure on $\Lambda^a$ (here $t_l(\lambda)$ with $l=1,2$ are the same as in Theorem \ref{prop:main}).

Fix an $m$-{tuple} $\bs z=(z_1,\ldots,z_m)$ of points in the interior of $\Lambda_{\ell_1,\ell_2}$, with $m\ge 2$. Theorem~\ref{prop:main} tells us that, 
for any $\bs l=(l_1,\ldots, l_m)\in\{1,2\}^m$, 
\begin{equation}\label{scalingtrunc}
	\begin{split}
		\media{\varepsilon_{l_1}^a(z_1); \dots ; \varepsilon_{l_m}^a(z_m)}_{\lambda,t_1(\lambda),t_2(\lambda);\Lambda^a}
		=&
		\big(Z_1(\lambda)\big)^{m_1}\big(Z_2(\lambda)\big)^{m_2}
		\media{\varepsilon_{l_1}^a(z_1); \dots ; \varepsilon_{l_m}^a(z_m)}_{0,t^*_1(\lambda),t^*_2(\lambda);\Lambda}
		\\
&+ R_{\bs l;\Lambda^a}(\bs z), 
	\end{split}
\end{equation}
where, if $\delta=\delta(\bs z)$ is the tree distance\footnote{$\delta(\bs z)$ is the same as the so-called Steiner diameter of $\bs z$. For $n$-ples of points in the plane, it is 
defined as $\delta(\bs z):=\min_{k\ge 0}\min_{\bs z'\in(\mathbb R^2)^k}\min_{\tau\in \cT(\bs z,\bs z')}L(\tau)$, where $\cT(\bs z,\bs z')$ is the set of  all possible trees with vertices
in the elements of the $(n+k)$-tuple $\bs z\oplus\bs z'\equiv(\bs z,\bs z')$, and $L(\tau)$ is the tree length of $\tau$, i.e., the sum of the Euclidean lengths of the edges of $\tau$. For $n$-ples of points in the cylinder the definition is the same, provided the Euclidean distance in $\mathbb R^2$ is replaced by its analogue in the cylinder.}
of $\bs z$, $d=\min_{1\le i<j\le m}d(z_i,z_j)$ and $d_\partial(\bs{z})=\min_{i=1}^m$ $\dist(z_i,\partial\Lambda_{\ell_1,\ell_2})$, 
\begin{equation} |R_{\bs l;\Lambda^a}(\bs z)|\le  C_{\theta,\varepsilon}^m  |\lambda| m!
\frac1{d^{m}}\left(\frac{d}{\delta}\right)^{2-2\varepsilon}\left(\frac{a}{\min\{d,d_\partial\}}\right)^\theta\;,\label{10b.vir}\end{equation}
where $C_{\theta,\varepsilon}$ is the same constant as in \eqref{10b}. Clearly, for any fixed $\bs z$, the right side of \eqref{10b.vir} goes to zero as $a\to 0$. 
Therefore, the scaling limit of the energy correlations is straightforwardly related to the corresponding quantity for the noninteracting model.
On a cylinder, this latter quantity can be given explicitly;  we formulate the result for the {\it non-truncated} energy correlations, rather than for truncated ones, since the resulting expression is now simpler in this form. 

\begin{corollary}\label{cor:main}
	Under the same assumptions as Theorem \ref{prop:main}, for any $m$-{tuple} $\bs z=(z_1,\ldots,z_m)$ of points in the interior of $\Lambda_{\ell_1,\ell_2}$, with $m\ge 2$,
and any $\bs l\in\{1,2\}^m$, 
\begin{eqnarray} && \lim_{a\to 0^+} \media{\varepsilon_{l_1}^a(z_1) \cdots  \varepsilon_{l_m}^a(z_m)}_{\lambda,t_1(\lambda),t_2(\lambda);\Lambda^a}=
\label{eq:cor.1}\\
&&\qquad=
\big(Z_1(\lambda)\big)^{m_1}\big(Z_2(\lambda)\big)^{m_2}\lim_{a\to 0^+}
		\media{\varepsilon_{l_1}^a(z_1) \cdots  \varepsilon_{l_m}^a(z_m)}_{0,t^*_1(\lambda),t^*_2(\lambda);\Lambda^a},\nonumber\end{eqnarray}
where $t_1(\lambda), t_2(\lambda), Z_1(\lambda), Z_2(\lambda)$ are the same as in Theorem \ref{prop:main}, and 
$\varepsilon_{l}^a(z)$ is defined as in \eqref{def_varepsilon}. The limit in the right side of \eqref{eq:cor.1} exists and equals
	\begin{equation} \lim_{a\to 0^+}
		\media{\varepsilon_{l_1}^a(z_1) \cdots  \varepsilon_{l_m}^a(z_m)}_{0,t^*_1(\lambda),t^*_2(\lambda);\Lambda^a}=(2t_2^*(\lambda))^{m_1}(1-(t_2^*(\lambda))^2)^{m_2}
		{\rm Pf}(\mathcal M(\bs z)),\label{scalinglim.free.1}\end{equation}
with $\mathcal M(\bs z)$ the $2m\times 2m$ anti-symmetrix matrix, whose elements, labelled by $(1,+),(1,-),\ldots$, $(m,+),(m,-)$, are equal to:
\begin{equation} \big[\mathcal M(\bs z)\big]_{(i,\omega),(j,\omega')}=\begin{cases} \big[\fg_\scal(z_i,z_j)\big]_{\omega\omega'} & \text{if $i\neq j$,}\\
0 & \text{otherwise,}\end{cases} \label{scalinglim.free.2}\end{equation} 
where $\fg_\scal$ is the scaling limit propagator in \eqref{eq:g_scal_cyl_def} below, computed at $t_1=t_1^*(\lambda)$, $t_2=t_2^*(\lambda)$. 
\end{corollary}
The analogue of \eqref{eq:cor.1} for the truncated correlations follows from \eqref{scalingtrunc}-\eqref{10b.vir}; 
since the non-truncated correlations are combinations of products of truncated ones (and viceversa), the statement for truncated 
correlations implies the one for non-truncated ones. The existence of the limit in the right side of \eqref{eq:cor.1} and the computation of its explicit form, 
\eqref{scalinglim.free.1}-\eqref{scalinglim.free.2}, follow from the exact solution of the nearest neighbor Ising model on the cylinder, which is reviewed 
in the companion paper \cite[Section~2]{AGG_part2}; in particular, for the proof of \eqref{scalinglim.free.1}-\eqref{scalinglim.free.2}, see 
\cite[Appendix~D]{AGG_part2}. It is easy to 
check that an analogue of Corollary \ref{cor:main} is valid for the energy correlations in the half-plane\footnote{The generalization of 
Corollary \ref{cor:main} to the case of the half-plane concerns the scaling limit of the energy correlations computed in the following way: $\lim_{a\to 0^+}\lim_{L,M\to\infty} \media{\varepsilon_{l_1}^a(x_1) \cdots  \varepsilon_{l_m}^a(x_m)}_{\lambda,t_1(\lambda),t_2(\lambda);\Lambda^a}$, i.e., $L,M\to\infty$ first, then $a\to 0$; here, as usual, $\Lambda^a=a\Lambda$, with $\Lambda$ the cylinder of sides $L$ and $M$, and the limit $L,M\to\infty$ is computed under the constraint that $L/M\in K$. 
The uniformity in $L,M$ of our bounds, valid as long as $L/M\in K$, implies that the limit computed this way is the same as the limit of \eqref{eq:cor.1}-\eqref{scalinglim.free.1} as $\ell_1,\ell_2\to\infty$, under the constraint that $\ell_1/\ell_2\in K$.}, as well. 

The explicit expression of the scaling limit shows that it is covariant under uniform rescalings of the cylinder 
$\Lambda_{\ell_1,\ell_2} \to \Lambda_{\xi\ell_1,\xi\ell_2}$, for any $\xi>0$, see \cite[Appendix~D]{AGG_part2}.
As commented there, uniform rescalings, translations and parity transformations are the only conformal transformations mapping finite cylinders $\Lambda_{\ell_1,\ell_2}$ to finite 
cylinders $\Lambda_{\ell_1',\ell_2'}$ (or translations thereof). 
In order to check conformal covariance of the scaling limit in a more complete sense, it would be desirable to extend the proof of Corollary \ref{cor:main}
to general finite domains with free boundary conditions; we discuss this further in {\it Generalizations and perspectives} below. 

\paragraph{Method of proof, motivations and comparison with previous works} As briefly mentioned above, the rigorous application of Wilsonian RG to interacting 
2D Ising models at the critical point was sparked by Spencer's proposal \cite{S00} of a rigorous strategy to compute the energy-energy critical exponent 
and by the related (unpublished) work of Pinson and Spencer \cite{PS}. 
The starting point of their approach is an exact representation of the partition and generating functions in terms of a non-gaussian Grassmann integral, 
a sort of fermionic $\phi^4_2$ theory, which can be 
studied via the constructive fermionic RG methods first developed in the mid `80s and early `90s \cite{BG,BGPS,FMRT,GK,Le} and later applied to several critical statistical mechanics models in two dimensions \cite{BFM02, BFM10, GM04,GM05, GMT17a, GMT19, M04}
and to condensed matter systems in one \cite{BM01,GM01},
two \cite{GM10,GMP12a,GMP12b} 
and higher dimensions \cite{GMP.Weyl,Mas.Weyl}. 
Dimensionally, the quartic interaction of the effective $\phi^4_2$ model which the interacting Ising model is equivalent to, is 
{\it marginal} in the RG jargon. The same is true for several other critical two-dimensional statistical mechanics systems, such as 
interacting dimer models, six- and eight-vertex models, and the Ashkin-Teller model (see \cite{BFM02, BFM10, GM04,GM05, GMT17a, GMT19, M04}
for a rigorous analysis of their critical behavior via fermionic RG methods). 
In all these systems, the marginal coupling is generically non-zero, and the analysis requires the use of subtle cancellations 
discovered by Benfatto, Gallavotti and Mastropietro in the context of interacting 1D fermions \cite{BGPS, BM01}. 
On the contrary, in the context of interacting Ising models, life is much simpler: the massless field in 
the fermionic formulation of the 2D critical Ising model is a two component Majorana fermion  \cite[Chapter~2]{ID}. This implies that there is no room 
for a local quartic coupling (see Section \ref{sec:cylinder_multiscale} for details, see in particular Eq.\eqref{pauli}) and, therefore, the fermionic quartic interaction 
is effectively {\it irrelevant}, rather than marginal. This is the key reason why one expects (and can prove in some cases, see \cite{GGM, GM13} and Corollary 
\ref{cor:main} above) that the infrared fixed point of 2D interacting Ising models is `trivial', that is, the scaling limit at the critical point coincides with the non-interacting 
one, up to a finite multiplicative renormalization of the energy observable.  

\medskip

All the rigorous RG results mentioned so far rely on translation invariance of the model. From a technical point of view, this guarantees, in particular, that the
relevant and marginal couplings are, in fact, constants, rather than functions depending on the position $z$ in $\Lambda$, the domain which the system is defined on. 
Constructive RG methods are not well developed yet in the case of critical theories in finite domains, where boundaries are present and affect the 
form of correlation functions in the scaling limit. This is a severe limitation for the rigorous construction of scaling limits in finite domains and for the study 
of their conformal covariance with respect to deformations of the domain. It is also a limitation for several other related problems, such as 
the understanding of interaction effects in systems with defects, which is an issue of relevance for, e.g., the Kondo problem \cite{Aff_rev, BGJ.Kondo, WiRMP}, the problem of many-body localization \cite{Alts,Mas17.loc,NH},
and even for the computation of monomer-monomer or spin-spin correlations in interacting dimer or Ising models (due to the fact that such correlations reduce to 
the computation of interacting Green functions in discrete Riemann surfaces with cuts, or `defects', at the locations of the monomers, or of the spins \cite{CHI,Dub11a}). At a theoretical physics level, RG methods in the presence of boundaries have been developed in the context of quantum wires \cite{FG, GM09, MEJ, MMSSSS} and of the Casimir effect \cite{Sym81, DDb, DDc}, but a systematic theory is still lacking. 

As discussed above, in the case of interacting 2D critical Ising models, the interaction is effectively irrelevant rather than marginal, contrary to many other 2D critical 
statistical mechanics systems. Therefore, this case looks like one of the easiest where to control boundary effects in the scaling limit. This is what we do 
in this paper; as far as we know, our work represents the first rigorous treatment of these effects in a critical theory. 
The methods we introduce may have an impact in related areas, such as the computation of boundary critical exponents in 
models of quantum wires or of quantum spin chains, the computation of the (universal?) sub-leading corrections to the critical free energy in models of interacting 
dimers, the Kondo problem, the Casimir effect in interacting systems, the computation of monomer-monomer or spin-spin correlations in interacting dimer or Ising models, etc. 

\medskip

Our strategy is roughly based on the following ideas: in the presence of a boundary, any contribution to the bulk thermodynamic functions, as well as to 
the generating function of correlations of observables located at points in the interior of the domain, 
can be decomposed into a bulk part (which is defined in a straightforward way based on its infinite plane counterpart), plus a remainder, which we call the `edge part'. 
One of the important results of this paper is that the edge part admits dimensional bounds that are dimensionally better by one scaling dimension, compared with their 
bulk counterparts: the edge part of a linearly relevant operator is marginal in the RG sense; the edge part of a marginal operator is irrelevant.  Fundamentally, this is because the local part of the edge part can be taken to be supported on the boundary (as in the definitions we introduce in \cref{sec:LeRcyl}), so that the sum or integral associated with these terms is over one coordinate rather than two.
This modified scaling dimension appears in Proposition~\ref{lm:W:scaldim_withm}, in particular as a dependence upon the label $E_v$, which is equal to $1$ for the edge 
contributions.

In the bulk theory of the 2D interacting Ising model studied in \cite{GGM} there is only one 
linearly relevant operator, corresponding to the `running temperature counterterm', denoted by $\nu_h$ both in \cite{GGM} and in this paper, which is used to 
fix the value of $\beta$ corresponding to the interacting critical inverse temperature. Its edge part can be localized at the boundary, and in this way one obtains a boundary marginal running 
coupling constant, whose flow is (potentially) logarithmically divergent. We expect that, in general, this logarithmic divergence is the one responsible for 
anomalous boundary critical exponents, like those expected in Luttinger liquids on the half-line \cite{FG,GM09,MEJ,MMSSSS}.

In our situation of interest, a remarkable cancellation, see \eqref{keycancellationedge}, implies that the boundary marginal running coupling constant is 
exactly zero. Therefore, the boundary terms are all effectively irrelevant, and they scale to zero in the infrared limit. Summarizing, 
if we start with the interacting Ising model on a cylinder, with open boundary conditions in the non-periodic direction, we tune the temperature at criticality, and take
the scaling limit, we get a limiting theory in the continuous cylinder with, again, open boundary conditions in the non-periodic direction. This result 
is in line with the CFT expectation that the scaling limit of the 2D Ising model supports only two independent 
conformal boundary conditions, open and fixed ($+/-$) boundary conditions \cite{Cardy.boundaries}\footnote{Although there is no exact duality relating different boundary conditions for the model with $\lambda\neq0$, 
we expect that the non-integrable model with fixed, say $+$, boundary conditions can be studied via a multiscale perturbation theory 
around a non-interacting model with renormalized, dressed, parameters and 
the same fixed boundary conditions. Such a reference non-interacting model \textit{is} dual to its counterpart with open boundary conditions, and the main features of the expansion should be the same as for the open boundary conditions we consider here: in particular boundary terms should be irrelevant
in this case as well.  Consequently, duality between open and fixed boundary conditions should 
be an emergent property in the scaling limit of the interacting model.}
The cancellation of the boundary marginal coupling 
is not related directly to the one of the bulk local quartic coupling, mentioned above: we could not anticipate it on the basis 
of the bulk analysis in \cite{GGM}. Rather, it is related to an approximate image rule satisfied by the propagator on the cylinder. 

\paragraph{Generalizations and perspectives} 

One of the limitations of this paper is the choice of cylindrical geometry\footnote{An additional, less consequential, limitation is the non-uniformity of our result,  
as the aspect ratio of the cylinder
 tends to zero or infinity. We expect to be able to overcome this limitation easily, by studying the regime of infrared scales corresponding lengths larger than $L$ 
but smaller than $M$, or viceversa, via a different multiscale scheme, taking into account the quasi-1D nature of the system at such scales. We decided not to do this explicitly in this 
paper, just in order to limit its length. We expect that such a refined scheme will allow us to prove the analogue of Theorem \ref{prop:main} and Corollary \ref{cor:main}
without any restriction on the aspect ratio. In particular, the result of Corollary \ref{cor:main} would generalize to infinite cylindrical strips, and would allow us to check conformal covariance of the limit from the strip to the half-plane.}: we expect analogous results to be true in finite domains of arbitrary shape with open 
boundary conditions, but we are currently unable to prove them. The generalization to rectangular domains should be straightforward (even if involved)
but extensions to more general domains appear to require new insights into the non-interacting model. In our approach, the choice of the domain is dictated by the availability of a 
sufficiently 
explicit exact solution for the reference non-interacting model. The partition function and the energy correlations of the non-interacting model 
exhibit a determinantal (or, more correctly, Pfaffian) structure in all 
domains, but the underlying matrix can be explicitly diagonalized only in very special cases, most notably the torus, the cylinder and the rectangle 
(explicit diagonalization of the relevant matrix on the cylinder is already quite involved, as compared to the torus, and is reviewed in the companion paper \cite{AGG_part2},
see Section \ref{sec:2.1} for a summary of the features of this exact diagonalization that are relevant for the present work;
diagonalization on the rectangle is known \cite{Hucht17a,Hucht17b} but even more involved\footnote{The diagonalization procedure on the rectangle may be simplified
substantially by using an exact image rule, following from $s$-holomorphicity, see \cite{CCK,Russkikh}, but this remains to be done in detail.}).
We use the explicit diagonalization of the relevant matrix in order to derive a Fourier representation of the propagator (the fermionic Green's function) and, 
correspondingly, a multiscale decomposition thereof, see \eqref{gch*h}; we also use it to write the propagator in Gram form, see \eqref{ddtthh},
which is needed for our technical estimates. If we were given the same inputs (in particular, the `right' decay bounds for the single-scale propagator, 
its Gram decomposition, and the cancellation of its appropriate components at the boundary) in a more general domain $\Omega$, 
then we would be able to construct the scaling limit in $\Omega$ as well: our multiscale RG construction, described in Section \ref{sec:renexp}, 
is insensitive to the geometric details of $\Omega$, provided the right inputs on the single-scale propagator are available. 

A natural idea for proving the desired properties for the propagator of the non-interacting theory in general domains is to use the 
results on the scaling limit of the fermionic Green's function in finite domains based on discrete holomorphicity, see \cite{HS13,Chelkak16.review}.  
The limiting propagator has all the desired properties; the hope is that, if the remainder (the difference between the rescaled finite-mesh propagator and 
its scaling limit) goes to zero sufficiently fast, then the desired properties can be proved for the finite-mesh propagator, as well. Unfortunately, 
the aforementioned results are not strong enough to 
provide us the desired inputs: the convergence to the scaling limit proved there is not quantitative\footnote{It is likely that an extension of these methods
would `easily' give non-optimal quantitative bounds \cite{Chelkak.private}, but these would not suffice for our purposes.}. 
The problem of computing the optimal convergence rate to the scaling limit for the planar Ising model is currently under investigation.

\medskip

If, instead, we stick to the same cylindrical geometry as in Theorem \ref{prop:main},
there are other, more straightforward, extensions of our main results:
(1) computation of the massive scaling limit of the energy correlations, (2)
computation of the scaling limit of the boundary energy and spin correlations, 
(3) computation of the universal sub-leading corrections to the critical free energy.

The solution of (1) for the massive scaling limit in the temperature direction
 is implicit in the proof of this paper: here we focus on the massless scaling limit only for simplicity, but 
our methods are flexible enough to allow us to control the massive one, see \cite{GGM}, where the massive scaling limit in the infinite plane was explicitly obtained. 
The computation of the massive scaling limit in the direction of the magnetic field is harder, and we do not know how to approach it at the moment. 
See  \cite{CGN15,CGN16} for recent progress on such scaling limit in the non-interacting case. 

Problem (2) has been solved in all its most essential aspects in \cite{Cava}, and we plan to present the full proof in a forthcoming paper. 

We expect that the solution of (3) will follow from a combination of the methods of this paper with 
those of \cite{GM13}. We hope to come back to this problem in a future publication. 

\paragraph{Summary and roadmap} \begin{itemize}
\item In Section \ref{sec:gen} we review the Grassmann representation of the generating function for the energy correlations of our model.
This representation originates from the well known Grassmann representation of the partition function of the nearest neighbor Ising model \cite{CCK,S80a}. 
While very similar to the one derived in \cite{GGM}, the specific form of the Grassmann generating function used in this paper is slightly different, and
we refer the reader to the companion paper \cite{AGG_part2} for its derivation. 
Moreover, in Section \ref{sec:gen}, 
we set up the stage for the multiscale computation of the Grassmann generating function, to be described in the following sections. In particular, 
we `add and subtract' quadratic terms to the Grassmann action, depending on two parameters $t_1^*,Z$, to be appropriately fixed in 
such a way that the reference non-interacting theory around which we expand is the correct one (i.e., the one with the correct dressed critical parameters
$t_1^*(\lambda), t_2^*(\lambda)$, see Theorem \ref{prop:main}, and with the correct asymptotics -- at the level of the finite prefactor in front of the dominant power law behavior 
at large distances -- of the fermionic Green's function at the critical point).
\item In Section \ref{sec:renexp} we describe the iterative, multiscale, computation of the Grassmann generating function. We begin with a general introduction to the 
iterative integration scheme to be followed, defining, in particular, the sequence of effective potentials $\cV^{(h)}$ (which define a 
sequence of coarse grained models obtained by averaging fluctuations on scales of the order $2^{-h}$ or smaller) and of `single-scale' contributions to the generating function
$\cW^{(h)}$ (that is, the contribution of the degrees of freedom eliminated in coarse-graining); in our conventions, the `scale label' $h$ is also equal to $h=1-j$, with $j$ the number of iteration steps performed. The goal of this section is to derive a 
uniformly convergent expansion for $\cV^{(h)}$ and $\cW^{(h)}$, with quantitative bounds on the norm of their kernels. After this general introduction, 
the exposition is organized as follows: in Section \ref{sec:cylinder_multiscale} we define the localization and renormalization operators $\cL$ and $\cR$, which allow us to 
isolate from $\cV^{(h)}$ the potentially divergent terms (the `local contributions' $\cL \cV^{(h)}$) and to rewrite the remainder in a convenient, interpolated, form, denoted
$\cR \cV^{(h)}$, which will be shown to satisfy improved dimensional bounds, 
compared to its local counterpart; in particular, we define the notion of localization on the boundary and 
exhibit the cancellation of the marginal boundary coupling, see \eqref{keycancellationedge}; compared to previous works, the definitions of $\cL$ and $\cR$ introduced here take 
into account boundary effects: correspondingly, the remainder $\cR \cV^{(h)}$ consists of two contributions, which we call `bulk' and `edge' contributions; on the contrary,
the local term $\cL \cV^{(h)}$ contains by construction only bulk contributions and is parametrized by five real parameters: 
the {\it running coupling constants} $\nu_h,\zeta_h,\eta_h$, and the {\it effective vertex renormalizations} $Z_{1,h}, Z_{2,h}$. 
In Section \ref{sec:treecyl} we define the so-called {\it Gallavotti-Nicol\`o} (GN) tree expansion for $\cV^{(h)}$ and $\cW^{(h)}$; compared to previous works, here we 
introduce the notions of bulk and edge vertices, and bulk and edge sub-trees, which are induced by the systematic decomposition of 
$\cR \cV^{(h)}$ into its bulk and edge contributions. In Section \ref{sec:informal_bounds} we prove 
$L^1$ weighted bounds for the tree values, and prove that each sourceless edge vertex 
comes with a dimensional gain, represented by the factors $2^{-E_v(h_v-h_{v'})}$ in Propositions \ref{lem:W_primitive_bound}, \ref{lem:W_primitive_with_m} (where $E_v$ is the `edge index', equal to $1$ for edge vertices, and $0$ otherwise) and the dependence upon 
$E_v$ in the scaling dimension in Proposition \ref{lm:W:scaldim_withm}; these are three of the 
main technical results of this paper. Finally, in Section \ref{sec:Zj}, we show the boundedness of the sequence of effective vertex renormalizations;
note that the boundedness (and asymptotic vanishing) of the running coupling constants follows from a standard fixed point argument, which requires the parameters 
$t_1^*,Z$ mentioned above to be fixed properly; see \cite[Section~4.5]{AGG_part2} for the proof in the specific setting required in this work. 
\item In Section \ref{sec:correlations} we conclude the proof of the main theorem, by adapting the bounds derived in Section \ref{sec:informal_bounds}
to the multipoint energy correlation functions. This section and the related Appendix \ref{app.proofboringb} contains the other key technical novelties of the work. 
The strategy for bounding the correlation functions 
parallels the analogous discussion in \cite[Section 4]{GGM}. However, in order to perform the sum over the scale labels we need to take into account that 
some branches of the GN trees have scaling dimension zero: this requires a modified procedure of summation over scales (which still leads to uniform bounds in the 
scaling limit, thanks to the fact that GN trees can be decomposed in a union of connected subtrees whose branches are all associated with negative scaling dimensions), 
described in Sections \ref{sec:5.1}, \ref{sec:corr_corrections} and Appendix \ref{app.proofboringb}. 
\end{itemize}

We emphasize that, in order to appreciate the technical novelties of the present paper, with respect to previous works on the multiscale construction of the bulk critical 
correlations of two-dimensional statistical mechanics models, such as \cite{GGM}, the reader should compare our definitions and technical estimates with the corresponding, 
simpler, ones, introduced and used for the treatment of the infinite plane critical correlation functions. In order to help comparison and to provide a self-contained reference for 
such an infinite plane construction, we summarized all the necessary material in the companion paper \cite{AGG_part2}, which contains, in addition to a derivation of  
of the exact solution of the nearest neighbor Ising model in cylindrical geometry using Grassmann variables, a review of the multiscale computation of the effective potentials 
in the infinite plane limit, in the same notation and via the same technical procedure used in the present work. 

\section{Grassmann representation of the generating function}\label{sec:gen}

In this section we rewrite the generating function of the energy correlations for the Ising model \eqref{eq:HM} with finite range interactions as an 
interacting, non-Gaussian, Grassmann integral, and we set the stage for the multiscale integration thereof, to be discussed in the following sections.
The estimates stated in this and in the following sections are uniform for $J_1/J_2, L/M\in K$ and $t_1,t_2\in K'$, but may depend upon the choice of $K,K'$, 
with $K,K'$ the compact sets introduced before the statement of Theorem \ref{prop:main}. 
As anticipated there, we will consider $K,K'$ to be fixed once and for all and, for simplicity, we will not track the 
dependence upon these sets in the constants $C,C', \ldots, c,c', \ldots, \kappa, \kappa', \ldots$, appearing below. Unless otherwise stated, 
the values of these constants may change from line to line. 

\medskip

The generating function of the energy correlations that we are interested in, in the notations introduced in Section \ref{sec:intro}, reads
\begin{equation}
	Z_\Lambda({\bs A}):=	\sum_{\sigma \in \Omega_\Lambda} \exp\left( 
		\sum_{x \in \fB_\Lambda} \left[ \beta J_{j(x)} +  A_x \right] \epsilon_x
	+ \beta\lambda \sum_{X \subset \Lambda} V( X)  \sigma_X\right),
	\label{eq:Ising_gen}
\end{equation}
where $j(x)$ is $1$, resp. $2$, if $x$ is the midpoint of a horizontal, resp. vertical,  bond. 
Before we formulate the general representation of $Z_\Lambda({\bs A})$ in terms of an interacting Grassmann integral, see Section \ref{sec:2.2} below,
let us briefly recall the {\it Gaussian} Grassmann representation 
of the partition function in the case $\lambda=0$; see \cite{AGG_part2} for additional details. 

\subsection{The non-interacting theory in the cylinder}\label{sec:2.1}

If $\lambda=0$ and ${\bs A}={\bs 0}$ we have \cite[Section~2]{AGG_part2}: 
\begin{equation}
	Z_\Lambda({\bs 0})\big|_{\lambda=0}= 2^{LM} (\cosh \beta J_1)^{LM}(\cosh \beta J_2)^{L(M-1)}\int  \mathcal D \phi\, \mathcal D\xi\,\, e^{\cS_{c}(\phi)+\cS_m(\xi)},
\label{eq:Z_pfaffian_grassmann}
\end{equation}
where $\phi=\{\phi_{\omega,z}\}_{\omega=\pm, z\in\Lambda}$ and $\xi=\{\xi_{\omega,z}\}_{\omega=\pm, z\in\Lambda}$ are two collections of $2LM$ Grassmann variables,
$\mathcal D \phi$ and $\mathcal D\xi$ denote the Grassmann `differentials', 
\begin{equation}\label{DphiDxi}\mathcal D \phi=\prod_{z\in\Lambda}d\phi_{+,z}\, d\phi_{-,z},\qquad \mathcal D \xi=\prod_{z\in\Lambda}d\xi_{+,z}\, d\xi_{-,z},\end{equation} 
and, letting  $\phi_{\omega,(z)_2}(k_1):= \sum_{(z)_1=1}^Le^{ik_1(z)_1}\phi_{\omega,((z)_1,(z)_2)}$ for any $k_1\in\mathcal D_L$, with 
\begin{equation}\label{defDL} \mathcal D_L:=\left\{\tfrac{\pi (2m-1)}{L}: \ m= -\tfrac{L}{2}+1,\cdots, \tfrac{L}2 \right\},\end{equation} 
and similarly for $\xi_{\omega,(z)_2}(k_1)$, 
\begin{eqnarray} 
 \cS_c(\phi)&=& \frac1L\sum_{k_1\in \mathcal D_L} \sum_{(z)_2=1}^M \Big[-b(k_1) \phi_{+,(z)_2}(-k_1) \phi_{-,(z)_2}(k_1)
+t_2  \phi_{+,(z)_2}(-k_1) \phi_{-,(z)_2+1}(k_1) \nonumber\\
&  & - \frac{i}2\Delta(k_1) \phi_{+,(z)_2}(-k_1) \phi_{+,(z)_2}(k_1)+\frac{i}2\Delta (k_1) \phi_{-,(z)_2}(-k_1) \phi_{-,(z)_2}(k_1)\big],\label{eq:2.1.13}\\
\cS_m(\xi)&=& \frac1L\sum_{k_1\in \mathcal D_L} \sum_{(z)_2=1}^M (1+t_1e^{-ik_1}) \xi_{+,(z)_2}(-k_1) \xi_{-,(z)_2}(k_1),\label{eq:2.1.12}
 \end{eqnarray}
with $\phi_{-,M+1}(k_1):= 0$, and, recalling that $t_l =\tanh \beta J_{l}$ for $l=1,2$, 
\begin{equation}b(k_1) := \frac{1-t_1^2 }{|1+t_1e^{ik_1}|^2},\qquad
\Delta(k_1) := \frac{2 t_1 \sin k_1}{|1+t_1e^{ik_1}|^2}. 
 \label{eq:b_def}
\end{equation}
\begin{remark} The Grassmann variables $\phi,\xi$ used here are different from the `standard' variables $\lis H,H,\lis V,V$ or $\psi,\chi$ used in \cite{GGM,ID}, even though
they are related to these other sets of coordinates via an (explicit) invertible linear transformation. 
The use of the coordinates $\phi,\xi$ is particularly convenient in our cylindrical geometry: in fact, the linear transformation relating $\phi,\xi$ with 
$\lis H,H,\lis V,V$ corresponds to a Schur reduction 
that block diagonalizes the quadratic Grassmann action in the cylinder, see \cite[Eq.~(2.1.9)]{AGG_part2}. 
The labels `c' and `m' attached to $\cS_c(\phi)$ and $\cS_m(\xi)$ stand for `critical' and `massive', and refer to the polynomial (resp. exponential) decay of the 
off-diagonal elements of the covariance matrix associated with $\cS_c(\phi)$ (resp. $\cS_m(\xi)$) on the critical line $t_2=(1-t_1)/(1+t_1)$. \end{remark}
\begin{remark}\label{rem:prop1} {\bf (Reflection symmetries)} The quadratic polynomials $\cS_c(\phi), \cS_m(\xi)$ are invariant under the following horizontal and vertical 
reflections: 
\begin{equation}
	\phi_{\omega,z} \to \Theta_1 \phi_{\omega,z} := i\omega \phi_{\omega, \theta_1 z}, \quad
	\xi_{ \omega,z} \to \Theta_1 \xi_{\omega,z} := i \xi_{-\omega,\theta_1 z}
	\label{eq:symm_horizontal_phixi}
\end{equation}
where $\theta_1z:=(L+1-(z)_1,(z)_2)$, and
\begin{equation}
	\phi_{\omega,z} \to \Theta_2  \phi_{\omega,z} :=  i \phi_{-\omega,\theta_2 z}, \quad
	\xi_{ \omega, z} \to \Theta_2 \xi_{ \omega, z} :=  -i\omega \xi_{\omega,\theta_2 z},
	\label{eq:symm_vertical_phixi}
\end{equation}
where $\theta_2z:=((z)_1,M+1-(z)_2)$. As discussed below, these symmetries are also present in the interacting theory, and will play a role in the multiscale computation of the 
generating function. 
\end{remark}
The quadratic forms $\cS_{c}(\phi)$ and $\cS_m(\xi)$ can be written as $\cS_{c}(\phi)=\frac12 (\phi, A_c\phi)$ and $\cS_{m}(\xi)=\frac12 (\phi, A_m\phi)$, respectively, for two 
$2LM\times 2LM$ anti-symmetric matrices $A_c=A_c(t_1,t_2)$ and $A_m=A_m(t_1,t_2)$. 
In terms of these matrices, \eqref{eq:Z_pfaffian_grassmann} can be rewritten as 
$$Z_\Lambda({\bf 0})\big|_{\lambda=0}=2^{LM} (\cosh \beta J_1)^{LM}(\cosh \beta J_2)^{L(M-1)}{\rm Pf}A_c\, {\rm Pf}A_m.$$ 
[We recall that the Pfaffian of a
$2n\times 2n$ antisymmetric matrix $A$ is defined as
\begin{equation}
{\rm Pf} A:=\frac1{2^n n!}\sum_\pi (-1)^\pi
A_{\pi(1),\pi(2)}...A_{\pi(2n-1),\pi(2n)}; \label{h1}
\end{equation}
where the sum is over permutations $\pi$ of $(1,\ldots,2n)$, with $(-1)^\pi$ denoting the signature.
One of the properties of the Pfaffian is that $({\rm Pf} A)^2={\rm det}A$.]
For later purpose, we also need to compute the averages of arbitrary monomials in the Grassmann variables $\phi_{\omega,z}$ and $\xi_{\omega,z}$, with $\omega\in\{+,-\}$
and $z\in\Lambda$. 
These can all be reduced to the computation of the inverses of $A_c$ and $A_m$, thanks to the `fermionic Wick rule': 
\begin{equation} \begin{split} & 
\int P_c(\mathcal D\phi) \,\phi_{\omega_1,z_1}\cdots \phi_{\omega_n,z_n}={\rm Pf}\,G_c\;,\\
& 
\int P_m(\mathcal D\xi) \,\xi_{\omega_1,z_1}\cdots \xi_{\omega_n,z_n} ={\rm Pf}\,G_m\;,\end{split}\label{2.2c}
\end{equation}
where $P_c(\cD\phi):=\cD\phi e^{\frac12(\phi,A_c\phi)}/{\rm Pf}(A_c)$, 
$P_m(\cD\xi):=\cD\xi e^{\frac12(\xi,A_m\xi)}/{\rm Pf}(A_m)$ are the Gaussian Grassmann integrations associated with $A_c$, $A_m$, respectively, and, 
if $n$ is even, $G_c$ and $G_m$ are the $n\times n$ matrices with entries 
\begin{equation}
  \label{eq:12} \begin{split}
[G_c]_{jk}&=\int P_c(\cD\phi)\,\phi_{\omega_j,z_j}\phi_{\omega_k,z_k}=-[A_c^{-1}]_{(\omega_j,z_j),(\omega_k,z_k)} \\
[G_m]_{jk}&=\int P_m(\cD\xi)\,\xi_{\omega_j,z_j}\xi_{\omega_k,z_k}=-[A_m^{-1}]_{(\omega_j,z_j),(\omega_k,z_k)}\end{split}
\end{equation}
(if $n$ is odd, the right sides of (\ref{2.2c}) should be interpreted as $0$). The two-point functions $\int P_c(\cD\phi)\phi_{\omega,z}\phi_{\omega',z'}$ and 
$\int P_m(\cD\xi)\xi_{\omega,z}\xi_{\omega',z'}$ are referred to as the {\it propagators} of the Grassmann fields $\phi$ and $\xi$, respectively. They have been explicitly 
computed via Fourier diagonalization of $A_c$ and $A_m$ in \cite[Section~2]{AGG_part2}. 
Let us summarize here the main properties of these propagators to be used in the remainder of this article. The massive and critical propagators are denoted 
\begin{equation}
  \label{eq:12bb} \begin{split}
\fg_m(z,z')&:=	\int P_m(\cD\xi)\begin{bmatrix}
		 \xi_{+,z} \xi_{+,z'}  &
		 \xi_{+,z} \xi_{-,z'}  \\
		 \xi_{-,z} \xi_{+,z'}  &
		 \xi_{-,z} \xi_{-,z'} \
		\end{bmatrix},\\
\fg_c(z, z')&:= \int P_c(\cD\phi)\begin{bmatrix}
		 \phi_{+,z} \phi_{+,z'}  &
		 \phi_{+,z} \phi_{-,z'}  \\
		 \phi_{-,z} \phi_{+,z'}  &
		 \phi_{-,z} \phi_{-,z'} 
		\end{bmatrix}
		\equiv \begin{bmatrix}
			g_{++}(z, z') &
			g_{+-}(z, z') \\
			g_{-+}(z, z') &
			g_{--}(z, z') 
		\end{bmatrix}. 		
		\end{split}
\end{equation}
The massive propagator, for all $t_1,t_2\in(0,1)$, is given by the explicit formula
\begin{equation}{\mathfrak g}_{m} (z, z') =\delta_{(z)_2,(z')_2} 
\begin{bmatrix} 0 & s_+((z-z')_1) \\
-s_-((z-z')_1) & 0\end{bmatrix}, \label{eq:propmassive}\end{equation}
where, for any $y\in\{1,\ldots,L\}$, $s_\pm(y):=\frac1{L}\sum_{k_1\in \mathcal D_L}\frac{e^{-ik_1y}}{1+t_1e^{\pm ik_1}}$. As for the critical propagator, we provide a detailed expression only on the critical line 
\begin{equation}
  t_1 t_2 + t_1 + t_2 = 1 \Leftrightarrow t_2= \frac{1-t_1}{1+t_1} \Leftrightarrow t_1= \frac{1-t_2}{1+t_2},
  \label{eq:anisotropic_critical_condition}
\end{equation}
which is the only case we need for the purposes of this paper: in this case, 
	\begin{equation}\begin{split}
		{\mathfrak g}_c (z, z') :=  &	\frac{1}{L}
		\sum_{k_1\in \mathcal D_L}
		\sum_{k_2 \in \mathcal{Q}_M(k_1)}
		\frac{1}{2N_M(k_1, k_2)}
		e^{-ik_1(z-z')_1}
		\\ & \times
		\left\{ e^{-ik_2(z-z')_2} \hat{\mathfrak{g}}(k_1, k_2)
		- e^{-ik_2(z+z')_2}
	        \begin{bmatrix}
		  \hat{g}_{++}(k_1, k_2) &
		 \hat{g}_{+-}(k_1,-k_2) \\
		  \hat{g}_{-+}(k_1, k_2) & e^{2ik_2(M+1)} 
		  \hat{g}_{--}(k_1, k_2)
		\end{bmatrix}
	      \right\}
	\end{split}
  \label{eq:g_finite_cyl}
\end{equation}
where $\cQ_M(k_1)$ is the set of solutions of {the following equation, thought of as an equation for $k_2$ at $k_1$ fixed}, in the interval $({-\pi,\pi})$:
\begin{equation}
\sin k_2(M+1)=B(k_1)\sin k_2M,
\label{eq:q_condition}
\end{equation}
with $B(k_1):= t_2\frac{|1+t_1 e^{ik_1}|^2}{1-t_1^2}$. Moreover, the normalization factor $N_M$ is defined as 
\begin{equation}
	\label{eq:N_M_def}
	N_M(k_1, k_2)=\frac{\frac{d}{dk_2}\left(B(k_1)\sin k_2M-\sin k_2(M+1) \right)}{B(k_1)\cos k_2M-\cos k_2(M+1)},
\end{equation}
and 
\begin{equation}
	\begin{split}
		\hat{\mathfrak g} (k_1, k_2)
		:=&
		\begin{bmatrix}
			\hat {g}_{++} (k_1, k_2) & 
			\hat {g}_{+-}(k_1, k_2) \\
			\hat {g}_{-+}(k_1, k_2) &
			\hat {g}_{--}(k_1, k_2) 
		\end{bmatrix}
		\\
		:= &
		\frac{1}{D(k_1, k_2)}
		 \begin{bmatrix}
			 {-}2 i t_1 \sin k_1 &
			- (1- t_1^2) (1 - B(k_1) e^{-ik_2}) \\
			 (1- t_1^2) (1 - B(k_1) e^{ik_2}) &
			{+} 2 i t_1 \sin k_1 
		\end{bmatrix},
		\label{eq:ghat}
	\end{split}
\end{equation}
with
\begin{equation}
D(k_1, k_2):= 2(1 - t_2)^2(1 - \cos k_1)+2(1 - t_1)^2(1 - \cos k_2). \label{defDk1k2}\end{equation}
Note that the symmetries detailed in \cref{rem:prop1} induce corresponding symmetry properties on the propagators.  In particular, for the critical 
propagator, these read 
\begin{equation}
\begin{split} & 		\hat g_{++} (k_1, k_2) = \hat g_{++}(k_1,-k_2) = - \hat g_{++}(-k_1, k_2) = \hat g_{--}(-k_1, k_2),\\
&		\hat g_{+-} (k_1, k_2) = \hat g_{+-} (-k_1, k_2) = - \hat g_{-+} (k_1,-k_2),\\
&		\hat g_{+ -}(k_1, k_2)
		=
		-
		e^{-2ik_2(M+1)} 
		\hat g_{- +} (k_1, k_2),\end{split}\label{symmpropa}
	\end{equation}
for all $k_1\in\mathcal D_L$ and $k_2 \in \cQ_M(k_1)$. 

\begin{remark} {\bf(Cancellation at the boundary)}\label{rem:cancell}	The definition~\eqref{eq:g_finite_cyl} can be extended to all $z, z' \in \bbR^2$;
in particular, using \eqref{symmpropa}, and letting $P_uz:=((z)_1,M+1)$ (resp. $P_dz:=((z)_1,0)$)
be the projection of $z$ on the row at height $M+1$ (resp. $0$), right above (resp. below) $\Lambda$ (the labels $u$ and $d$ stand for `up' and `down', respectively), 
	\begin{equation}\begin{split}
&		g_{++} \left( P_dz, z' \right)
=		g_{++} \left(z, P_dz'  \right)
		=
		g_{+-}  \left( P_dz,z' \right)
		=
		g_{-+} \left(z, P_dz' \right)
		=
		0,\\
&		g_{+-}\left(z, P_uz' \right)
		=
		g_{-+} \left( P_uz, z' \right)
		=
		g_{--}\left(  P_uz, z' \right)
		=
		g_{--} \left(z, P_uz' \right)
		=
		0
		,\end{split}
		\label{eq:g_cyl_boundary2}
	\end{equation}
for all $z, z'\in\mathbb R^2$. In the following, it will be convenient to introduce for any $z\in\Lambda$ four additional fictitious variables, $\phi_{\omega,P_{u}z}$ and $\phi_{\omega,P_{d}z}$, with $\omega\in\{+,-\}$,
to be understood as 	Grassmann fields that, if tested against another field $\phi_{\omega',z'}$ with respect to $P_c(\cD\phi)$, produce as output the propagators 
$[\fg_c(P_uz,z')]_{\omega,\omega'}$ and $[\fg_c(P_dz,z')]_{\omega,\omega'}$, respectively. In particular, thanks to \eqref{eq:g_cyl_boundary2}, 
the Grassmann fields $\phi_+$ located at $P_dz$ and the Grassmann fields $\phi_-$ located at $P_uz$ can be interpreted as being identically zero, 
and we shall do so in the following. Formally, the fact that $\phi_+$ and $\phi_-$ vanish at $P_dz$ and at $P_uz$, respectively, 
can be understood by noting that the system in the cylinder $\Lambda=\mathbb Z_L\times(\mathbb Z\cap [1,M])$ can be obtained from one in the larger domain 
$\mathbb Z_L\times(\mathbb Z\cap [0,M+1])$ 
by removing the vertical couplings connecting the two bottom-most and the two up-most rows;
equivalently, by setting the associated weights in the dimer representation to zero; or, again equivalently, by replacing the Grassmann variables associated with the `outside'
ends of these bonds (i.e., those at height $0$ and $M+1$, respectively) with zero.
For the setup we consider in this paper,
these are precisely the variables $\phi_+$ at $P_dz$ (also called $\overline V$ in the `standard' notation where the four types of Grassmann variables are denoted $\overline H, H, \overline V, V$, see \cite[Eq.(2.1.10)]{AGG_part2}) and $\phi_-$ at $P_uz$ (also called $V$, see \cite[Eq.(2.1.10)]{AGG_part2}).
\end{remark}

\begin{remark}\label{rem:scaling}{\bf(Scaling limit)}
The scaling limit propagator in \eqref{scalinglim.free.2} is defined as
\begin{equation}\fg_\scal(z, z'):=\lim_{a\to 0}a^{-1}\fg_c(\lfloor a^{-1} z\rfloor, \lfloor a^{-1} z'\rfloor),\label{eq:g_cyl_scaling_finite_a}
\end{equation}
and is given explicitly by the following expression: 
\begin{eqnarray} 
  \label{eq:g_scal_cyl_def}
  &&
\fg_\scal(z, z')=\sum_{\bs n\in\mathbb Z^2}(-1)^{\bs n} \Biggl\{ \fg^{\scal}_\infty(z-z'+(n_1\ell_1,2n_2\ell_2))\\
&&\hskip-.5truecm+
\begin{bmatrix} 
-g^{\scal}_{1}((z-z')_1+n_1\ell_1,(z+z')_2+2n_2\ell_2) & g^{\scal}_{2}((z-z')_1+n_1\ell_1,(z+z')_2+2n_2\ell_2) \\
-g^{\scal}_{2}((z-z')_1+n_1\ell_1,(z+z')_2+2n_2\ell_2) & g^{\scal}_{1}((z-z')_1+n_1\ell_1,(z+z')_2+2(n_2-1)\ell_2) \end{bmatrix}
\Biggr\},\nonumber\end{eqnarray}
where, letting, for $z=((z)_1,(z)_2)$, 
\begin{equation}
g^{\scal}(z):=\frac{1}{t_2(1-t_2)} \iint_{\mathbb R^2}\frac{\dd k_1\dd k_2}{(2\pi)^2} e^{-ik_1(z)_1-ik_2(z)_2}\frac{-i k_1}{k_1^2+k_2^2}
=-\frac1{{2\pi} t_2(1-t_2)}\frac{(z)_1}{\|z\|_2^2},
\label{eq:g_scale_scalar}
\end{equation}
we denoted $g_1^{\scal}(z):=g^{\scal}(\frac{(z)_1}{1-t_2},\frac{(z)_2}{1-t_1})$, $g_2^{\scal}(z):=g^{\scal}(\frac{(z)_2}{1-t_1},\frac{(z)_1}{1-t_2})$, and
\begin{equation}
  \fg_\infty^\scal(z)
    :=
	\begin{bmatrix}
    g_1^\scal (z) & g_2^\scal (z) \\
    g_2^\scal (z) & - g_1^\scal (z)
	\end{bmatrix}.
\end{equation}
For the computation leading to these explicit formulas and for constructive estimates on the 
speed of convergence of the right side of \eqref{eq:g_cyl_scaling_finite_a} to the limit, 
see \cite[Section~2.3 and Appendix~C]{AGG_part2}. 
\end{remark}

\subsection{The interacting theory in the cylinder}\label{sec:2.2}

Let $\mathcal O=\{+,-,i,-i\}$ and $\cM_{1,\Lambda}$ be the set %
of the tuples $\Psi=((\omega_1,z_1),\ldots,(\omega_n,z_n))\in (\cO\times\Lambda)^n$
for some $n\in 2\mathbb N$. Given a tuple $\Psi=((\omega_1,z_1),\ldots,(\omega_n,z_n))\in \cM_{1,\Lambda}$, 
we let $\phi(\Psi)$ be the following Grassmann monomial:
\begin{equation}\label{eq:2.2.1}
\phi(\Psi):=\phi_{\omega_1,z_1}\cdots \phi_{\omega_n,z_n},
\end{equation}
where $\phi_{\omega,z}$ denotes
$\phi_{+,z},\phi_{-,z},\xi_{+,z},\xi_{-,z}$ for $\omega=+,-,i,-i$, respectively. 
In the following, with some abuse of notation, any element 
$\Psi\in{\cM_{1,\Lambda}}$ of length $|\Psi|=n$ will be denoted indistinctly by $\Psi=((\omega_1,z_1),\ldots,(\omega_n,z_n))$ 
or $\Psi=(\bs \omega, \bs z)$, with the understanding that $\bs \omega=(\omega_1,\ldots,\omega_n)$ and $\bs z=(z_1,\ldots, z_n)$. 

Moreover, we let $\cX_\Lambda$ be the set of the tuples $\bs x=(x_1,\ldots,x_m)\in\fB_\Lambda^m$ for some $m\in\mathbb N_0$, with the 
understanding that, if $m=0$, then $\bs x$ is the empty set. Given a tuple $\bs x=(x_1,\ldots,x_m)\in\cX_\Lambda$, we let $\#\bs x:=m$ be its length, and 
$\bs A(\bs x):=\prod_{i=1}^mA_{x_i}$, with the understanding that, if $m=0$, then $\bs A(\emptyset)=1$. 

Given these definitions, we are ready to state the desired representation theorem for the (Taylor coefficients of the) generating function $Z_\Lambda(\bs A)$ in \eqref{eq:Ising_gen},
see \cite[Section~3]{AGG_part2} for the proof.

\begin{proposition}\label{prop:repr}
	For any translationally invariant, even, interaction $V$ of finite range, there exists $\lambda_0=\lambda_0 (V)$ 
	such that, for any $|\lambda| \le \lambda_0 (V)$,  the derivatives of $\log Z_\Lambda({\bs A})$ of order $2$ or more, with no repetitions, computed at ${\bs A}={\bs 0}$, are the same as those of $\log\Xi_\Lambda(\bs A)$, with
\begin{equation} 
	\Xi_\Lambda(\bs A)= 
	e^{\cW(\bs A)}
	\int  P_{c} (\cD\phi) P_{m} (\cD\xi) 
	\,
	e^{\cV(\phi,\xi,\bs A)
	}
	,	
	\label{eq:ZA_rescaled_tris}
\end{equation}
where: \begin{enumerate}
\item 
\begin{equation}
			\mathcal W(\bs A)=\sum_{\substack{\bs x\in\cX_\Lambda:\\ \#\bs x\ge 2\,\phantom{.}}}w_\Lambda(\bs x) \bs A(\bs x)	\label{eq:B_expansion_tris}
		\end{equation}
for suitable real functions $w_\Lambda$ such that, for any $m\ge 2$, and suitable positive constants $C,c_0,\kappa$\footnote{The factor $2$ in front of $c_0$ at exponent is 
chosen for convenience and uniformity with later notations: in fact, the bare Grassmann action will be identified with the bare effective potential on scale 
$h=1$, see \eqref{eq:startfrom}, \eqref{eq:Vcyl_N}, and following equations; the factor $2$ at exponent should be interpreted as being equal to $2^h$, with scale index $h=1$.},
		\begin{equation}
			\sup_{x_1\in\fB_\Lambda}\sum_{x_2,\ldots,x_m\in \fB_\Lambda}|w_\Lambda(\bs x)| e^{2c_0 \delta(\bs x)}\le
			C^{m}|\lambda|^{\max (1,\kappa m)},
			\label{eq:B_base_decay2}
		\end{equation}
where $\bs x=(x_1,\ldots,x_m)$ and $\delta(\bs x)$ is the tree distance of $\bs x$, that is, the cardinality of the smallest connected subset of $\fB_\Lambda$ containing the 
elements of $\bs x$.
		\label{it:B_cyl_base2}
\item  
\begin{equation}
	\cV (\phi,\xi,\bs A)= \cB^{\free}(\phi,\xi,\bs A)+ \cV^{\rm int}(\phi,\xi,\bs A)\label{expanV1.0}\end{equation}
where
\begin{equation} \begin{split} &\cB^{\free}(\phi,\xi,\bs A):=\sum_{x\in\fB_\Lambda} (1-t_{j(x)}^2) E_x A_x\\
& \cV^{\rm int}(\phi,\xi,\bs A):=\sum_{\Psi\in\cM_{1,\Lambda}}\sum_{\bs x\in\cX_\Lambda}
	W^{\rm int}_\Lambda(\Psi,\bs x) \phi(\Psi) \bs A(\bs x),\end{split}
	\label{expanV1}\end{equation}
	and:
\begin{itemize}
\item in the first line of \eqref{expanV1}, $j(x)=1$ (resp. $=2$) if $x$ is horizontal (resp. vertical), and: if $x$ is a horizontal edge 
with endpoints $z,z+\hat e_1$, then $E_x:=H_{+,z} H_{-,z+\hat e_1}$, with (recall that $s_\omega$ was defined after \eqref{eq:propmassive})
\begin{equation} H_{\omega,z}:= \xi_{\omega,z}+ 
\sum_{y=1}^L s_\omega((z)_1-y) \big(\phi_{+,(y,(z)_2)}-\omega \phi_{-,(y,(z)_2)}\big),\label{eq:sc_2}
\end{equation}
while, if $x$ is a vertical edge $x$ with endpoints $z,z+\hat e_2$, then $E_x:=\phi_{+,z}\phi_{-,z+\hat e_2}$;
\item in the second line of \eqref{expanV1}, $W_\Lambda^{\rm int}$ are suitable real functions such that, 
for any $n\in 2\mathbb N$, $m\in\mathbb N$ and the same $C,c_0,\kappa$ as in item \ref{it:B_cyl_base2}, 
\begin{equation}\begin{split}
&\sup_{\bs\omega\in\cO^n}\sup_{z_1\in\Lambda}\sum_{z_2,\ldots, z_{n} \in \Lambda}e^{2c_0 \delta(\bs z)}
\big|W_\Lambda^{{\rm int}}((\bs\omega,\bs z),\emptyset)\big|\le C^{n}|\lambda|^{\max (1, \kappa n)},\\
&\sup_{\bs\omega\in\cO^n}\sup_{x_1\in\fB_\Lambda}\sum_{x_2,\ldots, x_{m} \in \fB_\Lambda}\sum_{\bs z\in\Lambda^n}e^{2c_0\delta(\bs z,\bs x)}
\big|W_\Lambda^{{\rm int}}((\bs\omega,\bs z),\bs x)\big|\le C^{n+m}|\lambda|^{\max (1, \kappa(n+m))},\end{split}
	\label{eq:WL1bound}
\end{equation}
where, in the first line, $\bs z=(z_1,\ldots,z_n)$ and, in the second line, $\bs x=(x_1,\ldots,x_m)$; moreover,  
$\delta(\bs z,\bs x)$ is the tree distance of $(\bs z,\bs x)$, that is, the cardinality of the smallest connected subset of $\fB_\Lambda$ that 
includes $\bs x$ and touches the points in $\bs z$ (we say that an edge $x\in\fB_\Lambda$ touches a vertex $z\in\Lambda$, if $z$ is one of the endpoints of $x$), and $\delta(\bs z)\equiv \delta(\bs z,\emptyset)$; 
\end{itemize}
\item $w_\Lambda, W^{\rm int}_\Lambda$, 
considered as functions of $\lambda$, $t_1$, and $t_2$, can be analytically continued to any complex $\lambda, t_1, t_2$ such that $|\lambda| \le \lambda_0$ and $|t_1|, |t_2|\in K'$, with 
$K'$ the same compact set introduced before the statement of Theorem \ref{prop:main}, and the analytic continuation satisfies the same bounds above.
\end{enumerate}
\end{proposition}

A few remarks are in order. First of all, for uniformity and compactness of notation, we rewrite \eqref{expanV1.0}-\eqref{expanV1} as: 
\begin{equation} \cV(\phi,\xi,\bs A)=\sum_{\Psi\in\cM_{1,\Lambda}}\sum_{\bs x\in\cX_\Lambda}
W_\Lambda(\Psi,\bs x) \phi(\Psi) \bs A(\bs x),\label{eq:2.2.8}\end{equation}
where $W_\Lambda=B_\Lambda^{\rm free}+W_\Lambda^{\rm int}$, with $B_\Lambda^{\rm free}$ the kernel associated with the first term in the right side of \eqref{expanV1} and 
invariant under the symmetries described momentarily, which is supported on tuples $(\Psi, \bs x)$ such that $|\Psi|=2$ and $\#\bs x=1$, and satisfies
\begin{equation}
\sup_{\bs\omega\in\cO^2}\sup_{x\in\fB_\Lambda}\sum_{\bs z \in \Lambda^2}e^{2c_0 \delta(\bs z,x)}
\big|B_\Lambda^{{\rm free}}((\bs\omega,\bs z),x)\big|\le C.
	\label{eq:WfreeL1bound}
\end{equation}
The functions $w_\Lambda$ and $W_\Lambda$ satisfy the following properties, which will play an important role in the multiscale computation of the generating function. 

\medskip

\begin{enumerate}
\item {\it Symmetries.}\label{it1:dop} From the proof of Proposition \ref{prop:repr} in \cite[Section~3]{AGG_part2}, it follows that the functions $w_\Lambda(\bs x)$ can be chosen to be:
symmetric under permutations of $\bs x$, translationally invariant in the horizontal direction (with periodic boundary conditions in the horizontal direction), 
and invariant under the reflection symmetries induced by the transformations $A_x\to A_{\theta_l x}$, with $l=1,2$, where $\theta_1$, $\theta_2$ act on a midpoint of an edge $x=((x)_1,(x)_2)\in\fB_\Lambda$ as: 
$\theta_1x=(L+1-(x)_1,(x)_2)$, $\theta_2x=((x)_1,M+1-(x)_2)$. Similarly, $W_\Lambda((\bs\omega,\bs z),\bs x)$ can be chosen to be:
anti-symmetric under simultaneous permutations of $\bs \omega$ and $\bs z$, 
symmetric under permutations of $\bs x$, invariant under simultaneous translations of $\bs z$ and $\bs x$ in the horizontal direction (with anti-periodic and periodic boundary conditions in $\bs z$ and $\bs x$, 
respectively), invariant under the reflection symmetries induced by the transformations $A_x\to A_{\theta_l x}$ and $\phi_{\omega,z}\to \Theta_l\phi_{\omega,z}$, see \eqref{eq:symm_horizontal_phixi}-\eqref{eq:symm_vertical_phixi}.
Therefore, with no loss of generality, we assume that $w_\Lambda,W^{\rm int}_\Lambda$ are invariant under these symmetries. 

\item {\it Infinite volume limit.} \label{it2:dop}
From the discussion in \cite[Section~3]{AGG_part2}, it also follows that $w_\Lambda, W_\Lambda$ admit an infinite volume limit, in the following sense. 
Let $\Lambda_\infty:=\mathbb Z^2$ and let $\fB_\infty$ be the set of nearest neighbor edges of $\mathbb Z^2$. Then, for any fixed 
tuple $(\bs z,\bs x)\in \Lambda_\infty^n\times \fB_\infty^m$, with $n\in2\mathbb N$ and $m\in\mathbb N_0$, we let 
\begin{equation}\label{eq:WZ2} W_{\infty}((\bs\omega,\bs z),\bs x):=\lim_{L,M\to\infty}W_\Lambda((\bs \omega,\bs z+z_{L,M}),\bs x+z_{L,M})
,\end{equation} where $z_{L,M}=(L/2,\lfloor M/2\rfloor)$, $\bs z+z_{L,M}$ is the translate of $\bs z_\infty$ by $z_{L,M}$, and similarly for $\bs x+z_{L,M}$. The limiting 
kernel $W_\infty$, besides being anti-symmetric under simultaneous permutations of $\bs \omega$ and $\bs z$, and 
symmetric under permutations of $\bs x$, it is translationally invariant {\it in both coordinate directions}, and invariant under the following infinite plane reflection symmetries:
$A_x\to A_{(-(x)_1,(x)_2)}$, $\phi_{\pm,z}\to \pm i\phi_{\pm,(-(z)_1,(z)_2)}$, $\phi_{\pm i,z}\to i\phi_{\mp i,(-(z)_1,(z)_2)}$ ({\it horizontal reflection}), 
and 
$A_x\to A_{((x)_1,-(x)_2)}$, $\phi_{\pm,z}\to i\phi_{\mp,((z)_1,-(z)_2)}$, $\phi_{\pm i,z}\to \mp i\phi_{\pm i,((z)_1,-(z)_2)}$ ({\it vertical reflection}).
Moreover, the decomposition 
$W_\Lambda=B^{\rm free}_\Lambda+W^{\rm int}_\Lambda$ induces an analogous decomposition for $W_{\infty}$. Of course, $B^{\rm free}_\infty$ and $W^{\rm int}_\infty$ 
admit the same bounds \eqref{eq:WfreeL1bound} and \eqref{eq:WL1bound} as $B_\Lambda^{\rm free}$ and $W_\Lambda^{\rm int}$, respectively, and are invariant under the same symmetries as $W_\infty$. 
Similar considerations hold for $w_\Lambda$, whose infinite volume limit is denoted by $w_\infty$. 
\item {\it Bulk-edge decomposition.} \label{it3:dop}
Not only does the infinite volume limit of the kernels exist, but it is reached exponentially fast. This allows us to conveniently decompose the finite volume kernels into a 
bulk plus an edge part, with the edge part decaying exponentially fast to zero away from the 
boundary of the cylinder. For this purpose, note that any $n$-tuple $\bs z\in\Lambda^n$
with horizontal diameter smaller than $L/2$ can be identified (non uniquely, of course) with an $n$-tuple of points of $\mathbb Z^2$ with the same diameter and `shape' as $\bs z$; we let $\bs z_\infty\in (\mathbb Z^2)^n$ be one of these representatives, chosen arbitrarily\footnote{For instance, 
given $\bs z=(z_1,\ldots, z_n)\in \Lambda^n$, recalling that $(z_i)_1\in\{1,2,\ldots, L\}$ and $(z_i)_2\in\{1,2,\ldots, M\}$, 
we can let $\bs z_\infty=(y_1,\ldots,y_n)$ be the $n$-tuple of points of $\mathbb Z^2$ such that: (1) the vertical coordinates are the same as those of $\bs z$, i.e., $(y_i)_2=(z_i)_2$, $\forall i=1,\ldots, n$; (2)
the horizontal coordinate of $y_1$ is the same as $z_1$, i.e., $(y_1)_1=(z_1)_1$; (3) all the other horizontal coordinates are the same modulo $L$, i.e., $(y_i)_1=(z_i)_1$ mod $L$, $\forall i=2,\ldots, n$; (4) 
the specific values of $(y_i)_2$ for $i\ge 2$ are chosen in such a way that the horizontal distances between the corresponding pairs in $\bs z$ and $\bs z_\infty$ 
are the same, if measured on the cylinder $\Lambda$ or on $\mathbb Z^2$, respectively.}; we use an analogous convention for the elements of $\fB_\Lambda^m$ and of $\Lambda^n\times\fB_\Lambda^m$ 
with horizontal diameter smaller than $L/2$ (with some abuse of notation, given $(\bs z,\bs x)\in \Lambda^n\times\fB_\Lambda^m$ with horizontal diameter smaller than $L/2$, 
we denote by $(\bs z_\infty,\bs x_\infty)\equiv (\bs z,\bs x)_\infty$ its infinite volume 
representative). Given these definitions, we write $W_\Lambda=W_\B+W_\E$, with  
\begin{equation} W_\B((\bs\omega,\bs z),\bs x) :=(-1)^{\alpha(\bs z)}\,\mathds 1(\diam_1(\bs z,\bs x)\le L/3)\,W_\infty((\bs\omega,\bs z_\infty),\bs x_\infty), \label{WB+WE}
\end{equation}
and $W_\E((\bs\omega,\bs z),\bs x)):=W_\Lambda((\bs\omega,\bs z),\bs x))-W_\B((\bs\omega,\bs z),\bs x))$, 
where $\mathds 1(A)$ is the characteristic function of $A$, $\diam_1$ is the horizontal diameter on the cylinder and, for any $\bs z$ with $\diam_1(\bs z)\le L/3$, 
\begin{equation}
	\alpha(\bs z)= 	\begin{cases} \#\{z_i\in\bs z\ : \ (z_i)_1\le L/3\}, & \text{if}\quad \max_{z_i,z_j\in\bs z}\{(z_i)_1-(z_j)_1\}\ge 2L/3,\\
	0 \,, & \text{otherwise.} \end{cases}\label{eq:alphaz}
\end{equation}
Note that the factor $(-1)^{\alpha(\bs z)}$ guarantees that both $W_\B$ and $W_\E$ are invariant under simultaneous translations of $\bs z$ and $\bs x$ in the horizontal direction, with anti-periodic and periodic boundary conditions in $\bs z$ and $\bs x$, respectively. In addition to this, both kernels are 
invariant under the same reflection symmetries as $W_\Lambda$. 
Of course, in analogy with $W_\Lambda$ and $W_\infty$, $W_\B$ and $W_\E$ can also be decomposed as $W_\B=B^{\rm free}_\B+W^{\rm int}_\B$
and $W_\E=B^{\rm free}_\E+W^{\rm int}_\E$. While $B_\B^{\rm free}$ and $W_\B^{\rm int}$ satisfy the same  bounds \eqref{eq:WfreeL1bound} and \eqref{eq:WL1bound} as $B_\Lambda^{\rm free}$ and $W_\Lambda^{\rm int}$, respectively, 
$B_\E^{\rm free}$ and $W_\E^{\rm int}$ satisfy the following improved bounds: 
		\begin{equation}
			\frac1{L}\sup_{\bs \omega\in\mathcal O^2}\sum_{\substack{\bs z\in\Lambda^2\\ x\in\fB_\Lambda}} |B_{\E}^{\rm free}((\bs\omega,\bs z),x)| e^{2c_0\delta_\E(\bs z,x)}\le C, 
			\label{eq:WfreeE_base_decay}
		\end{equation}
and, for any $n\in\mathbb N$ and $m\in\mathbb N_0$, 
		\begin{equation}
			\frac1{L}\sup_{\bs \omega\in\mathcal O^n}\sum_{\substack{\bs z\in\Lambda^n\\ \bs x\in\fB_\Lambda^m}} |W_{\E}^{\rm int}((\bs\omega,\bs z),\bs x)| e^{2c_0\delta_\E(\bs z,\bs x)}\le C^{m+n} |\lambda|^{\max (1,\kappa(m+n))}, 
			\label{eq:WE_base_decay}
		\end{equation}
with the same $C,c_0,\kappa$ as in item \ref{it:B_cyl_base2} of Proposition \ref{prop:repr}, where $\delta_\E(\bs z,\bs x)$ is the `edge' tree distance of $(\bs z,\bs x)$,
\begin{equation}
	\delta_\E(\bs z, \bs x ) = \min_{T \in \fT_\E (\bs z, \bs x)}|T|
	\label{eq:delta_E_zx_def}
\end{equation}
where $\fT_\E(\bs z, \bs x)$ is the set of connected $T \subset \fB_\Lambda$ which include $\bs x$, touch all the points in $\bs z$, and also have at least one of the following properties:
\begin{enumerate}
	\item $T$ touches the boundary of the cylinder or
	\item $T$ includes two bonds $x$ and $y$ whose horizontal coordinates differ by more than $L/3$, taking periodicity into account, i.e.\ $\diam_1(T) > L/3$.
\end{enumerate}
For the proof of \cref{eq:WfreeE_base_decay,eq:WE_base_decay}, see 
\cite[Lemma~3.2 and Eqs.~(3.20)--(3.21)]{AGG_part2}
\end{enumerate}

Before we start the multiscale computation, let us make a final rewriting of the Grassmann 
generating function. It is expected (and it will be proved below) that the interaction has, among others, the 
effect of modifying (`renormalizing') the large distance behavior of the bare propagator, 
by effectively rescaling it by $1/Z$ (where $Z$ plays the role of the `wave function renormalization', in a QFT analogy), and by changing 
the bare couplings $t_1,t_2$ into dressed ones, $t_1^*,t_2^*$. In order to take this effect into account, it is convenient to write the Grassmann generating function 
in terms of a reference Gaussian Grassmann integration with dressed parameters. Since, of course, these dressed parameters $Z,t_1^*,t_2^*$ are 
a priori unknown, they will be fixed a posteriori via a self-consistent equation, whose solution requires the use of the implicit function theorem, see Remark \ref{remacon} below and the related discussion in 
\cite[Section~4.5]{AGG_part2}. For the time being, we let $t_1^*$ be any element of the compact set $K'$ defined before the statement of Theorem \ref{prop:main}, and $t_2^*:=(1-t_1^*)/(1+t_1^*)$. 

Motivated by the previous considerations, we rewrite the Grassmann generating function $\Xi(\bs A)$ in \eqref{eq:ZA_rescaled_tris} as follows: recalling that $P_c(\cD\phi)=\cD\phi e^{\cS_c(\phi)}/{\rm Pf}A_c$, with 
$\cS_c(\phi)\equiv \frac12(\phi,A_c\phi)$ as in \eqref{eq:2.1.13} 
we add and subtract at exponent the quantity $Z(\cS^*_c(\phi)+\cS^*_m(\xi))$, with $\cS^*_c(\phi)=\frac12(\phi,A_c^*\phi)$ and $\cS^*_m(\xi)=\frac12(\xi,A_m^*\xi)$, the same as in \eqref{eq:2.1.13} and \eqref{eq:2.1.12},
respectively, computed at $t_1^*,t_2^*$ instead of $t_1,t_2$; next we rescale the Grassmann fields as $\phi\to Z^{-1/2}\phi$, $\xi\to Z^{-1/2}\xi$, 
thus getting 
\begin{equation} \Xi_\Lambda(\bs A)\propto
	e^{\cW(\bs A)}	\int  P^*_{c} (\cD\phi) P_{m}^* (\cD\xi) \, e^{\cV^{(1)}(\phi,\xi,\bs A)	}, \label{eq:startfrom}\end{equation}
where: $\propto$  indicates that the ratio of the two expressions is independent of $\bs A$, so that they are generating functions of the same correlations;
$P_c^*(\cD\phi)=\cD\phi\, e^{\cS_c^*(\phi)}/{\rm Pf}A^*_c$; $P_m^*(\cD\xi)=\cD\xi\, e^{\cS_m^*(\xi)}/{\rm Pf}A^*_m$; 
\begin{equation} \cV^{(1)}(\phi,\xi,\bs A):=Z^{-1}(\cS_c(\phi)+\cS_m(\xi))-(\cS^*_c(\phi)+\cS^*_m(\xi))+\cV(Z^{-1/2}\phi,Z^{-1/2}\xi,\bs A).\label{cV1defin}\end{equation}
For later reference, we note that, in light of \eqref{eq:2.2.8}, $\cV^{(1)}(\phi,\xi,\bs A)$ can be written as:
\begin{equation}\cV^{(1)}(\phi,\xi,\bs A)= \sum_{\Psi\in \cM_{1,\Lambda}}\sum_{\bs x\in\cX_\Lambda}
W_\Lambda^{(1)}(\Psi,\bs x) \phi(\Psi) \bs A(\bs x)\label{eq:forma1}\end{equation}
where $W^{(1)}_\Lambda$ is analytic in $\lambda$, in the sense of item 3 of Proposition \ref{prop:repr}, it has the symmetry properties described in item \ref{it1:dop} after \eqref{eq:WfreeL1bound}, it admits an infinite volume limit, in the sense of 
item \ref{it2:dop} after \eqref{eq:WfreeL1bound}, denoted by $W_\infty^{(1)}$ (which is also analytic in $\lambda$), and admits a bulk-edge decomposition, in the sense of item \ref{it3:dop} after \eqref{eq:WfreeL1bound}. 
Eqs.\eqref{eq:startfrom}--\eqref{eq:forma1} will be the starting point of the multiscale analysis discussed in the next section. 

\section{The renormalized expansion}\label{sec:renexp}

In  this section, we will show that for every $J_1/J_2, L/M\in K$, $t_1,t_2\in K'$, with $K,K'$ the compact sets introduced before the statement of Theorem \ref{prop:main}, 
$|\lambda|$ sufficiently small, and an appropriate choice of $t_1^*$, $Z$, the derivatives of $\Xi_\Lambda(\bs A)$ of order $m\ge 2$, 
with no repetitions, at $\bs A=\bs 0$ admit an 
expansion as a uniformly convergent sum. Such an expansion is based on the following iterative evaluation of $\Xi_\Lambda(\bs A)$: 
starting from \eqref{eq:startfrom}, we first define
\begin{equation}
    e^{\cW^{(0)}(\bs A)+\cV^{(0)}(\phi,\bs A)}\propto 
 \int P^*_{m} (\cD\xi) \, e^{\cV^{(1)}(\phi,\xi,\bs A)},   
	\label{eq:Vcyl_N}
\end{equation}
where once again $\propto$ means `up to a multiplicative constant independent of $\bs A$', and 
$\cW^{(0)},\cV^{(0)}$ are normalized in such a way that $\cW^{(0)}(\bs 0)=\cV^{(0)}(0,\bs A)=0$. 
The function $\cW^{(0)}$ is the $h=0$ contribution to the generating function, and $\cV^{(0)}$ the effective potential on scale $0$. 

Next, we need to compute the integral $\int P_c^*(\mathcal D\phi) e^{\cV^{(0)}(\phi,\bs A)}$, where $P_c^*(\mathcal D\phi)$ is the 
Gaussian Grassmann integration with propagator $\fg_c^*$, whose explicit expression is 
given by \eqref{eq:g_finite_cyl} and following equations, computed at $t_1^*,t_2^*$ rather than at $t_1,t_2$. In order to obtain a convergent expansion for the logarithm 
of $\int P_c^*(\mathcal D\phi) e^{\cV^{(0)}(\phi,\bs A)}$, we use a multiscale procedure, based on the multiscale decomposition of the critical 
propagator discussed in \cite[Section~2.2]{AGG_part2}, whose most relevant features are recalled here. 

\medskip

{\it Multiscale decomposition of the critical propagator.} In view of \cite[eq.~(2.2.3)]{AGG_part2}, letting $h^*:=-{\lfloor}\log_2(\min\{L,M\}){\rfloor}$, 
we write 
\begin{equation} \fg_c^*(z,z')=\mathfrak g^{(\le h^*)}(z, z')+\sum_{h=h^*+1}^0\mathfrak g^{(h)}(z, z'),\label{gch*h}\end{equation}
where the {\it single-scale propagator} $\fg^{(h)}$ satisfies the same cancellation property at the boundary as \eqref{eq:g_cyl_boundary2} (see \cite[Eq.~(2.2.8)]{AGG_part2}), as well as the dimensional bounds stated in \cite[Proposition~2.3]{AGG_part2}, which 
we summarize here for the reader's convenience: 
for any $\bs r=(r_{1,1}, r_{1,2}, r_{2,1}, r_{2,2})\in\mathbb Z_+^4$, letting $\partial_{1,j}$ (resp. $\partial_{2,j}$) be the discrete derivative in direction $j$ with respect to the first (second) argument, defined by $\partial_{1,j}f(z,z'):=f(z + \hat{e}_j, z') - f(z, z')$, with $\hat{e}_j$ the $j$-th Euclidean basis vector, we have 
\begin{equation}\| \bs\partial^{\bs r}{\mathfrak g}^{(h)} (z, z')\|\le C^{1+|\bs r|_1}\bs r!\,2^{(1+|\bs r|_1)h} e^{-c_0 2^h \| z- z'\|_1}
\label{ghbounds1}\end{equation}
where $\bs\partial^{\bs r}=\prod_{i,j=1}^2 \partial_{i,j}^{r_{i,j}}$, $\bs r!=\prod_{i,j=1}^2 r_{i,j}!$, and $\|z\|_1=|\per_L(z_1)|+|z_2|$, with $\per_L (z_1)=z_1- L 
\floor{\frac{z_1}{L} + \frac12}$. Moreover, there exists a Hilbert space $\cH_{LM}$ with inner product $(\cdot,\cdot)$ including elements 
$\gamma^{(h)}_{\omega,\bs s, z}$, $\tilde{\gamma}^{(h)}_{\omega, \bs s, z}$, $\gamma^{(\le h)}_{\omega,\bs s, z}$, $\tilde{\gamma}^{(\le h)}_{\omega, \bs s, z}$  
(for $\bs s=(s_1,s_2)\in\mathbb Z_+^2$, $z\in\Lambda$) such that, whenever $h^*< h \le 0$, the following {\it Gram representation} holds for the elements 
$g_{\omega\omega'}^{(h)}(z,z')$ of $\fg^{(h)}(z,z')$, with $\omega,\omega'\in\{+,-\}$, and their derivatives:
\begin{equation}\bs \partial^{(\bs s,\bs s')}g_{\omega\omega'}^{(h)} (z, z') \equiv \left( \tilde \gamma^{(h)}_{\omega,\bs s, z},\gamma^{(h)}_{\omega', \bs s', z'} \right),
\label{ddtthh} \end{equation}
where $\partial^{(\bs s, \bs s')}:= \prod_{j=1}^2 \partial_{1,j}^{s_j} \partial_{2,j}^{s'_j}$ and, letting $|\cdot|$ be the norm generated by the inner product $(\cdot,\cdot)$, $\max\{\big|\gamma^{(h)}_{\omega,\bs s, z}\big|^2,\big|\tilde \gamma^{(h)}_{\omega, \bs s, z}\big|^2\} \le C^{1+|\bs s|_1}\bs s!\, 2^{h(1+2|\bs s|_1)}$. Similarly, $\bs \partial^{(\bs s,\bs s')}g_{\omega\omega'}^{(\le h^*)} (z, z') \equiv \big( \tilde \gamma^{(\le h^*)}_{\omega,\bs s, z},\gamma^{(\le h^*)}_{\omega', \bs s', z'} \big)$, with $\max\{\big|\gamma^{(\le h)}_{\omega,\bs s, z}\big|^2,\big|\tilde \gamma^{(\le h)}_{\omega, \bs s, z}\big|^2\}$ $\le C^{1+|\bs s|_1}\bs s!\, 2^{h^*(1+2|\bs s|_1)}$, so that, in particular, $\| \bs\partial^{\bs r}  {\mathfrak g}^{(\le h^*)} (z, z')\|\le C^{1+|\bs r|_1} \bs r! \, 2^{(1+|\bs r|_1)h^*}$ (we recall that $2^{h^*}\simeq L^{-1}$ by the very definition of $h^*$). We also recall that, by denoting $\fg_m^*\equiv \fg^{(1)}$ the massive propagator, i.e., the one associated with 
$P^*_m(\cD\xi)$, $\fg^{(1)}$ satisfies the same estimates and admits the same Gram decomposition spelled above for $\fg^{(h)}$, with the scale label $h$ replaced by the value $1$. 

Another important ingredient of the multiscale construction described in the following sections is the bulk-edge decomposition of the propagator, 
summarized here: for $h\le 1$, let $\fg_\infty^{(h)}(z-z')$ 
be the infinite volume limit of $\fg^{(h)}(z,z')$, in a sense analogous to \eqref{eq:WZ2}, that is, $\fg_\infty^{(h)}(z-z')=\lim_{L,M\to\infty}\fg^{(h)}(z+z_{L,M},y+z_{L,M})$, see also \cite[Eq.~(2.2.9)]{AGG_part2}. 
Of course, $\fg_\infty^{(h)}(x-y)$ satisfies the same bounds \eqref{ghbounds1} as $\fg^{(h)}(z,z')$, and admits an analogous Gram representation. Given this, 
we let the bulk part of $\fg^{(h)}$ be defined as the restriction of $\mathfrak{g}_\infty^{(h)}$ to the cylinder, with the appropriate (anti-periodic) boundary conditions in the horizontal direction: 
\begin{equation} \fg^{(h)}_{\B} (z, z'):=s_L \left( (z-z')_1 \right)\fg_\infty^{(h)}\left( \per_L\left( (z-z')_1\right), (z-z')_2\right)\end{equation}
where $\per_L$ was defined after \eqref{ghbounds1} and, recalling that $(z)_1,(z')_1\in\{1,\ldots, L\}$,  
\begin{equation}{s_L((z-z')_1):=\piecewise{+1, & |(z-z')_1|<L/2\\
\phantom{+}0, & |(z-z')_1|=L/2 \\ -1, & |(z-z')_1|>L/2}}\end{equation}
The edge part is, by definition, the difference between the full cutoff propagator and its bulk part: 
\begin{equation} \fg^{(h)}_{\E} (z, z') 
	:=\fg^{(h)} (z, z') -\fg^{(h)}_{\B} (z,z'),\label{cosacome}\end{equation}
and satisfies an improved dimensional bound, as compared with \eqref{ghbounds1}; namely, if $z, z'\in\Lambda$ are such that $|\per_L((z-z')_1)|< L/2-|r_{1,1}|-|r_{2,1}|$, 
\begin{equation}\| \bs\partial^{\bs r} {\mathfrak g}^{(h)}_{\E} (z, z')\|\le C^{1+|\bs r|_1} 	\bs r!\,2^{(1+|\bs r|_1)h} e^{-c_0 2^h {d_\E(z, z')}},\label{cosacomedove}\end{equation}
where $d_\E(z, z') :=\min\{ |\per_L((z-z')_1)| + \min\{(z+z')_2,2(M+1)-(z+z')_2\}, L-|\per_L((z-z')_1)|+|(z-z')_2|\}$; note the factor of $2^h$ in the exponent, so that this implies
\begin{equation}
	\left\| \sum_{h=-\infty}^0 \fg_\E^{(h)}(z,z') \right\|
	=
	\cO \left( \frac{1}{d_\E(z,z')} \right),
\end{equation}
that is to say that $\fg_\E$ has a similar decay to $\fg_\infty$ but in a different distance.
With no loss of generality we can assume that the constant $c_0$ in \eqref{ghbounds1} is  the same as in \eqref{cosacomedove}, and is also the same as the constant $c_0$ in Proposition \ref{prop:repr}, as well as in \eqref{eq:WfreeL1bound}, \eqref{eq:WfreeE_base_decay}, and \eqref{eq:WE_base_decay}.

\bigskip

The multiscale decomposition \eqref{gch*h} and the addition formula for Grassmann integrals (see e.g.\ \cite[Proposition~1]{GMT17a}), implies that 
$\int P_c^*(\mathcal D\phi) e^{\cV^{(0)}(\phi,\bs A)}$ can be rewritten as 
\begin{equation} \int P_c^*(\mathcal D\phi) e^{\cV^{(0)}(\phi,\bs A)}=\int P^{(\le h^*)}(\mathcal D\phi^{(\le h^*)})\int P^{(h^*+1)}(\mathcal D\phi^{(h^*+1)})
\cdots \int  P^{(0)}(\mathcal D\phi^{(0)})e^{\mathcal V^{(0)}(\phi^{(\le 0)}, \bs A)},\label{eq:3.9}\end{equation}
where $\phi^{(\le 0)}:=\phi^{(\le h^*)}+\sum_{h=h^*+1}^0\phi^{(0)}$, and $P^{(h)}(\mathcal D\phi^{(h)})$ (resp. $P^{(\le h^*)}(\mathcal D\phi^{(\le h^*)})$) 
is the Gaussian Grassmann integration with propagator $\fg^{(h)}$ (resp. $\fg^{(\le h^*)}$). The right side of \eqref{eq:3.9} will be computed one step at the time, 
by first integrating out $\phi^{(0)}$, then $\phi^{(-1)}$, etc. This iterative procedure induces the definition of the sequence of {\it effective potentials} $\cV^{(h)}$ and of single-scale contributions to the generating function $\cW^{(h)}$, via:
\begin{equation}
	e^{\cW^{(h-1)}(\bs A)+\cV^{(h-1)}(\phi,\bs A)}
	\propto \int   P^{(h)}(\cD\varphi)e^{\cV^{(h)}(\phi+\varphi,\bs A)},
	\label{eq:Vcyl_h_part1}
\end{equation}
for all $h^*<h\le 0$, and $\cV^{(h)}$, $\cW^{(h)}$ are fixed in such a way that $\cW^{(h)}(\bs 0)=\cV^{(h)}(0,\bs A)=0$. Finally, 
\begin{equation}e^{\cW^{(h^*-1)}(\bs A)}\propto \int P^{(\le h^*)} (\cD \phi) e^{\cV^{(h^*)}(\phi,\bs A)},\label{eq:Z_V_hbar_part1}\end{equation}
so that, eventually, 
\begin{equation}
	\Xi_\Lambda (\bs A)\propto \exp\left(\sum_{h=h^*-1}^1\cW^{(h)}(\bs A)\right).\label{eventually}
\end{equation}
where $\cW^{(1)}\equiv \cW$ is the same function as in \cref{eq:startfrom}.
The basic tool for the evaluation of \eqref{eq:Vcyl_N}, \eqref{eq:Vcyl_h_part1} and \eqref{eq:Z_V_hbar_part1} is the following formula, which we spell out in detail only 
for \eqref{eq:Vcyl_h_part1}, even though an analogous one is valid for \eqref{eq:Vcyl_N} and \eqref{eq:Z_V_hbar_part1}. Suppose that, for all $h\le 0$, 
the effective potential $\cV^{(h)}$ can be written 
in a way analogous to the one on scale $h=1$, see \eqref{eq:forma1}, namely 
\begin{equation}\cV^{(h)}(\phi,\bs A)= \sum_{\Psi\in \cM_{0,\Lambda}}\sum_{\bs x\in\cX_\Lambda}
W_\Lambda^{(h)}(\Psi,\bs x) \phi(\Psi) \bs A(\bs x),\label{expcVh}\end{equation}
where $\cM_{0,\Lambda}$ is the subset of $\cM_{1,\Lambda}$ consisting of tuples $\Psi\in(\{+,-\}\times \Lambda)^n$ 
for some $n\in 2\mathbb N$. 
Then $\cV^{(h-1)}$, as computed from \eqref{eq:Vcyl_h_part1}, admits an expansion analogous to \eqref{expcVh}, with $W_\Lambda^{(h)}$ replaced by 
\begin{equation}
	\begin{split}
		W^{(h-1)}_\Lambda(\Psi,\bs x) 
		= \sum_{s=1}^{\infty}\frac{1}{s!} &
		\sum_{\Psi_1,\ldots,\Psi_s\in \cM_{0,\Lambda}}^{(\Psi)}\ 
		\sum_{\bs x_1,\ldots,\bs x_s\in \cX_\Lambda}^{(\bs x)}\left(  \prod_{j=1}^s W^{(h)}_\Lambda(\Psi_j,\bs x_j)\right)
		\alpha(\Psi;\Psi_1,\ldots,\Psi_s)
		\\ & \times \mathbb E^{(h)}(\phi(\bar\Psi_1);\cdots;\phi(\bar\Psi_s)),
	\end{split}
	\label{Wh-1trun}
\end{equation}
	where: the superscript $(\Psi)$ on the sum over $\Psi_1,\ldots, \Psi_s$ indicates 
that the sum runs over all ways of representing $\Psi$ as an ordered sum of $s$ (possibly empty) tuples, 
$\Psi'_1+\cdots\Psi'_s=\Psi$, and over all tuples $\cM_{0,\Lambda}\ni\Psi_j\supseteq\Psi'_j$; for each such term, we denote by $\bar\Psi_j:=\Psi_j\setminus\Psi'_j$ and by $\alpha(\Psi;\Psi_1,\ldots,\Psi_s)$ the sign of the permutation from $\Psi_1\oplus\cdots\oplus \Psi_s$ to $\Psi\oplus\bar\Psi_1\oplus\cdots\oplus\bar\Psi_s$ (here $\oplus$ indicates concatenation of ordered tuples); the superscript $(\bs x)$ on the sum over $\bs x_1,\ldots, \bs x_s$ indicates the constraint ${\bs x}_1\oplus \cdots\oplus {\bs x}_s=\bs x$;
$\mathbb E^{(h)}(\phi(Q_1);\cdots;\phi(Q_s))$ is the {\it truncated expectation} of the Grassmann monomials $\phi(Q_1),\ldots,\phi(Q_s)$ with respect 
	to the Gaussian Grassmann integration with propagator $\fg^{(h)}$, with the understanding that, if $s>1$ and one of the $Q_i$s is the empty set, then $\mathbb E^{(h)}(\phi(Q_1);\cdots;\phi(Q_s))=0$, while, 
	 if $s=1$ and $Q_1=\emptyset$, then $\mathbb E^{(h)}(\phi(\emptyset))=1$. The single-scale contribution to the generating function admits a similar 
	representation, namely, 
\begin{equation}\cW^{(h-1)}(\bs A)= \sum_{\bs x\in\cX_\Lambda}^{\bs x\neq\emptyset}w_\Lambda^{(h-1)}(\bs x) \bs A(\bs x),\label{kernelw}
\end{equation}
where $w_\Lambda^{(h-1)}(\bs x)$ is defined by the counterpart of \cref{Wh-1trun} with $\Psi$ replaced by the empty set and, since we have not included $\mathcal W^{(h)}({\bf A})$ in the right hand side of \cref{eq:Vcyl_h_part1}, there is no term with $s=1$ and $\Psi_{1}=\emptyset$.
In \cref{Wh-1trun}, the truncated expectation can be evaluated explicitly
	in terms of the following Pfaffian formula, originally due to Battle, Brydges and Federbush
\cite{B,BrF},  later improved and simplified \cite{AR98,BK87} and rederived in several review papers \cite{GM01,Gi,GMR}, see e.g. \cite[Lemma 3]{GMT17a}: if $s>1$ and 
$Q_i\neq\emptyset$, $\forall i\in\{1,\ldots, s\}$, then 
	\begin{eqnarray} &&	\mathbb E^{(h)}(\phi(Q_1);\cdots;\phi(Q_s))=\sum_{T \in \cS(Q_1,\ldots,Q_s)}
	\fG_T^{(h)}(Q_1,\ldots,Q_s),\label{eq:BBF}\\
	& \text{with}&	\fG_T^{(h)}(Q_1,\dots,Q_s)
	:=
	\alpha_T(Q_1,\dots,Q_s)
		\left[ \prod_{\ell \in T} g^{(h)}_\ell \right]
		\int P_{Q_1,\dots,Q_s,T} (\dd \bs t)\, {\rm{Pf}} \big(G^{(h)}_{Q_1,\ldots,Q_s,T} (\bs t)\big) ,
\nonumber	\end{eqnarray}
where: 
\begin{itemize}
\item $\cS(Q_1,\ldots,Q_s)$ denotes the set of all the `spanning trees' on $Q_1,\ldots,Q_s$, that is, 
of all the sets $T$ of ordered pairs $(f,f')$, with $f \in Q_i$, $f' \in Q_j$ and $i < j$, whose corresponding graph $G_T=(V,E_T)$, 
with vertex set $V=\{1,\ldots,s\}$ and edge set $E_T=\{(i,j)\in V^2\,:\, \exists (f,f')\in T\ \text{with}\  f\in Q_i, f'\in Q_j\}$, is a tree graph;
\item $\alpha_T(Q_1,\dots,Q_n)$ is the sign of the permutation from $Q_1\oplus\cdots\oplus Q_s$ to $T\oplus(Q_1\setminus T)\oplus\cdots\oplus(Q_s\setminus T)$;
\item for $\ell=((\omega_i,z_i),(\omega_j,z_j))$, $g_\ell^{(h)}$ is a shorthand for $g^{(h)}_{\omega_i\omega_j}(z_i,z_j)$ (we recall that  
$g^{(h)}_{\omega\omega'}(z,z')$ are the components of the $2\times2$ matrix $\fg^{(h)}(z,z')$);
\item ${\bs t}=\{t_{i,j}\}_{1\le i,j \le s}$, and $P_{Q_1,\ldots,Q_s,T}(\dd \bs t)$
is a probability measure with support on a set of ${\bs t}$ such that
$t_{i,j}=\bs u_i\cdot\bs u_{j}$ for some family of vectors $\bs u_i=\bs u_i({\bs t})\in \mathbb R^s$ of
unit norm;
\item letting $2q=\sum_{i=1}^s|Q_i|$, $G^{(h)}_{Q_1,\ldots,Q_s,T}({\bs t})$ is an antisymmetric
  $(2q-2s+2)\times (2q-2s+2)$ matrix, whose off-diagonal elements are given by
  $\big(G^{(h)}_{Q_1,\ldots,Q_s,T} (\bs t)\big)_{f,f'}=t_{i(f),i(f')}g^{(h)}_{\ell(f,f')}$, where $f,
  f'$ are elements of the tuple $(Q_1\setminus T)\oplus\cdots\oplus(Q_s\setminus T)$, and $i(f)$ is the integer in $\{1,\ldots,s\}$ such that 
  $f$ is an element of $Q_i\setminus T$.
\end{itemize}
Similarly, if $s=1$ and $Q_1\neq\emptyset$, we let $\cS(Q_1)=\{\emptyset\}$ and we write the analogue of \eqref{eq:BBF} as $\mathbb E^{(h)}(\phi(Q_1))=\fG^{(h)}_\emptyset(Q_1):={\rm Pf}\big(G^{(h)}_{Q_1}\big)$,
where $\big(G^{(h)}_{Q_1}\big)_{f,f'}=g^{(h)}_{(f,f')}$ and $f,f'$ are elements of the tuple $Q_1$. 

\bigskip

The systematic use of \eqref{eq:BBF}, in combination with the decay bounds on $\fg^{(h)}$, see \eqref{ghbounds1}, 
allows one to get bounds on the kernels of $\cW^{(h)}$ and $\cV^{(h)}$ and, consequently, on those of $\log\Xi_\Lambda(\bs A)$. 
Such bounds are sufficient to show that the sums in terms of which these kernels are expressed are absolutely convergent for any fixed
$h^*$ (that is, for any fixed volume, recall the definition of $h^*$ given before \eqref{gch*h}); 
however, a priori, this convergence is not at all uniform in $h^*$ and so tells us nothing about the 
thermodynamic limit. For a more quantitative discussion of the reason why the bounds obtained via this `naive' 
procedure are non uniform in $h^*$, see, e.g., \cite[Sect.5.2.2]{GMT17a}. 

\medskip

The basic idea of the strategy we use to get past this is to find, at each step $h\le 0$ of the iterative scheme, a decomposition\footnote{Actually, in order to define 
a convergent expansion, it is not necessary to split the effective source term $\cB^{(h)}(\phi,\bs A ):=\cV^{(h)}(\phi, \bs A) - \cV^{(h)}(\phi, 0)$
into local plus irrelevant part, since this is not the source of any divergence. Nevertheless, in order to 
compute the scaling limit of the energy correlations, it is convenient to also decompose the effective source term into a local part plus a remainder, 
and we shall do so in Section \ref{sec:cylinder_multiscale} and following sections.}
\begin{equation}
	\cV^{(h)}(\phi,\bs A)
	=
	\cL
	\cV^{(h)}(\phi,\bs A)
	+
	\cR
	\cV^{(h)}(\phi, \bs A),
\label{cVLcVRcV}
\end{equation}
where: 
\begin{itemize}
\item $\cL \cV^{(h)}$ is the so-called {\it local part} of the effective potential, which includes terms that tend to `expand' under iterations, 
	usually called the {\it relevant} and {\it marginal} contributions in Renormalization Group (RG) terminology. In our case, $\cL \cV^{(h)}$ includes: three terms that are 
	quadratic in the Grassmann variables and independent of $\bs A$, depending on a sequence of $h$-dependent parameters which we denote $\allct=\{(\nu_h,\zeta_h,\eta_h)\}_{h\le 0}$ and call the {\it running coupling constants}; and two terms that are quadratic in the Grassmann variables and linear in $\bs A$, depending on another sequence of effective parameters, $\{Z_{1,h}, Z_{2,h}\}_{h\le 0}$, called the {\it effective vertex renormalizations}. 
\item $\cR \cV^{(h)}$ is the so-called {\it irrelevant}, or {\it renormalized}, part of the effective potential, which is not the source of any divergence.
\end{itemize}

\medskip

Such a decomposition, described in detail in Section \ref{sec:cylinder_multiscale} below, corresponds to a systematic reorganization, or `resummation', of the expansions arising from the multiscale computation of the 
generating function. We anticipate that the running coupling constants and the effective vertex renormalizations are defined in terms of the infinite volume limit 
of the kernels of the effective potential, whose flow has been studied and controlled in \cite{GGM} and 
reviewed in greater detail in \cite[Section~4]{AGG_part2}, exactly in the same setting needed for the present paper\footnote{The running coupling constants used here and in \cite{AGG_part2} are not strictly speaking equal to those of \cite{GGM} even in the isotropic case because of a number of changes in the quantities used to define them, most importantly the different decomposition of the propagators, but the general outline of the construction is the same.}.
As shown in \cite[Section~4.5]{AGG_part2}, by appropriately tuning the inverse temperature $\beta$ to a value $\beta_c(\lambda)$, which has the interpretation of interacting critical 
temperature, and by appropriately fixing the parameters $Z,t_1^*$ entering the definition of $\cV^{(1)}$, see \eqref{eq:startfrom}-\eqref{cV1defin}, 
the whole sequence $\allct$ is bounded and smaller than $C|\lambda|$. Under these conditions, we will be able to show that the resulting expansions for 
multipoint energy correlations in the cylinder $\Lambda$ are convergent, uniformly in $h^*$. Our estimates are based on writing the quantities involved as sums over terms indexed by Gallavotti-Nicol{\`o} (GN) trees~\cite{GN,GM01,Ga}, which emerge naturally from the multiscale procedure; the relevant aspects of the definitions of the GN trees 
will be reviewed in Section~\ref{sec:treecyl} below. Weighted $L^1$ bounds on the kernels of the effective potentials, based on their GN tree expansion, are derived in Section \ref{sec:informal_bounds}. 
As a corollary, in Section \ref{sec:Zj} we show how to control the flow of the effective vertex renormalizations, as well as their edge counterpart. 

\subsection{Localization and renormalization: the operators $\cL$ and $\cR$}
\label{sec:cylinder_multiscale}

As anticipated above, in order to define a convergent multiscale expansion for the effective potentials and the generating function, at each step $h\le 0$ of the iteration
we decompose the effective potential $\cV^{(h)}$ into a `local' part $\cL\cV^{(h)}$ plus a remainder $\cR\cV^{(h)}$ (the `irrelevant', or `renormalized' part). In 
this section we first give a number of necessary preliminaries and then define the action of the operators $\cL$ and $\cR$ on $\cV^{(h)}$, 
see Section \ref{sec:3.1.3} and in particular \eqref{eq:3.1.42(}, \eqref{eqfin1.}, and \eqref{RWdecomp.4} for the `final' formulas. The decomposition in 
\eqref{eq:3.1.42(}, \eqref{eqfin1.}, \eqref{RWdecomp.4} is defined in terms of other operators, $\cL_\B,\cR_\B$ and $\cL_\E,\cR_\E$ 
(the bulk- and edge-localization and -renormalization operators), which we introduce in Section \ref{sec:LeRcyl}, after having described in Section \ref{subsec:3.1.1} the general structure and properties 
of the kernels of the effective potential; Section \ref{subsec:normRBRE} collects a couple of norm bounds on $\cR_\B,\cR_\E$ that will be useful in the following. 

\subsubsection{Representation of the kernels of the effective potential}\label{subsec:3.1.1}

{\it Field multilabels.} Let $\bar\Lambda=\mathbb Z_L \times \left(\mathbb Z\cap [0, M+1]\right)$, be the {\it closure} of $\Lambda$, 
which can be thought of as the union of $\Lambda$ and of two additional rows at vertical heights $0$ and $M+1$, right below and above $\Lambda$, respectively. Recall also that 
for $\omega=\pm$ and $z\in\bar\Lambda\setminus\Lambda$, the symbol $\phi_{\omega,z}$ should be interpreted as explained in Remark \ref{rem:cancell}. 
Given these premises, we let $\cM_\Lambda$ be the set of {\it field multilabels}, which we define as tuples 
$$
\Psi=((\omega_1,D_1,z_1),\ldots,(\omega_n,D_n,z_n))\in (\{+,-\}\times\{0,1,2\}^2\times \bar\Lambda)^n
$$ 
for some $n\in 2\mathbb N$, satisfying the following constraints: $\|D_i\|_1\le 2$ and $z_i+D_i\in\bar\Lambda$.
Given $\Psi=((\omega_1,D_1,z_1),\ldots,(\omega_n,D_n,z_n))\in \cM_\Lambda$, we let $\phi(\Psi)$ be the Grassmann monomial
\begin{equation}\label{afterm}
\phi(\Psi)=\partial^{D_1}\phi_{\omega_1,z_1}\cdots \partial^{D_n}\phi_{\omega_n,z_n},
\end{equation}
where $\partial^{D_i}\phi_{\omega_i,z_i}:=\partial_1^{(D_i)_1}\partial_2^{(D_i)_2}\phi_{\omega_i,z_i}$, 
with $\partial_1$ and $\partial_2$ the right discrete derivatives in directions 1 and 2, respectively, defined by $\partial_jf(z)=f(z+\hat e_j)-f(z)$ (with the understanding that, if $(z)_1=L$, then $\partial_1\phi_{\omega,z}:=-\phi_{\omega,(1,(z)_2)}-\phi_{\omega,z}$, and similarly for the higher order derivatives, in agreement with the anti-periodic boundary conditions in the horizontal direction on the Grassmann fields).  
For later reference, we also let $\cM_\Lambda^\circ$ be the subset of $\cM_\Lambda$ consisting of tuples $((\omega_1,D_1,z_1),\ldots,(\omega_n,D_n,z_n))$
for some $n\in2\mathbb N$ 
such that $z_i,z_i+D_i\in\Lambda$ (rather than $z_i,z_i+D_i\in\bar\Lambda$).
In the following, with some abuse of notation, any element 
$\Psi\in{\cM_{\Lambda}}$ of length $|\Psi|=n$ will be denoted indistinctly by $\Psi=((\omega_1,D_1,z_1),\ldots,(\omega_n,D_n,z_n))$ 
or $\Psi=(\bs \omega, \bs D, \bs z)$, with the understanding that $\bs \omega=(\omega_1,\ldots,\omega_n)$, etc. 

\medskip

\noindent{\it Structure of the effective potential.} Recall that the effective potential $\cV^{(1)}$ on scale $1$ can be represented as in \eqref{eq:forma1}. 
For smaller scales, i.e., for $h\le0$, we will inductively prove that the effective potential can be represented in a similar way, namely as
\begin{equation}\cV^{(h)}(\phi,\bs A)= \sum_{\Psi\in \cM_{\Lambda}}\sum_{\bs x\in\cX_\Lambda}
W_\Lambda^{(h)}(\Psi,\bs x) \phi(\Psi) \bs A(\bs x),\label{eq:formalh}\end{equation}
for a suitable real {\it kernel function} (or, simply, kernel) $W_\Lambda^{(h)}:\mathcal M_\Lambda\times \cX_\Lambda\to\mathbb R$. Eq.\eqref{eq:formalh} is analogous to the `naive' expansion \eqref{expcVh}, with the important difference that the sum over $\Psi$ now ranges over $\cM_{\Lambda}$, 
rather than over $\cM_{0,\Lambda}$. It will be shown iteratively that the kernel $W_\Lambda^{(h)}$, labelled by $\Lambda$ and $h$, can be chosen in such a way that 
it satisfies the symmetry properties described in item \ref{it1:dop} after \eqref{eq:WfreeL1bound}. 

In addition to $W_\Lambda^{(h)}$, we iteratively assume the existence of an infinite volume kernel $W^{(h)}_\infty$
on $\mathcal M_\infty\times \cX_\infty$, where $\cM_\infty$ is the set of field multilabels in the infinite plane, defined in \cite[Section~4]{AGG_part2}, and $\cX_\infty$ is
the set of the tuples $\bs x=(x_1,\ldots,x_m)\in\fB_\infty^m$ for some $m\in\mathbb N_0$ (as usual, if $m=0$, then $\bs x$ is understood as the empty set). $W^{(h)}_\infty$ will be used momentarily to define the 
bulk-edge decomposition of $W_\Lambda^{(h)}$, and, a posteriori, it will turn out to be the infinite volume limit of $W_\Lambda^{(h)}$, in the sense of 
item \ref{it2:dop} after \eqref{eq:WfreeL1bound}; however, a priori, we define $W_\infty^{(h)}$ to be 
the solution to the iterative equations \eqref{eq:BBF0infty}-\eqref{eq:BBFh-1infty} below, with $W_\infty^{(1)}$ defined as in \eqref{eq:forma1} and following lines. 
Note that the restriction of $W_\infty^{(h)}$ at $\bs x=\emptyset$, denoted by $V_\infty^{(h)}$, 
is the same infinite volume translationally invariant kernel constructed and estimated in detail in \cite[Section~4]{AGG_part2}; the (straightforward) generalization of the construction of $W_\infty^{(h)}$, as well 
as the derivation of weighted $L^1$ bounds on it, to any $\bs x\in \cX_\infty$, goes along the same lines, and is given for completeness in Sections \ref{sec:treecyl}-\ref{sec:informal_bounds} below. 

Given $W_\Lambda^{(h)}$ and $W_\infty^{(h)}$ for some $h\le 0$, for any $(\bs\omega,\bs D,\bs z)\in\cM_\Lambda$ we let 
\begin{equation}\begin{split} &W_\B^{(h)}((\bs\omega,\bs D,\bs z),\bs x):=(-1)^{\alpha(\bs z)}\,I_\Lambda((\bs \omega,\bs D,\bs z),\bs x)\,
W_\infty^{(h)}((\bs\omega,\bs D,\bs z_\infty),\bs x_\infty),\\
&\qquad \text{with} \qquad I_\Lambda(\Psi,\bs x):=\mathds 1(\Psi\in\cM_\Lambda^\circ)\mathds 1(\diam_1(\bs z,\bs x)\le L/3),\end{split}
 \label{eq:3.1.4}\end{equation} 
where: $\alpha(\bs z)$ was defined in \eqref{eq:alphaz}; $\cM_\Lambda^\circ$ was defined as the subset of $\cM_\Lambda$ consisting of tuples 
$((\omega_1,D_1,z_1),\ldots , (\omega_n,D_n,z_n))$ for some $n\in2\mathbb N$ such that $z_i,z_i+D_i\in\Lambda$; $\diam_1$ is the horizontal diameter on $\Lambda$); $(\bs z_\infty, \bs x_\infty)\equiv 
(\bs z,\bs x)_\infty$ is the infinite volume representative of $(\bs z,\bs x)$, in the sense of item \ref{it3:dop} after \eqref{eq:WfreeL1bound}. Moreover, we let 
\begin{equation} W_\E^{(h)}:=W_\Lambda^{(h)}-W_\infty^{(h)}.\label{eq:3.1.4bis}\end{equation}
For later reference, we also introduce the following definitions: 
\begin{enumerate}
\item\label{itemi:dop}We denote by $V_\Lambda^{(h)}$ the restriction of $W_\Lambda^{(h)}$ to $\bs x=\emptyset$ (the `sourceless'  part of the effective potential,
which can be naturally thought of as a kernel on $\cM_{1,\Lambda}$ or on $\cM_\Lambda$, depending on whether $h=1$ or $h\le 0$), and by $B_\Lambda^{(h)}$ the restriction 
of $W_\Lambda^{(h)}$ to $\bs x\neq \emptyset$ (the `effective source term'), so that $W_\Lambda^{(h)}=V_\Lambda^{(h)}+B_\Lambda^{(h)}$.
\item\label{itemii:dop} Given a kernel $W$ on $\cM_\Lambda\times \cX_\Lambda$, 
we denote by $W_{n,p,m}$ its restriction to field multilabels of length $n$, whose derivative labels have $1$-norm equal to $p$, and to tuples $\bs x$ of length $m$. 
Similarly, given a kernel $V$ on $\cM_\Lambda$, we let $V_{n,p}$ be its restriction to  field multilabels of length $n$, whose derivative labels have $1$-norm equal to $p$. 
Often, with some abuse of notation, we shall identify $V_{n,p}$ with the kernel whose only non-zero component is $V_{n,p}$, and similarly for $W_{n,p,m}$.
\item
We say that two kernels $W$ and $\widetilde W$ on $\cM_\Lambda\times \cX_\Lambda$ 
are {\it equivalent}, and we write $W\sim \widetilde W$, 
if the corresponding potentials $$\cV(\phi,\bs A):=\sum_{\Psi\in\cM_\Lambda} \sum_{\bs x\in\cX_\Lambda}W(\Psi,\bs x)\phi(\Psi)\bs A(\bs x) \quad \text{and}
\quad 
\widetilde{\cV}(\phi,\bs A):=\sum_{\Psi\in\cM_\Lambda} \sum_{\bs x\in\cX_\Lambda}\widetilde W(\Psi,\bs x)\phi(\Psi)\bs A(\bs x)$$ are equal. For the notion 
of equivalence among infinite volume kernels, see \cite[Section~4]{AGG_part2}. 
\end{enumerate}

\subsubsection{The bulk- and edge-localization and -renormalization operators}\label{sec:LeRcyl}

In this subsection, we introduce bulk-localization and bulk-renormalization operators $\cL_\B$ and $\cR_\B$, as well as 
edge-localization and edge-renormalization operators  $\cL_\E$ and $\cR_\E$. These are {\it linear} operators acting on kernels of $\xi$-independent potentials on 
the cylinder $\Lambda$; if applied to kernels that are invariant under the symmetries described in item \ref{it1:dop} after \eqref{eq:WfreeL1bound}
(horizontal translations, permutations, and reflections), they act as projection operators, with $\cR_\B$ orthogonal to $\cL_\B$ and 
$\cR_\E$ orthogonal to $\cL_\E$. 

We will first define $\cL_\B$ and $\cR_\B$ on the sourceless part of the kernels, then $\cL_\E$ and $\cR_\E$ on the sourceless part of the kernels, 
and finally $\cL_\B$ and $\cR_\B$ on the effective source term (for the purpose of the following discussion, 
we will not need to define the edge-localization and -renormalization operators on the effective source term, because they are automatically dimensionally irrelevant). 
The definition of 
$\cL_\B$ and $\cR_\B$ on the sourceless part of the kernels generalizes the analogous definition on the infinite volume kernels discussed in 
\cite[Section~4.2]{AGG_part2}.
The action of the `full' localization and renormalization operators, $\cL$ and $\cR$, on the kernels of $\cV^{(h)}$ for $h\le 0$, 
will be defined in terms of $\cL_\B, \cR_\B, \cL_\E, \cR_\E$ in Section \ref{sec:3.1.3} below. 

\bigskip

\noindent{\bf The action of $\cL_\B$ and $\cR_\B$ on the sourceless part of the kernels.}

\noindent Consider a kernel $V:\mathcal M_\Lambda\to \mathbb R$ supported on $\cM_\Lambda^\circ$ (for the definition of $\cM_\Lambda^\circ$, see a few lines after 
\eqref{afterm}). First of all, recalling item \ref{itemii:dop} after \eqref{eq:3.1.4} for the definition of  $V_{n,p}$, we let
\begin{equation} \cL_\B (V_{n,p}):=0, \quad \text{if}\quad 2-\frac{n}2-p<0.\label{loc1}\end{equation}
In the RG jargon, the combination $2-\frac{n}2-p$ is the {\it scaling dimension} of $V_{n,p}$\footnote{\label{scaldim}On the basis of the general Wilsonian RG theory, the scaling dimension
of a kernel describes the way in which it transforms under rescalings of the spatial coordinates and of the fields or, equivalently, under the action of the linearized Wilsonian RG transformation. On general grounds, for a kernel of order $(n,p)$, it is equal to $d-n\Delta_\phi-p$, where $d$ is the spatial dimension, and $\Delta_\phi$ the scaling dimension of the field $\phi$, i.e., half the critical exponent of the two-point function of $\phi$: $\langle \phi_x\phi_y\rangle \sim |x-y|^{-2\Delta_\phi}$. In our case, $d=2$ and $\Delta_\phi=1/2$, 
which explains the expression $2-n/2-p$ in \eqref{loc1}.}; in this sense, \eqref{loc1} says that 
the local part of the terms with negative scaling dimension (the irrelevant terms, in the RG jargon) is set equal to zero. 

There are only a few cases for which $2-\frac{n}2-p\ge 0$, namely $(n,p)=(2,0), (2,1), (4,0)$. In these cases, the action of $\cL_\B$ on $V_{n,p}$ is non trivial, and will be defined in terms 
of other basic operators, the first of which is $\tcL$, which is defined as follows: for $(n,p)=(2,0), (2,1), (4,0)$, 
we let $(\tcL V)_{n,p}=\tcL (V_{n,p})\equiv \tcL V_{n,p}$ be the kernel on $\cM_\Lambda$ such that 
\begin{equation}
	\tcL V_{n,p} (\bs\omega,\bs D,\bs z)
	:=
	\left(\prod_{j=2}^n\delta_{z_j,z_1}\right)
	  \sum_{\bs y \in \Lambda^n}\delta_{y_1,z_1}(-1)^{\alpha(\bs y)}
	  V_{n,p}(\bs \omega,\bs D,\bs y).
	\label{eq:cLtilde_kernel_def.2}
\end{equation}
Note that this definition imples $\tcL \tcL V_{n,p} = \tcL V_{n,p}$.
If $(n,p)=(2,1), (4,0)$, we let 
\begin{equation} \cL_\B (V_{2,1}):=\cA (\tcL V_{2,1}), \qquad  \cL_\B (V_{4,0}) :=\cA (\tcL V_{4,0}), \label{eq:local2140}\end{equation}
where  $\cA$ is the operator that antisymmetrizes with respect to permutations {\it and} symmetrizes with respect to reflections in the horizontal and vertical directions,
more explicitly
\begin{equation}
	\cA V(\Psi)
	=
	\frac{1}{4 |\Psi|!}
	\sum_\pi \sum_{\Theta \in \left\{ 1,\Theta_1,\Theta_2,\Theta_1 \Theta_2 \right\}}
	(-1)^\pi \sigma_\Theta (\Psi)
	V\left(\Theta \pi \Psi  \right)
	\label{eq:cA_def}
\end{equation}
where the sum in $\pi$ is over permutations of sequences with $|\Psi|$ elements, and 
recalling the notation of \cref{eq:2.2.1},
$\sigma_\Theta (\Psi)$ and $\Theta \Psi$ are defined by
\begin{equation}
	\sigma_\Theta (\Psi) \phi(\Theta \Psi)
	=
	\Theta \phi (\Psi)
	\label{eq:symmetries_field_labels}
\end{equation}
where the action of $\Theta_1,\Theta_2$ on Grassmann monomials on the right hand side is given in
\cref{eq:symm_horizontal_phixi,eq:symm_vertical_phixi}.
Note that although $\cA$ and $\tcL$ do not commute (this is in fact part of the point of introducing $\cA$: $\tcL$ does not preserve antisymmetry of the kernels), one can verify that $\cA \tcL \cA V = \cA \tcL V$.

A first important remark, related to the definitions \eqref{eq:local2140}, is that the following key cancellation holds: 
\begin{equation}  \cL_\B (V_{4,0})\sim 0,\label{pauli}\end{equation}
simply because $\omega$ only assumes two values and $\tcL V_{4,0}$ is supported on $\bs z$ such that $z_1=z_2=z_3=z_4$. 

Let us now define the action of $\cL_\B$ on $V_{2,0}$. In this case we need to include in $\cL_\B(V_{2,0})$ a term of order $(2,1)$, see \eqref{eq:local20}-\eqref{defcL.1} below,
so we first need to rewrite $V_{2,0}- \tcL V_{2,0}$
in `interpolated form', in terms of an equivalent kernel supported on field multilabels $(\bs\omega,\bs D,\bs z)$ with $\|\bs D\|_1=1$. 
For each pair of distinct sites $z, z' \in \bar \Lambda$, 
let $\gamma(z, z')=(z, z_1, z_2, \ldots, z_n,z')$ be the shortest path obtained by going first vertically and then horizontally from $z$ to $z'$\footnote{
If $\diam_1(z,z')<L/2$, there is no ambiguity in the definition of such a {\it shortest} path. If $\diam_1(z,z')=L/2$, we choose 
$\gamma(z, z')=(z, z_1, z_2, \ldots, z_n,z')$ to be the path going first vertically and then horizontally from $z$ to $z'$ in such a way that the
$\min\{(z)_1,(z')_1\}\le (z_j)_1 \le \max\{(z)_1,(z')_1\}$, for all $j=1,\ldots, n$.}.
Note that $\gamma$ is covariant under the symmetries of the model, i.e., 
\begin{equation}
	S \gamma (z, z')
	=
	\gamma (S z, S z')
	\label{eq:gamma_symmetry}
\end{equation}
where $S: \bar\Lambda \to \bar\Lambda$ is some composition of horizontal translations, and of horizontal and vertical reflections.
Given $z,z'$ two distinct sites in $\bar\Lambda$, let $\INT (z,z')$ be the set of 
$(\sigma,(D_1,D_2),(y_1,y_2))\equiv ( \sigma, \bs D, \bs y)\in\{\pm\}\times\{0,\hat e_1,\hat e_2\}^2\times  \bar\Lambda^2$ such that: (1) $y_1=z$, (2) $D_1=0$, (3) $y_2, y_2 + D_2\in \gamma (z,z')$, (4)
$\sigma=+$ if $y_2$ precedes $y_2 + D_2$ in the sequence defining $\gamma (z,z')$, and $\sigma=-$ otherwise. In terms of this definition, one can easily check that 
(cf.\ \cite[Eqs.~(4.2.6) and (4.2.8)]{AGG_part2}):
\begin{equation}  \begin{split}\sum_{\bs z\in\Lambda^2}
[V_{2,0}-\tcL V_{2,0}](\bs \omega,\bs 0,\bs z)
\phi(\bs \omega, \bs 0, \bs z)
=&\sum_{\bs z\in\Lambda^2}(-1)^{\alpha(\bs z)}V_{2,0}(\bs \omega,\bs 0,\bs z)\sum_{(\sigma,\bs D,\bs y)\in\INT(\bs z)}(-1)^{\alpha(\bs y)}\sigma\phi(\bs \omega,\bs D,\bs y)\\
\equiv&\sum_{\bs y\in\Lambda^2}\sum_{\bs D}^{(1)}(\tcR V)_{2,1}(\bs \omega,\bs D,\bs y)\phi(\bs \omega,\bs D,\bs y), \end{split}
\label{juve.n} \end{equation}
where, if $\bs z=(z_1,z_1)$, the sum over $(\sigma,\bs D,\bs y)$ in the first line should be interpreted as being equal to zero (in this case, we let $\INT(z_1,z_1)$ be the empty set). In going from the first to the second line, we exchanged the order of summations over $\bs z$ and $\bs y$; moreover, $\sum_{\bs D}^{(p)}$ denotes the sum 
over the pairs $\bs D=(D_1,D_2)$ such that $\|\bs D\|_1=p$, and, for $\bs D$ such that $\|\bs D\|_1=1$, 
\begin{equation} (\tcR V)_{2,1}(\bs \omega,\bs D,\bs y)= (\tcR V_{2,0})(\bs \omega,\bs D,\bs y):=\sum_{\substack{\sigma=\pm,\, \bs z\in\Lambda^2:\\(\sigma,\bs D,\bs y)\in\INT(\bs z)}}(-1)^{\alpha(\bs z)+\alpha(\bs y)}\sigma V_{2,0}(\bs \omega,\bs 0,\bs z).\label{juve.m}\end{equation}
From the previous manipulations, it is clear that %
$(\tcR V)_{2,1}\sim V_{2,0}-\tcL V_{2,0}$. 
Since $\tcL V_{2,0}$ is supported on arguments with $\bs z = (z_1,z_1)$, and recalling that $\INT(z_1,z_1)=\emptyset$, we also have 
$\tcR (\tcL V_{2,0}) = 0$ and $\tcR (\cA \tcL V_{2,0})=0$.
Note that the action of $\tcR$ ``adds a derivative'', taking a kernel supported on multilabels without derivatives to one which is supported on multilabels containing a single derivative, and $\tcR$, unlike $\tcL$, does not commute with the operation of restricting to a certain order:
$\tcR V_{2,0}$, the result of $\tcR$ acting on the restriction of a kernel to order $(2,0)$, is equal to $(\tcR V)_{2,1}$, the restriction to terms of order $(2,1)$ of $\tcR V$.
We warn the reader to be alert to the positioning of the subscripts in the remainder of this section in particular.

We are finally ready to define the action of $\cL_\B$ on $V_{2,0}$: 
\begin{equation} (\cL_\B(V_{2,0}))_{n,p}:=\begin{cases} \cA(\tcL V_{2,0}) & \text{if}\ (n,p)=(2,0)\\
\cA (\tcL (\tcR V_{2,0})) & \text{if}\ (n,p)=(2,1)\\ 0 & \text{otherwise}.\end{cases}  
\label{eq:local20}\end{equation}
Combining this with the previous definitions and collecting terms of the same order, in view of \eqref{pauli}, 
\begin{equation}\label{defcL.1}
(\cL_\B V)_{n,p}=\begin{cases} \cA(\tcL V_{2,0}) &\text{if} \ (n,p)=(2,0),\\
	\cA(\tcL V_{2,1}+\tcL (\tcR V_{2,0})) & \text{if} \ (n,p)=(2,1),\\
0 & \text{otherwise}\end{cases}\end{equation}
\begin{remark}
	Recalling that $\tcL \tcL = \tcL$, $\cA \tcL \cA = \cA\tcL$, and $\tcR (\cA \tcL V_{2,0}) =0$, 
	\begin{equation}
		\begin{split}
(\cL_\B \cL_\B V)_{2,0} &= \cA\,\tcL (\cL_\B  V)_{2,0} = \cA\,\tcL\cA\,\tcL V_{2,0} = \cA\,\tcL V_{2,0} =(\cL_\B V)_{2,0}\\
(\cL_\B \cL_\B V)_{2,1} &=\cA\,\tcL (\cL_\B V)_{2,1}+\cA\,\tcL(\tcR (\cL_\B V)_{2,0}))\\
&=\cA\,\tcL \cA\tcL V_{2,1}+ \cA\,\tcL \cA\,\tcL(\tcR V_{2,0})+ \cA\,\tcL(\tcR (\cA \tcL V_{2,0}))\\
&=\cA\,\tcL V_{2,1}+ \cA\,\tcL(\tcR V_{2,0}) =(\cL_\B V)_{2,1},
\end{split}\end{equation}
and $(\cL_\B V)_{n,p}=0$ for all other values of $(n,p)$. In other words, $\cL_\B \cL_\B = \cL_\B$.
	\label{rem:LB_projection}
\end{remark}
\bigskip

Let us now define the action of $\cR_\B$ on $V$, in such a way that $\cR_\B V\sim \\cA V-\cL_\B V$; note that we will mainly apply these operators to kernels for which $\cA V \sim V$, and thus $\cR_\B \sim V - \cL_\B V$. 
First of all, we let 
\begin{equation} \cR_\B(V_{n,p})=(\cR_\B V)_{n,p}:=\cA V_{n,p},\quad \text{if:} \quad  n\ge 6,  \quad \text{or} \ n=4\ \text{and} \ p\ge 2, \quad \text{or} \ n=2\ \text{and} \ p\ge 3.\end{equation}
Moreover, we let 
\begin{equation} (\cR_\B V)_{2,0}=(\cR_\B V)_{2,1}=(\cR_\B V)_{4,0}:=0. \end{equation}
The only cases in which $(\cR_\B V)_{n,p}$ is non trivial are $(n,p)=(2,2), (4,1)$. For these values of $(n,p)$, 
$(\cR_\B V)_{n,p}$ is defined in terms of an interpolation generalizing the definition of $(\tcR V)_{2,1}$ in \eqref{juve.m}. 
As a preparation for the definition, we first introduce $(\tcR V)_{n,p}$ for $(n,p)=(2,2), (4,1)$. For this purpose, we start 
from the analogue of \eqref{juve.n} in the case that $(2,0)$ is replaced by $(n,p)=(2,1), (4,0)$: for such values of $(n,p)$ we write
(cf.\ \cite[Eqs.\ (4.2.16) and (4.2.18)]{AGG_part2})
\begin{equation}\label{supersplit}
	\begin{split}
		&\phantom{=}  \sum_{\bs z\in\Lambda^n}\sum_{\bs D}^{(p)}[V_{n,p}-\tcL V_{n,p}](\bs \omega,\bs D,\bs z)\phi(\bs \omega,\bs D,\bs z)
		\\
		&=\sum_{\bs z\in\Lambda^n}\sum_{\bs D}^{(p)}(-1)^{\alpha(\bs z)}V_{n,p}(\bs \omega,\bs D,\bs z)\sum_{(\sigma,\bs D',\bs y)\in\INT(\bs z)}(-1)^{\alpha(\bs y)}\sigma\phi(\bs \omega,\bs D+\bs D',\bs y)
		\\
		&= \sum_{\bs y\in\Lambda^n}\sum_{\bs D}^{(p+1)}(\tcR V)_{n,p+1}(\bs \omega,\bs D,\bs y)\phi(\bs \omega,\bs D,\bs y).
	\end{split}
\end{equation}	
In the second line, 
\begin{itemize}
\item if $(n,p)=(2,1)$, then $\INT(\bs z)$ is the same defined after \eqref{eq:gamma_symmetry}; 
\item if $(n,p)=(4,0)$, then, for any $\bs z\in\Lambda$, $\INT(\bs z)$ is the set of 
$(\sigma,(D_1,\ldots, D_4),(y_1,\ldots, y_4))\equiv ( \sigma, \bs D, \bs y)\in\{\pm\}\times\{0,\hat e_1,\hat e_2\}^4\times \Lambda^4$ such that: either
$y_1=y_2=y_3=z_1$, $D_1=D_2=D_3=0$, and $(\sigma, (0,D_4), (z_1,y_4))\in\INT(z_1,z_4)$; or $y_1=y_2=z_1$, $y_4=z_4$, $D_1=D_2=D_4=0$, and $(\sigma, (0,D_3), (z_1,y_3))\in\INT(z_1,z_3)$;
or $y_1=z_1$, $y_3=z_3$, $y_4=z_4$, $D_1=D_3=D_4=0$, and $(\sigma, (0,D_2), (z_1,y_2))\in\INT(z_1,z_2)$. 
\end{itemize} In the third line of \eqref{supersplit}, 
for $\bs D$ such that $\|\bs D\|_1={p+1}$, 
\begin{equation} (\tcR V)_{n,p+1}(\bs \omega,\bs D,\bs y)=(\tcR V_{n,p})(\bs \omega,\bs D,\bs y):=\sum_{\substack{\sigma=\pm,\, \bs z\in\Lambda^n,\\ \bs D'\in\{0,\hat e_1,\hat e_2\}^n:\\(\sigma,\bs D',\bs y)\in\INT(\bs z)}}(-1)^{\alpha(\bs z)+\alpha(\bs y)}\sigma V_{n,p}(\bs \omega,\bs D-\bs D',\bs z).\label{deftcR}\end{equation}
Recall that $\INT(\bs z)$ is empty if $\bs z$ consists of a single repeated point; as a result  $\tcR (\cA \tcL V_{n,p}) = 0$ also for $(n,p)=(2,1),(4,0)$ (on the case $(n,p)=(2,0)$ we already commented in Remark \ref{rem:LB_projection}).
We are now ready to define: 
\begin{equation}
(\cR_\B V)_{2,2}:= \cA(V_{2,2}+(\tcR V)_{2,2}+ (\tcR(\tcR V))_{2,2}), \qquad 
(\cR_\B V)_{4,1}:=\cA (V_{4,1}+(\tcR V)_{4,1}).\end{equation}
Summarizing and rewriting the order indices,
\begin{equation}
(\cR_\B V)_{n,p}=\begin{cases} 0 &\text{if} \ (n,p)=(2,0), (2,1), (4,0),\\
\cA(V_{2,2}+\tcR V_{2,1}+ \tcR (\tcR V_{2,0})) & \text{if} \ (n,p)=(2,2),\\
\cA(V_{4,1}+\tcR V_{4,0}) & \text{if} \ (n,p)=(4,1),\\
\cA V_{n,p} & \text{otherwise.}\end{cases}\label{defcR}\end{equation}
By construction, $\cA V\sim \cL_\B V+ \cR_\B V$ with $\cL_\B V$ as in \eqref{defcL.1}. 
if $V$ is invariant under the symmetries described in  item \ref{it1:dop} after \eqref{eq:WfreeL1bound}
(horizontal translations, permutations, and reflections), then  $V \sim \cA V$ and 
\begin{equation} \label{eq:3.1.20..}V\sim \cL_\B V+ \cR_\B V. \end{equation}

\begin{remark}
	Examining the orders of the kernels appearing in \cref{defcR,defcL.1}, we see that $\cL_\B \cR_\B V = 0$ and $\cR_\B \cR_\B V = \cA \cR_\B V = \cR_\B V$.  
	Recalling that $\tcR (\cA \tcL V_{n,p}) =0$ for $(n,p)=(2,0),(2,1)$, we also see that $\cR_\B \cL_\B V =0$.

	Together with \cref{rem:LB_projection}, we see that $\cL_\B$ and $\cR_\B$ are orthogonal projections.  These properties (and the corresponding ones for further related operators introduced below) are used in the iterative decomposition we introduce in \cref{sec:treecyl}, in particular in the structure and definition of the trees involved in the expansions
\eqref{treeexponcyl} and \eqref{treeexpinfty}: we refer, e.g., to the condition that endpoints with scale label $h_v$ smaller than $2$ are the children of vertices with scale label $h_v-1$, see the sixth dotted item after \eqref{treeexpinfty}; see also the analogous item in \cite{AGG_part2} (i.e., the last dotted one before  \cite[Eq.(4.3.3)]{AGG_part2}) and the related explanation given there. 
\label{rem:RB_LB_ortho}
\end{remark}

\begin{remark} \label{rem:3.1}As already mentioned, in the infinite volume limit, the operators $\cL_\B,\cR_\B$ reduce to those introduced in \cite[Section~4.2]{AGG_part2}, to 
be denoted 
by $\cL_\infty,\cR_\infty$ in this paper. The definition of these infinite plane operators is the same as the one for $\cL_\B,\cR_\B$, respectively, 
modulo a few trivial replacement (e.g., the operator $\cA$ should be interpreted 
as the one that, besides anti-symmetrizing with respect to permutations, symmetrizes with respect to horizontal and vertical reflections in the infinite plane; $\tcL$ is defined via 
an equation analogous to the right side of \eqref{eq:cLtilde_kernel_def.2}, with the factor $(-1)^{\alpha(\bs y)}$ replaced by $1$ and the sum over $\bs y\in\Lambda^n$ replaced 
by $\bs y\in\Lambda_\infty^n$, with $\Lambda_\infty\equiv \mathbb Z^2$; etc). 
\end{remark}

\noindent{\bf The action of $\cL_\E$ and $\cR_\E$ on the sourceless part of the kernels.}

\noindent Given a kernel $V$ on $\cM_\Lambda$, we first introduce the following operators: 
\begin{eqnarray} &&
{\tcL}_\E V_{2,0}(\bs\omega,\bs 0,\bs z)=\left(\prod_{j=1}^2\delta_{z_j,z_\partial(\bs z)}\right) 
\sum_{\substack{\bs y \in \bar\Lambda^2: \\ z_\partial(\bs y)=z_\partial(\bs z)}}(-1)^{\alpha(\bs y)} V_{2,0}(\bs \omega,\bs 0,\bs y),\\
&& (\tcR_\E V)_{2,1}(\bs \omega,\bs D,\bs z)=\sum_{\substack{\sigma, \bs y:\\(\sigma,\bs D,\bs z)\in\INT_\E(\bs y)}} (-1)^{\alpha(\bs y)+\alpha(\bs z)}\sigma\,
V_{2,0}(\bs \omega,\bs 0,\bs y),\label{atty}
\end{eqnarray}
where $$z_\partial(\bs z)=\begin{cases} ((z_1)_1,0) & \text{if $(z_1)_2\le \lfloor M/2\rfloor$}, \\
((z_1)_1,M+1) & \text{otherwise}. \end{cases}$$ 
Moreover, in \eqref{atty}, $\bs y$ is summed over $\bar\Lambda^2$, and $\INT_\E(z_1,z_2)$ is the set of 
$(\sigma,(D_1,D_2), (y_1,y_2))\equiv ( \sigma, \bs D, \bs y)\in\{\pm\}\times\{0,\hat e_1,\hat e_2\}^2\times  \bar\Lambda^2$ such that: either
$y_1=z_1$, $D_1=0$, and $(\sigma, (0,D_2), (z_\partial(\bs z),y_2))\in\INT(z_\partial(\bs z),z_2)$; or $y_2=z_\partial(\bs z)$, $D_2=0$, and $(\sigma, (0,D_1), (z_\partial(\bs z),y_1))\in\INT(z_\partial(\bs z),z_1)$. 

\begin{remark}\label{rem:cancellation_boundary}
In connection with the definition of $z_\partial(\bs z)$, we recall that the field $\phi_{\omega,z}$ at $z=z_\partial(\bs z)\in\bar\Lambda\setminus\Lambda$ 
is defined in the sense of Remark \ref{rem:cancell}. 
Moreover, as discussed in the same remark, $\phi_{+,z}$ (resp. $\phi_{-,z}$) is equivalent to $0$ if the vertical component of $z$ is equal to $0$ (resp. to $M+1$).
\end{remark}

Given the definitions of $\tcL_\E$ and $\tcR_\E$, we define the edge-localization and edge-renormalization operators as follows: 
\begin{eqnarray}
&& (\cL_\E V)_{n,p}=\piecewise{
\cA(\tcL_\E V_{2,0}) & \text{if $(n,p)=(2,0)$}\\
0 & \text{otherwise},}\label{defcLE.1}\\
&& (\cR_\E V)_{n,p}=\piecewise{
0 & \text{if $(n,p)=(2,0)$}\\
\cA(V_{2,1}+(\tcR_\E V)_{2,1}) & \text{if $(n,p)=(2,1)$}\\
\cA V_{n,p} & \text{otherwise}}\label{REdef}
\end{eqnarray}
which satisfy 
$\cA V\sim \cL_\E V+\cR_\E V.$; when $V \sim \cA V$ (i.e.\ with the symmetries mentioned in connection with \cref{eq:3.1.20..}),
\begin{equation} \label{eq:3.1.200..}V\sim \cL_\E V+\cR_\E V.\end{equation}
As we shall see in \cref{lem:W_primitive_bound}, the scaling dimension of $\E$ terms in the effective interaction is improved by one, so that $\cR_\E$ in fact consists of only irrelevent terms.
A further crucial observation is that 
\begin{equation} \label{keycancellationedge}\cL_\E V\sim 0,\end{equation}
because $\cL_\E V$ is supported on field multilabels of the form $\Psi=((\omega_1,0,z), (\omega_2,0,z))$ with $z$ having vertical component $(z)_2\in \{0,M+1\}$, so that $\phi(\Psi)$ is equivalent to $0$, either because $\omega_1=\omega_2$, or because of the considerations in Remark \ref{rem:cancellation_boundary}. 
Moreover, 
in the same way as in \cref{rem:RB_LB_ortho,rem:LB_projection}
$\cL_\E$ and $\cR_\E$ act on $V$ 
as projection operators, orthogonal among each other. 

\begin{remark} As discussed in Section \ref{sec:3.1.3} below, the operators $\cL_\E$ and $\cR_\E$ will be applied to the edge part of (the sourceless part of) the effective potential, whose effective scaling dimension is better by one as compared to the corresponding `bulk scaling dimension'.
The intuition behind this improvement in the edge vs bulk scaling dimension is very simple, and is based on the general expression 
$d-n\Delta_\phi-p$ for the scaling dimension discussed in footnote \ref{scaldim}. For edge contributions, which are localized along the one-dimensional boundary of the cylinder, 
the effective spatial dimension to be used is $d=1$ rather than $d=2$. Using the fact that the scaling dimension of $\phi$ is $\Delta_\phi=1/2$ also along the boundary, 
we find that the scaling dimension of an `edge kernel' of order $(n,p)$ is $1-n/2-p$, see \eqref{defscaldimwithm} in Remark \ref{importremmE} below.
The rationale behind 
the definition \eqref{defcLE.1} is that $\cL_\E$, when acting on an edge kernel of order $(n,p)$,
returns a non-trivial output only in the case that $1-n/2-p=0$, corresponding to the 
edge marginal contributions. The key cancellation \eqref{keycancellationedge} tells us that the edge marginal contributions vanish. 
\end{remark}

\noindent{\bf The action of $\cL_\B$ and $\cR_\B$ on the effective source term.}

\noindent Consider a kernel $B:\mathcal M_\Lambda\times\cX_\Lambda\to \mathbb R$ such that $B_{n,p,0}=0$ for all $n\in2\mathbb N$ and $p=0,1,2$ 
(see item \ref{itemii:dop} after \eqref{eq:3.1.4} for the definition of $B_{n,p,m}$ with $m\ge 0$).
We define the action of the `bulk-localization' and `bulk-renormalization' operators $\cL_\B$ and $\cR_\B$ on $B$ as follows:
\begin{equation}\label{defcL.100}
(\cL_\B B)_{n,p,m}=\begin{cases} \cA\,  (\tcL B_{2,0,1}) &\text{if} \ (n,p,m)=(2,0,1),\\
0 & \text{otherwise}\end{cases}\end{equation}
and 
\begin{equation}
(\cR_\B B)_{n,p,m}=\begin{cases} 0 &\text{if} \ (n,p,m)=(2,0,1),\\
\cA(B_{2,1,1}+(\tcR B)_{2,1,1}) & \text{if} \ (n,p,m)=(2,1,1),\\
\cA B_{n,p,m} & \text{otherwise}\end{cases}\label{defcR.200}\end{equation}
where we recall that $\cA$ is the operator that anti-symmetrizes with respect to permutations of $\Psi$ and symmetrizes with respect to reflections, and 
$\tcL$ and $\tcR$ are defined in analogy with \eqref{eq:cLtilde_kernel_def.2} and \eqref{deftcR}; namely, specializing to the only cases of interest, and 
denoting by $z_x$ the left/bottom vertex of $x\in\fB_\Lambda$ (i.e., $x=z_x+\hat e_j/2$, for $j$ equal either to $1$ or $2$), 
\begin{equation} \begin{split}&
\tcL B_{2,0,1} \big((\bs\omega,\bs 0,\bs z), x\big)=\Big(\prod_{j=1}^2\delta_{z_j,z_x}\Big)
	  \sum_{\bs y \in \bar\Lambda^2}(-1)^{\alpha(\bs y)}
	  B_{2,0,1}\big((\bs \omega,\bs 0,\bs y),x\big),\\
& (\tcR B)_{2,1,1}\big((\bs \omega,\bs D,\bs z),x\big)=\sum_{\substack{\sigma=\pm,\, \bs y\in\bar\Lambda^2:\\(\sigma,\bs D,\bs z)\in\INT_x(\bs y)}} (-1)^{\alpha(\bs y)+\alpha(\bs z)}\sigma
B_{2,0,1}\big((\bs \omega,\bs 0,\bs y),x\big),\end{split}\label{deftcR.2zzw}\end{equation}
where, in the second line, $\INT_x(\bs y)$ is the set of 
$(\sigma,(D_1,D_2), (y_1,y_2))\equiv ( \sigma, \bs D, \bs y)\in\{\pm\}\times\{0,\hat e_1,\hat e_2\}^2\times \bar\Lambda^2$ such that: either
$y_1=z_x$, $D_1=0$, and $(\sigma, (0,D_2), (z_x,y_2))\in\INT(z_x,z_2)$; or $y_2=z_2$, $D_2=0$, and $(\sigma, (0,D_1), (z_x,y_1))\in\INT(z_x,z_1)$. 
The reader can check that once again the definitions \eqref{defcL.100}-\eqref{defcR.200} are such that 
\begin{equation} \label{eq:3.1.2000..}B\sim \cL_\B B+ \cR_\B B\end{equation}
for suitably symmetric kernels,
and that $\cL_\B$ and $\cR_\B$ continue to act as orthogonal projections as in \cref{rem:RB_LB_ortho}.

\begin{remark} \label{rem:3.4}As in Remark \ref{rem:3.1}, we denote by $\cL_\infty,\cR_\infty$ the infinite volume counterparts of $\cL_\B,\cR_\B$; when acting on 
the effective source term (see item  \ref{itemi:dop} after \eqref{eq:3.1.4}) associated with an infinite volume kernel, $\cL_\infty,\cR_\infty$ are defined via the right sides of
\eqref{defcL.100},\eqref{defcR.200}, with $\cA$ to be interpreted as in Remark \ref{rem:3.1}, and $\tcL,\tcR$ given by the infinite plane analogues of the right sides 
of \eqref{deftcR.2zzw} (in the first line of \eqref{deftcR.2zzw}, the factor $(-1)^{\alpha(\bs y)}$ should be replaced by $1$ and the sum on $\bs y$ should run over 
$\Lambda_\infty^2$; in the second line of \eqref{deftcR.2zzw}, the factor $(-1)^{\alpha(\bs y)+\alpha(\bs z)}$ should be replaced by $1$, the sum on $\bs y$ should run over 
$\Lambda_\infty^2$, and $\INT_x(\bs y)$ is defined as explained after \eqref{deftcR.2zzw}, with the only difference that $( \sigma, \bs D, \bs y)$ belongs to 
$\{\pm\}\times\{0,\hat e_1,\hat e_2\}^2\times \Lambda_\infty^2$ rather than to $\{\pm\}\times\{0,\hat e_1,\hat e_2\}^2\times \bar\Lambda^2$). 
\end{remark}

\subsubsection{Norm bounds on $\cR_\B$ and $\cR_\E$} \label{subsec:normRBRE}

In this subsection we collect a couple of technical estimates on the norms of $\cR_\B$ and $\cR_\E$, which will be useful in the following. 
Consider two real kernels $V:\mathcal M_\Lambda\to \mathbb R$
and $B:\mathcal M_\Lambda\times \cX_\Lambda\to \mathbb R$, invariant 
under horizontal translations (with the usual anti-periodic and periodic boundary conditions in $\bs z$ and $\bs x$, respectively), and $B$ such that 
$B_{n,p,0}\equiv 0$. 
If $V$ is supported in $\cM_\Lambda^\circ$, then, for any $\kappa\ge 0$ and $\epsilon,\epsilon'>0$, the action of $\cR_\B$ on $V$ can be bounded as follows
(we formulate the bound only for the choices of $(n,p)$ for which the action of $\cR_\B$ is non trivial, see \eqref{defcR}):
\begin{eqnarray}
&& \| (\cR_\B V)_{2,2}\|_{(\kappa)}\le \|V_{2,2}\|_{(\kappa)}+\epsilon^{-1}\|V_{2,1}\|_{(\kappa+\epsilon)}+(\epsilon')^{-2}\|V_{2,0}\|_{(\kappa+2\epsilon')},\label{staz.ok0}\\
&& \| (\cR_\B V)_{4,1}\|_{(\kappa)}\le \|V_{4,1}\|_{(\kappa)}+3\epsilon^{-1} \|V_{4,0}\|_{(\kappa+\epsilon)}, \label{staz.ok} \end{eqnarray}
where the norm  $\|V_{n,p}\|_{(\kappa)}$ is defined as:
\begin{equation} \|V_{n,p}\|_{(\kappa)}:=\sup_{\bs \omega}\,\sup_{z\in\bar\Lambda}\sum_{\substack{\bs z\in\bar\Lambda^n:\\ z_1=z}}e^{\kappa\delta(\bs z)}\sup_{\bs D}^{(p)}\,|V_{n,p}(\bs \omega,\bs D,\bs z)|\label{eq:weightnorm}\end{equation}
(here the label $(p)$ on the sup over $\bs D$ indicates the constraint that $\|\bs D\|_1=p$, $\delta$ denotes the tree distance, defined after \eqref{eq:WL1bound}, and 
it is understood that the summand in the right side vanishes if $(\bs \omega, \bs D, \bs z)\not\in\cM_\Lambda$).
The definition \eqref{eq:weightnorm} is the finite volume analogue of \cite[Eq.~(4.2.21)]{AGG_part2}, and the 
proof of \eqref{staz.ok0} and \eqref{staz.ok} is the same as the one leading to \cite[Eqs.~(4.2.24), (4.2.25)]{AGG_part2}, which we refer the reader
to for additional details. A repetition of the same proof provides the following bounds on the action of $\cR_\E$ on $V$ and of $\cR_\B$ on $B$ (no need that $V$ is supported 
in the interior of $\Lambda$, here; again, 
we restrict our attention to the choice of the labels $n,p,m$ corresponding to a non-trivial action of the renormalization operators, see \eqref{REdef} and \eqref{defcR.200}): 
\begin{equation}
\|(\cR_\E V)_{2,1}\|_{(\E;\kappa)}\le \|V_{2,1}\|_{(\E;\kappa)}+ \frac2{\epsilon}\|V_{2,0}\|_{(\E;\kappa+\epsilon)},\label{eq:REbound}\end{equation}
\begin{equation}\label{eq:RBB}
\|(\cR_\B B)_{2,1,1}(x)\|_{(\kappa)}\le \|B_{2,1,1}(x)\|_{(\kappa)}+ \frac2{\epsilon}\|B_{2,0,1}(x)\|_{(\kappa+\epsilon)},\end{equation}
where, letting the symbol $* \parallel z_1$ on a sum indicate the constraint that the sum is taken with the horizontal coordinate of $z_1$ fixed, 
$\|V_{n,p}\|_{(\E;\kappa)}$ is defined as
\begin{equation}\|V_{n,p}\|_{(\E;\kappa)}:=\sup_{\bs \omega}\sum_{\bs z \in \bar\Lambda^n}^{* \parallel z_1}
		e^{\kappa \delta_\E(\bs z)}\sup_{\bs D}^{(p)}
		\left| V_{n,p}(\bs \omega, \bs D, \bs z) \right|,
		\label{eq:cRE_sumbound}\end{equation}
where $\delta_\E$ is the `edge' tree distance, defined after \eqref{eq:WE_base_decay}, and $\|B_{n,p,m}(\bs x)\|_{(\kappa)}$ is defined as 
\begin{equation}
\|B_{n,p,m}(\bs x)\|_{(\kappa)}:=\sup_{\bs \omega}\sum_{\bs z \in \bar\Lambda^n}e^{\kappa \delta(\bs z,\bs x)}\sup_{\bs D}^{(p)}
		\left| B_{n,p,m}\big((\bs \omega, \bs D, \bs z),\bs x\big) \right|.\label{normonsource}\end{equation}
Again, both in \eqref{eq:cRE_sumbound} and in \eqref{normonsource}, it is understood that the summand must be interpreted as zero if $(\bs \omega,\bs D,\bs z)\not\in\cM_\Lambda$. 
 In the following, we will use the bounds \eqref{staz.ok0}, \eqref{staz.ok}, \eqref{eq:REbound} and \eqref{eq:RBB} to evaluate the size of the renormalized part of the effective potential on scale $h$. In such a case, $\kappa$ will be chosen of the order $2^{h}$. Let us also anticipate that, in order to bound the size of the edge part of the effective source term, we will need an additional norm, 
which we introduce here for later reference: 
\begin{equation}\|B_{n,p,m}(\bs x)\|_{(\E;\kappa)}:=\sup_{\bs \omega}\sum_{\bs z \in \bar\Lambda^n}e^{\kappa \delta_\E(\bs z,\bs x)}\sup_{\bs D}^{(p)}
		\left| B_{n,p,m}\big((\bs \omega, \bs D, \bs z),\bs x\big) \right|,\label{normonsourceEdge}
		\end{equation}
where the summand is interpreted as zero if $(\bs \omega, \bs D, \bs z)\not\in\cM_\Lambda$. 

\begin{remark} If $B$ is the effective source term associated with an infinite volume kernel, the infinite volume analogue of the the bulk norm 
\eqref{normonsource} is given by the right side of \eqref{normonsource}, with the sum on $\bs z$ running over $\Lambda_\infty^n$, rather than over $\bar\Lambda^n$. 
In terms of this norm, the analogue of \eqref{eq:RBB} holds, namely:
\begin{equation}\label{eq:RBBinfty}
\|(\cR_\infty B)_{2,1,1}(x)\|_{(\kappa)}\le \|B_{2,1,1}(x)\|_{(\kappa)}+ \frac2{\epsilon}\|B_{2,0,1}(x)\|_{(\kappa+\epsilon)}.\end{equation}
\end{remark}

\subsubsection{The action of $\cL$ and $\cR$ on $\cV^{(h)}$ and the running coupling constants}\label{sec:3.1.3}

We finally have all the necessary ingredients for defining the desired decomposition of the kernel of the effective potential \eqref{eq:formalh} on scale $h\le 0$ into 
its local and renormalized parts. Recall that, as discussed after \eqref{eq:formalh}, $W_\Lambda^{(h)}$ is inductively 
assumed to satisfy the (analogue of the) symmetry property 
described in item \ref{it1:dop} after \eqref{eq:WfreeL1bound}; moreover, we assume the existence of an infinite volume kernel $W_\infty^{(h)}$, 
in term of which we define the bulk and edge parts of $W_\Lambda^{(h)}$ as in \eqref{eq:3.1.4} and \eqref{eq:3.1.4bis}, respectively (in particular, we recall that its sourceless 
part, $V_\infty^{(h)}$ is the same constructed in \cite[Section~4]{AGG_part2}). 

We first write $W_\Lambda^{(h)}=V_\Lambda^{(h)}+B_{\Lambda}^{(h)}$ (recall item \ref{itemi:dop} after \eqref{eq:3.1.4}) and 
decompose both $V_\Lambda^{(h)}$ and $B_\Lambda^{(h)}$ in their bulk and edge parts, 
\begin{equation} V_\Lambda^{(h)}=V_\B^{(h)}+V_\E^{(h)}, \qquad B_\Lambda^{(h)}=B_\B^{(h)}+B_\E^{(h)}.\label{eq:3.1.40;}\end{equation}
Next, by using the decompositions \eqref{eq:3.1.20..} 
and \eqref{eq:3.1.2000..} on $V_\B^{(h)}$ and $B_\B^{(h)}$, respectively, 
we rewrite
\begin{equation} V_\Lambda^{(h)}\sim \cL_\B V_\B^{(h)}+\cR_\B V_\B^{(h)}+V_\E^{(h)}, \qquad B_\Lambda^{(h)}\sim \cL_\B B_\B^{(h)}+\cR_\B B_\B^{(h)}+B_\E^{(h)}.\label{eq:3.1.41;}\end{equation}

We further decompose $\cL_\B V_\B^{(h)}$, $\cR_\B V_\B^{(h)}$, $\cL_\B B_\B^{(h)}$ and $\cR_\B B_\B^{(h)}$ in their bulk and edge parts, by letting, 
 for any $\Psi=(\bs\omega,\bs D,\bs z)\in \cM_\Lambda$,
\begin{equation}(\cL_\B V_\B^{(h)})_\B(\Psi):=(-1)^{\alpha(\bs z)}I_\Lambda(\Psi) (\cL_\infty V_\infty^{(h)})(\Psi), \qquad (\cL_\B V_\B^{(h)})_\E:= \cL_\B V_\B^{(h)}-(\cL_\B V_\B^{(h)})_\B,\end{equation}
and similarly for $(\cR_\B V_\B^{(h)})_\B, (\cR_\B V_\B^{(h)})_\E, (\cL_\B B_\B^{(h)})_\B, (\cL_\B B_\B^{(h)})_\E, (\cR_\B B_\B^{(h)})_\B, (\cR_\B B_\B^{(h)})_\E$.
Next, we define 
\begin{equation} \begin{split} & \cL V_\Lambda^{(h)}:=(\cL_\B V_\B^{(h)})_\B=2^h\nu_h F_{\nu,\B}+\zeta_h F_{\zeta,\B}+\eta_h F_{\eta,\B}\equiv \upsilon_h\cdot F_\B,\\
& \cL B_\Lambda^{(h)} := (\cL_\B B_\B^{(h)})_\B = Z_{1,h} F_{1,\B}+Z_{2,h} F_{2,\B}\equiv Z_h\cdot F_\B^A,\\
& \mathcal E V_\B^{(h)}:=(\cL_\B V_\B^{(h)})_\E+(\cR_\B V_\B^{(h)})_\E, \qquad \mathcal E B_\B^{(h)}:=(\cL_\B B_\B^{(h)})_\E+(\cR_\B B_\B^{(h)})_\E, \end{split}\label{eq:3.1.42(}\end{equation}
where:
\begin{itemize}
\item $F_{\nu,\B},F_{\zeta,\B},F_{\eta,\B}, F_{j,\B}$ are the $\cA$-invariant kernels associated with the following `local' potentials: 
\begin{eqnarray} 
&&\cF_{\nu,\B}:=\sum_{z\in\Lambda}\phi_{+,z}\phi_{-,z}, \quad 
\cF_{\zeta,\B}:=\sum_{\omega=\pm}\sum_{z\in\Lambda} \omega\phi_{\omega,z}\dd_1\phi_{\omega,z}, \quad 
\cF_{\eta,\B}:=\sum_{\omega=\pm}\sum_{z\in\Lambda}\phi_{\omega,z}\hdd_2\phi_{-\omega,z},\nonumber\\
&& \cF_{j,\B}:= \frac12\sum_{j=1,2}\sum_{x\in\fB_{j,\Lambda}} \sum_{\sigma=0,1}A_x \phi_{+,z_x+\sigma\hat e_j}\phi_{-,z_x+\sigma\hat e_j}, 
\label{defFjB}
\end{eqnarray}
where $\dd_1$, $\hdd_2$ are symmetric discrete derivatives, acting on Grassmann variables $\phi_{\omega,z}$ with $z\in\Lambda$ as
$\dd_1 \phi_{\omega,z}:=\frac12(\partial_1 \phi_{\omega,z}+\partial_1\phi_{\omega,z-\hat e_j})$\footnote{
It is understood that: if $(z)_1=1$, then $\partial_1\phi_{\omega,z-\hat e_1}\equiv -\partial_1\phi_{\omega,(L,(z)_2)}\equiv\phi_{\omega,z}+\phi_{\omega,(L,(z)_2)}$;
if $(z)_1=L$, then $\partial_1\phi_{\omega,z}\equiv -\phi_{\omega,(1,(z)_2)}-\phi_{\omega,z}$.},
and $\hdd_2 \phi_{\omega,z}:=\frac12(\mathds 1(z+\hat e_2\in\Lambda)\,\partial_2 \phi_{\omega,z}+\mathds 1(z-\hat e_2\in\Lambda)\,\partial_2\phi_{\omega,z-\hat e_j})$;
moreover, in the last line, $\fB_{1,\Lambda}$ and $\fB_{2,\Lambda}$ are the subsets of $\fB_\Lambda$ 
consisting of the horizontal and vertical edges, respectively.
\item The constants $\nu_h, \zeta_h, \eta_h$, called the {\it running coupling constants}, only depend on $V_\infty^{(h)}$ and are the same 
	defined in \cite[Eqs.~(4.2.30), (4.5.2)]{AGG_part2} 
and studied in \cite[Section~4.5]{AGG_part2}, see Remark \ref{remacon} below. 
The constants $Z_{j,h}$ with $j=1,2$ are defined as 
\begin{equation} \label{eeqq:3.1.42}Z_{j,h}=4(\cL_\infty B_\infty^{(h)})_{2,0,1}((\bs \omega,\bs 0,\bs 0),x)\Big|_{\bs \omega=(+,-)}^{x=\hat e_j/2}
=2\sum_{\bs y\in\Lambda_\infty^2}(B_\infty^{(h)})_{2,0,1}((\bs \omega,\bs 0,\bs y),x)\Big|_{\bs \omega=(+,-)}^{x=\hat e_j/2}\end{equation}
and are called the {\it effective vertex renormalizations}. 
\end{itemize}

Several remarks are in order: 

\begin{remark} 
The fact that $(\cL_\B V_\B)_\B$ and $(\cL_\B B_\B)_\B$ have the simple structure stated in the first two lines of \eqref{eq:3.1.42(} follows from the definitions
and the fact that $W_\infty^{(h)}$ is invariant\footnote{The stated invariance of $W_\infty^{(h)}$ follows from its definition, see \eqref{treeexpinfty} and following definitions.}
under translations and under the action of (the infinite volume analogue of) $\cA$, cf.\ \cite[Eq.~(4.2.30)]{AGG_part2}. 
\end{remark}

\begin{remark}\label{remacon} As mentioned above, the sequence $\allct=\{(\nu_h,\zeta_h,\eta_h)\}_{h\le 0}$ has been 
studied in \cite[Section~4.5]{AGG_part2}. In particular, in
\cite[Propositions~4.10 and~4.11]{AGG_part2} we proved that, for $\lambda$ small enough, there exist  
functions $\beta_c(\lambda)$, $t_1^*(\lambda)$ and $Z(\lambda)$, such that, fixing 
$t_1=\tanh(\beta_c(\lambda) J_1)$, $t_2=\tanh(\beta_c(\lambda) J_2)$, $t_1^*=t_1^*(\lambda)$ and $Z=Z(\lambda)$ in \eqref{eq:startfrom}, 
the corresponding sequence of running coupling constants satisfies the following bound, for all $0\le \vartheta<1$ and a suitable $K_\vartheta>0$:
\begin{equation} \epsilon_h:=\max\{|\nu_h|, |\zeta_h|, |\eta_h|\}\le K_\vartheta \left|\lambda\right| 2^{\vartheta h}.\label{eq:ct:short_memory}\end{equation}
The functions $\beta_c(\lambda), t_1^*(\lambda), Z(\lambda)$ are analytic in $\lambda$, and such that the differences $\beta_c(\lambda)-\beta_c(J_1,J_2)$, $t_1^*(\lambda)-t_1, Z(\lambda)-1$ are all of order $O(\lambda)$ (recall that $\beta_c(J_1,J_2)$ is the unperturbed critical temperature, defined a few lines before \eqref{media}). 
From now on, we assume $\beta=\beta_c(\lambda)$ and $t_1,t_2, t_1^*,Z$ to be fixed in this special way, so that \eqref{eq:ct:short_memory} holds. 
One of the key goals of the following sections is to show that, once that $\beta, t_1,t_2,t_1^*,Z$ are fixed in this special way, not only 
the running coupling constants are well defined uniformly in $h$ and infinitesimal as $h\to-\infty$, but the whole sequence of 
effective potentials is well defined and analytic in $\lambda$, uniformly in the scale label and in the volume, and their kernels satisfy natural 
dimensional bounds, proved in Section \ref{sec:informal_bounds} below.
\end{remark}

\begin{remark}\label{remar35}
If $(n,p)\neq(2,0),(2,1),(2,2),(4,0),(4,1)$, then $(\mathcal E V_\B^{(h)})_{n,p}=((V_\B^{(h)})_{n,p})_\E$, which is {\it zero}, because it is the edge part of the bulk part
of a kernel; moreover, $(\mathcal E V_\B^{(h)})_{4,0}=0$, by the very definition of $(\cL_\B V_\B^{(h)})_{4,0}$ and $(\cR_\B V_\B^{(h)})_{4,0}$.
Note also that, even if $(V_\B^{(h)})_\E=0$ and $V_\B^{(h)}\sim \cL_\B V_\B^{(h)}+ \cR_\B V_\B^{(h)}$, the kernel 
$\mathcal E V_\B^{(h)}=(\cL_\B V_\B^{(h)})_\E+ (\cR_\B V_\B^{(h)})_\E$ is not equivalent to zero, because the extraction of the edge part does not commute with 
the operators $\cL_\B, \cR_\B$. Similar considerations are valid for $\mathcal E B_\B^{(h)}$. In particular, if $(n,p,m)\neq(2,0,1),(2,1,1)$, then $(\mathcal E B_\B^{(h)})_{n,p,m}=((B_\B^{(h)})_{n,p,m})_\E=0$.
\end{remark}

\begin{remark} 
	\label{rem:belissemo}
	Using the decay bounds for $V_\infty^{(h)}$ proved in \cite[Lemma~4.8]{AGG_part2} and the bound on $\epsilon_h$ in \eqref{eq:ct:short_memory}, it follows that the kernel $(\mathcal E V_\B^{(h)})_{n,p}$, whenever it does not vanish (see previous remark), satisfies, for the same $c_0$ as in Eq.\eqref{cosacomedove}:
\begin{equation}
\label{belissemo} 
	\|(\cE V_\B^{(h)})_{n,p}\|_{(\E;\frac{3}4 c_02^h)}
	\le 
	C |\lambda| 2^{-h}2^{\theta h} 2^{(2-\frac{n}2-p)h},
\end{equation}
where the norm in the left side was defined in \eqref{eq:cRE_sumbound}, 
and, in the right side, $\theta=3/4$\footnote{The choice $\theta=3/4$ is arbitrary and made just for definiteness: the bound remains valid for any fixed $\theta$ smaller than $1$; the constant $C$ may diverge as $\theta\to1^-$.}.
See Appendix \ref{app:segunda} for the proof. Similar bounds on $\mathcal E B_\B^{(h)}$ will be proved in Sect.\ref{sec:Zj} below. 
\end{remark}
In light of \eqref{eq:3.1.41;} and \eqref{eq:3.1.42(}, we have: 
\begin{equation} \begin{split} & V_\Lambda^{(h)}\sim \upsilon_h\cdot F_\B+(\cR_\B V_\B^{(h)})_\B+\mathcal EV_\B^{(h)}+V_\E^{(h)},\\
& B_\Lambda^{(h)}\sim Z_h\cdot F^A_\B+
(\cR_\B B_\B^{(h)})_\B+\mathcal E B_\B^{(h)}+B_\E^{(h)}.\end{split}\label{eq:3.1.46;}\end{equation}
Finally, using the decomposition \eqref{eq:3.1.200..} on $\mathcal EV_\B^{(h)}+V_\E^{(h)}$ and the cancellation \eqref{keycancellationedge}, we write
$\mathcal E V_\B^{(h)}+V_\E^{(h)}\sim \cL_\E(\mathcal EV_\B^{(h)}+ V_\E^{(h)})+\cR_\E (\mathcal E V_\B^{(h)}+V_\E^{(h)})\sim \cR_\E (\mathcal E V_\B^{(h)}+V_\E^{(h)})$, so that, in conclusion, for all $h\le 0$, 
the kernel $W_\Lambda^{(h)}=V_\Lambda^{(h)}+B_\Lambda^{(h)}$ can be decomposed as 
\begin{equation} W_\Lambda^{(h)}\sim  \upsilon_h\cdot F_\B+Z_h\cdot F^A_\B + \cR W_\Lambda^{(h)},\label{eqfin1.}\end{equation}
where $\upsilon_h\cdot F_\B+Z_h\cdot F^A_\B\equiv \cL W_\Lambda^{(h)}$ and 
\begin{equation} \cR W_\Lambda^{(h)}=\big[(\cR_\B V_\B^{(h)})_\B+(\cR_\B B_\B^{(h)})_\B\big]+\big[\cR_\E(\mathcal E V_\B^{(h)}+V_\E^{(h)})+\mathcal E B_\B^{(h)}+B_\E^{(h)}\big].
\label{RWdecomp.4}\end{equation}
The expressions in the first and second brackets in the right side are, by definition, the bulk and edge parts of $\cR W_\Lambda$, respectively. 

\subsection{The tree expansion for the effective potential on the cylinder}\label{sec:treecyl}

Once that the action of the operators $\cL$ and $\cR$ on $\cV^{(h)}$ has been defined, we are ready to define the recursion formulas for the kernels of the effective potential. At the first step, recalling the representation \eqref{eq:forma1} for the effective potential $\cV^{(1)}$ on scale $h=1$ and the definition \eqref{eq:Vcyl_N} of 
the effective potential $\cV^{(0)}$ on scale $h=0$, we find that the kernel of $\cV^{(0)}$ admits a representation analogous to \eqref{Wh-1trun} (with $\mathbb E^{(1)}$ 
represented as in \eqref{eq:BBF} and $\fG_{T}^{(1)}$ defined as after \eqref{eq:BBF}, with $g^{(1)}_{\omega\omega'}(z,z')$ the elements of the $2\times2$ matrix 
$\fg^{(1)}(z,z')\equiv \fg_m(z,z')$); that is, for any $(\Psi,\bs x)\in\cM_{0,\Lambda}\times \cX_\Lambda$ (recall that $\cM_{0,\Lambda}$ is the subset of $\cM_{1,\Lambda}$ consisting of tuples $\Psi\in(\{+,-\}\times \Lambda)^n$ for some $n\in 2\mathbb N$):
\begin{eqnarray}
&&W^{(0)}_\Lambda(\Psi,\bs x) =\sum_{s=1}^{\infty}\frac{1}{s!}
	\sum_{\substack{\Psi_1,\ldots,\Psi_s\in \cM_{1,\Lambda}\\ \bs x_1,\ldots,\bs x_s\in \cX_\Lambda}}^{(\Psi, \bs x)}\
	\sum_{T \in \cS(\bar\Psi_1,\ldots,\bar\Psi_s)}
	\fG_{T}^{(1)}(\bar\Psi_1,\ldots,\bar\Psi_s)
	\cdot\nonumber\\
	&&\hskip1.9truecm\cdot\ \alpha(\Psi;\Psi_1,\ldots,\Psi_s)\left(  \prod_{j=1}^s W^{(1)}_\Lambda(\Psi_j, \bs x_j)\right)
.\label{eq:BBF0cyl}\end{eqnarray}
Recall that $W_\Lambda^{(1)}$ in the right side of \eqref{eq:BBF0cyl} can be decomposed in its bulk and edge parts: $W_{\Lambda}^{(1)}=W_\B^{(1)}+
W_\E^{(1)}$. The single-scale contributions to the generating function, $w_\Lambda^{(0)}(\bs x)$, 
is defined in the same way as \eqref{eq:BBF0cyl}, with $\Psi$ replaced by the empty set, except that there is no term with $s=1$ and $\Psi_{1}=\emptyset$.

The definition of the kernel of the effective potential for $h^*<h\le 0$ is similar: for any $(\Psi,\bs x)\in \cM_\Lambda\times\cX_\Lambda$, we let 
\begin{eqnarray}&&
\hskip-.6truecm
W^{(h-1)}_\Lambda(\Psi,\bs x) =\sum_{s=1}^{\infty}\frac{1}{s!}
	\sum_{\substack{\Psi_1,\ldots,\Psi_s\in \cM_{\Lambda}\\\bs x_1,\ldots,\bs x_s\in \cX_\Lambda}}^{(\Psi,\bs x)}\
	\sum_{T \in \cS(\bar\Psi_1,\ldots,\bar\Psi_s)}
	\fG_{T}^{(h)}(\bar\Psi_1,\ldots,\bar\Psi_s)\cdot\nonumber\\
&&\hskip-.4truecm\cdot\ \alpha(\Psi;\Psi_1,\ldots,\Psi_s)
\Biggl(  \prod_{j=1}^s \Big[\upsilon_h \cdot F_{\B}(\Psi_j)+Z_h\cdot F_\B^A(\Psi_j,\bs x_j)+\cR W^{(h)}_\Lambda(\Psi_j,\bs x_j)\Big]\Biggr),\label{eq:BBFh-1cyl}\end{eqnarray}
which is the analogue of \eqref{Wh-1trun}-\eqref{eq:BBF}, with $W_\Lambda^{(h)}$ decomposed as in \eqref{eqfin1.}. Recall that $\cR W_\Lambda^{(h)}$ in the right side of \eqref{eq:BBFh-1cyl} can be decomposed in its bulk and edge parts, as in \eqref{RWdecomp.4}. 
Note also that, with some abuse of notation, 
in \eqref{eq:BBFh-1cyl} we denoted by $\fG_{T}^{(h)}(Q_1,\ldots, Q_s)$ a function that is not exactly the same as the one defined in and after \eqref{eq:BBF},
but a generalized one, differing from it for one important feature: since now the field multilabels $Q_i$ have the form $(\bs \omega_i,\bs D_i,\bs z_i)$, with $\bs D_i$ different from $\bs 0$, in general, the propagators $g_\ell^{(h)}$ appearing in the second line of \eqref{eq:BBF}, with $\ell=((\omega_i,D_i,z_i),(\omega_j,D_j,z_j))$,
should now be interpreted as $\partial^{D_{i\phantom{j}}}_{z_i}\!\partial^{D_j}_{z_j}g^{(h)}_{\omega_i\omega_j}(z_i,z_j)$, and similarly 
for the propagators appearing in ${\rm Pf}(G_{Q_1,\ldots,Q_s,T}(\bs t))$. Of course, for any $h^*<h\le 0$, the single-scale contributions to the generating function, $w_\Lambda^{(h-1)}(\bs x)$, 
is defined in the same way as \eqref{eq:BBFh-1cyl}, with $\Psi$ replaced by the empty set, except that there is no term with $s=1$ and $\Psi_{1}=\emptyset$.

Finally, at the last step, we write $w^{(h_*-1)}_\Lambda(\bs x)$ as in \eqref{eq:BBFh-1cyl} with $\Psi=\emptyset$
(and the usual convention that there is no term with $s=1$ and $\Psi_1=\emptyset$), 
and the understanding that $\fG_{T}^{(h^*)}(\bar\Psi_1,\ldots,\bar\Psi_s)$ is the analogue of 
$\fG_{T}^{(h)}(\bar\Psi_1,\ldots,\bar\Psi_s)$, with $\fg^{(\le h^*)}$ replacing $\fg^{(h)}$. 

\medskip

For the following discussion, we will also need recursion formulas for the bulk and edge parts of the kernel of the effective potential: these are obtained, 
starting from \eqref{eq:BBF0cyl} or \eqref{eq:BBFh-1cyl}, by 
letting $W_\E^{(h-1)}(\Psi,\bs x)=W_\Lambda^{(h-1)}(\Psi,\bs x)-W_\B^{(h-1)}(\Psi,\bs x)$, where, for $\Psi=(\bs\omega,\bs D,\bs z)$, 
$W_\B^{(h-1)}(\Psi,\bs x)=(-1)^{\alpha(\bs z)} I_\Lambda(\Psi,\bs x)\, W_\infty^{(h-1)}(\Psi_\infty,\bs x_\infty)$ (here $I_\Lambda$ is defined as in \eqref{eq:3.1.4} and 
$\Psi_\infty=(\bs\omega,\bs D,\bs z_\infty)$ with $(\bs z_\infty,\bs x_\infty)\equiv(\bs z,\bs x)_\infty$ the infinite volume representative of $(\bs z,\bs x)$ in the sense of item \ref{it3:dop} after \eqref{eq:WfreeL1bound}) and  
$W_\infty^{(h-1)}(\Psi,\bs x)$ is the solution to the following infinite volume recursive equations: 
\begin{eqnarray}
&&W^{(0)}_\infty(\Psi,\bs x) =\sum_{s=1}^{\infty}\frac{1}{s!}
	\sum_{\substack{\Psi_1,\ldots,\Psi_s\in \cM_{1,\infty}\\ \bs x_1,\ldots,\bs x_s\in \cX_\infty}}^{(\Psi, \bs x)}\
	\sum_{T \in \cS(\bar\Psi_1,\ldots,\bar\Psi_s)}
	\fG_{T,\infty}^{(1)}(\bar\Psi_1,\ldots,\bar\Psi_s)
	\cdot\nonumber\\
	&&\hskip1.9truecm\cdot\ \alpha(\Psi;\Psi_1,\ldots,\Psi_s)\left(  \prod_{j=1}^s W^{(1)}_\infty(\Psi_j, \bs x_j)\right)
.\label{eq:BBF0infty}\end{eqnarray}
and 
\begin{eqnarray}&&
\hskip-.6truecm
W^{(h-1)}_\infty(\Psi,\bs x) = \sum_{s=1}^{\infty}\frac{1}{s!}
	\sum_{\substack{\Psi_1,\ldots,\Psi_s\in \cM_{\infty}\\\bs x_1,\ldots,\bs x_s\in \cX_\infty}}^{(\Psi,\bs x)}\
	\sum_{T \in \cS(\bar\Psi_1,\ldots,\bar\Psi_s)}
	\fG_{T,\infty}^{(h)}(\bar\Psi_1,\ldots,\bar\Psi_s)\nonumber\\
&& \hskip-.3truecm \cdot\ \alpha(\Psi;\Psi_1,\ldots,\Psi_s)\Biggl(  \prod_{j=1}^s \Big[\upsilon_h \cdot F_{\infty}(\Psi_j)+Z_h\cdot F_\infty^A(\Psi_j,\bs x_j)+\cR_\infty W^{(h)}_\infty(\Psi_j,\bs x_j)\Big]\Biggr), 
	\label{eq:BBFh-1infty}\end{eqnarray}
where $\cR_\infty W_\infty^{(h)}=\cR_\infty V_\infty^{(h)}+\cR_\infty B_\infty^{(h)}$ is defined as explained in Remarks \ref{rem:3.1} and \ref{rem:3.4}; similarly, all the other
quantities with the label $\infty$ are the infinite volume analogues of their finite-cylinder counterpart, obtained by replacing $\Lambda$ by $\mathbb Z^2$, $\fB_\Lambda$ by the set of nearest neighbor edges of the infinite square lattice, $\fg^{(h)}(z,z')$ by its infinite volume limit $\fg_\infty^{(h)}(z-z'):=
\lim_{L,M\to\infty}\fg^{(h)}(z+z_{L,M},z'+z_{L,M})$ (with $z_{L,M}$ defined as after \eqref{eq:WZ2}), etc. 
The infinite volume recursion equations 
have been introduced and studied in \cite[Section~4]{AGG_part2}, in the special case that $\bs x=\emptyset$ (i.e., for the sourceless 
part of the infinite plane effective potential), 
which we refer the reader to for additional details on the definitions of the sets and of the functions entering \eqref{eq:BBF0infty}-\eqref{eq:BBFh-1infty}, 
such as $\cM_\infty, \fG_{T,\infty}^{(h)}, F_\infty$.

\medskip

The systematic, iterative, use of \eqref{eq:BBF0cyl}-\eqref{eq:BBFh-1cyl} and of the bulk-edge decomposition, allows us to express $W_\Lambda^{(h)}$ purely in terms of 
$\{\nu_k, \zeta_k, \eta_k, Z_{1,k}, Z_{2,k}\}_{h<k\le 0}$, $\{\mathcal EV_\B^{(k)}\}_{h<k\le 0}$, 
$\{\mathcal EB_\B^{(k)}\}_{h<k\le 0}$, $W_\Lambda^{(1)}$ and of the propagators $\{\fg^{(k)}\}_{h<k\le 1}$, $\{\fg^{(k)}_\infty\}_{h<k\le 1}$. 
The result can be conveniently expressed in terms of a {\it Gallavotti-Nicol\`o} (GN) tree expansion, analogous to the one described in 
\cite[Section~4.3]{AGG_part2} for $V_\infty^{(h)}$. We refer the 
reader to \cite[Section~4.3]{AGG_part2} for details about the derivation of the GN tree expansion and the definition of GN trees in the infinite volume case. 

A repetition of the proof leading to \cite[Eqs.~(4.3.3) and~(4.3.4)]{AGG_part2} implies that 
\begin{equation} \begin{split} 
& W_\Lambda^{(h)}\sim \sum_{\tau\in \cT^{(h)}}W_\Lambda[\tau], \quad \text{with} \quad 
W_\Lambda[\tau]=\sum_{\substack{\ul P\in\cP(\tau): \\ P_{v_0}\neq\emptyset}}\ \sum_{\ul T\in\cS(\tau,\ul P)}\sum_{\ul D\in \cD(\tau,\ul P)}W_\Lambda[\tau,\ul P,\ul T, \ul D], \\
& w_\Lambda^{(h)}\sim \sum_{\substack{\tau\in \cT^{(h)}:\\ m_{v_0}>0}}^*w_\Lambda[\tau], \quad \text{with} \quad 
w_\Lambda[\tau]=\sum_{\substack{\ul P\in\cP(\tau):\\ P_{v_0}=\emptyset}}\ \sum_{\ul T\in\cS(\tau,\ul P)}\sum_{\ul D\in \cD(\tau,\ul P)}W_\Lambda[\tau,\ul P,\ul T, \ul D], \end{split}
\label{treeexponcyl}\end{equation}
where the $*$ in the first sum in the second line indicates the constraint that the root $v_0$ of $\tau$ is `dotted' (see below: all the involved notions, sets and functions will be defined shortly); moreover, $W_\infty^{(h)}$ is defined via the infinite volume analogue of the first line of \eqref{treeexponcyl}:
\begin{equation} W_\infty^{(h)}:=\sum_{\tau\in \cT^{(h)}}W_\infty[\tau], \quad \text{with} \quad 
W_\infty[\tau]=\sum_{\substack{\ul P\in\cP(\tau): \\ P_{v_0}\neq\emptyset}}\ \sum_{\ul T\in\cS(\tau,\ul P)}\sum_{\ul D\in \cD(\tau,\ul P)}W_\infty[\tau,\ul P,\ul T, \ul D].
\label{treeexpinfty} \end{equation}
In \eqref{treeexponcyl}, 
the set $\cT^{(h)}$ is a family of GN trees similar to the family $\cT_\infty^{(h)}$ defined in \cite[Section~4.3]{AGG_part2}, 
whose generic element, denoted $\tau\in\cT^{(h)}$, see Fig.\ref{fig:edge_tree} for an example, is characterized as follows: 

\begin{figure}
	\centering
	  \begin{tikzpicture}[scale=0.8]
      \foreach \x in {3,...,9}
      {
	  \draw[very thin] (\x ,-1) -- (\x , 8);
      }
      \draw (3,-1.5) node {$h+1$};
      \draw (9,-1.5) node {2};
      \draw (3,4) node[vertex,E, label=-135:{$v_0$}] (vbar) {};
      \draw (vbar) -- ++ (1,0.75) node[vertex,E] {} -- ++(1,0.75) node[vertex,E] (vt1) {} -- ++ (2,0.67) node[vertex] (vt11) {}
      -- ++(2,-0.5) node[vertex] {};
      \draw (vt11) -- ++(1,0.67) node[ctVertex] (t1e1) {}; 
      \draw (vt1) -- ++(1,-0.5) node[bareProbeEP] (t1e2) {};
      \draw (vbar) -- (5,2) node[vertex,E] (v1) {} -- (7,2.33) node[vertex,E] (vt2) {} -- (8,3) node[ctVertex,E] {};
      \draw (vt2) -- ++(1,-0.33) node[vertex] (t2e1) {} -- ++(1,-0.33) node[vertex] (t2e2) {};
      \draw (v1) -- ++(2,-1) node[vertex,E] (t3v1) {} -- +(2,-1) node[bareProbeEP,E] {};
    \end{tikzpicture}
    \caption{Example of a tree $\tau \in \cT^{(h)}$. Note that the white vertices are `hereditary', that is, if $w\in V(\tau)$ is white, then all the dotted vertices 
    preceding $w$ on $\tau$ are white.}
	\label{fig:edge_tree}
\end{figure}
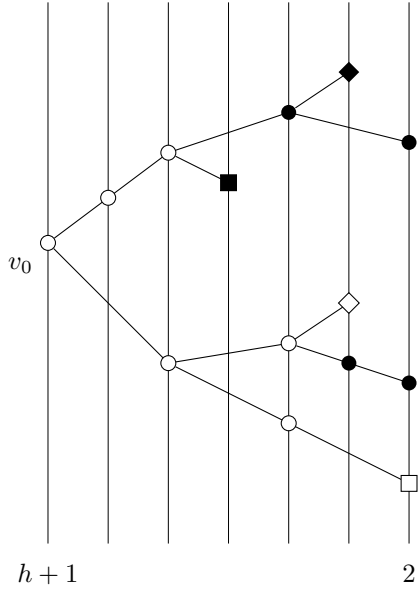

\begin{itemize}
\item $\tau$ is a rooted labelled tree, with vertex set $V(\tau)$ consisting of at least two elements; 
any vertex $v\in V(\tau)$ carries, in particular, a scale label $h_v\in [h+1,2]\cap\mathbb Z$. 
\item The root $v_0=v_0(\tau)$ of $\tau$ is the unique leftmost vertex of the tree, of scale $h_{v_0}=h+1$. The set of endpoints $V_e(\tau)$ is the subset of $V'(\tau):=V(\tau)\setminus \{v_0\}$ with degree (number of incident edges) equal to $1$. 
We let $V_0(\tau):=V(\tau)\setminus V_e(\tau)$ and $V'_0(\tau):=V_0(\tau)\setminus\{v_0\}$. 
\item Every vertex $v\in V_0'(\tau)$ is `dotted' and colored black or white, i.e., it is graphically represented either \tikzvertex{vertex} or \tikzvertex{vertex,E}. 
The root $v_0=v_0(\tau)$ may or may not be dotted; in order for $v_0$ not to be dotted, its degree must be $1$. The endpoints $v\in V_e(\tau)$ 
can be of six different types: either `bulk' interaction endpoints (represented \tikzvertex{vertex}), or `edge' interaction endpoints (represented \tikzvertex{vertex,E}),
or `bulk' counterterm endpoints (represented \tikzvertex{ctVertex}), 
or `edge' counterterm endpoints (represented \tikzvertex{ctVertex,E}), 
or `bulk' effective source endpoints (represented \tikzvertex{bareProbeEP}), or 
`edge' effective source endpoints (represented \tikzvertex{bareProbeEP,E}). 
\item if an endpoint $v\in V_e(\tau)$ is of type `bulk interaction' \tikzvertex{vertex}, or `edge interaction' \tikzvertex{vertex,E}, then $h_v=2$ necessarily; 
if an endpoint $v\in V_e(\tau)$ is of type `bulk counterterm' \tikzvertex{ctVertex}, or `edge counterterm' \tikzvertex{ctVertex,E},
then $h_v\in[h+2,1]\cap \mathbb Z$; if an endpoint $v\in V_e(\tau)$ is of type
`bulk effective source' \tikzvertex{bareProbeEP}, or `edge effective source' \tikzvertex{bareProbeEP,E}, then $h_v\in[h+2,2]\cap \mathbb Z$. 
\item Given two vertices $v,w\in V(\tau)$, $v \ge w$ or `$v$ is a successor of $w$' means that the (unique) path from $v$ to $v_0$ passes through $w$.
  Obviously, $v>w$ means that $v$ is a successor of $w$ and $v\neq w$.  Moreover, `$v$ is an immediate successor of $w$', denoted $v \successor w$, means that $v \ge w$, $v \neq w$, and $v$ and $w$ are directly connected. For any $v \in \tau$, $S_v$ is the set of $w \in \tau$ such that $w \successor v$.
Vertices, other than the root, with exactly one successor, are called `trivial'. For any $v\in V'(\tau)$, we denote by $v'$ the unique vertex in $V(\tau)$ such that $v \rhd v'$.  
 \item If $v\in V_e(\tau)$ is an endpoint with $h_v<2$, then $v'$ is non trivial and $h_{v'}=h_v-1$. 
 \item In addition to its scale label, every endpoint $v\in V_e(\tau)$ has an index $m_v\in\mathbb N_0$, which indicates the number of probe fields $\bs A$ associated with it; we allow $m_v>0$ if and only if $v$ is an effective source endpoint, \tikzvertex{bareProbeEP} or \tikzvertex{bareProbeEP,E}; if $v$ is an endpoint \tikzvertex{bareProbeEP} or \tikzvertex{bareProbeEP,E} with scale $h_v<2$, then $m_v=1$. 
If $v\in V_0(\tau)$, we let $m_v=\sum_{w\in V_e(\tau)}^{w>v}m_w$. 
\item If $v\in V(\tau)$ is colored white, then any other vertex such that $v<w$ is also colored white (including $v_0$, if it is dotted).
We attach a label $E_v\in\{0,1\}$ to all the vertices $v\in V(\tau)$, such that $E_v=0$ if $v$ is black and $E_v=1$ if $v$ is white (if $v_0$ is not dotted, we let $E_{v_0}$ be the same as $E_{v_0^*}$, with $v_0^*$ the immediate successor of $v_0$). 
\item Subtrees: for each $v \in V_0(\tau)$, we let $\tau_v \in \cT^{(h_v-1)}$ denote the subtree consisting of the vertices with $w \ge v$.
\end{itemize}

The sets $\cP(\tau)$, $\cS(\tau,\ul P)$, $\cD(\tau,\ul P)$ in the formulas for $W_\Lambda[\tau]$ and $w_\Lambda[\tau]$ in \eqref{treeexponcyl} are 
the sets of allowed field labels, allowed spanning trees, and allowed derivative maps, defined in a way analogous to those of \cite[Section~4.3]{AGG_part2}. More precisely, given $\tau\in \cT:=\cup_{h^*-1\le h\le 0}\cT^{(h)}$, any element $\ul P=\{P_v\}_{v\in V(\tau)}\in \cP(\tau)$ is characterized by the following properties:
\begin{itemize}
	\item $|P_v|$ is even and, if $v>v_0$, it is positive. 
		If $v\in V_e(\tau)$ is an endpoint of scale $h_v<2$ (therefore, of type  \tikzvertex{ctVertex}, \tikzvertex{ctVertex,E}, \tikzvertex{bareProbeEP}, or \tikzvertex{bareProbeEP,E}), then either $|P_v|=2$ or $|P_v|=4$, the case $|P_v|=4$ being allowed only for endpoints of type \tikzvertex{ctVertex,E}.
	\item We arbitrarily order the elements of $V_e(\tau)$, thus producing an ordered sequence out of it, 
	and, for every $v\in V_e(\tau)$, we let $j_v\in\{1,\ldots,|V_e(\tau)|\}$ be its position in such an ordered sequence
(e.g., if we graphically represent $\tau$ as in Fig.\ref{fig:edge_tree},
	we can let $j_v$ be the position of $v$ in the list of endpoints ordered from top to bottom). Then, for $v\in V_e(\tau)$, $P_v$ has the form $\left\{ (j_v,1,\omega_1),\dots,(j_v,2n,\omega_{2n}) \right\}$, where $\omega_l\in\{+,-,i,-i\}$, if $h_v=2$, while $\omega_l\in\{+,-\}$, if 
	$h_v<2$. Given $f=(j,l,\omega_l)$, we let $o(f)=(j,l)$ and $\omega(f):=\omega_l$.
	\item If $v$ is not an endpoint, $P_v \subset \bigcup_{w \in S_v} P_w$.
	\item If $v\in V_0(\tau)$, we let $Q_v:= \left( \bigcup_{w \in S_v} P_w \right) \setminus P_v$ be the set of {\it contracted fields}. If $v$ is dotted, then we require 
$|Q_v| \ge 2 $ and $|Q_v| \ge 2 (|S_v| - 1)$; and conversely $Q_v$ is empty if and only if $v=v_0$ and $v_0$ is not dotted. 
	\item If $v$ is not an endpoint and $h_v = 1$, then $Q_v = \bigcup_{w \in S_v} \condset{f \in P_w\, }{\, \omega(f) \in\{+i,-i\}}$, while, if $h_v<1$, $\omega(f)\in\{+,-\}$ for all $f\in Q_v$ (i.e., all and only the massive fields are integrated on scale $1$). As a consequence, any endpoint $v$ such that $v'$ has scale $h_{v'}<1$ 
is such that $\omega(f)\in\{+,-\}$ for all $f\in P_v$.\end{itemize}

We also denote by $\bs \omega_{v}$ the tuple of components $\omega(f)$, with $f\in P_v$. 
Note that the definitions 
imply that for $v,w \in \tau$ such that  neither $v \geq w$ or $v \leq w$ (for example when $v' = w'$ but $v \neq w$), $P_v$ and $P_w$ are disjoint,
as are $Q_v$ and $Q_w$.  

Next, given $\ul P\in \cP(\tau)$, we let $\cS(\tau,\ul P)$ be the whose elements $\ul T=\{T_v\}_{v\in V_0(\tau)}$ are characterized by the following properties: 
\begin{itemize}
\item for all $v\in V_0(\tau)$, 
$$T_v = \left\{ \left( f_1,f_2 \right),\dots, \left(f_{2 |S_v| - 3}, f_{2 |S_v| - 2} \right) \right\}\subset Q_v^2,$$
which is called the {\it spanning tree} associated with $v$;
\item if $w(f)$ denotes the (unique) $w \in S_v$ for which $f \in P_w$, then $(f,f')\in T_v\Rightarrow$ $w(f)\neq w(f')$ and $o(f)<o(f')$; moreover, 
$\left\{ \left\{ w(f_1), w(f_2) \right\} , \dots,  \left\{  w(f_{2 |S_v| - 3}), w(f_{2 |S_v| - 2})\right\}\right\}$
is the edge set of a tree with vertex set $S_v$.
\end{itemize}
For each $v\in V(\tau)$, we also denote by $\cD(\tau, \ul P)$ the set of families of maps $\ul D=\{D_v\}_{v\in V(\tau)}$ such that $D_v : P_v \to \{0,1,2\}^2$; the reader should think that a derivative operator 
$\partial^{D_v(f)}$ acts on the field labelled $f$. We also denote 
by $\bs D_v$ the tuple of components $D_v(f)$, with $f\in P_v$, and by $\bs D_v\big|_{Q}$ the restriction of $\bs D_v$ to any subset $Q\subset P_v$. 
Additionally, if a map $z: P_v\to \Lambda$ is assigned, we denote 
by $\bs z_v$ the tuple of components $z(f)$, with $f\in P_v$, and by $\Psi_v=(\bs\omega_v,\bs D_v,\bs z_v)$ the field multilabel associated with 
$\bs\omega_v,\bs D_v,\bs z_v$; moreover, if $v\in V_0(\tau)$ and also the maps $z: P_w\to \Lambda_\infty$, for all $w\in S_v$, are assigned, for each $w\in S_v$ 
we denote by 
$\bar\Psi_w=\Psi(P_w\setminus P_v)=(\bs \omega_w\big|_{P_w\setminus P_v}, \bs D_w\big|_{P_w\setminus P_v}, \bs z_w\big|_{P_w\setminus P_v})$
the restriction of $\Psi_w$ to $P_w\setminus P_v$ (here $\bs z_w\big|_Q$ is the restriction of $\bs z_w$ to the subset $Q\subset P_w$).  
Finally, if a tuple $\bs x_v\in \fB_\Lambda^{m_v}$ is assigned to a vertex 
$v\in V_0(\tau)$, for any $w\in S_v$ we let $\bs x_w\in\fB_\Lambda^{m_w}$ be the sub-tuple (of length $m_w$) of $\bs x_v$ such that 
$\bs x_v=\oplus_{w\in S_v}\bs x_w$, where the elements of $S_v$ are ordered in the way induced by the ordering of the endpoints (see the second item in the list of properties of $\underline P\in\cP(\tau)$).

\bigskip

We are now ready for the recursive definition of tree values. Let us start with the infinite volume tree values, which are defined via a slight extension of 
\cite[Eq.~(4.3.5)]{AGG_part2}, as follows: $W_\infty[\tau,\underline P, \underline T, \underline D]$ is non zero only if $E_{v_0}=0$ 
(i.e., if all the colored vertices of $\tau$ are black), in which case, letting $\bs D_{v_0}':=\oplus_{v\in S_{v_0}}\bs D_v\big|_{P_{v_0}}$ for $h_{v_0}<1$ and $\bs D_{v_0}':={\bf 0}$ for
$h_{v_0}=1$, and given $\bs x_{v_0}\in \fB_\infty^{m_{v_0}}$ with $\fB_\infty=\fB_{\mathbb Z^2}$ the set of nearest neighbor edges of the infinite square lattice,  
\begin{equation}
	\begin{split}
		&
		W_\infty [\tau, \ul P, \ul T, \ul D]((\bs \omega_0,\bs D_0,\bs z_0),\bs x_{v_0})=\mathds 1(\bs \omega_0=\bs \omega_{v_0})\mathds 1\Big(\bs D_0=\bs D_{v_0}=\bs D_{v_0}'\Big)\,\frac{\alpha_{v_0}}{ |S_{v_0}|!}
		\\
		&\qquad \times  \sum_{\substack{z: P_{v_0} \cup Q_{v_0}\to \mathbb Z^2\\ 
		\bs z_0= \bs z_{v_0}}}\hskip-.2truecm
		\fG_{T_{v_0},\infty}^{(h_{v_0})}(\bar\Psi_{v_1},\ldots,\bar\Psi_{v_{s_{v_0}}})
		\prod_{v \in S_{v_0}} K_{v,\infty}(\Psi_v,\bs x_v), 	
	\end{split}
	\label{eq:W8def:1.2}
\end{equation}
where $\alpha_{v_0}=\alpha(\Psi_{v_0};\Psi_{v_1},\ldots,\Psi_{v_{s_{v_0}}})$ (see \eqref{Wh-1trun} and following lines), and 
we recall that, if $|S_{v_0}|=1$, then $T_{v_0}=\emptyset$. In this case, if $\Psi_{v_1}=\Psi_{v_0}$, then $\fG_{\emptyset,\infty}^{(h_{v_0})}(\emptyset)$ should be interpreted as being equal to $1$; this latter case is the one in which, graphically, $v_0$ is not dotted. In the second line of \eqref{eq:W8def:1.2}, 
$K_{v,\infty}(\psi_v,\bs x_v)$ also depends on $\tau_v, \ul P, \ul T,$ and $\ul D$, but we leave this dependence implicit for brevity;  it is defined as follows:
if $h_{v_0}=1$, 
\begin{equation}
	K_{v,\infty}(\Psi_v, \bs x_v)= \piecewise{ V_\infty^{(1)}(\Psi_v)& \text{if $v$ is of type}\ \tikzvertex{vertex},\\
	B_\infty^{(1)}(\Psi_v,\bs x_v)& \text{if $v$ is of type}\  \tikzvertex{bareProbeEP}}\label{Kvinfty1}\end{equation}
(here, as usual, $V_\infty^{(1)}$ and $B_\infty^{(1)}$ are the restrictions of $W_\infty^{(1)}$ to the terms with $\bs x=\emptyset$ or $\bs x\neq \emptyset$, respectively)
while, if $h_{v_0}<1$, 
\begin{equation} 
	K_{v,\infty}(\Psi_v, \bs x_v)
	:=	
	\piecewise{\upsilon_{h_{v_0}}\cdot F_\infty(\Psi_v) 
		  	& \text{if $v\in V_e(\tau)$ is of type \tikzvertex{ctVertex} (so that $h_v = h_{v_0}+1$)}\\ 
			\cR_\infty V_\infty^{(1)}(\Psi_v) & \text{if $v\in V_e(\tau)$ is of type \tikzvertex{vertex} (so that $h_v=2$)} \\
			Z_{h_{v_0}}\cdot F_\infty^A(\Psi_v,\bs x_v) & \text{if $v\in V_e(\tau)$ is of type \tikzvertex{bareProbeEP} and $h_v = h_{v_0}+1$,}\\ 
			\cR_\infty B_\infty^{(1)}(\Psi_v, \bs x_v) & \text{if $v\in V_e(\tau)$ is of type \tikzvertex{bareProbeEP} and $h_v = 2$,}\\
			\lis W_\infty [\tau_v, \ul P_{v}, \ul T_{v}, \ul D_{v}](\Psi_v, \bs x_v)
		  	& \text{if $v\in V_0(\tau)$,}\label{defKv4.1}
		}
\end{equation}
where, in the last line of \eqref{defKv4.1}, letting $\ul P_v$ (resp. $\ul T_v$, resp. $\ul D_v$) be the restriction of $\ul P$ (resp. $\ul T$, resp. $\ul D$) 
to the subtree $\tau_v$, and $\ul D_v':=\{D_{v}'\}\cup\{D_w\}_{w\in V(\tau): w>v_0}$ (here $D_{v}'$ is the map such that 
$\bs D_{v}':=\bigoplus_{w\in S_{v}}\bs D_w\big|_{P_{v}}$), we denoted 
\begin{equation}
\label{lisWinfty}\lis W_\infty [\tau_v, \ul P_{v}, \ul T_{v}, \ul D_{v}]:=\cR_\infty W_\infty [\tau_v, \ul P_{v}, \ul T_{v}, \ul D_{v}']\Big|_{\bs D_{v}},
\end{equation}
where, if $\bs D_v=(D_1,\ldots,D_{n})$ with $n=|P_v|$ and $\|\bs D\|_1=p$, 
we denoted by $\cR_\infty W_\infty [\tau_v, \ul P_{v}, \ul T_{v}, \ul D_{v}']\Big|_{\bs D_{v}}$ the restriction of $(\cR_\infty W_\infty [\tau_v, \ul P_{v}, \ul T_{v}, \ul D_{v}'])_{n,p}$ to that 
specific choice of derivative label.

\bigskip

Let us now define the tree values in finite volume $\Lambda$. We distinguish three cases. 

\medskip

\noindent1) If $E_{v_0}=0$ (that is, all the colored vertices of $\tau$ are black), then, letting as usual 
$\Psi=(\bs \omega, \bs D, \bs z)$, we define 
\begin{equation}\label{eq:item1}W_\Lambda[\tau,\ul P,\ul T,\ul D](\Psi,\bs x):=(-1)^{\alpha(\bs z)}\, I_\Lambda(\Psi,\bs x)
W_\infty[\tau,\ul P,\ul T,\ul D](\Psi_\infty,\bs x_\infty),\end{equation} 
where $I_\Lambda$ was defined as in \eqref{eq:3.1.4}, 
$\Psi_\infty=(\bs\omega,\bs D,\bs z_\infty)$ with $(\bs z_\infty,\bs x_\infty)\equiv(\bs z,\bs x)_\infty$ the infinite volume representative of $(\bs z,\bs x)$ in the sense of item \ref{it3:dop} after \eqref{eq:WfreeL1bound}, and 
$W_\infty[\tau,\ul P,\ul T,\ul D](\Psi,\bs x)$ was defined in \eqref{eq:W8def:1.2} and following lines. Note that the fact that $W_\infty^{(h)}=\sum_{\tau\in\cT^{(h)}}^{E_{v_0}=0}W_\infty[\tau]$ readily implies that $W_\B^{(h)}= \sum_{\tau\in \cT^{(h)}}^{E_{v_0}=0}W_\Lambda[\tau]$, with $W_\Lambda[\tau]$ given by the second equation in the first line of \eqref{treeexponcyl}.

\medskip

\noindent2) If $E_{v_0}=1$ and $E_v=0$ for all $v>v_0$ (that is, all the vertices of $\tau$ but $v_0$ are black), then, letting $\tau_{\black}$ be the tree obtained from $\tau$ by changing 
the color of $v_0$ to black, we define 
\begin{equation}
W_\Lambda[\tau,\ul P,\ul T, \ul D]:= \widetilde W_\B[\tau_\black,\ul P,\ul T,\ul D]-W_\Lambda[\tau_{\black},\ul P,\ul T,\ul D].\label{Wtaublack-}\end{equation}
where $\widetilde W_\B[\tau_\black,\ul P,\ul T,\ul D]$ is defined by the finite volume analogue of \eqref{eq:W8def:1.2}, that is, 
\begin{eqnarray}
&&
	 \widetilde W_\B [\tau_\black, \ul P, \ul T, \ul D]\big((\bs \omega_0,\bs D_0,\bs z_0),\bs x_{v_0}\big)=\mathds 1(\bs \omega_0=\bs \omega_{v_0}) \mathds 1(\bs D_0=\bs D_{v_0}=\bs D_{v_0}')\, \frac{\alpha_{v_0}}{ |S_{v_0}|!}\cdot\nonumber\\
		&&\cdot  \sum_{\substack{z: \cup_{v\in S_{v_0}}\!P_{v}\to\bar\Lambda:\\ 
		\bs z_0= \bs z_{v_0}}}\hskip-.2truecm
		\fG_{T_{v_0}}^{(h_{v_0})}(\bar\Psi_{v_1},\ldots,\bar\Psi_{v_{s_{v_0}}})\, \prod_{v \in S_{v_0}}K_v(\Psi_v,\bs x_v),	
		\label{eq:W8def:1.2op}\end{eqnarray}
where in the second line the summand must be interpreted as zero if $\Psi_v\not\in\cM_\Lambda$ for some $v\in S_{v_0}$, and 
$K_v$ is defined by the `bulk' finite volume analogue of \eqref{Kvinfty1}-\eqref{defKv4.1}: more precisely, if $h_{v_0}=1$, then 
\begin{equation} K_v(\Psi_v,\bs x_v):=	
 \piecewise{ V_\B^{(1)}(\Psi_v)& \text{if $v$ is of type}\ \tikzvertex{vertex},\\
	B_\B^{(1)}(\Psi_v,\bs x_v)& \text{if $v$ is of type}\  \tikzvertex{bareProbeEP}}\end{equation}
while, if $h_{v_0}<1$, 
\begin{equation} K_v(\Psi_v,\bs x_v):=
	\piecewise{\upsilon_{h_{v_0}}\cdot F_\B(\Psi_v) 
		  	& \text{if $v\in V_e(\tau)$ is of type \tikzvertex{ctVertex} (so that $h_v = h_{v_0}+1$)}\\ 
			(\cR_\B V_\B^{(1)})_\B(\Psi_v) & \text{if $v\in V_e(\tau)$ is of type \tikzvertex{vertex} (so that $h_v=2$)} \\
			Z_{h_{v_0}}\cdot F_\B^A(\Psi_v,\bs x_v) & \text{if $v\in V_e(\tau)$ is of type \tikzvertex{bareProbeEP} and $h_v = h_{v_0}+1$,}\\ 
			(\cR_\B B_\B^{(1)})_\B(\Psi_v, \bs x_v) & \text{if $v\in V_e(\tau)$ is of type \tikzvertex{bareProbeEP} and $h_v = 2$,}\\
			\lis W_\Lambda [\tau_v, \ul P_{v}, \ul T_{v}, \ul D_{v}](\Psi_v, \bs x_v)
		  	& \text{if $v\in V_0(\tau)$,}\label{after2A}
		}
\end{equation}
where in the last line $\lis W_\Lambda [\tau_v, \ul P_{v}, \ul T_{v}, \ul D_{v}]$ is defined via the analogue of \eqref{lisWinfty}, with $\cR_\infty W_\infty$ in the right side replaced by $(\cR_\B W_\Lambda)_\B$, and $W_\Lambda$ defined as in item 1). 

\medskip

\noindent 3) If $E_v=1$ for at least one vertex $v>v_0$, then we define $ W_\Lambda [\tau, \ul P, \ul T, \ul D]\big((\bs \omega_0,\bs D_0,\bs z_0),\bs x_{v_0}\big)$ 
by the same expression as in the right side of \eqref{eq:W8def:1.2op}, where now some of the vertices $v\in S_{v_0}$ appearing in the product in the second line are white, i.e., 
$E_v=1$. In such cases, the definition of $K_v$ must be modified as follows: 
if $h_{v_0}=1$ and $E_v=1$, then  
\begin{equation} K_v(\Psi_v,\bs x_v):=	\piecewise{ 
		  V_\E^{(1)}(\Psi_v)& \text{if $v$ is of type}\ \tikzvertex{vertex,E},\\
		    B_\E^{(1)}(\Psi_v, \bs x_v)& \text{if $v$ is of type}\ \tikzvertex{bareProbeEP,E},}
\end{equation}
while, if $h_{v_0}<1$,
\begin{equation} K_v(\Psi_v,\bs x_v):=\piecewise{
		\cR_\E C_\E^{(h_{v_0})}(\Psi_v) & \text{if $v\in V_e(\tau)$ is of type \tikzvertex{ctVertex,E} (so that $h_v = h_{v_0}+1$)}\\ 
		\cR_\E (V^{(1)}_\E+\mathcal E V^{(1)}_\B)(\Psi_v) & \text{if $v\in V_e(\tau)$ is of type \tikzvertex{vertex,E} (so that $h_v = 2$)}\\ 
		D_\E^{(h_{v_0})}(\Psi_v,\bs x_v) & \text{if $v\in V_e(\tau)$ is of type \tikzvertex{bareProbeEP,E} and $h_v =h_{v_0}+1$}\\ 
		(B^{(1)}_\E+\mathcal E B^{(1)}_\B)(\Psi_v, \bs x_v) & \text{if $v\in V_e(\tau)$ is of type \tikzvertex{bareProbeEP,E} and $h_v = 2$,}\\ 
		\lis W_\Lambda [\tau_v, \ul P_{v}, \ul T_{v}, \ul D_{v}](\Psi_v, \bs x_v)  & \text{if $v\in V_0(\tau)$,}
		}\label{Kvwhitedef}
\end{equation}
where, in the first and third lines, $C_\E^{(h)}$ and $D_\E^{(h)}$ are defined inductively (in the scale $h$) as follows: 
\begin{equation} \label{defCeZe} 
C_\E^{(h)} :=
	\sum_{k=h}^0
	\sum_{\substack{\tau \in \cT^{(k)} \\ E_{v_0}=m_{v_0} = 0}}^*
	\cE W_\Lambda [\tau], \qquad 
	D_\E^{(h)}
	:=
	\sum_{k=h}^0
	\sum_{\substack{\tau \in \cT^{(k)} \\ E_{v_0}=0,\ m_{v_0}= 1}}^*
	\cE W_\Lambda [\tau],
\end{equation}
where the $*$ indicates the constraint that $\tau$ is required to have $v_0$ dotted, and, for any $\tau\in\cT^{(h)}$ such that $E_{v_0}=0$, 
$\mathcal E W_\Lambda[\tau]$ is defined, in analogy with the third line of \eqref{eq:3.1.42(}, as $\mathcal E W_\Lambda[\tau]:=\cL_\B W_\Lambda[\tau]-
(\cL_\B W_\Lambda[\tau])_\B+\cR_\B W_\Lambda[\tau]-(\cR_\B W_\Lambda[\tau])_\B\equiv (\cL_\B W_\Lambda[\tau])_\E+(\cR_\B W_\Lambda[\tau])_\E$. 
Moreover, in the last line of \eqref{Kvwhitedef}, 
if $v\in V_0(\tau)$ is white, then $\lis W_\Lambda [\tau_v, \ul P_{v}, \ul T_{v}, \ul D_{v}]$ is defined via the analogue of \eqref{lisWinfty}, with $\cR_\infty W_\infty$
in the right side replaced by $\cR_\E W_\Lambda$ (with the understanding that, if $E_v=1$ and $m_v>0$, then $\cR_\E W_\Lambda\equiv W_\Lambda$); if, instead, $v\in S_{v_0}$ is black, than $K_v$ is defined as in item (2).

\begin{remark} Noting that, in view of the definition of tree values, the bulk contribution to the sourceless part of the effective potential, 
$V_\B^{(h)}= \sum_{\tau\in\cT^{(h)}}^{E_{v_0}=m_{v_0}=0}W_\Lambda[\tau]$, can be equivalently rewritten as: 
\begin{equation} V_\B^{(h)}= \sum_{\substack{\tau\in\cT^{(h)}:\\E_{v_0}=m_{v_0}=0}}W_\Lambda[\tau]=\begin{cases} V_\B^{(1)}+\sum_{\substack{\tau\in\cT^{(0)}:\\E_{v_0}=m_{v_0}=0}}^* W_\Lambda[\tau] & \text{\quad if $h=0$}\\
\upsilon_{h+1}\cdot F_\B+(\cR_\B V_\B^{(1)})_\B+\sum_{k=h}^0\sum_{\substack{\tau\in\cT^{(k)}:\\E_{v_0}=m_{v_0}=0}}^* W_\Lambda[\tau] & \text{\quad if $h<0$}\end{cases}
\end{equation}
where, again, the $*$ on the sums over $\tau$ indicate the constraint that $\tau$ is required to have $v_0$ dotted, we see that 
$C_\E^{(h)} = \mathcal E(V_\B^{(0)}-V_\B^{(1)})$, and $C_\E^{(h)}= \mathcal E(V_\B^{(h)}-\upsilon_{h+1}\cdot F_\B-(\cR_\B V_\B^{(1)})_\B)$ for $h<0$. Moreover, 
by using the definition of $\mathcal E$, $\cL_\B$ and $\cR_\B$, it follows that $\mathcal E(\upsilon_{h+1}\cdot F_\B+(\cR_\B V_\B^{(1)})_\B)=0$, so that, in conclusion, 
\begin{equation} C_\E^{(h)}=\mathcal E(V_\B^{(h)})-\delta_{h,0}\mathcal E(V_\B^{(1)}).\label{betterdefCe}\end{equation}
Similarly, one finds that 
\begin{equation} D_\E^{(h)}=\mathcal E(B_\B^{(h)})-\delta_{h,0}\mathcal E(B_\B^{(1)}).\label{betterdefZe}\end{equation}
\end{remark}

Using the definitions in items (2)-(3) above, it follows that 
$W_\E^{(h)}\sim \sum_{\tau\in \cT^{(h)}}^{E_{v_0}=1}W_\Lambda[\tau]$, with $W_\Lambda[\tau]$ given by the second equation in the first line of \eqref{treeexponcyl};
for the proof, see Appendix \ref{app:itdec}. This, in combination with the fact that $W_\B^{(h)}= \sum_{\tau\in \cT^{(h)}}^{E_{v_0}=0}W_\Lambda[\tau]$, 
proves the first line of \eqref{treeexponcyl}. The proof of the second line of \eqref{treeexponcyl} is completely analogous. 

\begin{remark}\label{importremmE}
As in the case of the infinite plane, see \cite[Remark~4.4]{AGG_part2}, 
we let $R_v=\| \bs D_v\|_1-\sum_{w\in S_v}\| \bs D_w\big|_{P_v}\|_1\equiv \| \bs D_v\|_1-\| \bs D_v'\|_1$. From the definitions, it follows that $R_{v_0}=0$, while, if $v\in V_0'(\tau)$, any $\ul D\in \cD(\tau,\underline P)$ with $\underline P\in\cP(\tau)$ is such that: 
(1) if $m_v=E_v=0$, then 
	\begin{equation}
		R_v
		=
		\piecewise{
			2, & 
				|P_v| = 2 \textup{ and } \left\| \bs D_v' \right\|_1 = 0 \\ 
			1, & 
				|P_v| = 2 \textup{ and } \left\| \bs D_v' \right\|_1 = 1 \textup { or } 
				|P_v| = 4 \textup{ and } \left\| \bs D_v' \right\|_1 = 0 \\ 
			0, & \textup{otherwise};
		}
		\label{eq:Rv_def}
	\end{equation}
(2) if $m_v+E_v=1$, then 
	\begin{equation}
		R_v
		=
		\piecewise{
			1, & 
				|P_v| = 2 \textup{ and } \left\| \bs D_v' \right\|_1 = 0 \\ 
			0, & \textup{otherwise};
		}
		\label{eq:Rv_defEdge}
	\end{equation}
(3) if $m_v+E_v\ge 2$, then $R_v=0$. 
In analogy with the case of the infinite plane, see \cite[Eq.~(4.3.10)]{AGG_part2}, we let 
\begin{equation} d(P_v,\bs D_v,m_v,E_v):=2-\frac{|P_v|}{2}-\|\bs D_v\|_1-m_v -E_v\delta_{m_v,0}
\label{defscaldimwithm}\end{equation}
be the {\bf scaling dimension} of $v$, whose allowed values have important implications for the (uniform) convergence of the GN tree expansion, 
see Proposition \ref{lm:W:scaldim_withm} below. From the definitions, it follows that, for any allowed 
set $\ul D$ and $v\in V_0(\tau)\cup \{v\in V_e(\tau): h_{v'}<h_v-1\}$, then $d(P_v,\bs D_v,m_v,E_v)\le -1$, with the only exception of the case 
$(|P_v|, \|\bs D_v\|_1, m_v, E_v)=(2,0,1,1)$, which is allowed and corresponds to $d(P_v,\bs D_v,m_v,E_v)=0$. 
\end{remark}

\begin{remark}
	With the other arguments fixed, the number of choices of $\ul D$ for which $W_\Lambda[\tau,\ul P, \ul T, \ul D]$ or $W_\infty[\tau,\ul P, \ul T, \ul D]$ do 
	not vanish is no more than $10^{\left|V(\tau)\right|}$, cf.\ \cite[Remark~4.5]{AGG_part2}: in fact, 
	there is a choice of at most 10 possible values for each endpoint\footnote{The value 10 bounds, in particular,
	the number of different terms that the operators $\cR_\B$, $\cR_\E$ or $\cR_\infty$ produce when they act non-trivially on interaction or on effective source endpoints. In fact, the cases in which $\cR_\B$ acts non-trivially on the sourceless part of the kernels are those listed in the right sides of \eqref{defcR}, with $(n,p)=(2,2), (4,1)$ (similar considerations apply for the action of $\cR_\B$ on the effective source term, and for the actions of $\cR_\E$ and of $\cR_\infty$). If $(n,p)=(2,2)$, the number of possible values taken by $D_v$ is 10 (one derivative in direction $i\in\{1,2\}$ on the first Grassmann field and one derivative in direction $j\in\{1,2\}$ on the second Grassmann field, etc.); if $(n,p)=(4,1)$, the number of possible values taken by $D_v$ is 8, which is smaller than 10 (one derivative in direction $i\in\{1,2\}$ on the $k$-the Grassmann field, with $k\in\{1,2,3,4\}$).}, and then the other values are fixed except for a choice of up to 10 possibilities each time that $\cR_\B$ or $\cR_\E$ or $\cR_\infty$ act non trivially on a vertex $v\in V_0'(\tau)$, i.e., each time that, for such a vertex, $R_v>0$, see \eqref{eq:Rv_def}-\eqref{eq:Rv_defEdge}. 
	\label{rem:D_choice_p2}
\end{remark}

\subsection{Bounds on the kernels of the effective potential}
\label{sec:informal_bounds}

In this section, we prove two key bounds on the kernels $W_\Lambda[\tau,\ul P,\ul T,\ul D]$, which will be used later in order to estimate the logarithmic derivatives of the generating 
function of energy correlations, once we express them in the form of sums over GN trees. These bounds are the natural generalization of those 
described in \cite[Section~4.4]{AGG_part2} for the sourceless part of the infinite volume kernels of the effective potential (which we invite the reader to consult before reading this section) 
to the case of the finite volume tree values defined in the previous section. Compared with \cite[Section~4.4]{AGG_part2}, 
the main novelty here is the 
presence of edge vertices, which come with the `edge contribution' to the kernels associated with the given subtrees. As apparent from the definitions 
in the previous section, several ingredients enter the definition of the tree values of GN trees with edge (white) vertices; however, loosely speaking, 
the reader may simply think that the value of a subtree rooted in a white vertex is the same as the value of the corresponding tree with all the vertices 
re-colored black, modulo the fact that at least one of the functions $\fG_{T_{v}}^{(h_{v})}(\Psi_{v_1}\setminus\Psi_{v_0},\ldots,\Psi_{v_{s_{v}}}\setminus\Psi_{v_0})$ entering its definition
is replaced by the corresponding edge contribution, or at least one of the endpoints comes with an edge contribution, such as $\cR_\E C_\E^{(h_{v}-1)}$.
Of course, one of the ingredients that we need for estimating the kernel of a subtree rooted in a white vertex is a bound on the difference between 
$\fG_{T_{v}}^{(h_{v})}(\Psi_{v_1}\setminus\Psi_{v_0},\ldots,\Psi_{v_{s_{v}}}\setminus\Psi_{v_0})$ and the corresponding bulk contribution. The desired bound is summarized in the following lemma. 

\begin{lemma} There exists $C>0$ such that the following holds. Consider $s\ge 1$ non empty field multilabels $Q_i=(\bs \omega_i,\bs D_i, \bs z_i)$, $i=1,\ldots, s$, 
and let $Q= \oplus_{i=1}^s Q_i\equiv (\bs \omega,\bs D,\bs z)$ and $Q_{i,\infty}=(\bs \omega_i,\bs D_i, \bs z_{i,\infty})$. Then, if $\diam_1(\bs z)\le L/3$, 
	\begin{equation}
		\left| 
		\fG_T^{(h)} (Q_1,\ldots, Q_s)
		- (-1)^{\alpha(\bs z)}
		\fG_{T,\infty}^{(h)} (Q_{1,\infty},\ldots, Q_{s,\infty})
		\right|
		\le C^{|Q|}	2^{(\frac{|Q|}2 + \|\bs D\|_1)h} e^{-c_0 2^h \delta_\E (T,\bs z)},\label{eq_lem:1}
	\end{equation}
for any $h\le 1$,  where $c_0$ is the same constant as in \eqref{cosacomedove}, and 
\begin{equation}
	\delta_\E(T, \bs z):=\sum_{(f,f')\in T}\|z(f)-z(f')\|_1+\min\{\dist(\bs z,\partial\Lambda),L/3\}.
	\label{eq:dE_Tz_def}
\end{equation}
\label{lem:fG_E_diff}
\end{lemma}

\begin{remark} The proof of the lemma actually gives a slightly better bound than \eqref{eq_lem:1}, with $\sum_{(f,f')\in T}\|z(f)-z(f')\|_1+\min\{\dist(\bs z,\partial\Lambda),L\}$
replacing $\delta_\E(T,\bs z)$ in the exponent. The reason for defining $\delta_\E$ with $L/3$ rather than with $L$ is to simplify the form that some bounds in the 
following take. Similarly, the reason for stating the result of the lemma as \eqref{eq_lem:1} rather than with the improved exponent is to write the final 
bound in the precise form it will be used later.\end{remark}

\begin{proof} Let us recall, for the reader's convenience, the definition of $\mathfrak G_T^{(h)}$, see \eqref{eq:BBF} and following list:
	\begin{equation}\fG_T^{(h)}(Q_1,\dots,Q_s)=
	\alpha_T(Q_1,\dots,Q_s)
		\left[ \prod_{\ell \in T} g^{(h)}_\ell \right]
		\int P_{Q_1,\dots,Q_s,T} (\dd \bs t)\, {\rm{Pf}} \big(G^{(h)}_{Q_1,\ldots,Q_s,T} (\bs t)\big).\end{equation}
Moreover, $(-1)^{\alpha(\bs z)}\fG_{T,\infty}^{(h)} (Q_1,\ldots, Q_s)$ is given by an analogous formula, with $g_\ell^{(h)}$ replaced by its bulk counterpart
$g_{\ell,\B}^{(h)}$, 
and similarly for the propagators in the Pfaffian. Recall that the propagators $\fg^{(h)}(x,y)$ entering the definition of $\fG_{T}^{(h)} (Q_1,\ldots, Q_s)$ can be decomposed as $\fg^{(h)}(x,y)=
\fg_\B^{(h)}(x,y)+\fg_\E^{(h)}(x,y)$, see \eqref{cosacome}, with $\fg_\B^{(h)},\fg_\E^{(h)}$ bounded as in \eqref{ghbounds1} and \eqref{cosacomedove}, respectively;
moreover, if $\diam_1(x-y)\le L/3$, as in the case under consideration, 
$\fg^{(h)}, \fg_\B^{(h)},\fg_\E^{(h)}$ can all be represented in Gram form, as in \eqref{ddtthh}, with bounds of the same qualitative form as those described after \eqref{ddtthh};
see \cite[Remark~2.6]{AGG_part2}. 
We now write out the difference $\fG_{T}^{(h)} (Q_1,\ldots, Q_s)-(-1)^{\alpha(\bs z)}\fG_{T,\infty}^{(h)} (Q_1,\ldots, Q_s)$ as a telescopic sum of single factor differences.  
There are $|T|$ terms with a difference between spanning tree propagators, of the form (letting $\ell=(f,f')$)
	\begin{eqnarray}
		\left|g_{\ell}^{(h)} - g_{\ell,\B}^{(h)}\right|
		=\left|g_{\ell,\E}^{(h)}\right|
		&\le& C	2^{(1 + \|D(f)\|_1+\|D(f')\|_1)h}
		e^{- c_02^h d_\E(z(f),z(f'))}\label{eq:g_fG_diff}\\
		&\le & C	2^{(1 + \|D(f)\|_1+\|D(f')\|_1)h}
		e^{- c_02^h (\|z(f)-z(f')\|_1+\min\{\dist(\bs z,\partial\Lambda), L\})},\nonumber
	\end{eqnarray}
where in the second inequality we used the definition of $d_\E(z(f),z(f'))$, see the line after \eqref{cosacomedove}.

Let us now consider the difference ${\rm{Pf}} \big(G^{(h)}_{Q_1,\ldots,Q_s,T} (\bs t)\big)- {\rm{Pf}} \big(G^{(h)}_{Q_1,\ldots,Q_s,T;\B} (\bs t)\big)$,
 where in the second term the argument of the Pfaffian is the matrix obtained by replacing the elements  $G^{(h)}_{Q_1,\ldots,Q_s,T} (\bs t)$ 
by their bulk counterparts. 
Let $2n=|Q|-2|T|$ be the dimension of $G^{(h)}_{Q_1,\ldots,Q_s,T} (\bs t)$. 
We write the difference in telescopic form as
\begin{equation}\label{last}
{\rm{Pf}} \big(G^{(h)}_{Q_1,\ldots,Q_s,T} (\bs t)\big)- {\rm{Pf}} \big(G^{(h)}_{Q_1,\ldots,Q_s,T;\B} (\bs t)\big)=\sum_{1\le 1<j\le 2n}({\rm Pf}\, A^{(i,j)} -{\rm Pf}\, A^{(i,j)'}),\end{equation}
where $A^{(i,j)}$ is the anti-symmetrix matrix whose elements above the diagonal with label smaller or equal to (resp. larger 
than) $(i,j)$ in the lexicographic order are equal to the elements of $G^{(h)}_{Q_1,\ldots,Q_s,T} (\bs t)$ (resp. $G^{(h)}_{Q_1,\ldots,Q_s,T;\B} (\bs t)$), and $(i,j)'$ is the label immediately preceding $(i,j)$ in the lexicographic 
order (if $(i,j)=(1,2)$, we interpret $A^{(1,2)'}\equiv G^{(h)}_{Q_1,\ldots,Q_s,T;\B} (\bs t)$). Using the definitions, we find that  
\begin{equation}\label{paciuga}\big|{\rm Pf}\, A^{(i,j)} -{\rm Pf}\, A^{(i,j)'}\big|\le |g^{(h)}_{\ell_{i,j}}-g^{(h)}_{\ell_{i,j},\B}|\cdot \big|{\rm Pf}\, A_{\hat\imath\hat\jmath}\big|,
\end{equation}
where $\ell_{i,j}$ is the pair of field 
indices associated with the matrix element $(i,j)$, and $A_{\hat \imath\hat\jmath}$ denotes the matrix $A^{(i,j)}$ with both the $i$-th 
and $j$-th rows and columns removed. Now, the difference $g^{(h)}_{\ell_{i,j}}-g^{(h)}_{\ell_{i,j};\B}$ is an edge propagator, bounded as in \eqref{eq:g_fG_diff}. In order to 
bound $\big|{\rm{Pf}} A_{\hat\imath\hat\jmath}\big|$, recall that  both $G^{(h)}_{Q_1,\ldots,Q_s,T} (\bs t)$ and $G^{(h)}_{Q_1,\ldots,Q_s,T;\B} (\bs t)$ are Gram matrices
(see \eqref{ddtthh} and the fourth and fifth items after \eqref{eq:BBF}); 
in particular, they can be written as 
\begin{equation} \label{gramrepp}(G^{(h)}_{Q_1,\ldots,Q_s,T} (\bs t))_{k,l}=(f_{k},h_{l}), \qquad (G^{(h)}_{Q_1,\ldots,Q_s,T;\B} (\bs t))_{k,l}=(f_{\B;k},h_{\B;l}),\end{equation}
for appropriate vectors $f, h, f_\B, h_\B$ in two apriori different Hilbert spaces
$\mathcal H$ and $\mathcal H_\B$, such that $\|f_k\|^2\le C2^{h(1+2r_k)}$, with $r_k$ the $\ell^1$ norm of the derivative vector associated with the $k$-th field label
(see the lines after \eqref{ddtthh}), and similarly for $\|h_l\|^2, \|f_{\B,k}\|^2$ and $\|h_{\B,l}\|^2$. Remarkably, also 
$A_{\hat\imath\hat\jmath}$ is in Gram form, that is, for any $k,l\in\{1,\ldots,2n\}\setminus\{i,j\}$, we can write
$(A_{\hat\imath\hat\jmath})_{k,l}=(F_k,H_l)$, where $(\cdot,\cdot)$ denotes the scalar product in $\mathcal H\oplus \mathcal H\oplus \mathcal H_\B$, and $F_k, H_l$ are the 
following vectors in $\mathcal H\oplus \mathcal H\oplus \mathcal H_\B$: 
\begin{equation}\label{schiaccia}F_k=\begin{cases} (f_{k},0,0) & \text{if $k<i$,}\\ (0,f_k,f_{\B;k}) & \text{if $k>i$,}\end{cases} \quad \text{and} \quad  
H_l=\begin{cases} (h_{l},h_l,0) & \text{if $l<i$,}\\ (h_l,0,h_{\B;l}) & \text{if $l>i$.}\end{cases} \end{equation}
Therefore, $\big|{\rm Pf}\, A_{\hat\imath\hat\jmath}\big|=\sqrt{\big|\det\,A_{\hat\imath\hat\jmath}\big|}$ 
can be bounded from above via the Gram-Hadamard inequality \cite[Appendix~A.3]{GM01}, which states that $|\det A_{\hat\imath\hat\jmath}|\le \prod_{k}\|F_k\|\,\|H_k\|$; 
recalling the definition of $F_k,H_l$ in \eqref{schiaccia} and the bounds on the norms of $f_k, f_{\B,k}, h_l, h_{\B,l}$ stated after \eqref{gramrepp}, we find
\begin{equation}\big|{\rm Pf}\, A_{\hat\imath\hat\jmath}\big|\le C^{|Q|-2|T|}2^{h(\frac{|Q|}{2}-|T|-1+\|\bs D\|_1-\|\bs D(T)\|_1- \|D(f_i)\|_1-\|D(f_j)\|_1)},
\label{paciughina}
\end{equation}
where $\|\bs D(T)\|_1=\sum_{(f,f')\in T}(\|D(f)\|_1+\|D(f')\|_1)$ and $f_i,f_j$ are the two field labels such that $\ell_{i,j}=(f_i,f_j)$. 
Combining \eqref{last} and \eqref{paciuga} with \eqref{eq:g_fG_diff} and \eqref{paciughina}, we find
\begin{equation} 
	\begin{split}
		&
		\Big|{\rm{Pf}} \big(G^{(h)}_{Q_1,\ldots,Q_s,T} (\bs t)\big)- {\rm{Pf}} \big(G^{(h)}_{Q_1,\ldots,Q_s,T;\B} (\bs t)\big)\Big|
		\\ & \qquad \le 
		(|Q|-2|T|)^2C^{|Q|-2|T|}2^{h(\frac{|Q|}{2}-|T|+\|\bs D\|_1-\|\bs D(T)\|_1)}
		e^{-c_0 2^h\min\{\dist(\bs z,\partial\Lambda),L\}}.
	\end{split}
\end{equation}
Putting it all together, we find that
\begin{equation} 
	\begin{split}
		& \left| 
		\fG_T^{(h)} (Q_1,\ldots, Q_s)
		- 
		\fG_{T,\B}^{(h)} (Q_1,\ldots, Q_s)\right| 
		\\ &\ \le  
		\Big[|T|+(|Q|-2|T|)^2\Big]C^{|Q|} 2^{h\big(\frac{|Q|}2+\|\bs D\|_1\big)}
		\Big(\prod_{(f,f')\in T} e^{-c_0 2^h \|z(f)-z(f')\|_1} \Big)e^{-c_0 2^h\min\{\dist(\bs z,\partial\Lambda),L\}}.
	\end{split}
\end{equation}
Of course, the factor in brackets in the right side can be reabsorbed in $C^{|Q|}$, up to a redefinition of the constant $C$, and, therefore, we obtain the desired inequality, \eqref{eq_lem:1}.
\end{proof}

We are now ready to state and prove the bounds on $W_\Lambda[\tau, \ul P,\ul T,\ul D]$. In order to measure the size of $W_\Lambda[\tau, \ul P,\ul T,\ul D]$, 
we use the following norm, whose definition depends on the values of $E_{v_0}$ and $m_{v_0}$: given $\bs x_{v_0}\in \fB_\Lambda^{m_{v_0}}$, we let 
\begin{equation}\label{argum}\|W_\Lambda[\tau,\ul P, \ul T, \ul D](\bs x_{v_0})\|_{E_{v_0};h_{v_0}}:=\piecewise{
\|W_\Lambda[\tau,\ul P, \ul T, \ul D](\bs x_{v_0})\|_{(\frac{c_0}22^{h_{v_0}})} & \text{if $E_{v_0}=0$,}\\ 
\|W_\Lambda[\tau,\ul P, \ul T, \ul D](\bs x_{v_0})\|_{(\E;\frac{c_0}2 2^{h_{v_0}})} & \text{if $E_{v_0}=1$,}}\end{equation}
where $c_0$ is the same constant as in \eqref{cosacomedove} and: the norm in the first line is defined as in 
\eqref{eq:weightnorm} or as in \eqref{normonsource}, depending on whether $m_{v_0}=0$ or $m_{v_0}>0$; 
the norm in the second line is defined as in \eqref{eq:cRE_sumbound} or as in \eqref{normonsourceEdge}, depending on whether $m_{v_0}=0$ or $m_{v_0}>0$. If 
$m_{v_0}=0$, we shall drop the argument $(\bs x_{v_0})=(\emptyset)$ in \eqref{argum}. 

\medskip

We start by discussing the case in which $\tau$ has no endpoints 
with source field labels
(in particular, no endpoints of type \tikzvertex{bareProbeEP} or \tikzvertex{bareProbeEP,E}), 
that is, $m_{v_0}=0$; in this case, the basic bound is summarized in the following proposition. 
Next, we will discuss the case with $m_{v_0}>0$, see Proposition \ref{lem:W_primitive_with_m} below.

\begin{proposition} \label{lem:W_primitive_bound}
Let $W_\Lambda[\tau,\ul P,\ul T,\ul D]$ be inductively defined as in Section \ref{sec:treecyl}. There exist $C,\kappa,\lambda_0>0$ 
such that, for any $\tau\in \cT$,
with $m_{v_0}=0$, $\ul P\in \cP(\tau)$, $\ul T\in \cS(\tau,\ul P)$, $\ul D\in \cD(\tau,\ul P)$ and $|\lambda|\le \lambda_0$,
	\begin{eqnarray}
&&\hskip-.8truecm \|W_\Lambda[\tau,\ul P,\ul T,\ul D]\|_{E_{v_0};h_{v_0}}\le     C^{\sum_{v\in V_e(\tau)}|P_v|} 2^{-E_{v_0} h_{v_0}}
   \Big(\prod_{v \in V_0(\tau)}
	      \frac{2^{(\tfrac12 |Q_v|+ \sum_{w\in S_v}\| \bs D_w |_{Q_v} \|_1 - R_v + 2 - 2 \left|S_v\right|)h_v}}{ |S_v|!}
	    \Big) \nonumber
	    \\ &&
	   \qquad \times \Big(\prod_{v\in V'(\tau)} 2^{-E_v(h_v-h_{v'})}\Big)\,\Big(\prod_{v \in V_e(\tau)}|\lambda|^{\max\{1,\kappa|P_v|\}}
	      2^{h_v(2 - \tfrac12 |P_v| - \left\|\bs D_v\right\|_1 )} 2^{\theta h_v} \Big)
		\label{eq:W8_primitive_boundEdge}
	  \end{eqnarray}
where, in the first product in the right side, $R_v$ is defined as in Remark \ref{importremmE}, and, in the last product, $\theta=3/4$.
\end{proposition}

\medskip

\begin{remark}
	Proposition \ref{lem:W_primitive_bound} is a (significant) generalization of \cite[Proposition~4.6]{AGG_part2}, which states the validity of the
bound \eqref{eq:W8_primitive_boundEdge} for the (analogue of the) $\|\cdot\|_{0;h_{v_0}}$ norm (simply denoted $\|\cdot\|_{h_{v_0}}$ in \cite{AGG_part2}, see
\cite[Eq.~(4.4.1)]{AGG_part2}) of the infinite volume tree value $W_\infty[\tau,\ul P,\ul T,\ul D]$. In fact, the reader can easily check that, if $E_{v_0}=0$ (which is the only 
case of relevance for the infinite volume tree value: in fact, $W_\infty[\tau,\ul P,\ul T,\ul D]$ can be different from zero 
only if $\tau$ has $E_{v_0}=0$, in which case $2^{-E_{v_0} h_{v_0}}\prod_{v\in V'(\tau)} 2^{-E_v(h_v-h_{v'})}=1$), using the bound \eqref{eq:ct:short_memory} on $\epsilon_h$ 
with $\vartheta=\theta=3/4$, the right side of 
\cite[Eq.(4.4.2)]{AGG_part2} reduces to the right side of \eqref{eq:W8_primitive_boundEdge}. \end{remark}

\begin{proof} If $E_{v_0}=0$, then we use the fact that 
\begin{equation}\label{finiteinfinite} \|W_\Lambda[\tau, \ul P,\ul T,\ul D]\|_{0;h_{v_0}}\le \|W_\infty[\tau, \ul P,\ul T,\ul D]\|_{h_{v_0}},\end{equation}
where the norm in the right side is the same one defined in \cite[Eq.~(4.4.1)]{AGG_part2} in the infinite volume case. 
Thanks to \eqref{finiteinfinite}, in this case the desired bound, \eqref{eq:W8_primitive_boundEdge}, follows from 
\cite[Proposition~4.6]{AGG_part2} and from the bound \eqref{eq:ct:short_memory} on $\epsilon_h$. 
The only new case is $E_{v_0}=1$, which we focus on from now on.
We limit our discussion to the case $h_{v_0}<1$, the case $h_{v_0}=1$ being analogous (and, actually, simpler than the complementary one). 

\medskip

\noindent{\it{\underline{Case 1}}: $v$ is the only white vertex of $\tau$.}
In this case $W_\Lambda[\tau,\ul P,\ul T,\ul D]$ is given by \eqref{Wtaublack-}, that is, 
recalling that $\tau$ has no effective source endpoints, 
	\begin{eqnarray} &&
	\hskip-.3truecm
  W_\Lambda [ \tau, \ul P, \ul T, \ul D](\bs \omega_0,\bs D_0,\bs z_0)= \label{firstcase_1.01} \\
  &&\hskip-.3truecm=\frac{\mathds 1_{v_0}}{ |S_{v_0}|!}	\sum_{\substack{z: \cup_{v\in S_{v_0}}P_v \to \Lambda\\ 
			\bs z_0 = \bs z_{v_0}}}\Big[\fG_{T_{v_0}}^{(h_{v_0})} 
			\Big( \prod_{v \in S_{v_0}} K_{v}(\Psi_v)	\Big)-
			(-1)^{\alpha(\bs z_{v_0})}I_\Lambda(\Psi_{v_0})\fG_{T_{v_0},\infty}^{(h_{v_0})} 
			\Big(\prod_{v \in S_{v_0}} K_{v,\infty}(\Psi_{v,\infty})\Big)\Big]  	\nonumber
	\end{eqnarray}
	where: $\mathds 1_{v_0}$ is a shorthand for $\alpha_{v_0}\mathds 1(\bs \omega_0=\bs \omega_{v_0})\, \mathds 1(\bs D_0=\bs D_{v_0}=\bs D_{v_0}')$;
	for notational convenience, we dropped the arguments of $\fG_{T_{v_0}}^{(h_{v_0})}$ and $\fG_{T_{v_0},\infty}^{(h_{v_0})}$, which are both equal to 
	$(\bar\Psi_{v_1}, \ldots, \bar\Psi_{v_{s_{v_0}}})$; in the sum over $z$ we could freely assume that $z$ maps to $\Lambda$ rather than to $\bar\Lambda$, due to the support properties of the summand. We now rewrite the difference in the right side of \eqref{firstcase_1.01} as
	\begin{eqnarray} &&
  W_\Lambda [ \tau, \ul P, \ul T, \ul D](\bs \omega_0,\bs D_0,\bs z_0)=\label{firstcase_1}\\
  &&\hskip-.7truecm =\frac{\mathds 1_{v_0}}{ |S_{v_0}|!}	\sum_{\substack{z: \cup_{v\in S_{v_0}}P_v \to \Lambda\\ 
			\bs z_0 = \bs z_{v_0}}}\Big\{\Big[I_\Lambda(\Psi_{S_{v_0}})\big(\fG_{T_{v_0}}^{(h_{v_0})}-(-1)^{\alpha(\bs z_{S_{v_0}})-\alpha(\bs z_{v_0})}\fG_{T_{v_0},\infty}^{(h_{v_0})}\big)
			\Big(\prod_{v \in S_{v_0}} K_{v}(\Psi_v)\Big)\Big] \nonumber \\ &&\hskip2.6truecm+
			\Big[\big(1-I_\Lambda(\Psi_{S_{v_0}})\big)\fG_{T_{v_0}}^{(h_{v_0})}	\Big( \prod_{v \in S_{v_0}} K_{v}(\Psi_v)\Big)\Big]\nonumber\\
			&&\hskip2.6truecm-\Big[
			(-1)^{\alpha(\bs z_{v_0})}I_\Lambda(\Psi_{v_0})\big(1-I_\Lambda(\Psi_{S_{v_0}})\big)
			\fG_{T_{v_0},\infty}^{(h_{v_0})}
			\Big(	\prod_{v \in S_{v_0}} K_{v,\infty}(\Psi_v)\Big)
\Big]	  \Big\},	\nonumber
	\end{eqnarray}
	where $\Psi_{S_{v_0}}:=\oplus_{v\in S_{v_0}}\Psi_v$ (and, correspondingly, $\bs z_{S_{v_0}}:=\oplus_{v\in S_{v_0}}\bs z_v$), and 
	we used the fact that in the case under consideration all the vertices in $S_{v_0}$ are black so that, by the definition of bulk contributions, 
\begin{equation}\label{uff.512021} I_\Lambda(\Psi_{S_{v_0}}) (-1)^{\alpha(\bs z_{S_{v_0}})} \prod_{v \in S_{v_0}}K_{v}(\Psi_v)=I_\Lambda(\Psi_{S_{v_0}})\prod_{v \in S_{v_0}} K_{v,\infty}(\Psi_{v,\infty}).\end{equation}
By using the decomposition \eqref{firstcase_1}, we find $$\|W_\Lambda[\tau,\ul P,\ul T,\ul D]\|_{1;h_{v_0}}\le (I)+(II)+(III),$$ where $(I)$, $(II)$ and $(III)$
are contributions corresponding to the expressions in the first, second and third square brackets in the second, third and fourth lines of \eqref{firstcase_1}, respectively. 
Concerning $(I)$, we use Lemma~\ref{lem:fG_E_diff} to find (recall that the notation $*\parallel z(f_0)$ was introduced right before \eqref{eq:cRE_sumbound}): 
\begin{eqnarray} 
	(I)
	&\le& 
	\frac{C^{|Q_{v_0}|}}{ |S_{v_0}|!}	
	\sum_{z: \cup_{v\in S_{v_0}}P_v \to \Lambda}^{*\parallel z(f_0)}
	e^{\frac{c_0}2 2^{h_{v_0}}\delta_\E(\bs z_{v_0})} 	
	2^{h_{v_0}(\frac12|Q_{v_0}| + \sum_{v\in S_{v_0}}\|\bs D|_{Q_{v_0}}\|_1)} e^{-c_0 2^{h_{v_0}} \delta_\E (T_{v_0},\bs z({Q_{v_0}}))}
	\cdot\nonumber\\
	&\cdot & \Big(\prod_{v \in S_{v_0}} \big|K_{v}(\Psi_v)\big|\Big),
		\label{eq:W_cyl_Estar_part1}
	\end{eqnarray}
	where $\bs z(Q_{v_0})$ is the tuple with elements $z(f)$, $f\in Q_{v_0}$, and $f_0$ is an element of $Q_{v_0}$ such that $\dist(\bs z(Q_{v_0}),\partial\Lambda)=\dist(z(f_0),\partial\Lambda)$\footnote{Note that there is no need to write the suprema over $\bs \omega_{v_0}$ and $\bs D_{v_0}$ in the right side of 
	\eqref{eq:W_cyl_Estar_part1}, even though the definition of edge norm in \eqref{eq:cRE_sumbound} includes them: the reason is that, at $\ul P, \ul D$ fixed, these suprema 
	are trivial (the involved kernel at fixed $\ul P,\ul D$ is non zero only for a specific choice of $\bs \omega_{v_0}$ and $\bs D_{v_0}$). Similar comments hold for the 
norms appearing in the following discussion.}; recall that $\delta_\E(T,\bs z)$ was defined in \cref{eq:dE_Tz_def}.
Note that 	
\begin{equation}
	\delta_\E (\bs z_{v_0}) \le \delta_\E (\bs z_{S_{v_0}}) \le \delta_\E (T_{v_0}, \bs z(Q_{v_0})) + \sum_{v\in S_{v_0}} \delta (\bs z_v),
\end{equation}
since there is a set $S \subset \fB_\Lambda$ whose cardinality is the expression on the right hand side, which fulfills the requirements defining $\fT_\E(\bs z)$, see \cref{eq:delta_E_zx_def}.
As a result
\begin{eqnarray} 
(I)&\le& \frac{C^{|Q_{v_0}|}}{ |S_{v_0}|!}	\sum_{z: \cup_{v\in S_{v_0}}P_v \to \Lambda}^{*\parallel z(f_0)}
e^{-\frac{c_0}2 2^{h_{v_0}}\delta_\E(T_{v_0},\bs z(Q_{v_0}))} 	2^{h_{v_0}(\frac12|Q_{v_0}| + \sum_{v\in S_{v_0}}\|\bs D|_{Q_{v_0}}\|_1)} 
	\cdot\nonumber\\
	&\cdot &  \Big(\prod_{v \in S_{v_0}}e^{\frac{c_0}2 2^{h_{v_0}}\delta(\bs z_v)}\big|K_{v}(\Psi_v)\big|\Big)
		\label{eq:W_cyl_Estar_part1.1}
	\end{eqnarray}
In term $(II)$, noting that $\prod_{v \in S_{v_0}} K_{v}(\Psi_v)$ contains a factor $\prod_{v \in S_{v_0}} I_\Lambda(\Psi_v)\propto\mathds 1(\Psi_{S_{v_0}}\in\cM_\Lambda^\circ)$, 
the characteristic function $1-I_\Lambda(\Psi_{S_{v_0}})$ can be replaced by $\mathds 1(\diam_1(\bs z_{S_{v_0}})>L/3)$. Therefore, 
the term $(II)$ can be bounded from above by an expression analogous to the right side of \eqref{eq:W_cyl_Estar_part1}, 
modulo the facts that $\delta_\E (T_{v_0},\bs z({Q_{v_0}}))$ 
is replaced by $\delta(T_{v_0},\bs z(T_{v_0})):=\sum_{(f,f')\in T_{v_0}}\| z(f)-z(f')\|_1$, and that the summand is multiplied by 
$\mathds 1(\diam_1(\bs z_{S_{v_0}})>L/3)$. 
Because of the presence of this indicator function,
in all the nonvanishing terms $\bs z$ is such that 
$$\delta_\E (\bs z_{S_{v_0}}) > L/3 \ge \delta_\E (T_{v_0},\bs z(Q_{v_0})) - \delta(T_{v_0}, \bs z(T_{v_0})) $$ 
and
\begin{equation*}
	\delta_\E (\bs z_{S_{v_0}})
	=
	\delta (\bs z_{S_{v_0}})
	\le 
	\sum_{v\in S_{v_0}}\delta(\bs z_v)
	+
	\delta(T_{v_0}, \bs z(T_{v_0}))
	;
\end{equation*}
combining these two inequalities gives us
\begin{equation}
	\begin{split}
		\tfrac12 \delta_\E (\bs z_{S_{v_0}})
		-
		\delta(T_{v_0}, \bs z(T_{v_0}))
		&
		=
		\tfrac34
		\left[ 
			\delta_\E (\bs z_{S_{v_0}})
			-
			\delta(T_{v_0}, \bs z(T_{v_0}))
		\right]
		-
		\tfrac14 \delta_\E (\bs z_{S_{v_0}})
		-
		\tfrac14
		\delta(T_{v_0}, \bs z(T_{v_0}))
		\\ &
		\le 
		\tfrac34 \sum_{v \in S_{v_0}} \delta(\bs z_v)
		-
		\tfrac14 \delta_\E (T_{v_0},\bs z(Q_{v_0})) 
		,
	\end{split}
	\label{eq:delta_E_case_II}
\end{equation}
which together with the aforementioned bound (and $\delta_\E(\bs z_{v_0})\le \delta_\E(\bs z_{S_{v_0}})$) gives
\begin{eqnarray} 
(II)&\le& \frac{C^{|Q_{v_0}|}}{ |S_{v_0}|!}	\sum_{z: \cup_{v\in S_{v_0}}P_v \to \Lambda}^{*\parallel z(f_0)}
e^{-\frac{c_0}4 2^{h_{v_0}}\delta_\E(T_{v_0},\bs z(Q_{v_0}))} 	2^{h_{v_0}(\frac12|Q_{v_0}| + \sum_{v\in S_{v_0}}\|\bs D|_{Q_{v_0}}\|_1)} 
	\cdot\nonumber\\
	&\cdot & \Big( \prod_{v \in S_{v_0}}e^{\frac34{c_0}2^{h_{v_0}}\delta(\bs z_v)}\big|K_{v}(\Psi_v)\big|\Big).
		\label{eq:W_cyl_Estar_part2}
	\end{eqnarray}
Note that this bound is weaker than \eqref{eq:W_cyl_Estar_part1.1}.
Finally, $(III)$ is bounded from above by an expression analogous to the right side of \eqref{eq:W_cyl_Estar_part2}, 
with the only difference that $K_{v}$ is replaced by $K_{v,\infty}$. 

	\begin{figure}[h]
		\centering
		\begin{tikzpicture}[every node/.style={circle, inner sep = 2pt, outer sep = 0pt, fill=black},
				every label/.style={rectangle,fill=none, outer sep=0},
				every fit/.style={ellipse,draw, fill=none}
			]
			\node (f11) [label={[name=label1]90:$f_0 = \tilde f_{v_1}$}] {};
			\node (f12) [right=of f11] {}; 
			\node (f13) [below=of f11] { };
			\node (f14) [right=of f13] { }; 
			\node (w1) [fit = (label1) (f11) (f12) (f13) (f14),label={above:$v_1$}] {} ;
			\draw (f14) -- ++ (3,1.5) 
			node [fill=none,midway,above] {$\ell_1$}
			node[label={[name=label2]right:$\tilde f_{v_2}$}] (f21){ };
			\node (f22) [right=of f21] { };
			\node (f23) [below=of f21] { };
			\node (f24) [below=of f22] { };
			\node (w2) [fit = (label2) (f21) (f22) (f23) (f24), label= {above:$v_2 = \tilde v_{\ell_1}$}] { };
			\draw (f13) -- ++ (0,-3)
			node [fill=none,midway,left] {$\ell_2$}
			node[label={[name=label3]left:$\tilde f_{v_3}$}] (f31) {} ;
			\node (f32) [below=of f31] { };
			\node (w3) [fit = (label3) (f31) (f32), label={below:$v_3 = \tilde v_{\ell_2}$}] { };
			\draw (f23) -- ++ (0,-3)
			node [fill=none,midway,left] {$\ell_3$}
			node[label={[name=label4]right:$\tilde f_{v_4}$}] (f41) {};
			\node (f42) [below=of f41] { };
			\node (w4) [fit = (label4) (f41) (f42), label={below:$v_4 = \tilde v_{\ell_3}$}] { };
		\end{tikzpicture}
		\caption{Example of $\tilde f_v$, $\tilde v_\ell$ defined in the proof of Proposition~\ref{lem:W_primitive_bound}.  The numbering of the vertices and spanning tree edges is arbitrary.}
		\label{fig:f_and_w_8}
	\end{figure}
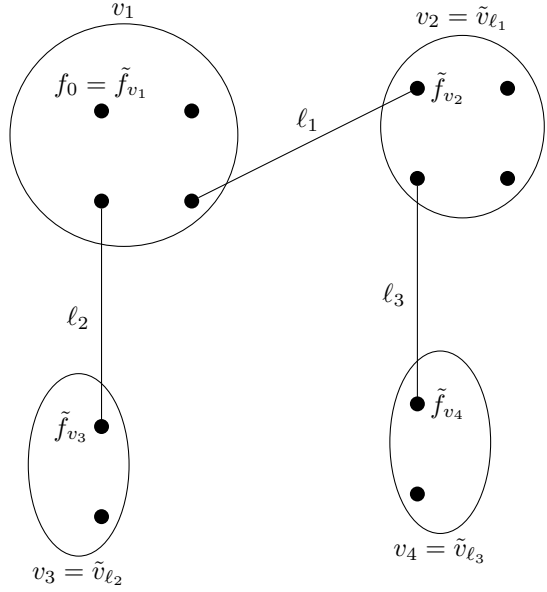

In order to sum over the coordinates, we proceed as follows. Given $f_0$ as above, let $\tilde f_v =  f_0$ for the $v\in S_{v_0}$ such that $f_0 \in P_v$; then for each $\ell \in T_{v_0}$ incident to $v$, let $\tilde v_\ell$ be the other vertex $\ell$ touches, and for this vertex let $\tilde f_{\tilde v_\ell}$ be the field label it shares with $\ell$; then do the same for all the other $\ell$ incident on the vertices mentioned so far, etc.  Figure~\ref{fig:f_and_w_8} gives an example.
	Having done so, and recalling that $\delta_\E(T_{v_0},\bs z(Q_{v_0}))=\min\{\dist(z(f_0),\partial\Lambda),L/3\}+\sum_{(f,f')\in T_{v_0}}\| z(f)-z(f')\|_1$, 
	we see that 
	\begin{eqnarray}
&&
\sum_{z: \cup_{v\in S_{v_0}}P_v \to \Lambda}^{*\parallel z(f_0)}
	    e^{-  \frac{c_0}{4}   2^{h_{v_0}}
	    \delta_\E (T_{v_0},\bs z(Q_{v_0}))}\Big(	\prod_{v \in S_{v_0}} e^{\frac34 c_0 2^{h_{v_0}}\delta(\bs z_v)}\big|K_{v}(\Psi_v)\big|\Big)
	    \nonumber\\ 
&& \qquad  \le   \Big(\sum_{(z)_2 = 1}^{M}   e^{- \frac{c_0}{4} 2^{h_{v_0}}\min\{(z)_2,M+1-(z)_2,L/3\}}\Big)\,\Big(\sum_{z \in \mathbb Z^2}  
		    e^{- \frac{c_0}{4} 2^{h_{v_0}} \|z\|_1}\Big)^{|S_{v_0}| - 1}\times		\label{eq:cyl_coord_sum_star}\\    
		    &&\qquad \qquad\times
		\prod_{v \in S_{v_0}}\Big(
		\sum_{z : P_v \to \Lambda}^{z(\tilde f_v)\ \text{fixed}}
		    e^{\frac34 c_0 2^{h_{v_0}}\delta(\bs z_v)}\big|K_{v}(\Psi_v)\big|\Big).
\nonumber	
\end{eqnarray}
The product in the last line can be rewritten as $\prod_{v \in S_{v_0}}\|K_{v}\|_{(\frac34c_02^{h_{v_0}})}$ and, recalling that the vertices $v\in S_{v_0}$ are black, 
$\|K_{v}\|_{(\frac34 c_02^{h_{v_0}})}\le \|K_{v,\infty}\|_{(\frac34 c_02^{h_{v_0}})}$, where the norm in the right side is the infinite volume norm (defined analogously to 
\eqref{eq:weightnorm}, with $\bar\Lambda$ replaced by $\Lambda_\infty$, see \cite[Eq.~(4.2.21)]{AGG_part2}). Now, the factors 
$\|K_{v,\infty}\|_{(\frac34 c_02^{h_{v_0}})}$ are bounded as discussed in \cite[Proposition~4.6]{AGG_part2}, see in particular 
the dotted list after \cite[Eq.~(4.4.9)]{AGG_part2} and \cite[Eq.~(4.4.13)]{AGG_part2}: using also \eqref{eq:ct:short_memory} with 
$\vartheta=\theta=3/4$, we get: 
\begin{equation}\|K_{v,\infty}\|_{(\frac34 c_02^{h_{v_0}})}\le \piecewise{C|\lambda|2^{(h_{v}-1)(2-\frac{|P_v|}2-\|\bs D_v\|_1)}2^{\theta h}& \text{if $v\in V_e(\tau)$ is of type \tikzvertex{ctVertex}}\\
			C^{|P_v|}|\lambda|^{\max\{1,\kappa |P_v|\}}& \text{if $v\in V_e(\tau)$ is of type \tikzvertex{vertex}} \\
			2^{-h_vR_v}\|W_\infty [\tau_v, \ul P_{v}, \ul T_{v}, \ul D_{v}']\|_{h_{v}}
		  	& \text{if $v\in V_0(\tau)$,}}\label{quest}\end{equation}
where, in the last line, $R_v$ is as in \eqref{eq:Rv_def}. Using \eqref{quest}, \eqref{eq:cyl_coord_sum_star} (or its analogue for the bound of the right side of $(III)$, 
 which is the same modulo a few trivial changes, due to replacing $\Lambda$ with $\Lambda_\infty$ and $K_{v}$ with $K_{v,\infty}$), and noting that 
 $\sum_{(z)_2 = 1}^{M}   e^{- \frac{c_0}{4} 2^{h_{v_0}}\min\{(z)_2,M+1-(z)_2,L/3\}}\le C 2^{-h_{v_0}}$ and $\sum_{z \in \mathbb Z^2}  
		    e^{- \frac{c_0}{4} 2^{h_{v_0}} \|z\|_1}\le C2^{-2h_{v_0}}$, we find:
\begin{eqnarray} \|W_\Lambda[\tau,\ul P,\ul T,\ul D]\|_{\E;h_{v_0}}&\le&\frac{C^{|Q_{v_0}|}}{ |S_{v_0}|!}2^{-h_{v_0}}	2^{h_{v_0}(\frac12|Q_{v_0}| + \sum_{v\in S_{v_0}}\|\bs D|_{Q_{v_0}}\|_1+2-2|S_{v_0}|)}\label{aquestami}\\
&\cdot&\prod_{v \in S_{v_0}} \piecewise{C|\lambda|2^{(h_{v}-1)(2-\frac{|P_v|}2-\|\bs D_v\|_1)}2^{\theta h}& \text{if $v\in V_e(\tau)$ is of type \tikzvertex{ctVertex}}\\
			C^{|P_v|}|\lambda|^{\max\{1,\kappa |P_v|\}}& \text{if $v\in V_e(\tau)$ is of type \tikzvertex{vertex}} \\
			2^{-h_vR_v}\|W_\infty [\tau_v, \ul P_{v}, \ul T_{v}, \ul D_{v}']\|_{h_{v}}.}\nonumber
\end{eqnarray}
Finally, using the bound of $\|W_\infty [\tau_v, \ul P_{v}, \ul T_{v}, \ul D_{v}']\|_{h_{v}}$ in 
 \cite[Proposition~4.6]{AGG_part2}, we obtain the desired bound, \eqref{eq:W8_primitive_boundEdge}, in the case under consideration, in 
which $E_{v_0}=1$ and $E_v=0$ for all $v>0$. 

\medskip

\noindent{\it{\underline{Case 2}}: $v$ is not the only white vertex of $\tau$.} In this case, $W_\Lambda [ \tau, \ul P, \ul T, \ul D]$ is defined as in item (3) after \eqref{after2A}, that is, 
	\begin{equation}  W_\Lambda [ \tau, \ul P, \ul T, \ul D](\bs \omega_0,\bs D_0,\bs z_0)=\frac{\mathds 1_{v_0}}{ |S_{v_0}|!}	\sum_{\substack{z: \cup_{v\in S_{v_0}}P_v \to \bar\Lambda\\ 
			\bs z_0 = \bs z_{v_0}}}\fG_{T_{v_0}}^{(h_{v_0})} 
			\Big( \prod_{v \in S_{v_0}} K_{v}(\Psi_v)	\Big)	 \label{secondcase_1} 
			\end{equation}
where the shorthand notations are the same as those used in \eqref{firstcase_1}.
If we now use the fact that 
	$$\delta_\E (\bs z_{v_0}) \le \delta_\E (\bs z_{S_{v_0}}) \le \delta(T_{v_0}, \bs z(T_{v_0})) 
	+ \sum_{v \in S_{v_0}} \delta_{E_v} (\bs z_v),$$
where $\delta_{E_v}(\bs z)=\delta_\E(\bs z)$ if $E_v=1$, and $\delta_{E_v}(\bs z)=\delta(\bs z)$ if $E_v=0$, we find
	\begin{eqnarray} 
&& \|W_\Lambda [\tau, \ul P, \ul T, \ul{D}]\|_{1;h_{v_0}}\le 
\frac{C^{|Q_{v_0}|}}{ |S_{v_0}|!}	2^{h_{v_0}(\frac12|Q_{v_0}| + \sum_{v\in S_{v_0}}\|\bs D|_{Q_{v_0}}\|_1)} \cdot 
\label{eq:W_cyl_main_bound1}\\
&&\qquad \cdot \sum_{z: \cup_{v\in S_{v_0}}P_v \to \bar\Lambda}^{*\parallel z(f_0)}\Big(\prod_{(f,f')\in T_{v_0}}e^{-\frac{c_0}2 2^{h_{v_0}}\|z(f)-z(f')\|_1}\Big)\, 
\Big( \prod_{v \in S_{v_0}} e^{\frac{c_0}2 2^{h_{v_0}}\delta_{E_v}(\bs z_v)}\big|K_{v}(\Psi_v)\big|\Big)\, ,
\nonumber	\end{eqnarray}
where $f_0$ is an arbitrary field label in $P_{w_0}$, with 
$w_0$ an arbitrary vertex in $S_{v_0}$ such that $E_{w_0}=1$. 
We define $\tilde f_v$ and $\tilde v_\ell$ in the same way as discussed after \eqref{eq:W_cyl_Estar_part2}, and rearrange the sum over positions as follows: 
each vertex $v$ with $E_v=0$ is associated with the sums over the coordinates of all of its field labels except for $\tilde f_v$, but when $E_v=1$ we also associate $v$ with the vertical coordinate of the position of $\tilde f_v$, thus obtaining a sum of the form appearing in the definition of the edge norm in \cref{eq:cRE_sumbound}.  The remaining sums are associated with the legs $\ell$ the spanning tree $T_{v_0}$; if $E_{\tilde v_\ell} = 0$ this sum is over two dimensions as before, but if $E_{\tilde v_\ell}=1$ the remaining sum is only over the horizontal coordinate. 
In this way, we bound the second line of \eqref{eq:W_cyl_main_bound1} as follows:
	\begin{eqnarray}
&& \sum_{z: \cup_{v\in S_{v_0}}P_v \to \bar\Lambda}^{*\parallel z(f_0)}\Big(\prod_{(f,f')\in T_{v_0}}e^{-\frac{c_0}2 2^{h_{v_0}}\|z(f)-z(f')\|_1}\Big) \Big( \prod_{v \in S_{v_0}}e^{\frac{c_0}2 2^{h_{v_0}}\delta_{E_v}(\bs z_v)}\big|K_{v}(\Psi_v)\big|\Big)\nonumber\\
&& \qquad \le\Biggl(\,\prod_{\substack{\ell \in T_v:\\ E_{\tilde v_\ell}=0}} \sum_{z\in \mathbb Z^2} e^{-\frac{c_0}2 2^{h_{v_0}} \|z\|_1}\Biggr)\, 
\Biggl(\,\prod_{\substack{\ell \in T_v:\\ E_{\tilde v_\ell}=1}} \sum_{(z)_1\in \mathbb Z} e^{-\frac{c_0}2 2^{h_{v_0}} |(z)_1|}\Biggr)\cdot \label{eq:cyl_coord_sum_main}\\
&&\qquad \quad \cdot\Biggl(\,\prod_{\substack{v\in S_{v_0}:\\ E_v=0}} \sum_{z : P_v \to \Lambda}^{z(\tilde f_v)\ \text{fixed}} e^{\frac{c_0}{2} 2^{h_{v_0}} \delta(\bs z_v)}
\big| K_v(\Psi_v)\big|\Biggr) \, \Biggl(\,\prod_{\substack{v \in S_{v_0}:\\ E_v=1}}	\sum_{z : P_v \to \bar\Lambda}^{* \parallel z (\tilde f_v) } e^{\frac{c_0}{2} 2^{h_{v_0}} \delta_\E(\bs z_v)}
\big| K_v(\Psi_v)\big|\Biggr)	\nonumber \\
&&\qquad \le C^{|S_{v_0}|}  2^{-h_{v_0}(2|S_{v_0}|-2)}\, 2^{-h_{v_0}} \Big(\,\prod_{v\in S_{v_0}} 2^{E_v h_{v_0}}\| K_v\|_{E_{v};h_{v_0}}\Big),
\nonumber	\end{eqnarray}
	where to obtain the last inequality we note that each $v\in S_{v_0}$ except for $w_0$ appears as $\tilde v_\ell$ for exactly one $\ell \in T_v$.
	
Now, the factors in the product over $v \in S_{v_0}$ with $E_v=0$ satisfy $\| K_v\|_{E_{v};h_{v_0}}\le \| K_{v,\infty}\|_{h_{v_0}}$ with $\| K_{v,\infty}\|_{h_{v_0}}$
bounded as in \eqref{quest}. 
We are left with bounding the factors in the last product, over the vertices $v\in S_{v_0}$ such that $E_v=1$. In this case, recalling the definition \eqref{Kvwhitedef}, 
if $v$ is an endpoint with $h_v<2$ (and, therefore, $h_v=h_{v_0}+1$), then 
$\| K_v\|_{1;h_{v_0}}=\|(\cR_\E C_\E^{(h_{v_0})})_{|P_v|,\|\bs D_v\|_1}\|_{1;h_{v_0}}$ with $C_\E^{(h)}$ defined as in \eqref{defCeZe}; now, using
 \eqref{betterdefCe}, we rewrite 
 \begin{equation}
	 \label{aggdet}\| K_v\|_{1;h_{v_0}}=\|(\cR_\E \mathcal E(V_\B^{(h_{v_0})}-\delta_{h_{v_0},0}V_\B^{(0)}))_{|P_v|,\|\bs D_v\|_1}\|_{1;h_{v_0}}
 \end{equation}
 that, in view of Remark \ref{remar35} and of the definition of $\mathcal R_\E$, see \eqref{REdef}, is non-zero only for $(|P_v|,\|\bs D_v\|_1)=(2,1),(2,2),(4,1)$;
 from this, using the bound  \eqref{eq:REbound} on the norm of $\cR_\E$, to be applied with $\kappa=\frac{c_0}2 2^{h_{v_0}}$ and 
$\kappa+\epsilon=\frac34 {c_0} 2^{h_{v_0}}$, and the bound \eqref{belissemo} on $\mathcal E V_\B^{(h)}$, we find 
\begin{equation}\label{3.3.28bis}\| K_v\|_{1;h_{v_0}}\le C|\lambda|2^{-h_v} 2^{h_v(2 - \tfrac12 |P_v| - \left\|\bs D_v\right\|_1 )} 2^{\theta h_v}.\end{equation}
Note that the scaling dimension $2 - \tfrac12 |P_v| - \left\|\bs D_v\right\|_1$ is $\le 0$ for all the values of $(|P_v|,\|\bs D_v\|_1)$ for which the left side is non-zero,
see the comment after \eqref{aggdet}.

On the other hand, if $v$ is an endpoint with $E_v=1$ and $h_v=2$, then $\| K_v\|_{1;h_{v_0}}=
\|(\cR_\E (V_\E^{(1)}+\mathcal E V_\B^{(1)}))_{|P_v|,\|\bs D_v\|_1,m_v}\|_{1;h_{v_0}}$; by using the 
the bound  \eqref{eq:REbound} on the norm of $\cR_\E$, the bound \eqref{eq:WE_base_decay} on the edge norm of $V_\E^{(1)}$ and the 
the bound \eqref{belissemo} on the norm of $\mathcal E V_\B^{(1)}$, we find that $\|(\cR_\E (V_\E^{(1)}+\mathcal E V_\B^{(1)}))_{|P_v|,\|\bs D_v\|_1,m_v}\|_{1;h_{v_0}}$
is bounded from above by $C^{|P_v|}|\lambda|^{\max\{1,\kappa|P_v|\}}$, which is also smaller than $C^{|P_v|}|\lambda|^{\max\{1,\kappa|P_v|\}}2^{-h_v}$ 
$2^{h_v(2 - \tfrac12 |P_v| - \left\|\bs D_v\right\|_1 )} 2^{\theta h_v}$, up to a redefinition of the constant $C$. 

Finally, if $v$ is a white vertex in $S_{v_0}\cap V_0(\tau)$, then $\|K_v\|_{1;h_{v_0}}=\|\lis W_\Lambda [\tau_v, \ul P_{v}, \ul T_{v}, \ul D_{v}]\|_{1;h_{v_0}}$, where we recall that 
$\lis W_\Lambda[\tau_v, \ul P_{v}, \ul T_{v}, \ul D_{v}]$ is defined via the analogue of \eqref{lisWinfty}, with $\cR_\infty W_\infty$
in the right side replaced by $\cR_\E W_\Lambda$. Once again, the norm of $\cR_\E$ is bounded as in \eqref{eq:REbound}, with $\kappa=\frac{c_0}2 2^{h_{v_0}}$ and 
$\kappa+\epsilon=c_0 2^{h_{v_0}} \le \tfrac{c_0}2 2^{h_v}$, so that, recalling that $R_v= \| \bs D_v\|_1-\| \bs D_v'\|_1$ (see Remark \ref{importremmE}), 
we find
$$\|\lis W_\Lambda [\tau_v, \ul P_{v}, \ul T_{v}, \ul D_{v}]\|_{1;h_{v_0}}\le C 2^{-R_v h_v}
\|W_\Lambda [\tau_v, \ul P_{v}, \ul T_{v}, \ul D_{v}']\|_{1;h_{v}}.$$
Putting things together, we get the analogue of \eqref{aquestami}, namely
\begin{eqnarray} 
&&\|W_\Lambda[\tau,\ul P,\ul T,\ul D]\|_{1;h_{v_0}}\le 
\frac{C^{|Q_{v_0}|}}{|S_{v_0}|!}  2^{h_{v_0}(\frac{|Q_{v_0}|}2+\sum_{v\in S_{v_0}}\|\bs D_v|_{Q_{v_0}}\|_1+2-2|S_{v_0}|)}\, 2^{-h_{v_0}}\cdot\nonumber\\
&&\cdot \prod_{v \in S_{v_0}}
		\piecewise{ C^{|P_v|}|\lambda|^{\max\{1,\kappa |P_v|\}} 2^{h_{v}(2-\frac{|P_v|}2-\|\bs D_v\|_1)} 2^{\theta h_v} 2^{-E_v(h_v-h_{v_0})}
		& \text{if $v\in V_e(\tau)$,}\\
		2^{-h_vR_v} 2^{E_v h_{v_0}}\|W_\Lambda [\tau_v, \ul P_{v}, \ul T_{v}, \ul D_{v}']\|_{E_v;h_{v}}  	& \text{if $v\in V_0(\tau)$.}
		}\nonumber\end{eqnarray}
Now, consider a factor 
$\|W_\Lambda [\tau_v, \ul P_{v}, \ul T_{v}, \ul D_{v}']\|_{E_v, h_{v}}$ appearing in the second case of the last product. If $E_v=0$, then we use 
\eqref{finiteinfinite} and bound it via \cite[Proposition~4.6]{AGG_part2}; 
if $E_v=1$, we iterate the bound, and we continue to do so until we are left only with endpoints or with vertices $v\in V_0'(\tau)$ such that $E_v=0$. 
By doing so, we obtain the desired bound, \eqref{eq:W8_primitive_boundEdge}.
\end{proof}

Let us now discuss the bound on the kernels associated with GN trees that have one or more effective source endpoints. 
The analogue of Proposition \ref{lem:W_primitive_bound} in the case of trees with $m_{v_0}\ge 1$ is summarized as follows. 

\medskip

\begin{proposition}
There exist $C,\kappa,\lambda_0>0$ 
such that, for any $\tau\in \cT_\cyl$ with $m_{v_0}\ge 1$,  $\bs x_{v_0}\in \fB_\Lambda^{m_{v_0}}$, 
$\ul P\in \cP(\tau)$, $\ul T\in \cS(\tau,\ul P)$, $\ul D\in \cD(\tau,\ul P)$ and $|\lambda|\le \lambda_0$,
\begin{eqnarray}
&&\hskip-.5truecm \|W_\Lambda[\tau,\ul P,\ul T,\ul D](\bs x_{v_0})\|_{E_{v_0};h_{v_0}}\le     C^{\sum_{v\in V_e(\tau)}(|P_v|+m_v)}\cdot\label{eq:W8_primitive_boundSource}  \\
&& \qquad \cdot \Big(\prod_{v \in V_0(\tau)}
	      \frac{2^{(\frac{|Q_v|}2+ \sum_{w\in S_v}\| \bs D_w |_{Q_v} \|_1 - R_v + 2 - 2 \left|S_v\right|)h_v}}{ |S_v|!}
	    \Big)\cdot\Big(\prod_{\substack{v\in V'(\tau):\\
	   m_v=0}}2^{-E_v(h_v-h_{v'})}\Big)\cdot\nonumber\\
	   &&\qquad\cdot\Big(\prod_{\substack{v\in V(\tau):\\
	   m_v\ge 1}} 2^{2[|S^*_v|-1]_+h_v} e^{-\frac{c_0}{12}2^{h_v}\delta_{E_v}(\bs x_v)}\Big) \cdot	
	      \Big(\prod_{\substack{v \in V_e(\tau):\\ m_v=0}}|\lambda|^{\max\{1,\kappa|P_v|\}}
	      2^{h_v(2 - \tfrac12 |P_v| - \left\|\bs D_v\right\|_1 )} 2^{\theta h_v} \Big)\cdot \nonumber\\
	   &&\qquad \cdot 
	      \Big(\prod_{\substack{v \in V_e(\tau):\\ m_v\ge 1}}\piecewise{
	      \tilde Z_{h_v-1}(E_v,\|\bs D_v\|_1,x_v) & \text{if $(|P_v|,m_v)=(2,1)$}\\
	       |\lambda|^{\max\{1,\kappa (|P_v|+m_v)\}} &  \text{otherwise} }\Big)
		      \nonumber	  \end{eqnarray}
where: in the first product in the third line, $S_v^*=\{w\in S_v: m_w \ge 1\}$, $[\cdot]_+$ is the positive part, and $c_0$ is the same constant as in \eqref{cosacomedove}; 
in the second product in the third line, $\theta=3/4$; in the product in the last line,
\begin{equation}  \tilde Z_{h_v-1}(E_v,\|\bs D_v\|_1,x_v) :=\piecewise{
\max_{j=1,2}\{|Z_{j,h_v-1}|\} & \text{if $E_v=0$ and $h_v\le 1$,}\\ 
\|(D_\E^{(h_v-1)})_{2,\|\bs D_v\|_1,1}(x_v)\|_{(\E;\frac38 c_02^{h_v})} & \text{if $E_v=1$ 
and $h_v\le 1$,}\\
1 & \text{if $E_v=0$ and $h_v=2$}\\
1+\max_{p=0,1}\|(\mathcal E B_\B^{(1)})_{2,p,1}\|_{(\E; c_0)} & \text{if $E_v=1$ and $h_v=2$}.}
\label{corpod}\end{equation}
with $c_0$ the same constant as in \eqref{cosacomedove}. 
Note that: 
in the product in the last line of \eqref{eq:W8_primitive_boundSource}, in
order for $(|P_v|,m_v)$ to be different from $(2,1)$, $h_v$ must be equal to $2$;
in the first line of the right side of \eqref{corpod}, $\|\bs D_v\|_1=0$ necessarily; in the second 
line of the right side of \eqref{corpod}, $\|\bs D_v\|_1\in\{0,1\}$ necessarily. 
\label{lem:W_primitive_with_m}
\end{proposition}

\begin{remark} The effective vertex renormalizations, $Z_{j,h}$, and the edge couplings $D_\E^{(h)}$, $\mathcal E B_\B^{(1)}$ will be bounded in Sect.\ref{sec:Zj} below.
\end{remark}

\begin{remark} Proposition \ref{lem:W_primitive_with_m} and its proof imply that the norm of the infinite volume effective source term, 
$\|W_\infty[\tau,\ul P,\ul T,\ul D](\bs x_{v_0})\|_{0;h_{v_0}}$ is bounded by the right side of 
\eqref{eq:W8_primitive_boundSource} with $E_v\equiv0$; note, in particular, that such a bound does not depend on the edge couplings $\{D_\E^{(k)}\}_{h_{v_0}\le k\le 0}$, 
while it does depend on the effective vertex renormalizations $\{Z_{j,k}\}_{h_{v_0}\le k\le 0}$. \end{remark} 

\begin{proof} As in the proof of Proposition \ref{lem:W_primitive_bound}, we limit our discussion to the case $h_{v_0}<1$, leaving the simpler case $h_{v_0}=1$ to 
the reader. We proceed as in Case 2 of the proof of Proposition \ref{lem:W_primitive_bound}: by using the fact that 
\begin{eqnarray}\delta_{E_{v_0}} (\bs z_{v_0}, \bs x_{v_0}) &\le& -\frac16\delta_{E_{v_0}}(\bs x_{v_0}) +\frac76
	\delta_{E_{v_0}} (\bs z_{S_{v_0}}, \bs x_{S_{v_0}})\\
	& \le &-\frac16\delta_{E_{v_0}}(\bs x_{v_0})+\frac76\big[\delta(T_{v_0}, \bs z(T_{v_0})) 
	+ \sum_{v \in S_{v_0}} \delta_{E_v} (\bs z_v)  \big], \end{eqnarray}
we obtain the analogue of \eqref{eq:W_cyl_main_bound1}:
	\begin{eqnarray} 
&& \|W_\Lambda [\tau, \ul P, \ul T, \ul{D}](\bs x_{v_0})\|_{E_{v_0};h_{v_0}}\le 
\frac{C^{|Q_{v_0}|}}{ |S_{v_0}|!}	2^{h_{v_0}(\frac12|Q_{v_0}| + \sum_{v\in S_{v_0}}\|\bs D|_{Q_{v_0}}\|_1)} \, e^{-\frac{c_0}{12}\delta_{E_{v_0}}(\bs x_{v_0})}\cdot 
\label{eq:W_cyl_main_bound1withm}\\
&&\qquad \cdot \sum_{z: \cup_{v\in S_{v_0}}P_v\to \bar\Lambda}\Big(\prod_{(f,f')\in T_{v_0}}e^{-\frac5{12}{c_0} 2^{h_{v_0}}\|z(f)-z(f')\|_1}\Big)\, 
\Big( \prod_{v \in S_{v_0}} e^{\frac7{12} {c_0} 2^{h_{v_0}}\delta_{E_v}(\bs z_v,\bs x_v)}\big|K_{v}(\Psi_v,\bs x_v)\big|\Big).
\nonumber	\end{eqnarray}
In order to sum over the coordinates, we introduce $\tilde v_\ell$ and $\tilde f_v$ be as follows (an example is given in Figure~\ref{fig:wtilde_ftilde_with_m}): 
	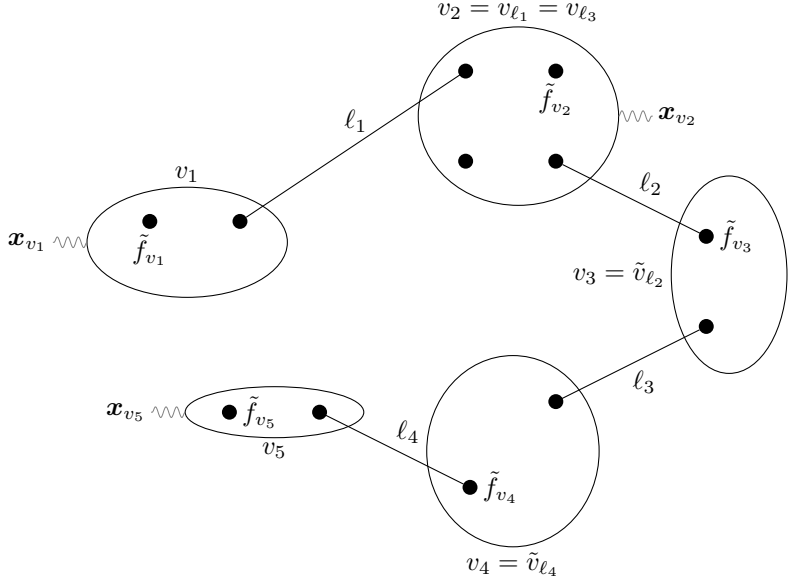
\begin{figure}[t]
		\centering
		\begin{tikzpicture}[every node/.style={circle, inner sep = 2pt, outer sep = 0, fill=black},
				every label/.style={rectangle,fill=none, outer sep=0},
				every fit/.style={ellipse,draw, fill=none, label distance=4pt, inner sep = 4pt},
				every pin/.style={rectangle,fill=none,outer sep=0,pin edge = {decorate,decoration={snake,amplitude=2pt,segment length =4pt}}}
			]
			\node (f11) [label={[name=label1]below:$\tilde f_{v_1}$}] { };
			\node (f12) [right=of f11]  { }; 
			\node (w1) [fit = (label1) (f11) (f12),pin={left:$\bs x_{v_1}$},label=above:$v_1$]  { };
			\draw (f12) -- ++ (3,2) 
			node [fill=none,midway,above] {$\ell_1$}
			node (f21) { };
			\node (f22) [right=of f21,label={[name=label2]below:$\tilde f_{v_2}$}] { };
			\node (f23) [below=of f21] { };
			\node (f24) [below=of f22] { };
			\node (w2) [fit= (label2) (f21) (f22) (f23) (f24),label={above:$v_2 = \tilde v_{\ell_1} = \tilde v_{\ell_3}$}, pin={right:$\bs x_{v_2}$}] { };
			\draw (f24) -- ++ (2,-1) 
			node [fill=none,midway,above right] {$\ell_2$}
			node (f31) [ label={[name=label3]right:$\tilde f_{v_3}$}] { };
			\node (f32) [below=of f31] {};
			\node (w3) [fit= (label3) (f31) (f32),label={left:$v_3 = \tilde v_{\ell_2}$}] {};
			\draw (f32) -- ++ (-2,-1) 
			node [fill=none,midway,below right] {$\ell_3$}
			node (f41) { };
			\node (f42) [below left =of f41, label={[name=label4]right:$\tilde f_{v_4}$}] {};
			\node (w4) [fit= (label4) (f41) (f42),label={below:$v_4=\tilde v_{\ell_4}$}] {};
			\draw (f42) -- ++ (-2,1) 
			node [fill=none,midway,above right] {$\ell_4$}
			node (f51) { };
			\node (f52) [left =of f51, label={right:$\tilde f_{v_5}$}] {};
			\node (w5) [fit= (f51) (f52),pin={left:$\bs x_{v_5}$},label={below:$v_5$}] {};
		\end{tikzpicture}
		\caption{Example of a selection of $\tilde f_v$, $\tilde v_\ell$ used in the proof Proposition~\ref{lem:W_primitive_with_m}. The clusters 
		with an external wiggly line correspond to vertices with $m_v\ge 1$, and $\bs x_v$ are the non-empty set of coordinates of the 
		corresponding probe fields. }
		\label{fig:wtilde_ftilde_with_m}
	\end{figure}
	\begin{enumerate}
		\item for each $v\in S_{v_0}$ with $m_v \ge 1$ let $\tilde f_v$ be an arbitrary element of $P_v$, and for each $\ell \in T_{v_0}$ which connects two such vertices let $\tilde v_\ell$ be one of them, chosen arbitrarily.
		\item pick another $\ell\in T_{v_0}$ which is incident on one of the vertices for which $\tilde f_v$ has already been designated:
		\begin{enumerate}
			\item If the other vertex $v'\in S_{v_0}$ which it touches does not yet have $\tilde f_v$ assigned, then let $\tilde v_\ell = v'$ and $\tilde f_{v'}$ be the unique element of $\ell \cap P_{v'}$.
			\item Otherwise, let $\tilde v_\ell$ be an arbitrarily chosen $v\in S_{v_0}$ with $m_v \ge 1$.
		\end{enumerate}
		\item Repeat the previous step until all $\tilde v_\ell$ with $\ell\in T_{v_0}$ have been assigned.
	\end{enumerate}

\begin{remark}
Note that, removing from $T_{v_0}$ the elements $\ell$ for which $\tilde v_\ell \in S_{v_0}^*$, separates $T_{v_0}$ into a collection of separate trees each of which contains exactly one 
$v \in S_{v_0}^*$, and so there are exactly $|S_{v_0}^*| -1$ such $\ell$: this is the key property we need from the definition of $\tilde v_\ell$. 
Note also that each $v \in S_{v_0} \setminus S_{v_0}^*$ is $\tilde v_\ell$ for exactly one $\ell$. \end{remark}

For the purpose of an upper bound, in the second line of \eqref{eq:W_cyl_main_bound1withm}, we 
replace the factors $e^{-\frac5{12}{c_0} 2^{h_{v_0}}\|z(f)-z(f')\|_1}$ associated with pairs $(f,f')\equiv \ell$ for which $\tilde v_\ell \in S_{v_0}^*$
by the factor $1$. By proceeding in this way, 
we bound the second line of \eqref{eq:W_cyl_main_bound1withm} as follows: 
  \begin{eqnarray}
  && \sum_{z: \cup_{v\in S_{v_0}}P_v \to \bar\Lambda}\Big(\prod_{(f,f')\in T_{v_0}}e^{-\frac5{12}{c_0} 2^{h_{v_0}}\|z(f)-z(f')\|_1}\Big)\, 
\Big( \prod_{v \in S_{v_0}} e^{\frac7{12}{c_0} 2^{h_{v_0}}\delta_{E_v}(\bs z_v,\bs x_v)}\big|K_{v}(\Psi_v,\bs x_v)\big|\Big)\le \nonumber\\
&&  \le\Big(\prod_{\substack{\ell \in T_v:\\ m_{\tilde v_\ell}=0, \ E_{\tilde v_\ell}=0}} \sum_{z\in \mathbb Z^2} e^{-\frac5{12}{c_0} 2^{h_{v_0}} \|z\|_1}\Big)\, 
\Big(\prod_{\substack{\ell \in T_v:\\ m_{\tilde v_\ell}=0, \ E_{\tilde v_\ell}=1}} \sum_{(z)_1\in \mathbb Z} e^{-\frac5{12}{c_0} 2^{h_{v_0}} |(z)_1|}\Big)\cdot \label{eq:eq:eq1}\\
&& \quad \cdot	\Big(\prod_{v\in S_{v_0}^*} \sum_{z : P_v \to \bar\Lambda} e^{\frac7{12}{c_0} 2^{h_{v_0}} \delta_{E_v}(\bs z_v,\bs x_v)}
\big| K_v(\Psi_v,\bs x_v)\big|\Big) \cdot\nonumber\\
&& \quad \cdot \Big(\prod_{\substack{v\in S_{v_0}\setminus S_{v_0}^*:\\ E_v=0}} \sum_{z : P_v \to \Lambda}^{z(\tilde f_v)\ \text{fixed}} e^{\frac7{12}{c_0} 2^{h_{v_0}}
\delta(\bs z_v)}\big| K_v(\Psi_v)\big|\Big) \, \Big(	\prod_{\substack{v \in S_{v_0}\setminus S_{v_0}^*:\\ E_v=1}}\ \
\sum_{z : P_v \to \bar\Lambda}^{* \parallel z (\tilde f_v) } e^{\frac7{12}{c_0} 2^{h_{v_0}} \delta_\E(\bs z_v)}\big| K_v(\Psi_v)\big|\Big).\nonumber 
\end{eqnarray}
Now, the products in the first, second, fourth and fifth parentheses in the right side are bounded as in Case 2 of the proof of Proposition \ref{lem:W_primitive_bound}, see \eqref{eq:cyl_coord_sum_main} and following discussion (the fact that the prefactor in the exponent is $7/12$ rather than $1/2$ causes some trivial changes). 
The product in the third parentheses in the right side is new, but it can be bounded via a similar procedure. We need to distinguish various cases:
\begin{itemize}
\item If $v\in S_{v_0}^*$ is in $V_0(\tau)$ and $E_v=0$, then $K_v(\Psi_v,\bs x_v)=
\lis W_\Lambda [\tau_v, \ul P_{v}, \ul T_{v}, \ul D_{v}](\Psi_v, \bs x_v)$, with $\lis W_\Lambda [\tau_v, \ul P_{v}, \ul T_{v}, \ul D_{v}]$ defined via the analogue of \eqref{lisWinfty}, with $\cR_\infty W_\infty$
replaced by $\cR_\B W_\Lambda$. By using \eqref{eq:RBB} (cf. also \cite[Eq.~(4.4.12)]{AGG_part2}), we find that
$$\sum_{z : P_v \to \bar\Lambda}e^{\frac7{12}{c_0} 2^{h_{v_0}} \delta(\bs z_v,\bs x_v)}\big| K_v(\Psi_v,\bs x_v)\big|\le C 2^{-h_v R_v} \| W_\Lambda [\tau_v, \ul P_{v}, \ul T_{v}, \ul D_{v}']\|_{0;h_v}.$$
\item If $v\in S_{v_0}^*$ is in $V_0(\tau)$ and $E_v=1$, then $K_v(\Psi_v,\bs x_v)=
W_\Lambda [\tau_v, \ul P_{v}, \ul T_{v}, \ul D_{v}](\Psi_v, \bs x_v)$, so that
$$\sum_{z : P_v \to \Lambda}e^{\frac7{12}{c_0} 2^{h_{v_0}} \delta_\E(\bs z_v,\bs x_v)}\big| K_v(\Psi_v,\bs x_v)\big|\le \| W_\Lambda [\tau_v, \ul P_{v}, \ul T_{v}, \ul D_{v}]\|_{1;h_v}.$$
\item If $v\in S_{v_0}^*$ is an endpoint of type \tikzvertex{bareProbeEP} and $h_v = h_{v_0}+1$, then 
$K_{v}(\Psi_v, \bs x_v)=Z_{h_{v_0}}\cdot F_\B^A(\Psi_v, \bs x_v)$, so that 
$$\sum_{z : P_v \to \bar\Lambda}e^{\frac7{12}{c_0} 2^{h_{v_0}} \delta(\bs z_v,\bs x_v)}\big| K_v(\Psi_v,\bs x_v)\big|\le \max_{j=1,2}\{|Z_{j,h_v-1}|\}.$$
\item If $v\in S_{v_0}^*$ is an endpoint of type \tikzvertex{bareProbeEP} and $h_v = 2$, then 
$K_{v}(\Psi_v, \bs x_v)=(\cR_\B B_\B^{(1)})_\B(\Psi_v, \bs x_v)$, so that: $K_v=0$ if $(|P_v|,m_v)=(2,1)$ and $\|\bs D_v\|_1=0$; and, otherwise, using 
\eqref{eq:WL1bound}, \eqref{eq:WfreeL1bound} and \eqref{eq:RBB}, 
\begin{equation}
\sum_{z : P_v \to \bar\Lambda}e^{\frac7{12}{c_0} 2^{h_{v_0}} \delta(\bs z_v,\bs x_v)}\big| K_v(\Psi_v,\bs x_v)\big|\le 
\piecewise{ C \hskip4.2truecm \text{if $(|P_v|,m_v)=(2,1)$,} & \\
C^{|P_v|+m_v}e^{-\frac{c_0}3\delta(\bs x_v)} |\lambda|^{\max\{1,\kappa(|P_v|+m_v)\}}\quad \text{otherwise.} & }\label{case.1.00}\end{equation}
\item If $v\in S_{v_0}^*$ is an endpoint of type \tikzvertex{bareProbeEP,E} and $h_v = h_{v_0}+1$, then 
$K_v=0$ unless $(|P_v|,\|\bs D_v\|_1,m_v)=(2,p,1)$ with $p=0,1$, in which case 
$K_{v}(\Psi_v, \bs x_v)=D_\E^{(h_{v_0})}(\Psi_v,\bs x_v)$, with $\bs x_v=x_v$ consisting of a single point, so that
\begin{equation}\begin{split}\sum_{z : P_v \to \bar \Lambda}e^{\frac7{12}{c_0} 2^{h_{v_0}} \delta_\E(\bs z_v,x_v)}\big| K_v(\Psi_v,x_v)\big|&\le
\| (D_\E^{(h_{v_0})})_{2,\|\bs D_v\|_1,1}(x_v)\|_{(\E;\frac7{12}{c_0}2^{h_{v_0}})}\\
&\le e^{-\frac{c_0}{12}2^{h_v}\delta_\E(x_v)}
\| (D_\E^{(h_{v_0})})_{2,\|\bs D_v\|_1,1}(x_v)\|_{(\E;\frac3{8}{c_0}2^{h_{v}})}.\end{split}\label{case.1.01}\end{equation}
\item If $v\in S_{v_0}^*$ is an endpoint of type \tikzvertex{bareProbeEP,E} and $h_v = 2$, then 
$K_{v}(\Psi_v, \bs x_v)=(B^{(1)}_\E+\mathcal E B_\B^{(1)})(\Psi_v, \bs x_v)$, so that
\begin{equation}\begin{split}& \sum_{z : P_v \to \bar\Lambda}e^{\frac{7c_0}{12} 2^{h_{v_0}} \delta_\E(\bs z_v,\bs x_v)}\big| K_v(\Psi_v,\bs x_v)\big|\le e^{-\frac{c_0}{3}\delta_\E(\bs x_v)}
\big(\delta_{\|\bs D_v\|_1,0}\|(B^{(1)}_\E)_{(|P_v|,0,m_v)}\|_{(\E;c_0)}\\
&\hskip2.truecm+ \delta_{|P_v|,2}\delta_{m_v,1}\mathds 1(\|\bs D_v\|_1\le 1)\max_{p=0,1}\|(\mathcal E B_\B^{(1)})_{(2,p,1)}\|_{(\E;c_0)}\big),\end{split}
\label{case.1.01bb}\end{equation}
where we used the fact that, by definition, $(B^{(1)}_\E)_{(n,p,m)}$ vanishes for $p=1$, and that, by Remark \ref{remar35}, $(\mathcal E B_\B^{(1)})_{(n,p,m)}$ 
vanishes unless $(n,p,m)\in\{(2,0,1), (2,1,1)\}$. 
Now, using \eqref{eq:WfreeE_base_decay} and \eqref{eq:WE_base_decay}, $\|(B^{(1)}_\E)_{n,0,m}\|_{(\E;c_0)}$ can be bounded from above by $C$, if $(n,m)=(2,1)$,
and by $C^{n+m}|\lambda|^{\max\{1,\kappa(n+m)\}}$, otherwise. In conclusion, 
\begin{equation}\begin{split} & \sum_{z : P_v \to \bar\Lambda}e^{\frac{7c_0}{12} 2^{h_{v_0}} \delta_\E(\bs z_v,\bs x_v)}\big| K_v(\Psi_v,\bs x_v)\big|\le \\
&\qquad \le e^{-\frac{c_0}{3}\delta_\E(\bs x_v)} \piecewise{ C+\max_{p=0,1}\|(\mathcal E B_\B^{(1)})_{2,p,1}\|_{(\E;c_0)} & \text{if $(|P_v|,m_v)=(2,1)$}, \\
C^{|P_v|+m_v}|\lambda|^{\max\{1,\kappa(|P_v|+m_v)\}} & \text{otherwise.} }\end{split}\end{equation}
\end{itemize}
Putting things together, we find 
\begin{eqnarray} 
&&\|W_\Lambda[\tau,\ul P,\ul T,\ul D]\|_{E_v;h_{v_0}}\le \frac{C^{|Q_{v_0}|}}{ |S_{v_0}|!}	2^{h_{v_0}(\frac12|Q_{v_0}| + \sum_{v\in S_{v_0}}\|\bs D|_{Q_{v_0}}\|_1+2|S_{v_0}^*|-2|S_{v_0}|)} \, e^{-\frac{c}{12}\delta_{E_{v_0}}(\bs x_{v_0})}
\cdot\nonumber\\
&&\cdot \prod_{v \in S_{v_0}\setminus S_{v_0}^*}
		\piecewise{ C^{|P_v|}|\lambda|^{\max\{1,\kappa |P_v|\}} 2^{h_{v}(2-\frac{|P_v|}2-\|\bs D_v\|_1)} 2^{\theta h_v} 2^{-E_v(h_v-h_{v_0})}
		& \text{if $v\in V_e(\tau)$}\\
		2^{-h_vR_v} 2^{E_{v}h_{v_0}}\|W_\Lambda [\tau_v, \ul P_{v}, \ul T_{v}, \ul D_{v}']\|_{E_v;h_{v}}  	& \text{if $v\in V_0(\tau)$,} %
		}\nonumber\\
&&\cdot \prod_{v \in S_{v_0}^*}
		\piecewise{e^{-\frac{c_0}{12}2^{h_v}\delta_{E_v}(x_v)} \tilde Z_{h_v-1}(E_v,\|\bs D_v\|_1,x_v) & \text{if $v\in V_e(\tau)$ and $(|P_v|,m_v)=(2,1)$,}\\
 C^{|P_v|+m_v} e^{-\frac{c_0}{3}\delta_{E_v}(\bs x_v)}|\lambda|^{\max\{1,\kappa (|P_v|+m_v)\}}	& \text{if $v\in V_e(\tau)$ and $(|P_v|,m_v)\neq(2,1)$,}\\
		2^{-h_vR_v}\|W_\Lambda [\tau_v, \ul P_{v}, \ul T_{v}, \ul D_{v}']\|_{E_v;h_{v}}  	& \text{if $v\in V_0(\tau)$.}} %
\nonumber\end{eqnarray}
Now, the factors associated with the vertices $v\in S_{v_0}\setminus S_{v_0^*}$ that are not endpoints are bounded as in Proposition \ref{lem:W_primitive_bound}. 
For the factors associated with the vertices $v\in S_{v_0^*}$ that are not endpoints, we iterate the bound, and we continue to do so until we are left only with endpoints, 
or with vertices in $V_0(\tau)$ with $m_v=0$. 
By doing so, we obtain the desired bound, \eqref{eq:W8_primitive_boundSource}.
\end{proof}

Next, we rearrange the bounds in Propositions \ref{lem:W_primitive_bound} and \ref{lem:W_primitive_with_m} 
in a different form, more suitable for summing over GN trees and their labels, 
and for deriving the desired dimensional bounds on the multipoint energy correlations, to be discussed in Section \ref{sec:correlations} below. The result is summarized in 
the following proposition. 

\begin{proposition} Under the same assumptions as Propositions \ref{lem:W_primitive_bound} and \ref{lem:W_primitive_with_m}, 
	\begin{eqnarray}
&& \|W_\Lambda[\tau,\ul P,\ul T,\ul D]\|_{E_{v_0};h_{v_0}}\le     \frac{C^{m_{v_0}+\sum_{v\in V_e(\tau)}|P_v|}}{|S_{v_0}|!}
2^{h_{v_0}d(P_{v_0}, \bs D_{v_0},m_{v_0},E_{v_0})} \cdot   \label{eq:W8_reworked_boundzaza}\\
&&\qquad \cdot   \Big(
	      \prod_{v \in V'(\tau)}\frac1{|S_{v}|!}
	      2^{(h_v-h_{v'})d(P_v,\bs D_v,m_v,E_v)}  \Big)\,\Big(\prod_{\substack{v\in V(\tau):\\
	   m_v\ge 1}} 2^{2[|S^*_v|-1]_+h_v} e^{-\frac{c_0}{12}2^{h_v}\delta_{E_v}(\bs x_v)}\Big)     \nonumber\\
	   &&\qquad \cdot 
	      \prod_{v \in V_e(\tau)}\piecewise{ 2^{h_v\|\bs D_v\|_1} \tilde Z_{h_v-1}(E_v,\|\bs D_v\|_1,x_v) &\text{if $(|P_v|,m_v)=(2,1)$}\\
	      	           |\lambda|^{\max\{1,\kappa(|P_v|+m_v)\}}2^{\theta h_v} & \text{otherwise}}\nonumber
	  \end{eqnarray}
	where $d(P_v,\bs D_v,m_v,E_v)=2-\frac{|P_v|}{2}-\|\bs D_v\|_1-m_v-E_v\delta_{m_v,0}$ is the scaling dimension of $v$, see \eqref{defscaldimwithm}. 
	\label{lm:W:scaldim_withm}
\end{proposition}

\begin{remark} As discussed in Remark \ref{importremmE}, the scaling dimension $d(P_v,\bs D_v,m_v,E_v)$ is negative (more precisely, it is $\le -1$) for all the allowed values of 
$P_v,\bs D_v,m_v,E_v$, with the only exception of the case $|P_v|=2$, $\bs D_v={\bf 0}$ and $m_v=E_v=1$, in which case it vanishes. A step-by-step 
repetition of the proof of \cite[Lemma~4.8]{AGG_part2} immediately implies that the sum over GN trees, restricted to allowed labels such that $(|P_v|,\|\bs D_v\|_1,m_v,E_v)\neq(2,0,1,1)$ for all the vertices in 
$V_0(\tau)\cup \{v\in V_e(\tau): h_{v'}<h_v-1\}$, is convergent, uniformly in the scale of the root. In particular, let us state the resulting bound on the GN tree expansion for $(B_\infty^{(h)})_{2,p,1}$, which will be used in 
the next section in order to bound $Z_{j,h}, D_\E^{(h)}$ and $\mathcal E B_\B^{(1)}$: for $h\le 0$, $p\in\{0,1,2\}$, and (say) $\theta=3/4$: 
\begin{equation} \begin{split}2^{-\theta h}\sum_{\substack{\tau\in \cT^{(h)}:\\ E_{v_0}=0,\, m_{v_0}=1}}^* &
\ \sum_{\substack{\ul P\in\cP(\tau)\\  |P_{v_0}|=2}}\ \sum_{\ul T\in \cS(\tau,\ul P)}\ \sum_{\substack{\ul D\in\cD(\tau,\ul P)\\ \|\bs D_{v_0}\|_1=p}}\|W_\infty[\tau,\ul P,\ul T,\ul D](\bs x_{v_0})\|_{0;h+1}\\ 
& \le 	C|\lambda| \Big(\delta_{h,0}+(1-\delta_{h,0})\max_{h<h'\le 0}\max_{j=1,2} |Z_{j,h'}|\Big) 2^{-ph},\end{split}
\label{eq:treesum}\end{equation}
provided that $|\lambda|$ is sufficiently small (here the $*$ on the sum indicates, as usual, the constraint that $v_0=v_0(\tau)$ is dotted; in the present situation, this implies that $\tau$ has at least one endpoint of type \tikzvertex{vertex}). Contrary to the trees 
contributing to $(B_\infty^{(h)})_{2,p,1}$ (or, more generally, to the bulk part of the effective source term, $B_\B^{(h)}$), 
trees contributing to the edge part of the effective source term, $B_\E^{(h)}$, will, in general, contain vertices with $(|P_v|,\|\bs D_v\|_1,m_v,E_v)=(2,0,1,1)$, for 
which the scaling dimension vanishes; the presence of such vertices has a significant impact on the proof of convergence of the sum over GN trees for the multipoint correlation functions, to be discussed in Section 
\ref{sec:correlations} and in Appendix \ref{app.proofboringb} below. 
\label{rem:4.21}
\end{remark}

\begin{proof} We start from \eqref{eq:W8_primitive_boundEdge} and \eqref{eq:W8_primitive_boundSource}. Recalling that 
	$R_v=\|\bs D_v\|_1-\sum_{w\in S_v}\|\bs D_w\big|_{P_v}\|_1$, and using \cite[Eq.~(4.4.23)]{AGG_part2} and the analogue of \cite[(4.4.24)]{AGG_part2}, we get 
	\begin{eqnarray}
&& \|W_\Lambda[\tau,\ul P,\ul T,\ul D]\|_{E_{v_0};h_{v_0}}\le     \frac{C^{m_{v_0}+\sum_{v\in V_e(\tau)}|P_v|}}{|S_{v_0}|!}
2^{h_{v_0}d(P_{v_0}, \bs D_{v_0})} \cdot\piecewise{2^{-E_{v_0}h_{v_0}} & \text{if $m_{v_0}=0$}\\ 1 &\text{if $m_{v_0}>0$}}\cdot		
\nonumber \\
&&\qquad \cdot   \Big(
	      \prod_{v \in V'(\tau)}\frac1{|S_{v}|!}
	      2^{(h_v-h_{v'})d(P_v,\bs D_v)}  \Big)\,\Big(\prod_{\substack{v\in V'(\tau):\\
	   m_v=0}}2^{-E_v(h_v-h_{v'})}\Big)\cdot\nonumber\\
	   &&\qquad \cdot  \Big(\prod_{\substack{v\in V(\tau):\\
	   m_v\ge 1}} 2^{2[|S^*_v|-1]_+h_v} e^{-\frac{c_0}{12}2^{h_v}\delta_{E_v}(\bs x_v)}\Big)\cdot\label{maquindi}\\
	   &&\qquad \cdot \Big( \prod_{v \in V_e(\tau)}\piecewise{     2^{-h_v d(P_v,\bs D_v)}\tilde Z_{h_v-1}(E_v,\|\bs D_v\|_1,x_v) & \text{if $(|P_v|,m_v)=(2,1)$,} \\
	      	           |\lambda|^{\max\{1,\kappa(|P_v|+m_v)\}}2^{\theta h_v} & \text{otherwise,}}\Big)\nonumber
	  \end{eqnarray}
where $d(P_v,\bs D_v)=2-\frac{|P_v|}2-\|\bs D_v\|_1$. If we now use the identity 
$$2^{-h_{v_0}m_{v_0}}\prod_{v\in V_0'(\tau)}2^{-(h_v-h_{v'})m_v}\prod_{v\in V_e(\tau)}2^{h_{v'}m_v}=1,$$
and note that, for any $v\in V_e(\tau)$ such that $(|P_v|,m_v)=(2,1)$ one has $2^{-h_v d(P_v,\bs D_v)}=2^{h_v\|\bs D_v\|_1}2^{-h_vm_v}$, after rearranging the resulting factors, we get 
\eqref{eq:W8_reworked_boundzaza}.
\end{proof}

\subsection{Beta function equation for $Z_{j,h}$ and bounds on $D_\E^{(h)}$, $\mathcal E B_\B^{(1)}$}\label{sec:Zj}

In this section we bound the effective vertex renormalizations $Z_{j,h}$, with $j=1,2$ and $h\le 0$, and the kernels $D_\E^{(h)}$, $\mathcal E B_\B^{(1)}$ involved in the definition of 
$\tilde Z_{h_v-1}(E_v,\|\bs D_v\|_1,x_v)$ in \eqref{corpod}. The bound the effective vertex renormalizations is based on the control of their flow equation, known as the 
`beta function equation' in the RG jargon, and is discussed in Section \ref{ebbZj}. 
The bounds on $D_\E^{(h)}$ and $\mathcal E B_\B^{(1)}$  follow from an adaptation of Appendix \ref{app:segunda}, and are discussed in Section \ref{sec.3.4.2}. 

\subsubsection{Beta function equation for $Z_{j,h}$}\label{ebbZj} The definition of the effective vertex renormalizations $Z_{1,h}$ and $Z_{2,h}$, see \eqref{eeqq:3.1.42}, 
combined with the GN tree expansion for the infinite volume effective potentials, see \eqref{treeexpinfty},
implies the following equation, for all $h\le 0$: 
\begin{equation}  Z_{j,h} =4\sum_{\tau\in \cT^{(h)}}\cL_\infty W_\infty[\tau]((\bs \omega,\bs 0,\bs 0),x)\Big|_{\bs \omega=(+,-)}^{x=\hat e_j/2}\equiv
2\sum_{\tau\in \cT^{(h)}}\sum_{\bs y\in\Lambda_\infty^2}
W_\infty[\tau]((\bs \omega,\bs 0,\bs y),x)\Big|_{\bs \omega=(+,-)}^{x=\hat e_j/2}
\label{defcLZj}\end{equation}
(recall $\Lambda_\infty := \bZ^2$).
Note that the trees contributing to \eqref{defcLZj} have exactly one endpoint of type \tikzvertex{bareProbeEP}, and all the others (if any)
of type \tikzvertex{vertex} or \tikzvertex{ctVertex}. Note also that sum over $\tau$ in \eqref{defcLZj} is absolutely convergent, with 
the contribution from the trees with $v_0$ dotted bounded as in \eqref{eq:treesum}. The case in which $v_0$ is not dotted corresponds to 
the contribution from the `trivial' tree with exactly one endpoint (which is, therefore, of type \tikzvertex{bareProbeEP}) on scale $h_v=h+2$ (in fact, if $h<0$, 
the contribution to \eqref{defcLZj} from the tree with exactly one endpoint $v$ on scale $h_v=2$ and $v_0$ not dotted vanishes,
because $\cL_\infty(\cR_\infty B^{(1)}_\infty)=0$, by the definition of $\cL_\infty$ and $\cR_\infty$); moreover, the contribution from such a trivial tree equals
$Z_{j,h+1}$, if $h<0$, and $4(\cL_\infty B_\infty^{(1)})_{2,0,1}(((+,-),\bs 0,\bs 0),\hat e_j/2)$, if $h=0$.

Thanks to these considerations, recalling also the fact that $W_\infty^{(1)}$ is analytic in $\lambda$, see the comments after \eqref{eq:forma1}, 
and the bound on the norm of $W_\infty^{(1)}$ following from \eqref{eq:WL1bound}, \eqref{eq:WfreeL1bound} and following discussion,
we find that, if $h=0$, then $Z_{j,0}$, with $j=1,2$, are analytic in $\lambda$, 
and bounded as $|Z_{j,0}|\le C_0$ for all $|\lambda|\le \lambda_0$ and some $C_0>0$. More explicitly, recalling the definition of $B_\infty^{(1)}$ following from 
\eqref{cV1defin} and \eqref{expanV1.0}, the bound \eqref{eq:WL1bound}, and the facts that $Z=1+O(\lambda)$, $t_1^*-t_1=O(\lambda)$ and $t_2^*-t_2=O(\lambda)$ with 
$t_2^*=(1-t_1^*)/(1+t_1^*)$ (see Remark \ref{remacon}), after a straightforward computation we find that 
\begin{equation}Z_{j,0}=4(\cL_\infty B_\infty^{(1)})_{2,0,1}(((+,-),\bs 0,\bs 0),\hat e_j/2)+O(\lambda)=Z^*_j+O(\lambda),\end{equation}
with
\begin{equation} Z^*_j:=\begin{cases} 2t_2^* & \text{if $j=1$}\\
1-(t_2^*)^2 & \text{if $j=2$}\end{cases}\label{behZ*}\end{equation}
If $h<0$, distinguishing the contribution from the trivial tree with one endpoint on scale $h_v=2$ from the rest, we can write 
\begin{equation}
\label{betafunZj} Z_{j,h}=Z_{j,h+1}+B^j_{h+1},\end{equation}
where the {\it beta function} $B^j_{h+1}$ is given by the analogue of the right side of \eqref{defcLZj}, modulo the constraint that trees contributing to $B^j_{h+1}$ must have $v_0$ dotted. Thanks to \eqref{eq:treesum},
\begin{equation}
\label{betafunZjbound}
|B^{j}_{h}|\le C |\lambda| 2^{\theta h}\max_{h'\ge h}\{|Z_{j,h'}|\}, \end{equation}
for $\theta=3/4$. Eqs.\eqref{betafunZj} and \eqref{betafunZjbound}, in combination with the absolute convergence of the expansion for $B^j_{h+1}$ and the 
analyticity of $W_\infty^{(1)}$, imply that $\{Z_{j,h}\}_{h\le 0}$, with $j=1,2$, are two Cauchy sequences, 
whose elements are real analytic in $\lambda$ for $|\lambda|\le \lambda_0$. We let $Z_{j,-\infty}=Z_{j,-\infty}(\lambda)=Z^*_j+O(\lambda)$ be the limit as 
$h\to -\infty$ of $Z_{j,h}$, which is also real analytic in $\lambda$. Note also that, for $\theta=3/4$ (say),
\begin{equation}\label{Zj-Zj-infty}\big| Z_{j,h}-Z_{j,-\infty}\big|\le C|\lambda| 2^{\theta h}.\end{equation}

\subsubsection{Bounds on $D_\E^{(h)}$ and $\mathcal E B_\B^{(1)}$}\label{sec.3.4.2}

In this subsection we derive bounds on $\|(D_\E^{(h)})_{2,p,1}(x)\|_{(\E;\frac34c_02^h)}$ and 
$\|(\mathcal E B_\B^{(1)})_{2,p,1}(x)\|_{(\E;c_0)}$, with $p=0,1$, 
which are the norms entering the definition \eqref{corpod}. Let us first consider $(D_\E^{(h)})_{2,p,1}$. 
Recall that, by the definition \eqref{defCeZe}, $D_\E^{(h)}=\sum_{k=h}^0 \cE \tilde B_\B^{(k)}$, with 
\begin{equation} \tilde B^{(k)}_\B:=\sum_{\substack{\tau \in \cT^{(k)} \\ E_{v_0}=0,\ m_{v_0}=1}}^*
W_\Lambda [\tau]=I_\Lambda \sum_{\substack{\tau \in \cT^{(k)} \\ E_{v_0}=0,\ m_{v_0} = 1}}^*
W_\infty [\tau]\equiv I_\Lambda \tilde B_\infty^{(k)},\end{equation}
with $I_\Lambda$ the kernel in \eqref{eq:3.1.4}, so that 
$\|(D_\E^{(h)})_{2,p,1}(x)\|_{(\E;\frac34c_02^h)}\le \sum_{k=h}^0\|(\mathcal E \tilde B_\B^{(k)})_{2,p,1}(x)\|_{(\E;\frac34c_02^k)}$. In order to bound the norm of 
$\mathcal E \tilde B_{\text{B}}^{(k)}$, we proceed in a way analogous to Appendix \ref{app:segunda}. Recall the definitions of $\cL_\B$ and $\cR_\B$ 
in \eqref{defcL.100}-\eqref{defcR.200}, from which it follows, in particular, that 
$(\mathcal E \tilde B_{\text{B}}^{(k)})_{2,0,1}=((\cL_\B \tilde B_\B^{(k)})_{2,0,1})_\E=(\cA(\tcL (\tilde B_\B^{(k)})_{2,0,1}))_\E$ and 
$(\mathcal E \tilde B_{\text{B}}^{(k)})_{2,1,1}=((\cR_\B \tilde B_\B^{(k)})_{2,1,1})_\E=(\cA(\tcR \tilde B_\B^{(k)})_{2,1,1})_\E$ (in writing the last identity, we used the fact that 
$((\tilde B_\B^{(k)})_{2,1,1})_\E=0$, because it is the edge part of the bulk part of a kernel). From these formulas and the definition of $\tcR$ in \eqref{deftcR.2zzw}, 
it is apparent that $(\mathcal E \tilde B_{\text{B}}^{(k)})_{2,p,1}$ with $p=0,1$ only depends on $(\tilde B_\infty^{(k)})_{2,0,1}$, which, in view of 
\eqref{eq:treesum} and the boundedness of $Z_{j,h}$, see \eqref{Zj-Zj-infty} and preceding lines, satisfies 
\begin{equation}\label{2+3=6}\|(\tilde B_\infty^{(k)})_{2,0,1}(x)\|_{(c_0 2^k)}\le C|\lambda|2^{\theta k}\qquad \forall k\le 0.\end{equation}
We start with bounding the edge norm of $(\mathcal E \tilde B_{\text{B}}^{(k)})_{2,0,1}=(\cA(\tcL (\tilde B_\B^{(k)})_{2,0,1}))_\E$, which satisfies 
$$\|(\mathcal E \tilde B_{\text{B}}^{(k)})_{2,0,1}(x)\|_{(\E;\kappa)}\le \|(\tcL (\tilde B_\B^{(k)})_{2,0,1})_\E(x)\|_{(\E;\kappa)}$$ with 
\begin{equation}
	\begin{split}
		&
		\big( 
			\tcL (\tilde B_\B^{(k)})_{2,0,1} 
		\big)_\E(	(\bs \omega, \bs 0, \bs z),x)
		=\tcL (\tilde B_\B^{(k)})_{2,0,1} ((\bs \omega, \bs 0, \bs z),x)
		-\big( \tcL (\tilde B_\B^{(k)})_{2,0,1} \big)_\B
		((\bs \omega, \bs 0, \bs z),x)
		\\ & \quad =
		\Big(\prod_{j=1}^2\delta_{z_j,z_x}\Big)\Big[ 
		\sum_{\substack{\bs y \in \Lambda^2 \\ \diam_1(\bs y,x)\le L/3}}
		(\tilde B_\infty^{(k)})_{2,0,1}((\bs \omega, \bs 0, \bs y_\infty),x_\infty)-
		\sum_{\bs y \in \Lambda_\infty^2}
		(\tilde B_\infty^{(k)})_{2,0,1}((\bs \omega, \bs 0, \bs y_\infty),x_\infty)
		\Big];
	\end{split}\label{rhs}
\end{equation}
the terms in the first sum on the right-hand-side are entirely cancelled by a corresponding set of terms in the second sum, leaving only those 
for which $\diam_1(\bs y, x) > L/3$ or $y_j\notin \Lambda$ for $j$ equal either to $1$ or $2$, 
and for these remaining terms we have $\delta (\bs y, x) \ge  \delta_\E(\bs z, x)$. 
As a result, for any $\kappa,\epsilon>0$,
\begin{equation}
	\sum_{\bs z \in \bar\Lambda^2}
		e^{\kappa \delta_\E (\bs z,x)}
		\Big|\big( \tcL (\tilde B_\B^{(k)})_{2,0,1} \big)_\E ((\bs\omega,\bs 0,\bs z),x)\Big|
			\le e^{-\epsilon \delta_\E(x)}\sum_{\bs y \in\Lambda_\infty^2}
		e^{(\kappa+\epsilon) \delta (\bs y,x)}
		\big|(\tilde B_\infty^{(k)})_{2,0,1}((\bs \omega, \bs 0, \bs y),x)\big|,
\label{eqref}\end{equation}
that is, recalling the definition \eqref{normonsourceEdge} of the edge norm, 
\begin{equation} \|(\tcL (B_\B^{(k)})_{2,0,1})_\E(x)\|_{(\E;\kappa)}	\le e^{-\epsilon\delta_\E(x)}	\|(\tilde B_\infty^{(k)})_{2,0,1}(x)\|_{(\kappa+\epsilon)}\end{equation}
where the norm in the right side is the infinite volume analogue of \eqref{normonsource}. 

Next, we bound the edge norm of $(\mathcal E \tilde B_{\text{B}}^{(k)})_{2,1,1}=(\cA(\tcR \tilde B_\B^{(k)})_{2,1,1})_\E$, where, recalling the defintion of $\tcR$ in \eqref{deftcR.2zzw}, 
\begin{equation}
	\begin{split}
		& ((\tcR \tilde B_\B^{(k)})_{2,1,1})_\E((\bs \omega, \bs D, \bs z),x)
		=(\tcR \tilde B_\B^{(k)})_{2,1,1}( (\bs \omega, \bs D, \bs z),x)
		-
		\big((\tcR \tilde B_\B^{(k)})_{2,1,1}\big)_\B((\bs \omega, \bs D, \bs z),x)
		\\ & \qquad =
		(-1)^{\alpha(\bs z)}
		\Big[ \sum_{\substack{\sigma = \pm 1, \bs y \in \Lambda^2: 
						\\ \diam_1 (\bs y,x) \le L/3,
						\\ (\sigma, \bs D,\bs z) \in \INT_x(\bs y)}}
					\sigma (\tilde B_\infty^{(k)})_{2,0,1} ((\bs \omega, \bs 0, \bs y_\infty),x_\infty)
					\\ &\hskip3.truecm
					- I_\Lambda((\bs\omega,\bs D,\bs z), x)
					\sum_{\substack{\sigma = \pm 1, \bs y \in \Lambda_\infty^2: 
						\\ (\sigma, \bs D,\bs z_\infty) \in \INT_{x_\infty}(\bs y)}}
					\sigma (\tilde B_\infty^{(k)})_{2,0,1}((\bs \omega, \bs 0, \bs y),x_\infty)
			\Big]
			;
		\end{split}\label{2+2=25}
\end{equation}
where the interpolation path $\INT$ in the last term in the right side is the infinite volume analogue of 
the one defined after \eqref{deftcR.2zzw}; the construction of $\INT_x$ is such that the first sum on the right hand side is over an empty set whenever either of the indicator functions before the second sum vanish, so again the first sum cancels all terms of the second sum which do not satisfy $\delta(\bs y,x) \ge  \delta_\E (\bs z,x)$, giving, 
for any $\kappa,\epsilon>0$,
\begin{equation}
	\begin{split}
		&
		\sum_{\bs z \in \bar\Lambda^2}e^{\kappa \delta_\E (\bs z,x)}\sup_{\bs D}^{(1)}
		\big|((\tcR \tilde B_\B^{(k)})_{2,1,1})_\E((\bs \omega, \bs D, \bs z),x)\big|
		\\ &
		\le e^{-\epsilon\delta_\E(x)}\sum_{\bs y \in \Lambda_\infty^{2}}
		e^{(\kappa+\epsilon) \delta(\bs y,x)}
		\sum_{(\sigma, \bs D, \bs z) \in \INT_x(\bs y)}
		| (\tilde B_\infty^{(k)})_{2,0,1} ((\bs \omega, \bs 0, \bs y),x_\infty) |
		\\ &
		\le 
		\frac{2}{\epsilon}e^{-\epsilon\delta_\E(x)}
		\sum_{\bs y \in \Lambda_\infty^2}
		e^{(\kappa + 2\epsilon) \delta(\bs y,x)}
		| (\tilde B_\infty^{(k)})_{2,0,1}((\bs \omega, \bs 0, \bs y),x_\infty) |
	\end{split}\label{2+2=28}
\end{equation}
(for the last inequality, proceed analogously to \cite[Equations~(4.2.26) to~(4.2.28)]{AGG_part2}). In conclusion, 
\begin{equation}
 \|((\tcR \tilde B_\B^{(k)})_{2,1,1})_\E(x)\|_{(\E;\kappa)}\le 
 	\frac2{\epsilon}e^{-\epsilon\delta_\E(x)}\|(\tilde B_\B^{(k)})_{2,0,1}(x)\|_{(\kappa + 2\epsilon)} \end{equation}
Putting things together, choosing $\kappa=\frac34c_02^k$ and $\kappa+2\epsilon=c_02^{k}$, and using \eqref{2+3=6}, 
we find, for $p=0,1$, 
\begin{equation}\|(\mathcal E \tilde B_{\text{B}}^{(k)})_{2,p,1}(x)\|_{(\E;\frac34c_02^k)}\le  C|\lambda|2^{-pk}2^{\theta k}e^{-\frac{c_0}82^k\delta_\E(x) }. \end{equation}
Summing over $k$ from $h$ to $0$, and bounding $2^{-pk}\le 2^{-ph}$, we find
\begin{equation} \|(D_\E^{(h)})_{2,p,1}(x)\|_{(\E;\frac34c_02^h)}\le C2^{-ph}|\lambda|\sum_{k=h}^0 2^{\theta k} e^{-\frac{c_0}82^k\delta_\E(x)}
\le C'2^{-ph}|\lambda|(\delta_\E(x))^{-\theta}.
\end{equation}
Similar considerations, in combination with the norm bound on $B_\infty^{(1)}$, which follows from the infinite volume analogues of 
\eqref{eq:WL1bound} and \eqref{eq:WfreeL1bound}, imply that 
\begin{equation} \max_{p=0,1}\|(\mathcal E B_\B^{(1)})_{2,p,1}(x)\|_{(\E;c_0)}\le C e^{-c_0\delta_\E(x)}.
\end{equation}
Summarizing, using the bounds of this subsection and recalling the definition \eqref{corpod} of $\tilde Z_{h_v-1}(E_v,$ $|\bs D_v\|_1,x_v)$, 
we have that the combination $2^{h\|\bs D_v\|_1}\tilde Z_{h_v-1}(E_v,|\bs D_v\|_1,x_v)$ entering \eqref{maquindi} satisfies 
\begin{equation} 2^{h\|\bs D_v\|_1}\tilde Z_{h_v-1}(E_v,|\bs D_v\|_1,x_v)\le  \begin{cases} 
C|\lambda|(\delta_\E(x))^{-\theta}& \text{if $E_v=1$ and $h_v\le 1$,}\\
C & \text{otherwise.}
\end{cases}
\label{boundtildeZEv}\end{equation} 

\section{Correlation functions}
\label{sec:correlations}

Recall that the truncated energy correlation functions of $\epsilon_{x_1}, \ldots, \epsilon_{x_m}$, with 
$\bs x=(x_1,\ldots,$  $x_m)$ an $m$-tuple consisting of $m\ge 2$ distinct edges (or, equivalently, of edge midpoints) in $\mathfrak B_\Lambda$,
can be computed as the derivative of $\log \Xi_\Lambda(\bs A)$ with respect to $A_{x_1},\ldots, A_{x_m}$ at $\bs A=\bs 0$; and that 
$\Xi_\Lambda(\bs A)$ is given by \eqref{eventually}, with $\mathcal W^{(1)}(\bs A)\equiv \mathcal W(\bs A)=\sum_{\bs x\in\cX_\Lambda}w_\Lambda(\bs x) \bs A(\bs x)$ as in  \eqref{eq:B_expansion_tris}, and 
$\mathcal W^{(h)}(\bs A)$ with $h\le 0$ as in \eqref{kernelw}, with the kernel $w_\Lambda^{(h)}$ expressed via the GN tree expansion \eqref{treeexponcyl}.
Consequently, we have arrived at 
\begin{equation}
	\langle\epsilon_{x_1}; \dots; \epsilon_{x_m}\rangle^{T}_{\lambda,t_1,t_2;\Lambda}
	=\sum_{\pi}\Big[w_\Lambda(\pi(\bs x))+\sum_{h=h^*-1}^{0}
	\sum_{\substack{\tau \in \cT^{(h)}:\\ m_{v_0}=m}}^*w_\Lambda[\tau](\pi(\bs x))\Big],	\label{eq:correlation_base_representation}
\end{equation}
where $t_1=\tanh(\beta_c(\lambda)J_1)$ and $t_2=\tanh(\beta_c(\lambda) J_2)$ are fixed as explained in Remark \ref{remacon}, 
the first sum in the right side runs over the $m!$ permutations $\pi$ 
of an $m$-tuple of distinct elements, and the $*$ on the last sum in the right side indicates the constraint that the vertex $v_0=v_0(\tau)$ is dotted. 
In \eqref{eq:correlation_base_representation}, $w_\Lambda$ comes from the transformation from spin to Grassmann variables summarized in Proposition \ref{prop:repr},
and vanishes at $\lambda=0$; the remaining terms come from the iterative integration of the massive and massless Grassmann variables 
described in the previous section, and may or may not vanish at $\lambda=0$, depending on the choice of $\tau$.

Let $m_1$ and $m_2$ be the numbers of horizontal and vertical edges in $\bs x$, respectively. 
We want to compare \eqref{eq:correlation_base_representation} with the rescaled, critical, non-interacting correlation function 
\begin{equation}\label{eq:rescaledfree}
\Big(\frac{Z_{1,-\infty}}{Z_{1}^{*}}\Big)^{m_1}\Big(\frac{Z_{2,-\infty}}{Z_{2}^{*}}\Big)^{m_2}
\langle\epsilon_{x_1}; \dots; \epsilon_{x_m}\rangle^{T}_{0,t_1^*,t_2^*;\Lambda}, \end{equation}
where $Z_{j,-\infty}=Z_{j,-\infty}(\lambda)$, with $j=1,2$, is the limiting value of $Z_{j,h}$ introduced in Section \ref{sec:Zj}, 
and $Z^*_j$ are defined as in \eqref{behZ*}. Recall that $Z_{j,-\infty}/Z_j^*=1+O(\lambda)$ for $j=1,2$. We intend to prove 
that the difference between \eqref{eq:correlation_base_representation} and \eqref{eq:rescaledfree} is bounded as the remainder term in Theorem \ref{prop:main}, 
see \eqref{10b}. This will in fact prove Theorem \ref{prop:main}.

Eq.\eqref{eq:rescaledfree} can be computed by deriving with respect to $\bs A$ the generating function 
\begin{equation} W^{\qf}(\bs A)\equiv \sum_{\substack{\bs x\in \cX_\Lambda:\\ \bs x\neq\emptyset}}w^\qf_\Lambda(\bs x)\bs A(\bs x):=\log\int  P^*_{c} (\cD\phi) P_{m}^* (\cD\xi) \, e^{\cB^{\qf}(\phi,\xi,\bs A)} \label{eq:startfromfree}\end{equation}
with\footnote{The label `qf' on the tree values stands for `quasi-free'; the `quasi' accounts for the presence of the dressed parameters $t_1^*, t_2^*$ and of the 
rescaling factors $Z_{j,-\infty}/Z^*_j$.}
\begin{equation} \label{Bqfree}\cB^{\qf}(\phi,\xi,\bs A):=
\sum_{x\in\fB_{1,\Lambda}}\frac{Z_{1,-\infty}}{Z^*_1} (1-(t_1^*)^2) H^*_{+,z_x} H^*_{-,z_x+\hat e_1}A_x+  \sum_{x\in\fB_{2,\Lambda}}
\frac{Z_{2,-\infty}}{Z^*_2} (1-(t_2^*)^2) \phi_{+,z_x}\phi_{-,z_x+\hat e_2} A_x,\end{equation}
where, letting $s^*_\omega(z):=\frac1{L}\sum_{k_1\in\mathcal D_L} \frac{e^{-ik_1z}}{1+t_1^*e^{\pm ik_1}}$ (and $\mathcal D_L$ defined as stated after \eqref{DphiDxi}),
\begin{equation} H^*_{\omega,z}:=\xi_{\omega,z}+ \sum_{y=1}^L s_\omega^*((z)_1-y) \big(\phi_{+,(y,(z)_2)}-\omega \phi_{-,(y,(z)_2)}\big).\label{eq:sc_2000}
\end{equation}
For later reference, we let $B^\qf_\Lambda$ be the $\cA$-invariant kernel associated with \eqref{Bqfree}, $B^\qf_\infty$ be its infinite volume limit, 
and $B^\qf_\B, B^\qf_\E$ be its bulk and edge parts, respectively. 
In order to compare \eqref{eq:correlation_base_representation} with \eqref{eq:rescaledfree}, it is convenient to rewrite the kernel $w^\qf_\Lambda(\bs x)$ in 
\eqref{eq:startfromfree} as a sum over GN trees, in analogy with the (sum over scales of the) second line of \eqref{treeexponcyl}, i.e., 
\begin{equation}\label{wqftree}w^\qf_\Lambda\sim \sum_{h=h^*-1}^0 \sum^*_{\substack{\tau \in\cT_\free^{(h)}:\\ m_{v_0}=m}}
w^\qf_\Lambda[\tau],\end{equation} 
where $\cT_{\free}^{(h)}$ is the subset of $\cT^{(h)}$ consisting of trees such that: (1) the endpoints are all either of type \tikzvertex{bareProbeEP} 
or of type \tikzvertex{bareProbeEP,E} and are all associated with a label $m_v=1$; (2) there are no endpoints $v$ of type \tikzvertex{bareProbeEP,E} on scale $h_v<2$;
(3) the vertices $v\in V_0'(\tau)$ have $|S_v|=|S_{v}^*|\ge 2$. Moreover, the $*$ on the sum over GN trees in the right side indicates the constraint that $v_0=v_0(\tau)$ is dotted. 
The definition of $w_\Lambda^\qf[\tau]$ will be given momentarily. Recalling that \eqref{eq:rescaledfree} is the $m$-th derivative of 
\eqref{eq:startfromfree} at the origin, we thus get 
\begin{equation}\Big(\frac{Z_{1,-\infty}}{Z_{1}^{*}}\Big)^{m_1}\Big(\frac{Z_{2,-\infty}}{Z_{2}^{*}}\Big)^{m_2}
 \langle\epsilon_{x_1}; \dots; \epsilon_{x_m}\rangle^{T}_{0,t_1^*,t_2^*;\Lambda}=\sum_\pi\sum_{h=h^*-1}^{0}\sum_{\substack{\tau \in \cT_\free^{(h)}:\\ m_{v_0}=m}}^*w_{\Lambda}^{\qf}[\tau](\pi(\bs x)).\label{eq:rescaledfreetree}\end{equation}
In \eqref{wqftree} and \eqref{eq:rescaledfreetree}, 
\begin{equation} \label{galeno}w_{\Lambda}^{\qf}[\tau]=
\sum_{\substack{\ul P\in\cP(\tau):\\ P_{v_0}=\emptyset}}\
\sum_{\ul T\in\cS(\tau,\ul P)}\sum_{\ul D\in \cD(\tau,\ul P)}W_\Lambda^\qf[\tau,\ul P,\ul T, \ul D],\end{equation}
with $W_\Lambda^\qf[\tau,\ul P,\ul T, \ul D]$ obtained via the same recursive definition as $W_\Lambda[\tau,\ul P,\ul T, \ul D]$, see 
Section \ref{sec:treecyl}, up to the following differences: 
\begin{itemize}
\item If $\tau\in \cT^{(h)}_\free$ and $E_{v_0}=0$, then $W_\Lambda^\qf[\tau,\ul P,\ul T, \ul D]$ is defined as in \eqref{eq:item1}, with the function $W_\infty$ in the right side replaced by $W_\infty^\qf$; the latter is defined as described after \eqref{eq:item1}, modulo the fact that $K_{v,\infty}$ should be replaced by $K_{v,\infty}^\qf$, 
with $K_{v,\infty}^\qf=B^\qf_\infty$ is $h_{v_0}=1$, and 
$$K_{v,\infty}^\qf=\piecewise{Z_{-\infty}\cdot F^A_\infty & \text{if $v\in V_e(\tau)$ is of type \tikzvertex{bareProbeEP} and $h_v=h_{v_0}+1$},\\
\cR_\infty B^\qf_\infty & \text{if $v\in V_e(\tau)$ is of type \tikzvertex{bareProbeEP} and $h_v=2$,}\\
{\lis W}^\qf_\infty[\tau_v,\ul P_v, \ul T_v, \ul D_v] & \text{if $v\in V_0(\tau)$},}
$$
otherwise. Here: $Z_{-\infty}\cdot F^A_\infty:= Z_{1,-\infty} F_{1,\infty}+
Z_{2,-\infty} F_{2,\infty}$, with $F_{j,\infty}$  the infinite volume analogue of $F_{j,\B}$, see \eqref{defFjB} and preceding lines, 
and ${\lis W}^\qf_\infty$ is defined recursively in a way analogous to ${\lis W}_\infty$. 
\item If $\tau\in \cT^{(h)}_\free$ and $E_{v_0}=1$, then $W_\Lambda^\qf[\tau,\ul P,\ul T, \ul D]$ is defined as in items (2) and (3) of Section \ref{sec:treecyl}, see \eqref{Wtaublack-} and following paragraphs, modulo the fact that $K_v$ must be replaced by $K_{v}^\qf$, where,  if $h_{v_0}=1$, then 
$K_{v}^\qf=\piecewise{B_{\B}^\qf & \text{if $v\in V_e(\tau)$ is of type \tikzvertex{bareProbeEP}}\\
B_{\E}^\qf & \text{if $v\in V_e(\tau)$ is of type \tikzvertex{bareProbeEP,E}}}$, while, if $h_{v_0}<1$, then 
$$K_v^\qf=\piecewise{Z_{-\infty}\cdot F^A_\B & \text{if $v\in V_e(\tau)$ is of type \tikzvertex{bareProbeEP} and $h_v=h_{v_0}+1$},\\
(\cR_\B B^\qf_\B)_\B & \text{if $v\in V_e(\tau)$ is of type \tikzvertex{bareProbeEP} and $h_v=2$,}\\
		B^{\qf}_\E+\mathcal E B^{\qf}_\B & \text{if $v\in V_e(\tau)$ is of type \tikzvertex{bareProbeEP,E} and $h_v = 2$,}\\ 
			\lis W_\Lambda^\qf [\tau_v, \ul P_{v}, \ul T_{v}, \ul D_{v}]  & \text{if $v\in V_0(\tau)$.}
		}$$
\end{itemize}
The proof that the kernel $w^\qf_\Lambda$ of $W^\qf(\bs A)$ can be computed via the GN tree expansion \eqref{wqftree}, \eqref{galeno}, 
is analogous to (and simpler than) the one in Appendix \ref{app:itdec}, and left to the reader. 
\begin{remark} The condition that 
the trees contributing to \eqref{wqftree} have no endpoints $v$ of type \tikzvertex{bareProbeEP,E} on scale $h_v<2$
is due to the following: the value $D_\E^{\qf;(h_v-1)}$, which would be naturally (by mimicking the proof in Appendix \ref{app:itdec}) associated with such vertices, 
should be defined as
\begin{equation}\label{wouldbe} D_\E^{\qf;(h)}:=\sum_{k=h}^0	\sum_{\substack{\tau \in \cT_\free^{(k)} \\ E_{v_0}=0,\ m_{v_0}= 1}}^*\cE W^\qf_\Lambda [\tau],\end{equation}
with the $*$ on the sum indicating the constraint that $v_0=v_0(\tau)$ is dotted. However, \eqref{wouldbe} 
is identically zero, because the constraint $m_{v_0}=1$ is incompatible with the requirement that 
$\tau\in\cT_\free^{(h)}$ and $v_0$ is dotted, as the reader can easily convince herself.
\end{remark}
Note that, due to the definitions above, the sum over $\ul P$ in the right side of \eqref{galeno} can be freely restricted to the set 
$\cP_\free(\tau)\subset \cP(\tau)$ for which $|P_v|=2$ for all $v\in V'(\tau)$: in fact, whenever 
$|P_v|\ge 4$ for some $v\in V'(\tau)$, $W_\Lambda^\qf[\tau,\ul P,\ul T, \ul D]=0$.
In particular $B^\qf$ vanishes when its argument has more than two field labels, whereas $B^{(1)}$, which it replaces in these expresssions, includes terms
with four or more Grassmann fields, which arise from the transformation from spin to Grassmann variables summarized in Proposition \ref{prop:repr}.

\bigskip

We now write
\begin{eqnarray} &&	\langle\epsilon_{x_1}; \dots; \epsilon_{x_m}\rangle^{T}_{\lambda,t_1,t_2;\Lambda}
-	\Big(\frac{Z_{1,-\infty}}{Z_{1}^{*}}\Big)^{m_1}\Big(\frac{Z_{2,-\infty}}{Z_{2}^{*}}\Big)^{m_2}\langle\epsilon_{x_1}; \dots; \epsilon_{x_m}\rangle^{T}_{0,t_1^*,t_2^*;\Lambda}=\sum_\pi\Big\{w_\Lambda(\pi(\bs x))\nonumber\\
&&\!\!\!+\sum_{h=h^*-1}^0\Big[ \sum^*_{\substack{\tau\in \cT^{(h)}_\free:\\ m_{v_0}=m}} \
\sum_{\substack{\ul P\in \cP_\free(\tau):\\ P_{v_0}=\emptyset}}\ \sum_{\ul T\in \cS(\tau,\ul P)}\sum_{\ul D\in \cD(\tau,\ul P)}
\Big(W_\Lambda[\tau,\ul P,\ul T,\ul D](\pi(\bs x))- W^\qf_\Lambda[\tau,\ul P,\ul T,\ul D](\pi(\bs x))\Big)\nonumber\\
&&\!\!\!+\ \sum^*_{\substack{\tau\in\cT^{(h)}:\\ m_{v_0}=m}}\ \sum_{\substack{\ul P\in \cP_*(\tau):\\ P_{v_0}=\emptyset}}\ \sum_{\ul T\in \cS(\tau,\ul P)}\sum_{\ul D\in \cD(\tau,\ul P)} W_\Lambda[\tau,\ul P,\ul T,\ul D](\pi(\bs x))\Big]
\Big\},
\end{eqnarray}
where the set $\cP_*(\tau)$ in the second sum in the last line is defined as 
$$\cP_*(\tau):=\piecewise{
\cP(\tau)\setminus \cP_\free(\tau) & \text{if $\tau\in \cT^{(h)}_\free$}\\
 \cP(\tau) & \text{if $\tau\not\in \cT^{(h)}_\free$}.}
$$
Note that for $\tau \in \cT_\free^{(h)}$, $\cP^*(\tau)$ is nonempty because \tikzvertex{bareProbeEP} and \tikzvertex{bareProbeEP,E} endpoints on scale $2$ include factors of $B^{(1)}_\B$ or $B^{(1)}_\E$ which, unlike $B^\free$, are not restricted to arguments with only two Grassmann fields.

We denote by $R_1(\bs x)$ and $R_2(\bs x)$ the contributions from the sums in the second and third lines, respectively, so that 
\begin{eqnarray}\label{eccocisi}
&&\langle\epsilon_{x_1}; \dots; \epsilon_{x_m}\rangle^{T}_{\lambda,t_1,t_2;\Lambda}
-	\Big(\frac{Z_{1,-\infty}}{Z_{1}^{*}}\Big)^{m_1}\Big(\frac{Z_{2,-\infty}}{Z_{2}^{*}}\Big)^{m_2}\langle\epsilon_{x_1}; \dots; \epsilon_{x_m}\rangle^{T}_{0,t_1^*,t_2^*;\Lambda}=\\
&&=\sum_\pi\Big[w_\Lambda(\pi(\bs x))+R_1(\pi(\bs x))+R_2(\pi(\bs x))\Big].\nonumber\end{eqnarray}
The contribution from $w_\Lambda$ can be bounded via \eqref{eq:B_base_decay2}, which implies 
\begin{equation}\label{eq:boundR0}\sum_{\pi}\big|w_\Lambda(\pi(\bs x))\big|\le C^m|\lambda| e^{-c\delta(\bs x)},\end{equation}
which is smaller than the right side of \eqref{10b}. We are left with the contributions from $R_1$ and $R_2$, which are discussed in the following two subsections. 

\subsection{The remainder term $R_1$}\label{sec:5.1}
Recall that 
$$R_1(\bs x)=\sum_{h=h^*-1}^0 \sum^*_{\substack{\tau\in \cT^{(h)}_\free:\\ m_{v_0}=m}}\ %
\sum_{\substack{\ul P\in \cP_\free(\tau):\\ P_{v_0}=\emptyset}}\ \sum_{\ul T\in \cS(\tau,\ul P)}\sum_{\ul D\in \cD(\tau,\ul P)}\Big(W_\Lambda[\tau,\ul P,\ul T,\ul D](\bs x)- 
W^\qf_\Lambda[\tau,\ul P,\ul T,\ul D](\bs x)\Big).$$
For each term in the sum, we write the difference $W_\Lambda[\tau,\ul P,\ul T,\ul D]- W^\qf_\Lambda[\tau,\ul P,\ul T,\ul D]$ 
as a telescopic sum of $m$ terms, each of which has one endpoint associated with the value $K_v-K^\qf_v$ or $K_{v,\infty}-K^\qf_{v,\infty}$.
These differences can be rewritten and bounded as follows (we restrict our attention to trees with $h_{v_0}<1$ and to $K_v-K^\qf_v$, the 
other cases, namely $h_{v_0}=1$ and $K_{v,\infty}-K^\qf_{v,\infty}$, being analogous). 
\begin{enumerate}
\item If $v\in V_e(\tau)$ is of type \tikzvertex{bareProbeEP} and $h_v<2$, then   
$K_v-K^\qf_v=(Z_{h_{v}-1}-Z_{-\infty})\cdot F^A_\B$, so that $\|(K_v-K_v^\qf)(x)\|_{0;h_v-1}\le C|\lambda|2^{\theta h_v}$, 
see \eqref{Zj-Zj-infty}. %
\item If $v\in V_e(\tau)$ is of type \tikzvertex{bareProbeEP} and $h_v=2$, then 
$K_v-K^\qf_v=\big(\cR_\B(B^{(1)}_\B-B^\qf_\B)_{2,0,1}\big)_{\B}$, where
$(B^{(1)}_\B-B^\qf_\B)_{2,0,1}=(Z^{-1}B^{\free}_\B-B^\qf_\B)_{2,0,1}+Z^{-1}(W^{\rm int}_\B)_{2,0,1}$, see \eqref{expanV1} and following paragraphs for the
definitions of the kernels $B^{\free}_\Lambda$, $W^{\rm int}_\Lambda$ and of their bulk counterparts. By the very definitions of $B^{\free}_\B$ and $B^\qf_\B$ and the 
facts that $Z-1=O(\lambda)$, $t_1-t^*_1=O(\lambda)$, $t_2-t_2^*=O(\lambda)$ (see Remark \ref{remacon}) and $Z_{j,-\infty}/Z_j^*-1=O(\lambda)$,
we find that $\|(Z^{-1}B^{\free}_\B-B^\qf_\B)_{2,0,1}(x)\|_{(2c_0)}\le C|\lambda|$, with $c_0$ the same constant as in Proposition \ref{prop:repr}. Moreover, by using \eqref{eq:WL1bound}, 
$\|(W^{\rm int}_\B)_{2,0,1}(x)\|_{(2c_0)}\le C|\lambda|$. Therefore, using also \eqref{eq:RBB}, we find that, for any $c'<2c_0$, $\|(K_v-K_v^\qf)(x)\|_{(c')}\le C|\lambda|/(2c_0-c')$. 
\item If $v\in V_e(\tau)$ is of type \tikzvertex{bareProbeEP,E} and $h_v=2$, then  
$K_v-K^\qf_v=(B^{(1)}_\E-B^\qf_\E)_{2,0,1}+\sum_{p=0}^1(\mathcal E B^{(1)}_\B-\mathcal E B^\qf_\B)_{2,p,1}$, where, similarly to the decomposition used in the previous item, 
we write
$(B^{(1)}_\E-B^\qf_\E)_{2,0,1}=(Z^{-1}B^{\free}_\E-B^\qf_\E)_{2,0,1}+Z^{-1}(W^{\rm int}_\E)_{2,0,1}$ and 
$(\mathcal E B^{(1)}_\B-\mathcal E B^\qf_\B)_{2,p,1}=(Z^{-1}\mathcal E B^{\free}_\B-\mathcal E B^\qf_\E)_{2,p,1}+Z^{-1}(\mathcal E W^{\rm int}_\E)_{2,p,1}$. By the same 
considerations as in the previous item and the bounds \eqref{eq:WfreeE_base_decay}, \eqref{eq:WE_base_decay}, we find  that, for any $c'<2c_0$, 
$$\|(B^{(1)}_\E-B^\qf_\E)_{2,0,1}(x)\|_{(\E;c')}\le C|\lambda| e^{-(2c_0-c')\delta_\E(x)}.$$ 
Similarly, using also considerations analogous to those of Section \ref{sec.3.4.2}, we find that, for any $c'<2c_0$, 
$$\|(\mathcal E B^{(1)}_\B-\mathcal E B^\qf_\B)_{2,p,1}\|_{(\E;c')}\le C|\lambda|e^{-(c_0-c'/2)\delta_\E(x)}/(2c_0-c').$$ 
\end{enumerate}
By using the definition of the norm $\|\cdot\|_{E_{v_0};h_{v_0}}$, we find, for $\bs x_{v_0}\equiv \bs x$, 
\begin{eqnarray} |R_1(\bs x)|&\le & \sum_{h=h^*-1}^0 \ \sum_{\substack{\tau\in \cT^{(h)}_\free:\\ m_{v_0}=|V_e(\tau)|=m}}e^{-\frac{c_0}{2}2^{h_{v_0}}\delta_{E_{v_0}}(\bs x)} \times
\label{eqqq:qq} \\
&\times& \sum_{\substack{\ul P\in \cP_\free(\tau): \\ P_{v_0}=\emptyset}}\
\sum_{\ul T\in \cS(\tau,\ul P)}\sum_{\ul D\in \cD(\tau,\ul P)}\|(
W_\Lambda[\tau,\ul P,\ul T,\ul D]- W^\qf_\Lambda[\tau,\ul P,\ul T,\ul D])(\bs x_{v_0})\|_{E_{v_0};h_{v_0}},\nonumber\end{eqnarray}
where, in view of the previous considerations, of the proof of Proposition \ref{lem:W_primitive_with_m}, and of Proposition \ref{lm:W:scaldim_withm}, 
the norm of the difference $W_\Lambda[\tau,\ul P,\ul T,\ul D]- W^\qf_\Lambda[\tau,\ul P,\ul T,\ul D]$ in the second line is bounded as 
in \eqref{eq:W8_reworked_boundzaza} (with the factor in the last line simplifying to $C^m$\footnote{The reason for this simplification is the following. 
Consider a tree $\tau\in\cup_{h^*-1}^0\cT_\free^{(h)}$ contributing to the right side of \eqref{eqqq:qq}. By definition, each endpoint $v\in V_e(\tau)$ has $(|P_v|,m_v)=(2,1)$: 
therefore, in the last factor in \eqref{eq:W8_reworked_boundzaza} one can neglect the second case, corresponding to $(|P_v|,m_v)\neq(2,1)$. Moreover, 
looking back at the definition of $\tilde Z_{h_v-1}(E_v,\|\bs D_v\|_1,x_v)$ in \eqref{corpod}, and recalling that $\tau$ has no endpoints of type \tikzvertex{bareProbeEP,E} on 
scales smaller than $2$, we see that we can neglect the second case in the right side of \eqref{corpod}; this, in light also of the fact that $\max_{j=1,2}\{|Z_{j,h}|\}\le C$, uniformly 
in $h$ (see Section \ref{ebbZj}) leads to $2^{h_v\|\bs D_v\|_1}\tilde Z_{h_v-1}(E_v,\|\bs D_v\|_1,x_v)\le C$ for all the cases of relevance for the current bound.})
times an additional factor $|\lambda| 2^{\theta h_M}$, which accounts for 
the bounds on $K_v-K_v^\qf$ discussed in items 1 to 3 above, where $h_M=h_M(\tau):=
\max_{v\in V_e(\tau)}h_v$. 
 
For $\tau \in \cT_\free^{(h)}$ and $\underline P\in\cP_\free(\tau)$, all endpoints $w\in V_e(\tau)$ are such that $|P_w|=2$ and $m_w=1$, which implies that $|P_{v}|=2$ and $m_v \ge 1$ for all $v\in V'(\tau)$;
therefore, recalling the definition $d(P_v, \bs D_v,m_v,E_v)=2-|P_v|/2-\|\bs D_v\|_1-m_v-E_v\delta_{m_v,0}$ 
for all such vertices $d(P_v,\bs D_v,m_v,E_v)=1-m_v-\|\bs D_v\|_1$.
Noting also $|P_{v_0}|=0$ (so that $d(P_{v_0}, \bs D_{v_0},m_{v_0},E_{v_0})=2-m$),
we find 
	\begin{eqnarray}
&& \|(W_\Lambda[\tau,\ul P,\ul T,\ul D]-W_\Lambda^\qf[\tau,\ul P,\ul T,\ul D])(\bs x_{v_0})\|_{E_{v_0};h_{v_0}}\le     \frac{C^m}{|S_{v_0}|!}
2^{h_{v_0}(2-m)}\, |\lambda| 2^{\theta h_M}
\label{eq:pkyut}
\\
&&\qquad \times   \Big(
	      \prod_{v \in V'(\tau)}\frac1{|S_{v}|!}
	      2^{(h_v-h_{v'})(1-m_v-\|\bs D_v\|_1)}  \Big)\,
	       \Big(\prod_{v\in V(\tau)} 2^{2[|S_v|-1]_+h_v} e^{-\frac{c_0}{12}2^{h_v}\delta_{E_v}(\bs x_v)}\Big).\nonumber
	  \end{eqnarray}
This equation is analogous to \cite[Eq.(4.5)]{GGM}. 
If we now plug \eqref{eq:pkyut} into \eqref{eqqq:qq}
and sum over $\ul P,\ul T,\ul D$ (see the proof of \cite[Lemma~4.8]{AGG_part2} for additional details), we get 
\begin{eqnarray}&& |R_1(\bs x)|\le  C^m\,|\lambda|\, \sum_{h=h^*-1}^0 \ \sum_{\substack{\tau\in \cT^{(h)}_\free\\ |V_e(\tau)|=m}}e^{-\frac{c_0}{2}2^{h_{v_0}}
\delta_{E_{v_0}}(\bs x)} 2^{h_{v_0}(2|S_{v_0}|-m)} 2^{\theta h_M}\cdot\label{5.201}\\
	      &&\cdot    \Big(\prod_{v \in V'_0(\tau)} 2^{(h_v-h_{v'})(1-m_v)} 2^{2(|S_v|-1)h_v} e^{-\frac{c}{12}2^{h_v}\delta_{E_v}(\bs x_v)}\Big)\, 
	      \Big(\prod_{\substack{v \in V_e(\tau):\\ E_v=0,\, h_v=2}} 2^{h_{v'}}\Big) 
	      \Big(\prod_{\substack{v \in V_e(\tau):\\ E_v=1,\, h_v=2}} e^{-\frac{c_0}{3}\delta_\E(x_v)}\Big),\nonumber\end{eqnarray}
where the last two products should be interpreted as $1$ if they run over an empty set, and the 
factors $2^{h_{v'}}$ associated with the endpoints $v$ with $E_v=0$ and $h_v=2$ come from the factors 
$2^{-(h_v-h_{v'})\|\bs D_v\|_1}$ in the second line of \eqref{eq:pkyut}: recall, in fact, that, due to the action of the $\cR_\B$ operator, 
the kernel associated with such endpoints is different from zero only if $\|\bs D_v\|_1=1$. 

We now manipulate this expression further. Write the factors $2^{2(|S_v|-1)h_v}$ as $2^{(|S_v|-1)h_v}\cdot 2^{(|S_v|-1)h_v}$. Keep one of these factors on a side, 
and rewrite the product of the other as 
$$\prod_{v \in V'_0(\tau)}2^{(|S_v|-1)h_v}=\prod_{v \in V'_0(\tau)}2^{(|S_v|-1)h_{v_0}}\cdot 2^{(|S_v|-1)\sum_{w\in V_0'(\tau)}^{w\le v} 
(h_w-h_{w'})}.$$
By using the identity \begin{equation}\label{telsv}\sum_{v\in V_0(\tau)}(|S_v|-1)=|V_e(\tau)|-1,\end{equation} which can be easily proved by induction, 
we can rewrite this product as 
$$2^{(m-|S_{v_0}|)h_{v_0}}\prod_{v \in V'_0(\tau)}2^{(m_v-1)(h_v-h_{v'})}.$$
Plugging this into \eqref{5.201} and letting, for short, $\delta_v:=\delta_{E_v}(\bs x_v)$,
we find
\begin{eqnarray}&& 
      \nonumber\end{eqnarray}
\begin{equation}
	\begin{split}
		|R_1(\bs x)|
		\le  
		C^m\,|\lambda|&\sum_{h=h^*-1}^0 \ 
		 \sum_{\substack{\tau\in \cT^{(h)}_\free\\ |V_e(\tau)|=m}}e^{-\frac{c_0}{2}2^{h_{v_0}}
	\delta_{v_0}} 2^{h_{v_0}|S_{v_0}|} 2^{\theta h_M}
	\\ & \times
	      \Big(\prod_{v \in V'_0(\tau)} 2^{(|S_v|-1)h_v} e^{-\frac{c}{12}2^{h_v}\delta_v}\Big)\, 
	      \Big(\prod_{\substack{v \in V_e(\tau):\\ E_v=0,\, h_v=2}} 2^{h_{v'}}\Big) 
	      \Big(\prod_{\substack{v \in V_e(\tau):\\ E_v=1,\, h_v=2}} e^{-\frac{c_0}{3}\delta_v}\Big).
	\end{split}
	\label{5.202}
\end{equation}
This is the analogue of \cite[Eq.(4.13)]{GGM}. Let $\mathfrak d=\mathfrak d(\bs x)$ be the minimal pairwise distance between the points in $\bs x$. 
Using the fact that, for any $n>1$, 
 \begin{equation}
    \delta(x_1,\dots,x_n)
    \ge
    \frac12 \min_{\pi \in \Pi(n)}
    \sum_{k=1}^{n-1} 
    \delta(\bs x_{\pi(k)},\bs x_{\pi(k+1)}), 
  \end{equation}
see \cite[Lemma~3.4]{BCO.TreeDecay}, we can bound from below $\delta_v\ge\frac12 (m_v-1)\mathfrak d\ge\frac12 (|S_v|-1)\mathfrak d$; we use this estimate in \eqref{5.202} for all vertices in $V_0(\tau)$ such that $v>v_0^*$, where $v_0^*=v_0$, if $|S_{v_0}|>1$, and $v_0^*$ equal to the only element of $S_{v_0}$, otherwise;
for $v\in\{v_0,v_0^*\}$ we use that $\delta_{v}=\delta(\bs x)\ge \frac13(\delta(\bs x)+(|S_v|-1)\mathfrak d)$. 
Next, in the sum over $\tau$, we distinguish two cases: either $\tau$ has all the endpoints  
on scales $<2$, in which case 
$2^{\theta h_M}=2^{\theta(h_{v^*}+1)}$ for some $v^*\in V_0'(\tau)$, or it has at least one endpoint on scale $2$, 
in which case, letting $\mathfrak{d}_\partial= \mathfrak{d}_\partial(\bs x)$ be the minimum of $\delta_\E(x)$ over the elements $x$ of $\bs x$, 
$\Big(\prod_{v \in V_e(\tau)}^{E_v=0,\, h_v=2} 2^{h_{v'}}\Big)\,\Big(\prod_{v \in V_e(\tau)}^{E_v=1,\, h_v=2} e^{-\frac{c_0}{3}\delta_v}\Big)\le 
2^{h_{v^*}}+e^{-\frac{c_0}{3}\mathfrak d_\partial}$ for some $v^*\in V_0'(\tau)$. 
In view of these considerations, letting $v^*$ be the vertex in $V_0'(\tau)$ such
that  $h_{v^*}=\max_{v\in V_0'(\tau)}\{h_v\}$, \eqref{5.202} implies that 
\begin{equation} 
	\begin{split}
		|R_1(\bs x)|
		\le  
		C^m\,|\lambda|\, \sum_{h=h^*-1}^0  
		&
		\ \sum_{\substack{\tau\in \cT^{(h)}_\free \\ |V_e(\tau)|=m}}
		2^{h_{v_0}-h_{v_0^*}}
		e^{-\frac{c_0}{36}2^{h_{v_0^*}}
		(\delta(\bs x)+(|S_{v_0^*}|-1)\mathfrak d)} 2^{h_{v_0^*}|S_{v_0^*}|} 
		\\ & \times \Big(\prod_{\substack{v \in V_0(\tau)\\ v>v_0^*}} 2^{(|S_v|-1)h_v} e^{-\frac{c_0}{24}2^{h_v}(|S_v|-1)\mathfrak d}\Big)\cdot (2^{\theta h_{v^*}}+e^{-\frac{c_0}{3}\mathfrak d_\partial})\equiv (I)+ (II), 
	\end{split}
	\label{eq:5.17a}
\end{equation}
where $(I)$ is the term proportional to $2^{\theta h_{v^*}}$, while $(II)$ is the one proportional to $e^{-\frac{c_0}{3}\mathfrak d_\partial}$. 
Term $(I)$ is bounded exactly as described in \cite[Eqs.(4.13)--(4.17)]{GGM}; by proceeding as discussed there, we find the analogue of 
\cite[Eq.(4.17)]{GGM}, namely 
\begin{equation}\label{boundI}(I)\le C^m |\lambda|  \Big(\frac1{\delta(\bs x)}\Big)^2\Big(\frac1{\mathfrak d}\Big)^{m-2+\theta}.\end{equation}
Term $(II)$ is bounded via an analogous strategy. For completeness, let us describe it explicitly: 
we split the sum over trees $\tau$ into a sum over 
the scale labels $\{h_v\}_{v\in V(\tau)}$, and a sum over the remaining structure of the tree, its `skeleton'. For fixed skeleton, we 
sum over the scale labels $h_v$, neglecting all the constraints but $h_{v_0}=h+1\le h_{v_0^*}$. By summing  
$2^{h_{v_0}-h_{v_0^*}}$ in the right side of \eqref{eq:5.17a} over $h_{v_0}\equiv h+1$, for $h_{v_0}\le h_{v_0^*}$, we get a factor $\sum_{j\le 0}2^j=2$. 
Next, by summing $e^{-\frac{c_0}{36}2^{h_{v_0^*}}
(\delta(\bs x)+(|S_{v_0^*}|-1)\mathfrak d)} 2^{h_{v_0^*}|S_{v_0^*}|}$ over $h_{v_0^*}$ we get a factor 
$$\Big[\sum_{h\in\mathbb Z}e^{-\frac{c_0}{36}2^{h}(\delta(\bs x)+(|S_{v_0^*}|-1)\mathfrak d)} 2^{h|S_{v_0^*}|}\Big]\le 
\frac1{\log 2}\Big(\frac{(72/c_0)}{\delta(\bs x)+(|S_{v_0^*}|-1)\mathfrak d}\Big)^{|S_{v_0^*}|}(|S_{v_0}|-1)!,$$
where we used the fact that, 
for any $\alpha,\delta >0$,
	\begin{equation}
	  \sum_{h\in\mathbb Z} 2^{\alpha h} e^{-2^h \delta}\le   
	  2^\alpha \int_{-\infty}^{\infty} 2^{\alpha x} e^{-2^x \delta} d x
	  =
	  \frac{(2/\delta)^\alpha}{\log 2} \Gamma(\alpha).\label{5.2.7}
	\end{equation}
Similarly, by summing $2^{(|S_v|-1)h_v} e^{-\frac{c_0}{24}2^{h_v}(|S_v|-1)\mathfrak d}$ over $h_v$ we get a factor 
$$\Big[\sum_{h\in\mathbb Z}  2^{(|S_{v}|-1)h}e^{-\frac{c_0}{24}2^{h}(|S_{v}|-1)\mathfrak d}\Big]\le 
\frac1{\log 2}\Big(\frac{(48/c_0)}{(|S_{v}|-1)\mathfrak d}\Big)^{|S_{v}|-1}(|S_{v}|-2)!.$$
Putting things together, and using the fact that
the number of skeletons with $m$ endpoints is bounded by (const.)$^m$, see \cite[Lemma A.1]{GM01}, we find that, 
up to a further redefinition of the constant $C$, 
\begin{eqnarray} (II)&\le&  C^m\,|\lambda|\, \,e^{-\frac{c_0}{3}\mathfrak d_\partial} \sup
 \frac{(|S_{v_0^*}|-1)!}{\big[\delta(\bs x)+(|S_{v_0^*}|-1)\mathfrak d\big]^{|S_{v_0^*}|}}
\Big(\prod_{\substack{v\in V_0(\tau)\\ v> v_0^*}} \frac{(|S_v|-2)!}{\big[(|S_v|-1)\mathfrak d\big]^{|S_v|-1}}\Big) \label{IcomeII}\\
&\le& C^m\,|\lambda|\, \, e^{-\frac{c_0}{3}\mathfrak d_\partial} \sup \Big(\frac1{\delta(\bs x)}\Big)^2\Big(
 \frac{|S_{v_0^*}|-1}{(|S_{v_0^*}|-1)\mathfrak d}\Big)^{|S_{v_0^*}|-2}
\Big(\prod_{\substack{v\in V_0(\tau)\\ v> v_0^*}} \Big(\frac{|S_v|-2}{(|S_v|-1)\mathfrak d}\Big)^{|S_v|-1}\Big),\nonumber\end{eqnarray}
where the sup is over the choices of $|S_v|$ compatible with the trees in $\cT^{(h)}_\free$ with $m$ endpoints, and in the second inequality 
we used the fact that $|S_{v_0^*}|\ge 2$. By using the fact that $\sum_{v\in V_0(\tau)}(|S_v|-1)=m-1$, see \eqref{telsv}, we finally get
\begin{equation}\label{boundII}(II)\le  C^m\,|\lambda|\, \Big(\frac1{\delta(\bs x)}\Big)^2\Big(\frac1{\mathfrak d}\Big)^{m-2}e^{-\frac{c_0}{3}\mathfrak d_\partial}.\end{equation}
Combining \eqref{boundI} and \eqref{boundII}, we obtain the desired bound on $R_1(\bs x)$,
\begin{equation}\label{summR1}|R_1(\bs x)|\le C^m\,|\lambda|\, \Big(\frac1{\delta(\bs x)}\Big)^2\Big(\frac1{\mathfrak d}\Big)^{m-2}\Big(\mathfrak d^{-\theta}+
e^{-\frac{c_0}{3}\mathfrak d_\partial}\Big).\end{equation}

\subsection{The remainder term $R_2$}
\label{sec:corr_corrections}

Let us now focus on 
\begin{equation} R_2(\bs x)
=\sum_{h=h^*-1}^0\sum^*_{\substack{\tau\in\cT^{(h)}:\\ m_{v_0}=m}}\ \sum_{\substack{\ul P\in \cP_*(\tau):\\ P_{v_0}=\emptyset}}\ \sum_{\ul T\in \cS(\tau,\ul P)}
\sum_{\ul D\in \cD(\tau,\ul P)} W_\Lambda[\tau,\ul P,\ul T,\ul D](\bs x),
\end{equation}
which, letting $\bs x_{v_0}\equiv \bs x$, we bound as
\begin{equation} |R_2(\bs x)|\le \sum_{h=h^*-1}^0\sum_{\substack{\tau\in\cT^{(h)} \\ m_{v_0}=m}}^*
e^{-\frac{c_0}{2}2^{h_{v_0}}\delta_{E_{v_0}}(\bs x)} 
\sum_{\substack{\ul P\in \cP_*(\tau) \\ P_{v_0}=\emptyset}}\ \sum_{\ul T\in \cS(\tau,\ul P)}\sum_{\ul D\in \cD(\tau,\ul P)} 
\|W_\Lambda[\tau,\ul P,\ul T,\ul D](\bs x_{v_0})\|_{E_{v_0};h_{v_0}}.\end{equation}
In view of Proposition \ref{lm:W:scaldim_withm}, the norm of $W_\Lambda[\tau,\ul P,\ul T,\ul D]$ is bounded as 
in \eqref{eq:W8_reworked_boundzaza}, with $\tilde Z_{h_v-1}(E_v,$ $\|\bs D_v\|_1,x_v)$ bounded as in \eqref{boundtildeZEv}, so that, after summing over $\ul T$ (using, in particular, 
\cite[Eq.~(4.4.29)]{AGG_part2})  
	\begin{eqnarray}
|R_2(\bs x)|&\le& C^m \sum_{h=h^*-1}^0\sum_{\substack{\tau\in\cT^{(h)} \\ m_{v_0}=m}}^*
e^{-\frac{c_0}{2}2^{h_{v_0}}\delta_{E_{v_0}}(\bs x)} 
\sum_{\substack{\ul P\in \cP_*(\tau) \\ P_{v_0}=\emptyset}}
 C^{\sum_{v\in V_e(\tau)}|P_v|}
2^{h_{v_0}(2-m)}\cdot 
\nonumber\\
&\cdot& \sum_{\ul D\in \cD(\tau,\ul P)}   \Big(
\prod_{v \in V'(\tau)} 2^{d_v(h_v-h_{v'})} \Big)\Big(\prod_{\substack{v\in V(\tau):\\ m_v\ge 1}} 2^{2[|S^*_v|-1]_+h_v} e^{-\frac{c_0}{12}2^{h_v}\delta_{E_v}(\bs x_v)}\Big)\cdot
\label{corona.v}\\
&\cdot& \Big(  \prod_{v \in V_e(\tau)}^{**}  |\lambda|^{\max\{1,\kappa(|P_v|+m_v)\}}2^{\theta h_v} \Big)\Big(\prod_{v \in V_e(\tau)}^{***}  |\lambda|(\delta_{\E}(x_v))^{-\theta}\Big)
,\nonumber\end{eqnarray}
where, in the first product  in the second line, $d_v$ is a shorthand for $d(P_v,\bs D_v,m_v,E_v)$,
and, in the products in the last line, $**$ indicates  the constraint that $(|P_v|,m_v)\neq(2,1)$, while ${*\!*\!*}$ indicates the constraint that 
$m_v=E_v=1$ and $h_v\le 1$ (which implies that $|P_v|=2$).
Note that the restriction to $\ul P \in \cP_*(\tau)$ implies that at least one of the two products in the last line runs over a non-empty set (and, as usual, whenever one of the products 
runs over the empty set, it should be interpreted as 1).
Recall also the following properties of $d_v$: for all the allowed choices of $P_v, \bs D_v, m_v, E_v$ 
in \eqref{corona.v}, $d_v\le -1$, with the following exceptions:
\begin{enumerate}
\item if $(|P_v|, \|\bs D_v\|_1, m_v, E_v)=(2,0,1,1)$, then $d_v=0$;
\item if $v$ is an endpoint such that $h_v-h_{v'}=1$, in which case $d_v$ may be equal to $0$ 
(if $(|P_v|, \|\bs D_v\|_1$, $m_v, E_v)=(4,0,0,0), (2,1,0,0), (2,0,0,1), (2,0,1,0), (2,0,1,1)$), or to $1$
(if $(|P_v|, \|\bs D_v\|_1, m_v$, $E_v)=(2,0,0,0)$).
\end{enumerate}
Note, incidentally, that, from the definition of tree values, the choice $(|P_v|, \|\bs D_v\|_1$, $m_v, E_v)=(4,0,0,0)$ (corresponding to a vanishing scaling dimension) 
is not allowed; the choices 
$(2,1,0,0)$, $(2,0,0,1)$, $(2,0,1,0)$, $(2,0,0,0)$ 
(corresponding to scaling dimensions equal to $0$ or $1$) are allowed, but, as just remarked, they require $h_{v'}=h_v-1$; the only 
allowed choice for which the scaling dimension is non-negative and $h_{v'}$ may be smaller than $h_v-1$ is $(|P_v|, \|\bs D_v\|_1$, $m_v, E_v)=(2,0,1,1)$.

The presence of cases for which $d_v\ge 0$ 
means that the sum must be handled differently than in the otherwise similar sums considered in \cite[Proof of Lemma~4.8]{AGG_part2}
(see Remark \ref{rem:4.21}), since the corresponding factors 
$2^{d_v(h_v-h_{v'})}$ are not exponentially small in $h_v-h_{v'}$. Clearly, the second case listed above is not problematic, because $h_v-h_{v'}=1$;
in that case, if desired, the factors $2^{d_v(h_v-h_{v'})}$ with $d_v=0,1$ can be freely replaced by $2^{-(h_{v}-h_{v'})}$, at the cost of a factor smaller than 
$2^{2N_{4}}$; here $N_{4}$ is the number of endpoints with $|P_v|\le 4$, which can be reabsorbed in the factor $ C^{\sum_{v\in V_e(\tau)}|P_v|}$ in the right 
side of \eqref{corona.v}, up to a redefinition of $C$. To summarize, the bound \eqref{corona.v} remains valid, up to a redefinition of the constant $C$, 
even if we replace the factors $2^{d_v(h_v-h_{v'})}$ in the 
first product  in the second line by the smaller factors $2^{d_v'(h_v-h_{v'})}$, where 
\begin{equation}\label{eq:defdv'}
d'_v=\piecewise{\min\{d_v,-1\} & \text{if $(|P_v|, \|\bs D_v\|_1,m_v,E_v)\neq (2,0,1,1)$}\\ 
0 & \text{if $(|P_v|, \|\bs D_v\|_1,m_v,E_v)=(2,0,1,1)$.}}\end{equation}

We now further manipulate \eqref{corona.v} (or, better, its rewriting with $d_v$ replaced by $d_v'$), by proceeding in a way similar to what it was done in \cite{GGM} after \cite[Eq.(4.20)]{GGM}. 
We let $V_e^{A}(\tau)=\{v\in V_e(\tau): \ m_v\ge 1\}$, and $\tau^*(\tau)$ be the minimal subtree of $\tau$ containing $v_0$ and all the endpoints in $V_e^A(\tau)$. Our first step consists in `pruning' the tree $\tau$ 
of the branches that are not in $\tau^*(\tau)$. For this purpose, we make the following rearrangement. 
Let $\mathcal{T}^{*}_{h,h_0^*;m}$ be the set of labelled trees with endpoints all of type \tikzvertex{bareProbeEP} or \tikzvertex{bareProbeEP,E}, 
$m_{v_0}=m$, no trivial vertices, dotted root $v_0$ on scale $h+1$, and with the 
leftmost  branching point, which we denote by $v_0^*$, on scale $h_0^*\ge h+1$ 
(if $h_0^*=h+1$, then $v_0=v_0^*$, otherwise $v_0^*$ is the vertex immediately following $v_0$ on $\tau^*$; if $\tau^*$ has no branching points, 
then $h_0^*=2$ and $v_0^*$ is the only endpoint of $\tau^*$); `as usual', the endpoints may be on scale $2$ or smaller; however, if $v$ is an endpoint on scale $h_v<2$, 
we do not assume that the vertex immediately preceding it on $\tau^*$ is on scale $h_{v}-1$, we allow the case that $h_{v'}<h_v-1$. 
Given $\tau^*\in \mathcal{T}^{*}_{h,h_0^*;m}$ and $v\in V_0(\tau^*)$, we denote 
by $s^*_v$ the number of vertices immediately following $v$ on $\tau^*$ (note that, since by definition $\tau^*$ does not have trivial vertices, $s^*_v>1$ for all $v\in V_0'(\tau^*)$), 

The idea is to rewrite the sum over $\tau$ in \eqref{corona.v} as a sum over $\tau^*\in \cT^*_{h,h_0^*;m}$ times a sum 
over trees $\tau$ compatible with $\tau^*$, i.e., such that $\tau^*(\tau)=\tau^*$. Having done this, we will perform the sums in the following order: 
first we sum over $\tau$ and $\ul P$ at $\tau^*$ fixed, and then over $\tau^*$. 

Fix $\theta=3/4$, as above, let $\epsilon:=(1-\theta)/3$, and define
\begin{equation} \tilde d_v:= \left\{ 
\begin{array}{ll}
1+\epsilon -m_v, &  m_v > 1 \\
-\theta, & \text{$m_v = 1$ and $E_v=0$} \\
0, & \text{otherwise}.\label{eq:5.2.5}
\end{array}\right. \end{equation}
Using the definition of $d_v'$ in \eqref{eq:defdv'} and of $d_v=d_v(P_v,\bs D_v,m_v,E_v)$ in \eqref{defscaldimwithm}, it is easy to check that
\begin{equation}  d_v'-\tilde d_v\le -\frac{\epsilon}{2}|P_v|-\theta \delta_{m_v,0}, \qquad \text{if $(|P_v|,\|\bs D_v\|_1,m_v,E_v)\neq (2,0,1,1)$},
\label{dv'-tildedv}\end{equation}
while $d_v'=\tilde d_v=0$, if $(|P_v|,\|\bs D_v\|_1,m_v,E_v)= (2,0,1,1)$. We now: neglect the factor $e^{-\frac{c_0}{2}2^{h_{v_0}}\delta_{E_{v_0}}(\bs x)}$
 in \eqref{corona.v}, since a comparable one is contained in the second product in the second line of \eqref{corona.v}, and, recalling that $h_{v_0}=h+1$, 
 we rearrange the result as follows:
\begin{eqnarray}
  |R_2(\bs x)| &\le& C^m \sum_{h=h^*-1}^0 2^{h(2-m)} \sum_{h<h_0^*\le 2}\ \sum_{\tau^* \in \mathcal{T}^*_{h,h_0^*;m}}
\Big(\prod_{v\in V(\tau^*)} 2^{2[s^*_v-1]_+h_v} e^{-\frac{c_0}{12}2^{h_v}\delta_{E_v}(\bs x_v)}\Big)\cdot\nonumber\\
&\cdot& \Big(\prod_{v \in V'(\tau^*)} 2^{\tilde d_v(h_v-h_{v'})} \Big)\,\cdot\Big\{ \Big(\prod_{v \in V_e(\tau^*)}^{***}  |\lambda|(\delta_{\E}(x_v))^{-\theta}\Big)
\sum_{\tau\in \cT(\tau^*)}^* \sum_{\substack{\ul P\in \cP_*(\tau) \\ |P_{v_0}|=0}}
\sum_{\ul D\in \cD(\tau,\ul P)}\cdot\nonumber\\
&\cdot &\Big(\prod_{v \in V'(\tau)} 2^{(d_v'-\tilde d_v)(h_v-h_{v'})} \Big) \Big(  \prod_{v \in V_e(\tau)}^{**} (C|\lambda|)^{\max\{1,\kappa(|P_v|+m_v)\}}2^{\theta h_v} \Big)
\Big\}, \label{corona.vi}\end{eqnarray}
where, given $\tau^*\in \cT^*_{h,h_0^*;m}$, $\cT(\tau^*)$ is the subset of trees in $\cT^{(h)}$ such that $\tau^*(\tau)=\tau^*$. Eq.\eqref{corona.vi}
is the analogue of the un-numbered equation after \cite[Eq.(4.22)]{GGM}. We now perform the sums in braces and, in analogy with \cite[Eq.(4.23)]{GGM}, we get 
\begin{eqnarray} &&\hskip-.3truecm \Big(\prod_{v \in V_e(\tau^*)}^{***}  |\lambda|(\delta_{\E}(x_v))^{-\theta}\Big)\sum_{\tau\in \cT(\tau^*)}^* 
\sum_{\substack{\ul P\in \cP_*(\tau) \\ |P_{v_0}|=0}}
\sum_{\ul D\in \cD(\tau,\ul P)} \Big(\prod_{v \in V'(\tau)} 2^{(d_v'-\tilde d_v)(h_v-h_{v'})} \Big)\cdot\nonumber\\
&&\hskip1.truecm \cdot 
\Big(  \prod_{v \in V_e(\tau)}^{**} (C|\lambda|)^{\max\{1,\kappa(|P_v|+m_v)\}}2^{\theta h_v} \Big) \le C^{m} |\lambda|(2^{\theta' h^*_M}+\mathfrak{d}_\partial^{-\theta}),\label{corona.vir}\end{eqnarray}
where $h^*_M=h_M(\tau^*)=\max\{h_v:\ v\in V_e(\tau^*)\}$, 
$\theta'=\theta-\epsilon/2$, and $\mathfrak d_\partial=\mathfrak d_\partial(\bs x)=\min_{v\in V_e(\tau^*)}\delta_\E(x_v)$.
The proof of \eqref{corona.vir} is given in Appendix \ref{app.proofboringb}, and is one of the important novelties of the current work, 
as compared to \cite{GGM}. Of course, the new source of difficulty, as compared to the proof of \cite[Eq.(4.23)]{GGM}, is the  
possible presence of vertices such that $d_v'=\tilde d_v=0$, for which the factor $2^{(d_v'-\tilde d_v)(h_v-h_{v'})}$ equals $1$, rather
then being exponentially small in $(h_v-h_{v'})$ and in $|P_v|$. 

Plugging \eqref{corona.vir} in \eqref{corona.vi} gives, letting again $\delta_{E_v}(\bs x_v)\equiv \delta_v$:
\begin{equation}
	\begin{split}
		|R_2(\bs x)| 
		\le& 
		C^m \sum_{h=h^*-1}^0 2^{h(2-m)} \sum_{h<h_0^*\le 2}\ \sum_{\tau^* \in \mathcal{T}^*_{h,h_0^*;m}}
		|\lambda|\, (2^{\theta' h^*_M}+\mathfrak{d}_\partial^{-\theta})
		\\ & \times
		\Big(\prod_{v\in V(\tau^*)} 2^{2[s^*_v-1]_+h_v} e^{-\frac{c_0}{12}2^{h_v}\delta_{v}}\Big)\,\Big(\prod_{v \in V'(\tau^*)} 2^{\tilde d_v(h_v-h_{v'})} \Big)
		\\ \le &
		(C')^m\, \sum_{h\le 0}\sum_{h<h_0^*\le 2} 2^{(h-h_0^*)(1-\epsilon)} \sum_{\tau^* \in \mathcal{T}^*_{h,h_0^*;m}} |\lambda|\,
		(2^{\theta' h^*_M}+\mathfrak{d}_\partial^{-\theta})2^{(2s^*_{v_0^*}-m)h_0^*}
		e^{-\frac{c_0}{12}2^{h_0^*}\delta_{v_0^*}} 
		\\ & \times
		\Big(\prod_{\substack{v \in V_0(\tau^*)\\ v>v_0^*}} 2^{2(s^*_v-1)h_v} e^{-\frac{c_0}{12}2^{h_v}\delta_v}2^{(1+\epsilon-m_v)(h_v-h_{v'})} \Big)\,
		\Big(\prod_{\substack{v \in V_e(\tau^*):\\ E_v=0,\, m_v=1}} 2^{-\theta(h_v-h_{v'})}\Big)
		\\ & \times
		\Big(\prod_{\substack{v \in V_e(\tau^*):\\ E_v=1,\, m_v=1}} e^{-\frac{c_0}{12}2^{h_v}\delta_v}\Big)\, 
		\Big(\prod_{\substack{v \in V_e(\tau^*):\\ m_v>1}} 2^{(m_v-1-\epsilon)h_{v'}} e^{-\frac{c}{3}\delta_v}\Big)=:(I)+(II)+(III),
	\end{split}
\label{daichecci}
\end{equation}
where in the second inequality we used the definition of $\tilde d_v$ explicitly, and we distinguished the terms in the products associated with 
$v_0^*$, from those associated with the vertices in $V_0(\tau^*)\setminus\{v_0^*\}$, from those 
associated with the endpoints. 

In the last identity in \eqref{daichecci}, the contributions denoted by $(I)$, $(II)$, $(III)$ are defined as follows. Consider first the contribution to the right side of \eqref{daichecci}
proportional to $2^{\theta' h^*_M}$, and let $\bar v$ be the endpoint that realizes the maximum in the definition of $h^*_M$, i.e., $h^*_M=h_{\bar v}$. 
We further distinguish two cases: (1) $(E_{\bar v},m_{\bar v})=(0,1)$ 
(which means that $\bar v$ is included in the product $\prod_{v \in V_e(\tau^*)}^{E_v=0,\, m_v=1} 2^{-\theta(h_v-h_{v'})}$); (2) 
$(E_{\bar v},m_{\bar v})\neq(0,1)$ (which means that $\bar v$ is included in one of the two products in the last line of \eqref{daichecci}). 
In terms of these definitions, we denote by $(I)$ (resp. $(II)$) the contribution to the right side of \eqref{daichecci} proportional to $2^{\theta' h^*_M}$ and associated with trees realizing case (1) (resp. (2)). 
Finally, we denote by $(III)$ the contribution to the right side of \eqref{daichecci} proportional to $\mathfrak{d}_\partial^{-\theta}$.

The definition of  $(I)$ is completely analogous to \cite[Eq.(4.24)]{GGM}; by proceeding as discussed there, we get the analogue of \cite[Eq.(4.33)]{GGM}, i.e., 
\begin{equation}\label{nonsicont.vir} (I)\le C^m |\lambda|  \Big(\frac1{\mathfrak d}\Big)^{m+\theta'}\Big(\frac{\mathfrak d}{\delta(\bs x)}\Big)^{2-2\epsilon}.
\end{equation}
Let us now consider $(II)$, which we manipulate as follows. First, we rewrite the product of the factors $2^{(1+\epsilon-m_v)(h_v-h_{v'})}$ as 
\begin{equation}\prod_{\substack{v \in V_0(\tau^*)\\ v>v_0^*}} 2^{(1+\epsilon-m_v)(h_v-h_{v'})}=2^{h^*_0[-(1+\epsilon)s^{*,2}_{v_0^*}+m^{*,2}_{v_0^*}]}
\prod_{\substack{v \in V_0(\tau^*)\\ v>v_0^*}} 2^{h_v[(1+\epsilon)(1-s^{*,2}_v)-m_v+m^{*,2}_v]},\label{rewtel.vir}\end{equation}
where $s^{*,2}_v=|S^{*,2}_v|$, with $S^{*,2}_v=S^*_v\cap V_0(\tau^*)$, and $m^{*,2}_v=\sum_{w\in S^{*,2}_v}m_w$. Recall the notation $s^{*}_v=|S^*_v|$, 
where, since $\tau^*$ has no endpoints with $m_v=0$, $S^*_v$ is the same thing as $S_v$, i.e., it is the set of vertices immediately following $v$ on $\tau^*$. 
For later reference, we also let
$s^{*,1}_v=\{$number of endpoints with $m_v=1$ immediately following $v$ on $\tau^*\}$, 
$s^{*,>}_v=\{$number of endpoints with $m_v>1$ immediately following $v$ on $\tau^*\}$, and $m^{*,>}_v=m_v-s^{*,1}-m^{*,2}_v$ (note that 
$m^{*,>}$ is the sum of $m_w$ over the endpoints $w$ with $m_w>1$ immediately following $v$ on $\tau^*$). If we use \eqref{rewtel.vir}
in \eqref{daichecci} and we associate each factor $2^{(m_v-1-\epsilon)h_{v'}}$ in the last product of \eqref{daichecci} with the vertex $v'$ immediately preceding 
the endpoint $v$ with $m_v>1$, we find: 
\begin{eqnarray} (II)&\le& C^m\, \sum_{h\le 0}\sum_{h<h_0^*\le 2} 2^{(h-h_0^*)(1-\epsilon)} \sum_{\tau^* \in \mathcal{T}^*_{h,h_0^*;m}}^{(II)} 
|\lambda|\,2^{\theta' h_{\bar v}}
  \Big(\prod_{\substack{v \in V_0(\tau^*)\\ v\ge v_0^*}} 2^{\alpha_v h_v} e^{-\frac{c_0}{12}2^{h_v}\delta_v} \Big) \cdot\nonumber\\
&\cdot&  
\Big(\prod_{\substack{v \in V_e(\tau^*):\\ (E_v,m_v)=(0,1)}} 2^{-\theta(h_v-h_{v'})}\Big)\,
\Big(\prod_{\substack{v \in V_e(\tau^*):\\ (E_v,m_v)\neq(0,1)}} e^{-\frac{c_0}{12}2^{h_v}\delta_v}\Big)\, 
\label{daicheccisi}\end{eqnarray}
where the label $(II)$ on the sum indicates the constraint that the endpoint $\bar v$ such that $h_{\bar v}=\max_{v\in V_e(\tau^*)}h_v$ 
is such that $(E_{\bar v},m_{\bar v})\neq(0,1)$ (in particular, the second product in the second line is not empty), and 
\begin{equation}\alpha_v = \left\{
\begin{array}{ll}
(1-\epsilon)s_{v}^*+\epsilon s_v^{*,1}, & v=v_0^* \\
(1-\epsilon)(s_{v}^* -1) + \epsilon s_v^{*,1} & \text{otherwise.}\end{array}
\right.\end{equation}
In order to bound \eqref{daicheccisi} we proceed in a way similar to the one used to bound term $(II)$ in \eqref{eq:5.17a}. 
Recalling that $\mathfrak d$ is the minimal pairwise distance between the points in $\bs x$ and that $\mathfrak d_\partial$ is the minimum of $\delta_\E(x)$ over the points $x$ in $\bs x$,  
we bound 
\begin{equation} 
\begin{split} 2^{\theta' h_{\bar v}}\Big(\prod_{\substack{v \in V_e(\tau^*):\\ (E_v,m_v)\neq(0,1)}} e^{-\frac{c_0}{12}2^{h_v}\delta_v}\Big) &\le
2^{\theta' h_{\bar v}}e^{-\frac{c_0}{24}2^{h_{\bar v}}\delta_{\bar v}}\Big(\prod_{\substack{v \in V_e(\tau^*):\\ m_v>1}} e^{-\frac{c_0}{6}\delta_v}\Big)\\
&\le \Big(\sum_{h\le 2} 2^{\theta' h} e^{-\frac{c_0}{24}2^{h}\mathfrak d_\partial}+2^{2\theta'}e^{-\frac{c_0}{6}\mathfrak{d}}\Big)
\Big(\prod_{\substack{v \in V_e(\tau^*):\\ m_v>1}} e^{-\frac{c_0}{6}\delta_v}\Big)\\
&\le C\Big(\mathfrak d_\partial^{-\theta'}+ e^{-\frac{c_0}{6}\mathfrak d}\Big)
\Big(\prod_{\substack{v \in V_e(\tau^*):\\ m_v>1}} e^{-\frac{c_0}{6}\delta_v}\Big),\end{split}\end{equation}
Next, we bound
$\delta_v$ from below by $\frac12(m_v-1)\mathfrak d\ge 
\frac12(s_v-1)\mathfrak d$ for all vertices of $\tau^*$ such that $v>v_0^*$; for $v=v_0^*$, we use 
$\delta_{v_0^*}\ge \frac13(\delta(\bs x)+(s_{v_0^*}-1)\mathfrak d)$. We 
perform the sum over $\tau^*$ by first summing over the scale labels, and then over the remaining structure (`skeleton') of the tree. 
By summing the factor $2^{(h-h_0^*)(1-\epsilon)}$ over $h<h_0^*$ with the other scales fixed, we get $\sum_{j\le -1}2^{j(1-\epsilon)}$. 
By summing the factors $2^{-\theta(h_v-h_{v'})}$ over the scale labels of the endpoints with $m_v=1$ and $E_v=0$ with the other scales fixed, we get 
a factor $\sum_{k\ge 1}2^{-\theta k}$ for each such endpoint.  By summing the factor 
$2^{\alpha_{v_0^*} h_0^*} e^{-\frac{c_0}{36}2^{h_0^*}(\delta(\bs x)+(m-1)\mathfrak d}$ over $h_0^*$ we get 
$[{(\rm const.)}/(\delta(\bs x)+(m-1)\mathfrak d)]^{\alpha_{v_0^*}}\Gamma(\alpha_{v_0^*})$ (see \eqref{5.2.7}). 
Similarly, by summing the factors $2^{\alpha_{v} h_v} e^{-\frac{c_0}{24}2^{h_0^*}(m_v-1)\mathfrak d}$ over $h_v$ 
we get $[({\rm const.})/((m_v-1)\mathfrak d)]^{\alpha_{v}}\Gamma(\alpha_{v})$. Finally, using the fact that the number of skeletons with $\le m$ endpoints is 
smaller than $({\rm const.})^m$, we find the analogue of \eqref{IcomeII}, namely
\begin{equation} \begin{split}(II)&\le  
C^m\,|\lambda|\Big(\mathfrak d_\partial^{-\theta'}+ e^{-\frac{c_0}{6}\mathfrak d}\Big) \sup_{\tau^*}\Biggl\{ \frac{(s^*_{v_0^*}-1)!}{\big[\delta(\bs x)+(s^*_{v_0^*}-1)\mathfrak d\big]^{(1-\epsilon)s_{v_0^*}^*+\epsilon s_{v_0^*}^{*,1}}}
\cdot\\
&\cdot \Big(\prod_{\substack{v\in V_0(\tau)\\ v> v_0^*}} \frac{(s^*_v-2)!}{\big[(s^*_v-1)\mathfrak d\big]^{(1-\epsilon)(s_{v}^* -1) + \epsilon s_v^{*,1}}}\Big)\cdot 
\Big(\prod_{\substack{v \in V_e(\tau^*):\\ m_v>1}} e^{-\frac{c_0}{12}(m_v-1)\mathfrak d}\Big)\Biggr\},
 \end{split}\label{daicheccisia}\end{equation}
 where the sup is over the trees $\tau^*$ involved in the sum in \eqref{daicheccisi} (and, therefore, over the corresponding
 choices of $s^*_v, s^{*,1}_v, m_v$). By using the fact that $s^*_{v_0^*}\ge 2$, that 
 $\sum_{v}(s^*_v-1)=N-1$, with $N=N(\tau^*)=|V_e(\tau^*)|$, and that $\sum_{v\in V_0(\tau)}s^{*,1}_v=m_1$, with $m_1=m_1(\tau^*)=\{$number of endpoints of $\tau^*$ with $m_v=1\}$, we get:
\begin{equation} (II)\le  
C^m\,|\lambda|\Big(\mathfrak d_\partial^{-\theta'}+ e^{-\frac{c_0}{6}\mathfrak d}\Big)  \sup_{\tau^*}\Biggl\{ \Big(\frac{\mathfrak d}{\delta(\bs x)}\Big)^{2-2\epsilon}\Big(\frac1{\mathfrak d}\Big)^{m_1}
\Big(\prod_{\substack{v \in V_e(\tau^*):\\ m_v>1}}\frac{e^{-\frac{c_0}{12}(m_v-1)\mathfrak d}}{\mathfrak d^{1-\epsilon}}\Big)\Biggr\}. \label{daicheccisiam}
\end{equation}
Each of the factors $\frac{e^{-\frac{c_0}{12}(m_v-1)\mathfrak d}}{\mathfrak d^{1-\epsilon}}$ can be bounded from above by $C^{m_v}\mathfrak d^{-m_v}$ for some constant $C>0$, so that, in conclusion, 
\begin{equation} (II)\le  
C^m\,|\lambda|\,  \Big(\frac{\mathfrak d}{\delta(\bs x)}\Big)^{2-2\epsilon}\Big(\frac1{\mathfrak d}\Big)^{m}\Big(\mathfrak d_\partial^{-\theta'}+ e^{-\frac{c_0}{6}\mathfrak d}\Big). \label{daicheccisiamo}
\end{equation}
Finally, a repetition of the proof leading to \eqref{daicheccisiamo} implies that, similarly, 
\begin{equation} (III)\le  
C^m\,|\lambda|\,  \Big(\frac{\mathfrak d}{\delta(\bs x)}\Big)^{2-2\epsilon}\Big(\frac1{\mathfrak d}\Big)^{m}\mathfrak d_\partial^{-\theta}. \label{daicheccisiamo!}
\end{equation}
Combining \eqref{daicheccisiamo} and \eqref{daicheccisiamo!} with \eqref{nonsicont.vir}, we find that 
\begin{equation}|R_2(\bs x)|\le C^m\,|\lambda|(\mathfrak d/\delta(\bs x))^{2-2\epsilon}(1/\mathfrak d)^{m}(\min\{\mathfrak d,\mathfrak d_\partial\})^{-\theta'}.\end{equation}
Finally, plugging this bound, together with the analogous bound on $R_1$, Eq.\eqref{summR1}, and \eqref{eq:boundR0}, into \eqref{eccocisi}, proves 
\eqref{10b} and concludes the proof of the main result of this paper, Theorem \ref{prop:main}.

\appendix

\section{Norm bounds for $\cE V_\B^{(h)} $}
\label{app:segunda}

In this appendix we prove \eqref{belissemo}. We recall that $\mathcal E V_\B^{(h)}=\cL_\B V_\B^{(h)}-(\cL_\B V_\B^{(h)})_\B+
\cR_\B V_\B^{(h)}-(\cR_\B V_\B^{(h)})_\B\equiv (\cL_\B V_\B^{(h)})_\E+(\cR_\B V_\B^{(h)})_\E$ and that $(\mathcal E V_\B^{(h)})_{n,p}$ vanishes unless
$(n,p)\in \{(2,0),(2,1)$, $(2,2),(4,1)\}$ (see Remark \ref{remar35}). From the definitions of $\cL_\B$ and $\cR_\B$ 
in \eqref{defcL.1} and \eqref{defcR} and the definition of $\mathcal E V_\B^{(h)}$, it follows, in particular, that:
$(\mathcal E V_\B^{(h)})_{2,0}=(\cA(\tcL (V_\B^{(h)})_{2,0}))_\E$; 
$(\mathcal E V_\B^{(h)})_{2,1}=(\cA(\tcL (V_\B^{(h)})_{2,1}+\tcL (\tcR V_\B^{(h)})_{2,1}))_\E$; 
$(\mathcal E V_\B^{(h)})_{2,2}=(\cA((\tcR V_\B^{(h)})_{2,2}+(\tcR (\tcR V_\B^{(h)}))_{2,2}))_\E$;
$(\mathcal E V_\B^{(h)})_{4,1}=(\cA((\tcR V_\B^{(h)})_{4,1}))_\E$ (in writing the last two identities, we used the fact that 
$((V_\B^{(h)})_{2,2})_\E$ and $((V_\B^{(h)})_{4,1})_\E$ vanish, because they are the edge part of the bulk part of a kernel). Therefore, 
\begin{equation} \begin{split} 
& \|(\mathcal E V_\B^{(h)})_{2,0}\|_{(\E;\kappa)}\le \|(\tcL (V_\B^{(h)})_{2,0})_\E\|_{(\E;\kappa)}, \\
&  \|(\mathcal E V_\B^{(h)})_{2,1}\|_{(\E;\kappa)}\le \|(\tcL (V_\B^{(h)})_{2,1})_\E\|_{(\E;\kappa)}+\|(\tcL (\tcR V_\B^{(h)})_{2,1})_\E\|_{(\E;\kappa)}, \\
&  \|(\mathcal E V_\B^{(h)})_{2,2}\|_{(\E;\kappa)}\le \|((\tcR V_\B^{(h)})_{2,2})_\E\|_{(\E;\kappa)}+\|((\tcR (\tcR V_\B^{(h)}))_{2,2})_\E\|_{(\E;\kappa)}, \\
& \|(\mathcal E V_\B^{(h)})_{4,1}\|_{(\E;\kappa)}\le \|((\tcR V_\B^{(h)})_{4,1})_\E\|_{(\E;\kappa)}. \end{split}\label{A.1}\end{equation}
From the previous formulas and the definitions of $\tcL$, $\tcR$, it follows that both the left and right sides of these inequalities can be written in terms of 
$(V_\infty^{(h)})_{n,p}$, with $(n,p)=(2,0),(2,1),(4,0)$, which, for $\theta=3/4$ and $|\lambda|$ small enough, satisfies
\begin{equation}\|(V_\infty^{(h)})_{n,p}\|_{(c_02^h)}
	\le C	|\lambda|^{\max\{1,\kappa n\}}
	2^{h (2-n/2-p)}2^{\theta h},\label{eq:cited_V8_bound}
\end{equation}
see \cite[Eq.~(4.4.27)]{AGG_part2} (in our cases of interest, in which $n\le 4$, the factor $|\lambda|^{\max\{1,\kappa n\}}$ can be simply replaced by 
$|\lambda|$). 

We start with bounding $\|(\tcL (V_\B^{(h)})_{2,p})_\E\|_{(\E;\kappa)}$ with $p=0,1$, where
\begin{equation}
	\begin{split}
		&
		\big( 
			\tcL (V_\B^{(h)})_{2,p} 
		\big)_\E(\bs \omega, \bs D, \bs z)
		=\tcL (V_\B^{(h)})_{2,p} (\bs \omega, \bs D, \bs z)
		-\big( \tcL (V_\B^{(h)})_{2,p} \big)_\B
		(\bs \omega, \bs D, \bs z)
		\\ & \quad =
\delta_{z_2,z_1}\Big[ 
		\sum_{\substack{\bs y \in \Lambda^2:\\ \bs y+\bs D\in\Lambda^2\\ \diam_1(\bs y)\le L/3}}\delta_{y_1,z_1}
		(V_\infty^{(h)})_{2,p}(\bs \omega, \bs D, \bs y_\infty)\\
		&\quad -\mathds 1(\bs z,\bs z+\bs D\in\Lambda^2)
		\sum_{\bs y \in \Lambda_\infty^2}\delta_{y_1,z_1}
		(V_\infty^{(h)})_{2,p}(\bs \omega, \bs D, \bs y)
		\Big].
	\end{split}\label{rhsApp}
\end{equation}
In order for this expression not to vanish, we must have $z_1=z_2\in\Lambda$ and $z_1+D_1\in\Lambda$, which we assume to hold. 
If $z_1+D_2\not\in\Lambda$, then the second term in square brackets vanishes, and $z_1=z_2$ belongs to the upper boundary of $\Lambda$;
if $z_1+D_2\in\Lambda$, then the terms in the first sum in square brackets are entirely cancelled by a corresponding set of terms in the second sum, leaving only those 
for which $\diam_1(\bs y) > L/3$ or $y_2\notin \Lambda$ or $y_2+D_2\notin\Lambda$; in all these cases, we have $\delta (\bs y) \ge  \delta_\E(\bs z)$. 
As a result, for any $\kappa,\epsilon>0$,
\begin{equation}\begin{split}
	\sum_{\substack{\bs z \in \bar\Lambda^2\\ (z_1)_1 \text{fixed}}}
		e^{\kappa \delta_\E (\bs z)}\sup_{\bs D}^{(p)}
		\Big|\big( \tcL (V_\B^{(h)})_{2,p} \big)_\E (\bs\omega,\bs D,\bs z)\Big|&\le \sum_{(z_1)_2=1}^Me^{-\epsilon\min\{(z_1)_2-1,M-(z_1)_2\}}\cdot\\
&\cdot\sum_{\substack{\bs y \in\Lambda_\infty^2:\\ y_1 \text{fixed}}}
		e^{(\kappa+\epsilon) \delta (\bs y)}\sup_{\bs D}^{(p)}
		\big|(V_\infty^{(h)})_{2,p}(\bs \omega, \bs D, \bs y)\big|,\end{split}
\label{eqref}\end{equation}
which, recalling the norm defined in \eqref{eq:cRE_sumbound} and the infinite volume analogue of \eqref{eq:weightnorm}, implies
\begin{equation}\label{eppA5} \|(\tcL (V_\B^{(h)})_{2,p})_\E\|_{(\E;\kappa)}\le \frac{2}{1-e^{-\epsilon}}\|(V_\infty^{(h)})_{2,p}\|_{(\kappa+\epsilon)}.\end{equation}
Let us consider next $\|(\tcR (V_\B^{(h)})_{n,p})_\E\|_{(\E;\kappa)}$ with $(n,p)=(2,2), (4,1)$, where, recalling the definition of $\tcR$ in \cref{deftcR},
\begin{equation}
	\begin{split}
		& ((\tcR V_\B^{(h)})_{n,p})_\E(\bs \omega, \bs D, \bs z)
		=(\tcR V_\B^{(h)})_{n,p}(\bs \omega, \bs D, \bs z)
		-
		\big((\tcR V_\B^{(h)})_{n,p}\big)_\B(\bs \omega, \bs D, \bs z)
		\\ & \qquad =
		(-1)^{\alpha(\bs z)}
		\Big[ \sum_{\substack{\sigma = \pm,\, \bs y \in \Lambda^n,\, \bs D'\in\{0,\hat e_1,\hat e_2\}^n: 
						\\ (\sigma, \bs D',\bs z) \in \INT(\bs y)}}I_\Lambda(\bs \omega,\bs D-\bs D',\bs y)
					\sigma (V_\infty^{(h)})_{n,p-1} (\bs \omega, \bs D-\bs D', \bs y_\infty)
					\\ &\hskip3.truecm
					- I_\Lambda(\bs\omega,\bs D,\bs z)
					\sum_{\substack{\sigma = \pm,\, \bs y \in \Lambda_\infty^n,\, \bs D'\in\{0,\hat e_1,\hat e_2\}^n: 
						\\ (\sigma, \bs D',\bs z_\infty) \in \INT(\bs y)}}
					\sigma (V_\infty^{(h)})_{n,p-1}(\bs \omega, \bs D-\bs D', \bs y)
			\Big]
			;
		\end{split}\label{2+2=25App}
\end{equation}
with $I_\Lambda$ the indicator function in \eqref{eq:3.1.4}, and the interpolation path $\INT$ in the second sum in square brackets the infinite volume analogue of 
the one defined after \eqref{eq:gamma_symmetry} and after \eqref{supersplit} (see the definitions given after \cite[Eq.~(4.2.7)]{AGG_part2}
and after \cite[Eq.~(4.2.16)]{AGG_part2}); the construction of $\INT$ is such that the first sum in square brackets is over an empty set whenever 
$I_\Lambda(\bs\omega,\bs D,\bs z)=0$, so again the first sum cancels all terms of the second sum which do not satisfy $\delta(\bs y) \ge  \delta_\E (\bs z)$,  
giving, for any $\kappa,\epsilon>0$,

\begin{equation}
	\begin{split}
		&
		\sum_{\substack{\bs z \in \bar\Lambda^2\\ (z_1)_1 \text{fixed}}}e^{\kappa \delta_\E (\bs z)}\sup_{\bs D}^{(p)}
		\big|((\tcR V_\B^{(h)})_{n,p})_\E(\bs \omega, \bs D, \bs z)\big|
		\\ &
		\le \sum_{(z_1)_2=1}^Me^{-\frac\epsilon2\max\{(z_1)_2-1,M-(z_1)_2\}}\sum_{\substack{\bs y \in \Lambda_\infty^{n}\\ y_1 \text{fixed}}}
		e^{(\kappa+\frac\epsilon2) \delta(\bs y)}\sup_{\bs D}^{(p)}
		\sum_{(\sigma, \bs D', \bs z) \in \INT(\bs y)}
		| (V_\infty^{(h)})_{n,p-1} (\bs \omega, \bs D-\bs D', \bs y) |
		\\ &
		\le 
		\frac{4(n-1)}{\epsilon(1-e^{-\frac\epsilon2})}\sum_{\substack{\bs y \in \Lambda_\infty^n\\ y_1 \text{fixed}}}
		e^{(\kappa + \epsilon) \delta(\bs y)}\sup_{\bs D}^{(p-1)}
		| (V_\infty^{(h)})_{n,p-1}(\bs \omega, \bs D, \bs y)|
	\end{split}\label{2+2=28APP}
\end{equation}
(for the last inequality, see \cite[Equations~(4.2.26) to~(4.2.28)]{AGG_part2}). In conclusion, for $\epsilon$ smaller than (say) $c_0$ and 
a suitable $C>0$,
\begin{equation}\label{eppA8}
 \|((\tcR V_\B^{(h)})_{n,p})_\E\|_{(\E;\kappa)}\le 
 	\frac{C}{\epsilon^2}\|(V_\B^{(h)})_{n,p-1}\|_{(\kappa + \epsilon)}. \end{equation}

Let us now briefly discuss the remaining two cases: consider $\|(\tcL (\tcR V_\B^{(h)})_{2,1})_\E\|_{(\E;\kappa)}$, with 
\begin{equation}
	\begin{split}
		& (\tcL(\tcR V_\B^{(h)})_{2,1})_\E(\bs \omega, \bs D, \bs z)
		=\tcL(\tcR V_\B^{(h)})_{2,1}(\bs \omega, \bs D, \bs z)
		-
		\big(\tcL(\tcR V_\B^{(h)})_{2,1}\big)_\B(\bs \omega, \bs D, \bs z)
		\\ & \qquad =
\delta_{z_2,z_1}\Big[ \sum_{\bs w\in\Lambda^2}\delta_{w_1,z_1}\sum_{\substack{\sigma = \pm,\, \bs y \in \Lambda^2,\, \bs D'\in\{0,\hat e_1,\hat e_2\}^2: \\
\diam_1(\bs y)\le L/3 	\\ (\sigma, \bs D',\bs w) \in \INT(\bs y)}}\sigma (V_\infty^{(h)})_{2,0} (\bs \omega, \bs 0, \bs y_\infty)\\ 
&\qquad - \mathds 1(\bs z,\bs z+\bs D\in\Lambda^2)\sum_{\bs w\in\Lambda_\infty^2}\delta_{w_1,z_1}
\sum_{\substack{\sigma = \pm,\, \bs y \in \Lambda_\infty^2,\, \bs D'\in\{0,\hat e_1,\hat e_2\}^2: \\ (\sigma, \bs D',\bs w) \in \INT(\bs y)}}
\sigma (V_\infty^{(h)})_{2,0}(\bs \omega, \bs 0, \bs y)\Big].
		\end{split}\label{2+2=425App}
\end{equation}
A cancellation similar to the one discussed in the previous items leads, for $\epsilon$ smaller than $c_0$ and a suitable $C>0$, to 
\begin{equation}\label{eppadieci}
 \|(\tcL(\tcR V_\B^{(h)})_{2,1})_\E\|_{(\E;\kappa)}\le 
 	\frac{C}{\epsilon^2}\|(V_\B^{(h)})_{2,0}\|_{(\kappa + \epsilon)}. \end{equation}
Similarly (details left to the reader), 
\begin{equation}\label{eppaundici}
 \|(\tcR(\tcR V_\B^{(h)})_{2,2})_\E\|_{(\E;\kappa)}\le 
 	\frac{C}{\epsilon^3}\|(V_\B^{(h)})_{2,0}\|_{(\kappa + \epsilon)}. \end{equation}
If we now choose $\kappa=\frac34 c_0 2^h$ and $\kappa+\epsilon=c_0 2^h$, and plug Eqs.\eqref{eppA5}, \eqref{eppA8}, \eqref{eppadieci}, \eqref{eppaundici}, 
recalling \eqref{eq:cited_V8_bound}, we obtain the desired bound, \eqref{belissemo}. 

\section{Proof of the validity of the tree expansion}\label{app:itdec}

In this appendix we prove that, for all $h\le 0$, 
$W_\E^{(h)}\sim \sum_{\tau\in \cT^{(h)}}^{E_{v_0}=1}W_\Lambda[\tau]$, with $W_\E^{(h)}:=W_\Lambda^{(h)}-I_\Lambda W_\infty^{(h)}$, 
$W_\Lambda^{(h)}$ the solution to the recursive equations \eqref{eq:BBF0cyl}-\eqref{eq:BBFh-1cyl}, and $W_\infty^{(h)}$ defined by \eqref{treeexpinfty}. In view of the fact 
that, from the definition of $W_\infty^{(h)}$ and the fact that $W_\B^{(h)}=I_\Lambda W_\infty^{(h)}$, it readily follows that 
$W_\B^{(h)}=\sum_{\tau\in \cT^{(h)}}^{E_{v_0}=0}W_\Lambda[\tau]$, the proof of $W_\E^{(h)}\sim \sum_{\tau\in \cT^{(h)}}^{E_{v_0}=1}W_\Lambda[\tau]$ is equivalent
to the proof of $W_\Lambda^{(h)}\sim \sum_{\tau\in \cT^{(h)}}W_\Lambda[\tau]$, which we now discuss. 

We proceed inductively in $h$. For $h=0$, in the sum of $W_\Lambda[\tau](\Psi,\bs x)$ over $\tau\in\cT^{(0)}$, we distinguish three kinds of contributions: 
\begin{enumerate}
\item the one from the trees in which $v_0$ is not dotted (note that, in this case, these trees have necessarily only one endpoint), 
which gives $(V^{(1)}_\B+B_\B^{(1)}+V^{(1)}_\E+B_\E^{(1)})(\Psi,\bs x)$;
\item the one from the trees in which $v_0$ is dotted (and is either black or white) and all the endpoints are black, which, thanks
to the definitions in \eqref{eq:item1}, \eqref{Wtaublack-}, and following lines, gives 
\begin{equation} \begin{split}& \sum_{s=1}^{\infty}\frac{1}{s!}
		\sum_{\substack{\Psi_1,\ldots,\Psi_s\in \cM_{1,\Lambda}:\\
		\text{if $s=1$ then $\Psi_1\neq \Psi$,}\\ 
		\bs x_1,\ldots,\bs x_s\in \cX_\Lambda}}^{(\Psi, \bs x)}\
		\sum_{T \in \cS(\bar\Psi_1,\ldots,\bar\Psi_s)}
		\fG_{T}^{(1)}(\bar\Psi_1,\ldots,\bar\Psi_s)
		\\
		&\qquad \cdot\ \alpha(\Psi;\Psi_1,\ldots,\Psi_s)  
		\prod_{j=1}^s \left(
			V^{(1)}_\B (\Psi_j, \bs x_j)
			+
			B^{(1)}_\B (\Psi_j, \bs x_j)
				\right);
	\end{split}
	\label{eq:W0_check}
\end{equation}
\item the one from the trees in which $v_0$ is dotted and white, and at least one endpoint is white, which gives 
\begin{equation} \begin{split}& \sum_{s=1}^{\infty}\frac{1}{s!}
		\sum_{\substack{\Psi_1,\ldots,\Psi_s\in \cM_{1,\Lambda}:\\
		\text{if $s=1$ then $\Psi_1\neq \Psi$,}\\ 
		\bs x_1,\ldots,\bs x_s\in \cX_\Lambda}}^{(\Psi, \bs x)}\
		\sum_{T \in \cS(\bar\Psi_1,\ldots,\bar\Psi_s)}
		\fG_{T}^{(1)}(\bar\Psi_1,\ldots,\bar\Psi_s)\alpha(\Psi;\Psi_1,\ldots,\Psi_s)  
		\\
		&\qquad \cdot\Big[\prod_{j=1}^s \left(
			V^{(1)}_\B (\Psi_j, \bs x_j)
			+
			B^{(1)}_\B (\Psi_j, \bs x_j)+
			V^{(1)}_\E (\Psi_j, \bs x_j)
			+
			B^{(1)}_\E (\Psi_j, \bs x_j)\right)\\ &\qquad -
\prod_{j=1}^s \left(
			V^{(1)}_\B (\Psi_j, \bs x_j)
			+
			B^{(1)}_\B (\Psi_j, \bs x_j)\right)\Big].
	\end{split}
	\label{eq:W0_checkk}
\end{equation}
\end{enumerate}
Summing these three contributions up, and using that $V^{(1)}_\B+B_\B^{(1)}+V^{(1)}_\E+B_\E^{(1)}=W_\Lambda^{(1)}$, we find that 
$\sum_{\tau\in\cT^{(0)}}W_\Lambda[\tau](\Psi,\bs x)$ is equal to the right side \eqref{eq:BBF0cyl}, which proves $W_\Lambda^{(0)}(\Psi,\bs x)=
\sum_{\tau\in\cT^{(0)}}$ $W_\Lambda[\tau](\Psi,\bs x)$. 

For $h\le -1$, we inductively assume that, for all $h\le k\le 0$, $W_\Lambda^{(k)}[\tau]\sim \sum_{\tau\in\cT^{(k)}}W_\Lambda[\tau]$, which we equivalently rewrite (distinguishing 
the contributions from the trees with $v_0$ not dotted from the rest) as 
\begin{equation} W_\Lambda^{(k)}[\tau]\sim W_\Lambda^{(1)}+\sum_{k'=k}^0 \sum^*_{\tau\in\cT^{(k')}}W_\Lambda[\tau],\label{eqforkappa}\end{equation}
for all $h\le k\le 0$, where, as usual, the $*$ on the sum indicates the constraint that $v_0$ is dotted. Under this inductive assumption, we intend to prove \eqref{eqforkappa} with $k$ replaced by 
$h-1$. From the definition of tree values, noting that we can choose an element of $\cT^{(h)}$ by first choosing the number of elements of $S_{v_0}$ and then, independently for 
each, choosing their type and scale and (if type \tikzvertex{vertex} or \tikzvertex{vertex,E} at scale <2) the subtree of which they are the root, we find
\begin{equation}
	\begin{split}
		&\sum_{\tau \in \cT^{(h-1)}}
		W_\Lambda [\tau](\Psi,\bs x)=
		\sum_{s=1}^{\infty}\frac{1}{s!}
		\sum_{\substack{\Psi_1,\ldots,\Psi_s\in \cM_{\Lambda}\\\bs x_1,\ldots,\bs x_s\in \cX_\Lambda}}^{(\Psi,\bs x)}\
		\sum_{T \in \cS(\bar\Psi_1,\ldots,\bar\Psi_s)}
		\fG_{T}^{(h)}(\bar\Psi_1,\ldots,\bar\Psi_s)\cdot
		\\ & \cdot
		\alpha(\Psi;\Psi_1,\ldots,\Psi_s)
	\Biggl(  \prod_{j=1}^s \Big[
			\begin{aligned}[t]
				&
				\upsilon_h \cdot F_{\B}
				+ ( \cR_\B V^{(1)}_\B)_\B
				+Z_h\cdot F_\B^A
				+ ( \cR_\B B^{(1)}_\B)_\B
				+ \cR_\E C_\E^{(h)}
				+ \cR_\E V^{(1)}_\E
				\\ &
				+ \cR_\E( \cE V_\B^{(1)})
				+D_\E^{(h)}
				+ B_\E^{(1)}
				+ \cE B_\B^{(1)}
				+ 
				\sum_{k_s=h}^0 \sum_{\tau_s \in \cT^{(k)}}^*
				\lis W_\Lambda[\tau_s]
				\Big](\Psi_j,\bs x_j)\Biggr),
			\end{aligned}
	\end{split}
	\label{eq:W-1_trees_base}
\end{equation}
where
\begin{equation}
	\lis W_\Lambda [\tau]
	=
	\begin{cases}
		( \cR_\B W_\Lambda[\tau])_\B
		, &
		\text{if $E_{v_0(\tau_s)}=0$},
		\\
		\cR_\E W_\Lambda [\tau]
		, &
		\text{if $E_{v_0(\tau_s)}=1$}. 
	\end{cases}
	\label{eq:bar_W_tau}
\end{equation}
The proof of this identity follows from the observation that each term in the square brackets corresponds to one choice of the type of the corresponding vertex of $S_{v_0}$, 
and thus of one of the cases from the definition of $K_v$, \cref{after2A,Kvwhitedef}, except that \tikzvertex{vertex} and \tikzvertex{vertex,E} at intermediate scales are combined 
in a single sum in the last term. Now, recalling (or noting) that:
\begin{itemize}
\item $\upsilon_h \cdot F_{\B}+Z_h\cdot F_\B^A=(\cL_\B W_\B^{(h)})_\B=I_\Lambda \cL_\infty W_\infty^{(h)}$, with $W_\infty^{(h)}=\sum_{\tau\in \cT^{(h)}}^{E_{v_0}=0}W_\infty[\tau]=W^{(1)}_\infty+\sum_{k=h}^0\sum^{\tikzvertex{vertex}}_{\tau\in\cT^{(k)}}W_\infty[\tau]$ (where the symbol \tikzvertex{vertex} on the sum denotes the constraint that $v_0$ is dotted and black), we have
\begin{equation}
	\upsilon_h \cdot F_{\B}
	+Z_h\cdot F_\B^A
	= ( \cL_\B W^{(1)}_\B)_\B	+\sum_{k=h}^0
	\sum_{\tau \in \cT^{(k)}}^{\tikzvertex{vertex}}( \cL_\B W_\Lambda [\tau])_\B;
	\label{eq:local_tree_decomp}
\end{equation}
\item $\cR_\E W \sim W$ for all kernels;
\item $C_\E^{(h)}+D_\E^{(h)}=
	\sum_{k=h}^0
	\sum_{\tau \in \cT^{(k)}}^{\tikzvertex{vertex}}\cE W_\Lambda [\tau]$, where, for any $\tau$ with $v_0$ dotted and black,  
\begin{equation}\begin{split}
\cE W [\tau]&=\cL_\B W_\Lambda [\tau]-( \cL_\B W_\Lambda [\tau])_\B+\cR_\B W_\Lambda [\tau]-( \cR_\B W_\B [\tau])_\B\\
&\sim 	W_\Lambda [\tau]-( \cL_\B W_\Lambda [\tau])_\B-( \cR_\B W_\B [\tau])_\B;\end{split}
	\label{eq:cE_W_tau}
\end{equation}
\item $\cE V_\B^{(1)}+\cE B_\B^{(1)}=\cL_\B W_\B^{(1)}-(\cL_\B W_\B^{(1)})_\B+\cR_\B W_\B^{(1)}-(\cR_\B W_\B^{(1)})_\B\sim 
W_\B^{(1)}-(\cL_\B W_\B^{(1)})_\B-(\cR_\B W_\B^{(1)})_\B$;
\end{itemize}
we find that 
\begin{equation}
	\begin{split}
& 	\upsilon_h \cdot F_{\B}+ ( \cR_\B V^{(1)}_\B)_\B+Z_h\cdot F_\B^A+ ( \cR_\B B^{(1)}_\B)_\B+ \cR_\E C_\E^{(h)}	+ \cR_\E( \cE V_\B^{(1)})+D_\E^{(h)}	+ \cE B_\B^{(1)}\\
&	+ \sum_{k_s=h}^0 \sum_{\tau_s \in \cT^{(k)}}^{\tikzvertex{vertex}}\lis W_\Lambda[\tau_s]\sim W_\B^{(1)}+\sum_{k_s=h}^0 \sum_{\tau_s \in \cT^{(k)}}^{\tikzvertex{vertex}}W_\Lambda[\tau_s].
	\end{split}
\end{equation}
Using this equivalence, recalling again that $\cR_\E W \sim W$ for all kernels, and noting that $W_\B^{(1)}+W_\E^{(1)}=W_\Lambda^{(1)}$, we find that the expression in square brackets in the right side of 
\eqref{eq:W-1_trees_base} is equivalent to $W_\Lambda^{(1)}+\sum_{k_s=h}^0\sum^*_{\tau_s\in\cT^{(k)}}W_\Lambda[\tau]$. This, in view of the inductive assumption \eqref{eqforkappa} for $k=h$, of 
 \eqref{eq:BBFh-1cyl} and of \eqref{eqfin1.}, proves \eqref{eqforkappa} for $k=h-1$, as desired. 
 
 \section{Proof of \eqref{corona.vir}} \label{app.proofboringb}

In this section we prove that the left side of \eqref{corona.vir}, which for the reader's convenience we recall is
\begin{equation}\begin{split} \Big(\prod_{v \in V_e(\tau^*)}^{***}  |\lambda|(\delta_{\E}(x_v))^{-\theta}\Big) &\sum_{\tau\in \cT(\tau^*)}^*
\sum_{\substack{\ul P\in \cP_*(\tau) \\ |P_{v_0}|=0}}
\sum_{\ul D\in \cD(\tau,\ul P)} \Big(\prod_{v \in V'(\tau)} 2^{(d_v'-\tilde d_v)(h_v-h_{v'})} \Big)\cdot\\
&\cdot  \Big(  \prod_{v \in V_e(\tau)}^{**} (C|\lambda|)^{\max\{1,\kappa(|P_v|+m_v)\}}2^{\theta h_v} \Big), \end{split}\label{eq:appG.1}\end{equation}
is bounded from above by $C^{m}|\lambda|(2^{\theta' h^*_M}+\mathfrak{d}_\partial^{-\theta})$,
where $\tau^*\in\cT^{*}_{h,h_0^*;m}$, $h^*_M=\max\{h_v:\ v\in V_e(\tau^*)\}$, and 
$\mathfrak d_\partial=\mathfrak d_\partial(\bs x)$ is the minimal distance between the points in $\bs x$ and $\partial \Lambda$.
We remind the reader that $\theta'=\theta-\epsilon/2=\theta-(1-\theta)/6$;
in this appendix, we let $\alpha:=\epsilon/2=(1-\theta)/6$. As already observed, see \eqref{dv'-tildedv}, $d_v'-\tilde d_v\le -\alpha|P_v|-\theta\delta_{m_v,0}$, 
with the exception of the vertices with $m_v=1$, $|P_v|=2$, $\|\bs D_v\|_1=0$, $E_v=1$, for which $d_v'-\tilde d_v=0$. These are the only potentially 
dangerous vertices as far as the sum over the scale labels is concerned. We will shortly see that, in fact, they do not create any problem, provided we  
devise a suitable summation procedure, which is slightly different from the `standard' one, i.e., the one explained in \cite[Appendices A.1 and A.3]{GM01}
and in \cite[Proof of Lemma~4.8]{AGG_part2}. 
We denote by $N(\tau)\subset V'(\tau)$ the set of `null vertices' of $\tau$, that is, those that can have $d_v'-\tilde d_v=0$ for some allowed choice of the labels $\ul P,\ul D$.
Using these definitions, and recalling that the number of allowed choices of $\ul D$ is smaller than $({\rm const.})^{|V(\tau)|}$, see Remark 
\ref{rem:D_choice_p2}, we find that 
\begin{eqnarray} && \eqref{eq:appG.1}\le \Big(\prod_{v \in V_e(\tau^*)}^{***}  |\lambda|(\delta_{\E}(x_v))^{-\theta}\Big)
 \sum_{\tau\in \cT(\tau^*)}^* C^{|V(\tau)|} 
\sum_{\substack{\ul P\in \cP_*(\tau) \\ |P_{v_0}|=0}}
\Big(\prod_{\substack{v \in V'(\tau) \\ v\not\in N(\tau)}} 2^{-(\alpha |P_v|-\theta\delta_{m_v,0})(h_v-h_{v'})} \Big)\cdot\nonumber\\
&&\quad \cdot
\Big(\prod_{v\in N(\tau)} 2^{(1-\frac{|P_v|}2)(h_v-h_{v'})} \Big)\,\Big(  \prod_{v \in V_e(\tau)}^{**} (C|\lambda|)^{\max\{1,\kappa(|P_v|+m_v)\}}2^{\theta h_v} \Big),
\label{eq:appG.2}\end{eqnarray}
where we also used the fact that, if $v\in N(\tau)$, then necessarily $m_v=E_v=1$ and, therefore, $d_v'-\tilde d_v=d_v\le 1-|P_v|/2$. 
Next, if $\prod_{v \in V_e(\tau)}^{**}$ runs over a non-empty set, we rewrite the factors $2^{\theta h_v}$ in the corresponding product as $2^{(\theta-\alpha) h_v}2^{\alpha h_v}$ (note that $\theta-\alpha=(7\theta-1)/6$ is positive for our choice of $\theta$), keep the factors $2^{\alpha h_v}$ on a side, and 
bound $(\prod^{**}_{v\in V_e(\tau)} 2^{(\theta-\alpha) h_v})(\prod_{v \in V'(\tau)}^{m_v=0} 2^{-\theta(h_v-h_{v'})})$ by $2^{(\theta-\alpha) h^*_M}\equiv 2^{\theta' h^*_M}$ (up to a redefinition of 
the constant $C$ in the product $\prod^{**}_{v\in V_e(\tau)}$). If $\prod_{v \in V_e(\tau^*)}^{***}$ runs over a non-empty set, we bound 
$(\prod_{v \in V_e(\tau^*)}^{***} (\delta_{\E}(x_v))^{-\theta})$ by $C\mathfrak{d}_\partial^{-\theta}$. 
Moreover, we bound the factor $C^{|V(\tau)|}$ by $({\rm const.})^m$, up to a further redefinition of the constant $C$ in the product $\prod^{**}_{v\in V_e(\tau)}$. 
Therefore, recalling that at least one of the two products $\prod_{v \in V_e(\tau)}^{**}$ and $\prod_{v \in V_e(\tau^*)}^{***}$ runs over a non-empty set, we find
\begin{eqnarray}
\eqref{eq:appG.2}&\le & (C')^m(2^{\theta' h_M^*}+\mathfrak{d}_\partial^{-\theta}) \Big(\prod_{v \in V_e(\tau^*)}^{***}|\lambda|\Big)\sum_{\tau\in \cT(\tau^*)}^* \sum_{\substack{\ul P\in \cP_*(\tau) \\ |P_{v_0}|=0}}
\Big(\prod_{\substack{v \in V'(\tau) \\ v\not\in N(\tau)}} 2^{-\alpha |P_v|(h_v-h_{v'})} \Big)\cdot \label{eq:appG.3}\\
&\cdot&\Big(\prod_{v\in N(\tau)} 2^{(1-\frac{|P_v|}2)(h_v-h_{v'})} \Big)
\Big(  \prod_{v \in V_e(\tau)}^{**} (C'|\lambda|)^{\max\{1,\kappa(|P_v|+m_v)\}}2^{\alpha h_v} \Big).\nonumber
\end{eqnarray}
In order to perform the sum over the trees, and in particular over the scale labels, it is convenient to characterize more precisely the set of null vertices
of $\tau\in \cT(\tau^*)$. First of all, note that the null vertices of $\tau$ are all contained in the branches of $\tau^*$ connecting an 
endpoint of type \tikzvertex{bareProbeEP,E} such that $m_v=1$ with the vertex immediately preceding it on $\tau^*$. 
Consider one such branch, and call it $(w',w)$, where $w$ is the endpoint with $m_v=1$ and $E_v=1$, and $w'$ the vertex immediately 
preceding $w$ on $\tau^*$; note that $w'$ must be a branching point, $s^*_{w'}>1$, and that $m_{w'}>1$. Let $w_0, \ldots w_n$, $n\ge 1$, be the vertices of $\tau$ 
contained in the branch $(w',w)$, labelled with the convention that $w_0=w'$, $w_k<w_{k+1}$, for all $k=0,\ldots, n-1$, and $w_n=w$. 
If the sequence $(w_1,\ldots, w_n)$ does not contain any non-trivial vertex, then the reader can easily 
convince herself that the only null vertex of $\tau$ contained in $(w',w)$ is $w_1$. 
If, on the contrary, $(w_1,\ldots, w_n)$ contains the non-trivial vertices $w_{k_1}, \ldots, w_{k_a}$, with $a\ge 1$ and $w_{k_1}<\cdots<w_{k_a}$, 
then the null vertices of $\tau$ contained in $(w',w)$ are $w_1$, $w_{k_1+1}$, $\ldots$, $w_{k_a+1}$. From this explicit construction 
of the set of null vertices, it follows, in particular, that their number can be bounded as
\begin{equation} \label{appG.boundNtau} |N(\tau)|\le m+ \kappa'\sum_{\substack{v\in V_e(\tau)\\ m_v=E_v=1}}|P_v|,\end{equation}
for some $\kappa'>0$.
If we erase from $\tau$ the edges $(v',v)$, with $v\in N(\tau)$ and $v'$ the vertex immediately preceding $v$ on $\tau$, 
the tree graph $\tau$ is disconnected into a certain number of maximal connected components,
some of which may consist of isolated vertices (we call such connected components `trivial'). By construction, any such trivial connected component 
either consists of the root $v_0$, or of an endpoint with $m_v=1$ and $E_v=1$. On the other hand, 
any non-trivial connected component consists of a subtree of $\tau$ 
that either contains the root $v_0$, or contains a counterterm endpoint (i.e., an endpoint of type \tikzvertex{ctVertex} or \tikzvertex{ctVertex,E}), or contains an endpoint on scale $2$; we denote these connected subtrees by $\tau_1,\ldots,\tau_{n_0}$, which we will think of as being rooted in their vertex with smallest scale label. 

\medskip

Let us now go back to \eqref{eq:appG.3}. Using the fact that, for all $v\in V'(\tau)$, $h_v-h_{v'}\ge 1$ and $|P_v|\ge 2(1+t_v)$, where 
$t_v$ is the number of trivial vertices immediately preceding $v$ on $\tau$ (i.e., those preceding $v$, but not preceding any non-trivial vertex $w<v$), 
we can bound the factor $2^{-\alpha |P_v|(h_v-h_{v'})}$ by $2^{-\frac{\alpha}2 |P_v|} \cdot 2^{-\alpha (1+t_v)(h_v-h_{v'})}$. Next, note that 
\begin{eqnarray}
 \Big(\prod_{\substack{v \in V'(\tau) \\ v\not\in N(\tau)}} 2^{-\frac{\alpha}2 |P_v|}\Big)\,\Big(\prod_{v\in N(\tau)} 2^{(1-\frac{|P_v|}2)(h_v-h_{v'})} \Big)
&\le& \Big(\prod_{\substack{v \in V'(\tau) \\ v\not\in N(\tau)}} 2^{-\frac{\alpha}2 |P_v|}\Big)\,\Big(\prod_{v\in N(\tau)} 2^{(1-\frac{|P_v|}2)} \Big)\nonumber\\
&\le& 2^{\alpha|N(\tau)|}\prod_{v \in V'(\tau)} 2^{-\frac{\alpha}2 |P_v|}.\end{eqnarray}
Thanks to \eqref{appG.boundNtau}, the factor $2^{\alpha|N(\tau)|}$ can be reabsorbed into a redefinition of the constant $C'$ in \eqref{eq:appG.3}. Therefore, 
up to a redefinition of this constant,
\begin{eqnarray}
\eqref{eq:appG.3}&\le & (C')^m(2^{\theta' h_M^*}+\mathfrak{d}_\partial^{-\theta})\Big(  \prod_{v \in V_e(\tau^*)}^{***} |\lambda| \Big) \sum_{\tau\in \cT(\tau^*)}^* \sum_{\substack{\ul P\in \cP_*(\tau) \\ |P_{v_0}|=0}}
\Big(\prod_{\substack{v \in V'(\tau) \\ v\not\in N(\tau)}} 2^{-\alpha (1+t_v)(h_v-h_{v'})} \Big)\cdot\nonumber\\
&\cdot&\Big(\prod_{v\in V'(\tau)} 2^{-\frac{\alpha}{2}|P_v|}\Big)\,
\Big(  \prod_{v \in V_e(\tau)}^{**} (C'|\lambda|)^{\max\{1,\kappa(|P_v|+m_v)\}}2^{\alpha h_v} \Big). \label{eq:appG.4}
\end{eqnarray}
Now, the sum 
$ \sum_{\ul P\in \cP_*(\tau)}^{\dagger}\Big(\prod_{v\in V'(\tau)} 2^{-\frac{\alpha}{2}|P_v|}\Big)$
where the $\dagger$ on the sum indicates the constraints that $p_v:=|P_v|$ is fixed for all $v\in V_e(\tau)$, and $p_{v_0}=0$, 
can be performed exactly as in \cite[Appendix A.6.1]{GM01}, and gives
$$ \sum_{\ul P\in \cP_*(\tau)}^{\dagger}\Big(\prod_{v\in V'(\tau)} 2^{-\frac{\alpha}{2}|P_v|}\Big)\le 
\prod_{v\in V_e(\tau)}\Big(\frac1{1-2^{-\frac\alpha2}}\Big)^{p_v}.$$
Note that this constant can be reabsorbed into a further redefinition of the constant $C'$ in \eqref{eq:appG.4}. 
Moreover, $\prod_{v \in V_e(\tau)}^*2^{\alpha h_v}$ can be bounded from above by $\prod_{v \in V_e(\tau)}^{m_v=0} 2^{\alpha h_v}$ times a factor
that can be again reabsorbed into a redefinition of $C'$. Therefore, up to these additional redefinitions, 
\begin{eqnarray}
\eqref{eq:appG.4}&\le & (C')^m(2^{\theta' h_M^*}+\mathfrak{d}_\partial^{-\theta})\Big(  \prod_{v \in V_e(\tau^*)}^{***}|\lambda|\Big) \sum_{\tau\in \cT(\tau^*)}^*\Big(\prod_{\substack{v \in V'(\tau) \\ v\not\in N(\tau)}} 2^{-\alpha (1+t_v)(h_v-h_{v'})} \Big)
\, \Big(\prod_{\substack{v\in V_e(\tau) \\ m_v=0}}2^{\alpha h_v}\Big)\cdot \nonumber\\
&\cdot&\sum_{\{p_v\}_{v\in V_e(\tau)}}^{\dagger\dagger}
\Big(  \prod_{v \in V_e(\tau)}^{**} (C'|\lambda|)^{\max\{1,\kappa (m_v+p_v)\}}\Big),\label{eq:appG.7}
\end{eqnarray}
where $\sum_{\{p_v\}_{v\in V_e(\tau)}}^{\dagger\dagger}$ indicates the sum over the labels $p_v$ in the positive even integers, with the constraint that, if $\tau\in \cT^{(h)}_\free$, 
then the configuration of labels $p_v\equiv2$ for all the endpoints is not allowed (i.e., at least one of the $p_v$'s must be $\ge 4$). 
By performing the sum in the second line, we find, up to a new redefinition of $C'$, 
\begin{eqnarray}
\eqref{eq:appG.7}&\le&  (C')^m (2^{\theta' h_M^*}+\mathfrak{d}_\partial^{-\theta})\Big(  \prod_{v \in V_e(\tau^*)}^{***}|\lambda|\Big)
 \Big(\prod_{\substack{v \in V_e(\tau^*) \\ m_v> 1}}|\lambda|^{\max\{1,\kappa m_v\}}\Big)\cdot
\label{eq:appG.8}\\
&\cdot&\sum_{\tau\in \cT(\tau^*)}^*|\lambda|^{\mathds 1(\tau\in \cT^{(h)}_\free)}(C'|\lambda|)^{|V_e^0(\tau)|}
\Big(\prod_{\substack{v \in V'(\tau) \\ v\not\in N(\tau)}} 2^{-\alpha (1+t_v)(h_v-h_{v'})} \Big)\, 
\, \Big(\prod_{v\in V_e^0(\tau)}2^{\alpha h_v}\Big),\nonumber
\end{eqnarray}
where $V_e^0(\tau)=\{v\in V_e(\tau)\,:\, m_v=0\}$. 
In order to perform the sum over $\tau$, we shall think of it as a sum over: the skeleton $\mathfrak t$ of $\tau$ (by `skeleton' here we mean a rooted tree, with 
nodes associated only with the root, the endpoints and the branching points, and no scale labels associated with the nodes); 
the insertion of additional, trivial, vertices in the branches of $\mathfrak t$; 
the types of endpoints; the choice of the scale labels of the vertices of $\mathfrak t$ and of the additional trivial vertices. 

We first perform the sum over the scale labels with the skeleton fixed and a given choice of insertion of trivial vertices. For this purpose, 
it is convenient to rearrange the last two products in the second line of \eqref{eq:appG.8}
in a form that is factored over $\tau_1,\ldots, \tau_{n_0}$, which are, as discussed above, 
the non-trivial maximal connected subtrees obtained from $\tau$ by erasing the edges $(v',v)$, with $v\in N(\tau)$. We rewrite: 
\begin{equation}\Big(\prod_{\substack{v \in V'(\tau) \\ v\not\in N(\tau)}} 2^{-\alpha (1+t_v)(h_v-h_{v'})} \Big)\, 
\, \Big(\prod_{v\in V_e^0(\tau)}2^{\alpha h_v}\Big)=\prod_{j=1}^{n_0}\Big[
\Big(\prod_{v \in V'(\tau_j)} 2^{-\alpha (1+t_v)(h_v-h_{v'})} \Big)\, 
\, \Big(\prod_{v\in V_e^0(\tau_j)}2^{\alpha h_v}\Big)\Big],\label{appG.scalef}\end{equation}
where $V'(\tau_j)$ is the set of vertices of $\tau_j$, with the exception of its root; $V_e^0(\tau_j)$ 
is the set of endpoints of $\tau_j$ with $m_v=0$ (by construction, $V_e^0(\tau_j)$ is contained in $V_e^0(\tau)$). 
Not all the factors in the right side of \eqref{appG.scalef} are really needed for summing over the scales; let us drop the un-necessary ones. 
For each $\tau_j$, we define $V_{\min}(\tau_j)$, the `minimal set of useful vertices' (for the purpose of the sum over scales), as follows: 
if the root of $\tau_j$ is the root $v_0$ of $\tau$, we let $V_{\Min}(\tau_j)=V(\tau_j)\setminus V_e^0(\tau_j)$;
if the root of $\tau_j$ is not the root $v_0$ of $\tau$ and $\tau_j$ contains at least one endpoint on scale $2$, 
we let $V_{\Min}(\tau_j)$ be the union of $V(\tau_j)\setminus V_e^0(\tau_j)$ and one of its endpoints on scale $2$;
if the root of $\tau_j$ is not the root $v_0$ of $\tau$ and $\tau_j$ does not contain any endpoint on scale $2$, 
we let $V_{\Min}(\tau_j)$ be the union of $V(\tau_j)\setminus V_e^0(\tau_j)$ and one of its endpoints of type \tikzvertex{ctVertex} or \tikzvertex{ctVertex,E}. 
[Note that these three cases exhaust the cases to consider, see above, right before the definition of $\tau_1,\ldots, \tau_{n_0}$.] 
We also let $\tau_j'$ be the minimal subtree of $\tau_j$ containing $V_\Min(\tau_j)$. For the purpose of an upper bound, 
in the right side of \eqref{appG.scalef} we neglect all the factors $2^{-\alpha (1+t_v)(h_v-h_{v'})}$ associated with edges $(v',v)$ that are not 
in $\tau_j'$; moreover, we neglect all the factors $2^{\alpha h_v}$ associated with endpoints $v\in V_e^0(\tau_j)$ that are not 
in $\tau_j'$ (up to a constant $2^{2\alpha|V_e^0(\tau)|}$). That is, we bound
\begin{equation} \eqref{appG.scalef}\le 2^{2\alpha|V_e^0(\tau)|}\prod_{j=1}^{n_0}\Big[
\Big(\prod_{v \in V'(\tau_j')} 2^{-\alpha (1+t_v)(h_v-h_{v'})} \Big)\, 2^{\alpha h^*_j}\Big],\label{appG.scalefvi}\end{equation}
where, if the root of $\tau_j$ does not coincide with $v_0$ and $\tau_j$ does not contain any endpoint on scale $2$, 
$h^*_j$ is the scale of the endpoint of type \tikzvertex{ctVertex} or in \tikzvertex{ctVertex,E} $\tau_j$, and is equal to zero otherwise. The reader can convince herself that the 
right side of \eqref{appG.scalefvi} is summable over the scale labels $\{h_v\}_{v\in V_0'(\tau)}$\footnote{Given $\{h_v\}_{v\in V_0'(\tau)}$, the sum
over the scales of the endpoints is `trivial', since it contains at most $2^{|V_e(\tau)|}$ terms: each factor 2 corresponds to the fact that an endpoint 
$v$ immediately following a vertex $v'$ at scale $h_{v'}$ can be either on scale $h_{v'}+1$ or on scale $2$.}; by performing the sum, we get 
a constant smaller than 
\begin{equation} 2^{2\alpha|V_e^0(\tau)|}\Big(\frac{2^{2\alpha}}{1-2^{-\alpha}}\Big)^{|V_e^{\ct}(\tau)|}\prod_{j=1}^{n_0}\Big[
\Big(\prod_{v \in V'(\tau_j')} \frac{2^{-\alpha (1+t_v)}}{1-2^{-\alpha(1+t_v)}} \Big)\, \Big],\label{appG.scalefvir}\end{equation}
where $V_e^{\ct}(\tau)=\{v\in V_e(\tau)\,:\, \text{$v$ is of type \tikzvertex{ctVertex} or \tikzvertex{ctVertex,E}}\}$. This concludes the discussion of the sum over the scale labels. 

Next, we sum over the insertions of trivial vertices. This means to sum the products of factors $\frac{2^{-\alpha (1+t_v)}}{1-2^{-\alpha(1+t_v)}}$
over the number of consecutive trivial vertices that one can insert, independently, on each branch of the skeleton. This produces a factor smaller than 
\begin{equation}\label{branchsk}1+\sum_{k\ge 0}\prod_{t=0}^k\Big( \frac{2^{-\alpha (1+t)}}{1-2^{-\alpha(1+t)}} \Big)\le 
1+\frac{\sum_{k\ge 1} 2^{-\frac\alpha2 k(k+1)}}{\prod_{k\ge 1}(1-2^{-\alpha k})}\end{equation}
for each branch of the skeleton. Noting that the number of branches of the skeleton is smaller than twice the number of its endpoints, we see that the product of 
the factors in \eqref{branchsk} over the branches of the skeleton is smaller than $({\rm const.})^N$, with $N$ the number of endpoints.

Next, we sum over the types of endpoints, and over the skeletons compatible with $\tau^*$ with $N$ endpoints: recalling that 
the number of un-labelled rooted trees with $N$ endpoints is smaller than $4^N$, see \cite[Lemma A.1]{GM01}, these sums produce an additional 
factor $({\rm const.})^N$. Finally, we sum over $N$. Putting things together, we find that the second line of \eqref{eq:appG.8} is bounded from above by 
$$(C')^m\sum_{N\ge |V_e(\tau^*)|}(C'|\lambda|)^{\max\{\mathds 1(\tau^*\in\cT^{(h)}_\free),N-|V_e|\}}, $$
for a suitable constant $C'$. %
By plugging this into \eqref{eq:appG.8} gives the desired bound, \eqref{corona.vir}.\qed

\subsection*{Acknowledgements}
We thank Hugo Duminil-Copin for several inspiring discussions. 
This work has been supported by the European Research Council (ERC) under the European Union's Horizon 2020 research and innovation programme (ERC CoG UniCoSM, grant agreement No.\ 724939 for all three authors and also ERC StG MaMBoQ, grant agreement No.\ 802901 for R.L.G.). 
G.A.\ acknowledges financial support from the Swiss Fonds National.
A.G.\ acknowledges financial support from MIUR, PRIN 2017 project MaQuMA PRIN201719VMAST01.

\printbibliography[heading=bibintoc]
\end{document}